\documentclass{nature}
\usepackage{times}  %Required
\usepackage{helvet}  %Required
\usepackage{courier}  %Required
\usepackage{url}            % simple URL typesetting
\usepackage{subfigure}
\usepackage[subfigure]{tocloft}
\usepackage{booktabs}       % professional-quality tables
\usepackage{amsfonts}       % blackboard math symbols
\usepackage{nicefrac}       % compact symbols for 1/2, etc.
\usepackage{microtype}      % microtypography
\usepackage{subfigure}
\usepackage{multirow}
\usepackage{enumitem}
\usepackage{amsmath}
\usepackage{enumerate}
\usepackage{bm}
\usepackage{graphicx}
\usepackage{subfigure}
\usepackage{color}
\usepackage{colortbl}
\usepackage{chngpage}
\usepackage{import}
\usepackage{amssymb,amsfonts}
\usepackage{textcomp}
\usepackage{flushend}
\usepackage{hhline}
\usepackage{gensymb}
\usepackage{rotating}
\usepackage{caption}
\usepackage{xr}

\usepackage{tabularx}

\usepackage{threeparttable}
\usepackage{array}
\usepackage{lineno}

\usepackage{epsfig}
\usepackage{algorithm2e}

\usepackage{algpseudocode}

\usepackage{lscape}
\usepackage[table,xcdraw]{xcolor}
\usepackage[hypertexnames=false]{hyperref}

\usepackage{amsmath,amsfonts,bm}

\def\eqref#1{equation~\ref{#1}}
\def\1{\bm{1}}

\def\rvx{{\mathbf{x}}}

\DeclareMathAlphabet{\mathsfit}{\encodingdefault}{\sfdefault}{m}{sl}
\SetMathAlphabet{\mathsfit}{bold}{\encodingdefault}{\sfdefault}{bx}{n}

\usepackage{acro}
\DeclareAcronym{ddp}{
short=DDP,
long=Distributed Data Parallel,
foreign-plural={}
}

\DeclareAcronym{ct}{
short=CT,
long=Computed Tomography,
foreign-plural={}
}

\DeclareAcronym{sam}{
short=SAM,
long=Segment Anything Model,
foreign-plural={}
}

\DeclareAcronym{css}{ % keep in supplementary
short=CSS,
long=Continual Semantic Segmentation,
foreign-plural={}
}

\DeclareAcronym{cs}{ % use in main paper
short=continual segmentation,
long=Continual Segmentation, 
foreign-plural={},
first-style=short
}

\DeclareAcronym{fls}{
short=FL-Supporting,
long=Feature Level Supporting,
foreign-plural={}
}

\DeclareAcronym{pl}{ % use in both main paper and supp
short=universal segmentation,
long=Universal Segmentation, % Partial Label Segmentation
foreign-plural={},
first-style=short
}

\DeclareAcronym{ema}{
short=EMA,
long=Exponential Moving Average,
foreign-plural={}
}

\DeclareAcronym{ge}{
short=GE,
long=General Encoder,
foreign-plural={}
}

\DeclareAcronym{clnet}{
short=CL-Net,
long=Continual Learning Network,
foreign-plural={}
}

\DeclareAcronym{ssl}{
short=SSL,
long=Self-Supervised Learning,
foreign-plural={}
}

\DeclareAcronym{lth}{
short=LTH,
long=Lottery Ticket Hypothesis,
foreign-plural={}
}

\DeclareAcronym{dsc}{
short=DSC,
long=Dice Similarity Coefficient,
foreign-plural={}
}

\DeclareAcronym{asd}{
short=ASD,
long=Average Surface Distance,
foreign-plural={}
}

\DeclareAcronym{bpr}{
short=BPR,
long=Body Part Regression,
foreign-plural={}
}

\DeclareAcronym{dsn}{ % dataset number
short=36,
long=36,
foreign-plural={},
first-style=short
}

\DeclareAcronym{orgn}{ % target organ number
short=235,
long=235,
foreign-plural={},
first-style=short
}

\DeclareAcronym{decn}{ % decoder number
short=101,
long=101,
foreign-plural={},
first-style=short
}

\DeclareAcronym{totalseg}{ % TotalSeg V2
short=TotalSegmentator V2,
long=TotalSegmentator V2,
foreign-plural={},
first-style=short
}

\DeclareAcronym{flare}{ % FLARE
short=FLARE22,
long=FLARE22,
foreign-plural={},
first-style=short
}

\DeclareAcronym{segthor}{ % SegThor
short=SegThor,
long=SegThor,
foreign-plural={},
first-style=short
}

\DeclareAcronym{structseg}{ % StructSeg
short=StructSeg19,
long=StructSeg19,
foreign-plural={},
first-style=short
}

\DeclareAcronym{kits}{ % KiTS
short=KiTS21,
long=KiTS21,
foreign-plural={},
first-style=short
}

\DeclareAcronym{cln_c5}{
short=\ac{clnet}$^{\mathcal{C}_5}$,
long=\ac{clnet}$^{\mathcal{C}_5}$,
foreign-plural={},
first-style=short
}

\DeclareAcronym{cln_unprn}{
short=\ac{clnet}$^{\mathcal{C}_5}_{\text{unpruned}}$,
long=\ac{clnet}$^{\mathcal{C}_5}_{\text{unpruned}}$,
foreign-plural={},
first-style=short
}

\DeclareAcronym{cln_u5}{
short=\ac{clnet}$^{\mathcal{U}_5}$,
long=\ac{clnet}$^{\mathcal{U}_5}$,
foreign-plural={},
first-style=short
}

\DeclareAcronym{nnu_e5}{
short=nnUNet$^{E_{5}}$,
long=nnUNet$^{E_{5}}$,
foreign-plural={},
first-style=short
}

\DeclareAcronym{nnu_e36}{
short=nnUNet$^{E_{36}}$,
long=nnUNet$^{E_{36}}$,
foreign-plural={},
first-style=short
}

\DeclareAcronym{cln_c36}{
short=\ac{clnet}$^{\mathcal{C}_{36}}$,
long=\ac{clnet}$^{\mathcal{C}_{36}}$,
foreign-plural={},
first-style=short
}

\DeclareAcronym{cln_u36}{
short=\ac{clnet}$^{\mathcal{U}_{36}}$,
long=\ac{clnet}$^{\mathcal{U}_{36}}$,
foreign-plural={},
first-style=short
}

\DeclareAcronym{cln_u36_unprn}{
short=\ac{clnet}$^{\mathcal{U}_{36}}_{\text{unpruned}}$,
long=\ac{clnet}$^{\mathcal{U}_{36}}_{\text{unpruned}}$,
foreign-plural={},
first-style=short
}

\DeclareAcronym{cln_u36_ft}{
short=\ac{clnet}$^{\mathcal{U}_{36}}_{\text{fine-tune}}$,
long=\ac{clnet}$^{\mathcal{U}_{36}}_{\text{fine-tune}}$,
foreign-plural={},
first-style=short
}

\DeclareAcronym{cln_c36_nofls}{
short=\ac{clnet}$^{\mathcal{C}_{36}}_{\text{noFL-Supporting}}$,
long=\ac{clnet}$^{\mathcal{C}_{36}}_{\text{noFL-Supporting}}$,
foreign-plural={},
first-style=short
}

\DeclareAcronym{cln_u36_nofls}{
short=\ac{clnet}$^{\mathcal{U}_{36}}_{\text{noFL-Supporting}}$,
long=\ac{clnet}$^{\mathcal{U}_{36}}_{\text{noFL-Supporting}}$,,
foreign-plural={},
first-style=short
}

\DeclareAcronym{gtv}{
short=lesion,
long=lesion,
foreign-plural={},
first-style=short
}

\DeclareAcronym{npc}{
short=NPC,
long=Nasopharyngeal Carcinoma,
foreign-plural={},
first-style=short
}

\DeclareAcronym{hn-zju-15}{
short=NPC\_FAH-ZU\_OAR,
long=NPC\_FAH-ZU\_OAR,
foreign-plural={},
first-style=short
}

\DeclareAcronym{hn-cgmh-42}{
short=HN\_CGMH\_OAR,
long=HN\_CGMH\_OAR,
foreign-plural={},
first-style=short
}

\DeclareAcronym{ch-zju-lungcan-12}{
short=LungCancer\_FAH-ZU\_OAR,
long=LungCancer\_FAH-ZU\_OAR,
foreign-plural={},
first-style=short
}

\DeclareAcronym{ch-zju-esocan-35}{
short=EsoCancer\_FAH-ZU\_OAR,
long=EsoCancer\_FAH-ZU\_OAR,
foreign-plural={},
first-style=short
}

\DeclareAcronym{ch-zju-rtog-13}{
short=LungCancer\_HHA-FU\_OAR,
long=LungCancer\_HHA-FU\_OAR,
foreign-plural={},
first-style=short
}

\DeclareAcronym{ab-td-5}{
short=Cervix\_LinkingMed\_OAR,
long=Cervix\_LinkingMed\_OAR,
foreign-plural={},
first-style=short
}

\DeclareAcronym{hn-lns-18}{
short=HN\_FAH-ZU\_LNS,
long=HN\_FAH-ZU\_LNS,
foreign-plural={},
first-style=short
}

\DeclareAcronym{ch-lns-15}{
short=Chest\_FAH-ZU\_LNS,
long=Chest\_FAH-ZU\_LNS,
foreign-plural={},
first-style=short
}

\DeclareAcronym{body-linkmed-1}{
short=Body\_LinkingMed\_OAR,
long=Body\_LinkingMed\_OAR,
foreign-plural={},
first-style=short
}

\DeclareAcronym{hn-zju-npc-1}{
short=NPCCancer\_FAH-ZU\_Lesion,
long=NPCCancer\_FAH-ZU\_Lesion,
foreign-plural={},
first-style=short
}

\DeclareAcronym{hn-fd-ln-1}{
short=HN\_ZH-FU\_LN,
long=HN\_ZH-FU\_LN,
foreign-plural={},
first-style=short
}

\DeclareAcronym{ch-wpy-lungcan-1}{
short=LungCancer\_SCH\_Lesion,
long=LungCancer\_SCH\_Lesion,
foreign-plural={},
first-style=short
}

\DeclareAcronym{ch-wpy-esocan-2}{
short=EsoCancer\_SCH\_Lesion,
long=EsoCancer\_SCH\_Lesion,
foreign-plural={},
first-style=short
}

\DeclareAcronym{ch-zju-lungln-1}{
short=LungCancer\_FAH-ZU\_LN,
long=LungCancer\_FAH-ZU\_LN,
foreign-plural={},
first-style=short
}

\DeclareAcronym{ch-zju-esoln-1}{
short=EsoCancer\_FAH-ZU\_LN,
long=EsoCancer\_FAH-ZU\_LN,
foreign-plural={},
first-style=short
}

\DeclareAcronym{ab-livercan-2}{
short=LiverCancer\_SHCH\_Lesion,
long=LiverCancer\_SHCH\_Lesion,
foreign-plural={},
first-style=short
}

\DeclareAcronym{hn-gz-12}{
short=NPC\_SMU\_OAR,
long=NPC\_SMU\_OAR,
foreign-plural={},
first-style=short
}

\DeclareAcronym{ch-zju-14}{
short=LungCancer\_NLH\_OAR,
long=LungCancer\_NLH\_OAR,
foreign-plural={},
first-style=short
}

\DeclareAcronym{hn-cgmh-npc-1}{
short=NPC\_CGMH\_Lesion,
long=NPC\_CGMH\_Lesion,
foreign-plural={},
first-style=short
}

\DeclareAcronym{ab-cgmh-livercan-1}{
short=LiverCancer\_CGMH\_Lesion,
long=LiverCancer\_CGMH\_Lesion,
foreign-plural={},
first-style=short
}

\DeclareAcronym{ch-cgmh-esocan-1}{
short=EsoCancer\_CGMH\_Lesion,
long=EsoCancer\_CGMH\_Lesion,
foreign-plural={},
first-style=short
}

\DeclareAcronym{ab-ch-kidneycan-1}{
short=KidneyCancer\_SHCH\_Lesion,
long=KidneyCancer\_SHCH\_Lesion,
foreign-plural={},
first-style=short
}

\usepackage{tocloft} % 用于自定义目录格式
\usepackage{lipsum} 

\usepackage[font=footnotesize]{caption} %footnotesize,small

\usepackage[capitalise]{cleveref}

\graphicspath{{./figs/}}

\newcommand{\acll}[1]{\acl[format/long=\MakeLowercase]{#1}}
\newcommand{\acfl}[1]{\acf[format/long=\MakeLowercase]{#1}}
\newcolumntype{P}[1]{>{\raggedright\arraybackslash}p{#1}}
\newcolumntype{M}[1]{>{\raggedright\arraybackslash}m{#1}}
\newcolumntype{C}[1]{>{\centering\arraybackslash}m{#1}}

\nolinenumbers
\pdfoutput=1 
\title{A Continual Learning-driven Model for Accurate and Generalizable Segmentation of Clinically Comprehensive and Fine-grained Whole-body Anatomies in CT}

\author{Dazhou Guo$^{1\dagger}$,
Zhanghexuan Ji$^{1\dagger}$,
Yanzhou Su$^{1,3\dagger}$, 
Dandan Zheng$^{2\dagger}$,
Heng Guo$^{1,3}$,
Puyang Wang$^{4}$,
Ke Yan$^{1,3}$,
Yirui Wang$^{1}$,
Qinji Yu$^{1}$,
Zi Li$^{1,3}$, 
Minfeng Xu$^{1,3}$,
Jianfeng Zhang$^{1,3}$,
Haoshen Li$^{1}$,
Jia Ge$^{2}$,
Tsung-Ying Ho$^{5}$,
Bing-Shen Huang$^{6}$,
Tashan Ai$^{7}$,
Kuaile Zhao$^{7}$,
Na Shen$^{8}$,
Qifeng Wang$^{9}$,
Yun Bian$^{10}$,
Tingyu Wu$^{11}$,
Peng Du$^{11}$,
Hua Zhang$^{12}$,
Feng-Ming Kong$^{13}$,
Alan L. Yuille$^{14}$,
Cher Heng Tan$^{15}$,
Chunyan Miao$^{16,17}$,
Perry J. Pickhardt$^{18}$, \,
Senxiang Yan$^{2*}$, \,
Ronald M. Summers$^{19*}$, \, 
Le Lu$^{1*}$,
Dakai Jin$^{1*}$,
Xianghua Ye$^{2*}$}

\begin{document}
\maketitle

%\vspace{-2mm}

\begin{affiliations}
 \item DAMO Academy, Alibaba Group
 \item Department of Radiation Oncology, The first Affiliated Hospital, Zhejiang University, Hangzhou, China
 \item Hupan Laboratory, Hangzhou, China
 \item School of Medicine, Johns Hopkins University, Baltimore, United States of America
 \item Department of Nuclear Medicine, Chang Gung Memorial Hospital, LinKou, ROC
 \item Department of Radiation Oncology, Proton and Radiation Therapy Center, Chang Gung Memorial Hospital, LinKou, ROC
 \item Department of Radiation Oncology, Fudan University Shanghai Cancer Center, Shanghai, China
 \item Department of Otolaryngology-Head \& Neck Surgery, Zhongshan Hospital, Fudan University, Shanghai, China
 \item Department of Radiation Oncology, Sichuan Cancer Hospital, Sichuan, China
 \item Department of Radiology, Shanghai Institution of Pancreatic Disease, Shanghai, China
 \item Department of Colorectal and Anal Surgery, Xinhua Hospital Affiliated to Shanghai Jiaotong University School of Medicine, Shanghai, China
 \item Linking Med Inc., Beijing, China
 \item Department of Clinical Oncology, School of Clinical Medicine, LKS Faculty of Medicine, The University of Hong Kong and University of Hong Kong–Shenzhen Hospital, Hong Kong, China
 \item Department of Computer Science, Johns Hopkins University, Baltimore, United States of America
 \item Department of Diagnostic Radiology, Tan Tock Seng Hospital, Singapore
 \item Center of AI in Medicine, Nanyang Technological University, Singapore
 \item Alibaba-NTU Global e-Sustainability CorpLab (ANGEL), Nanyang Technological University, Singapore
 \item  Radiology \& Medical Physics, School of Medicine \& Public Health, University of Wisconsin, Madison, United States of America
 \item Radiology and Imaging Sciences, National Institutes of Health Clinical Center, Bethesda, United States of America
 
% \item Department of Diagnostic Imaging and Interventional Radiology, Moffitt Cancer Center, Tampa, United States of America 

$\dagger$ Daggers indicate first coauthors, they contributed equally \\
$*$ Asterisks indicate co-corresponding authors 
 %\item xxxxxxxx xxxxxxxxxxxxxx xxxxxxxx xxxxxxxx xxxxxxxxx xxxxxxxxxxxxxxxxx xxxxxxxxx \\
\end{affiliations}

% \section*{Abstract}
\vspace{6pt}
\begin{abstract}
\textbf{\textit{Abstract}.} 
Precision medicine in the quantitative management of chronic diseases and oncology would be greatly improved if the Computed Tomography (CT) scan of any patient could be segmented, parsed and analyzed in a precise and detailed way. However, there is no such fully annotated CT dataset with all anatomies (e.g., up to 235 organs and sub-organs) delineated for training because of the exceptionally high manual cost, the need for specialized clinical expertise, and the time required to finish the task.  To this end, we proposed a novel continual learning-driven CT model that can segment complete anatomies presented using dozens of previously partially labeled datasets, dynamically expanding its capacity to segment new ones without compromising previously learned organ knowledge. Existing multi-dataset approaches are not able to dynamically segment new anatomies without catastrophic forgetting and would encounter optimization difficulty or infeasibility when segmenting hundreds of anatomies across the whole range of body regions. Our single unified CT segmentation model, CL-Net, can highly accurately segment a clinically comprehensive set of 235 fine-grained whole-body anatomies (including 73 new anatomies not available elsewhere, but with good clinical utility). Composed of a universal encoder, multiple optimized and pruned decoders, and a body-part-guided merging module, CL-Net is developed using 13,952 CT scans from 20 public (4,855) and 16 private (9,097) high-quality partially labeled CT datasets of various vendors, different contrast phases, and pathologies. Extensive evaluation demonstrates that CL-Net consistently outperforms the upper limit of an ensemble of 36 specialist nnUNets trained per dataset with the complexity of $\sim$5\% model size and significantly surpasses the segmentation accuracy of recent leading ``segment anything''-style medical image foundation models by large margins ($>$9.9\% Dice score). Our continual learning-driven CL-Net model (with 55.5MB parameters) would lay a solid foundation to facilitate many downstream tasks of oncology and chronic diseases using the most widely adopted CT imaging. %and will be made publicly available. %precision management clinical

\end{abstract}
\newpage 
\setcounter{tocdepth}{3}
\tableofcontents
\vspace{2em}

\begin{table}[h!]
\centering
\resizebox{0.98\textwidth}{!}{
\begin{tabular}{lllll}
\cline{1-2} \cline{4-5}
\multicolumn{1}{c}{Abbreviation} & \multicolumn{1}{c}{Definition} &  & \multicolumn{1}{c}{Abbreviation} & \multicolumn{1}{c}{Definition} \\ \cline{1-2} \cline{4-5} 
\acs{asd} & \acl{asd} &  & \acs{fls} & \acl{fls} \\
\acs{bpr} & \acl{bpr} &  & \acs{ge} & \acl{ge} \\
\acs{clnet} & \acl{clnet} &  & \acs{lth} & \acl{lth} \\
\acs{css} & \acl{css} &  & LN & Lymph Node \\
\acs{ct} & \acl{ct} &  & LNS & Lymph Node Station \\
\acs{ddp} & \acl{ddp} &  & \acs{npc} & \acl{npc} \\
\acs{dsc} & \acl{dsc} &  & \acs{sam} & \acl{sam} \\
\acs{ema} & \acl{ema} &  & \acs{ssl} & \acl{ssl} \\ \cline{1-2} \cline{4-5} 
\end{tabular}
}
\caption*{\textbf{List of Abbreviation}~\textbar~All abbreviations used in the manuscript are presented in alphabetical order.}
\label{tab:abbr}
\end{table}

\newpage 
\newpage 
\section{Introduction}
\label{sec:intro}

% CT -> most popular (75% usage? double check + citation) -> segmentation on CT > multi-modality -> solving seg on CT == solved
% SUNSeg structure -> comprehensive + openly available + adapting to new class + non-forgetting
% SUNSeg performance -> compare to all other methods -> we are the best -> weidi xie's SAT + nnUNet + MultiTalent + Peter D LEE?
% SUNSeg -> 
% comparing  to SUNSeg ...  
% final dsc, asd, pruning, runing time
% table 

% CT 用的最广 --- 在CT上的seg解决了，其他也就解决了
% Seg的过程里的困难有：数据集大小，数据集是否可以见，label种类，器官update困难。我们都解决了
% Seg的分割yyds是什么：nnUNet，我们可以超过他，并且超过当前现有的SAT（rull them all) multitalent，nnunet ensemble
% 我们claim seg解决了吗？

%background, single dataset model advantage and limitation
Semantic image segmentation parses medical scans into spatially coherent and meaningful anatomical compartments that serve as an essential functionality to facilitate numerous downstream clinical tasks. Among all advanced cross-sectional imaging modalities, computed tomography (CT) is the most widely used in radiology and oncology due to its wide availability and effectiveness in cancer screening~\cite{ardila2019end,lotter2021robust}, disease diagnosis~\cite{wang2024data}, treatment planning and monitoring~\cite{mcintosh2021clinical}. CT is widely used for a broad range of non-oncologic clinical indications, including unexplained abdominal pain and trauma, etc. For example, harness the rich cardiometabolic data embedded within CT scans that can add great clinical value and describe the rapidly growing field of ``opportunistic CT screening''~\cite{pickhardt2022value,pickhardt2024harnessing}. 3D organ or \ac{gtv} segmentation in CT has been a well-studied problem. With the emergence of many dedicatedly labeled organ imaging datasets~\cite{antonelli2022medical} and the rapid advancement of deep learning segmentation techniques, particularly the introduction of nnUNet~\cite{isensee2021nnu}, 3D segmentation network models trained in specific datasets have achieved quantitative performance comparable to that of human experts~\cite{tang2019clinically,ye2022comprehensive,shi2022deep}, potentially suitable for real clinical practice. However, these models so far are restricted to segmenting only the types of anatomies present in the specific training datasets and training separate models would be required to handle any additional anatomies or datasets, which may be (highly) inconvenient and lack the scalability for wide clinical task deployment.

%multi-datasets approaches (non-SAM based), its limitation
Segmenting all anatomies semantically and accurately from multiple or up to dozens of different training datasets using a single unified model faces the challenge of partial labeling problem, since all datasets are annotated for a specific organ/\ac{gtv} or a limited group of organs/\ac{gtv}s, given the laborious nature of 3D volume annotation. To solve this issue, recent approaches have explored in-context learning~\cite{zhang2021dodnet,xie2023learning,liu2023clip}, various loss designs (marginal and exclusive losses)~\cite{fang2020multi,shi2021marginal,liu2023clip}, or re-mapping all labels with a unique ID across datasets and simply extending the last convolutional layer to produce output channels for all re-mapped labels~\cite{ulrich2023multitalent}. Although partially effective, these methods are often applied to a specific body region and can only segment a limited number of anatomies~\cite{zhang2021dodnet,xie2023learning,liu2023clip,fang2020multi,shi2021marginal} (normally 3$\sim$30). Expanding the scope to cover a broader range of body regions and more anatomy types may lead to optimization difficulty or infeasibility and a tendency to be overfitted onto dominant classes. For example, the recent MultiTalent~\cite{ulrich2023multitalent} method, as a straightforward extension of nnUNet to handle multi-dataset scenarios, experiences (extremely) slow convergence and takes over 1,000 GPU hours when training on using only five datasets of 1,471 CT scans. Moreover, this approach does not produce a unique output for the same organs present in multiple datasets and would require post-merging or dataset ID to generate the final output.

%MultiTalent first re-maps all labels across datasets by assigning each label a unique ID, even if it overlaps with labels from other datasets, then simply expands the final convolutional layer (output) to produce output channels for all re-mapped labels with sigmoid activation function to avoid partial labeling issue. 
%However, this approach does not produce a unique output for same organs present in multiple datasets (requiring post-merging or dataset ID to generate final output). It also lacks sufficient scalability, since training on a large number of CT scans leads to optimization difficulty with extremely slow convergence. For instance, training on only 1,471 CT scans from five datasets can take over 1,000 GPU hours to converge~\cite{ulrich2023multitalent}. 

%multi-dataset approaches, sam-based, advantage and limitation 
With the recent emergence of the Segment Anything Model (SAM)~\cite{kirillov2023segment} in the natural 2D image, a growing number of SAM-style medical universal segmentation models have been developed~\cite{ma2024segment,wu2023medical,chen2024ma,Zhao2023one,Ouyang2024towards,zhao2024foundation,he2024vista3d}. Benefiting from SAM's core design principles of class-agnostic segmentation, multi-modality prompt encoding, and iterative training, SAM~\cite{kirillov2023segment} can learn from large and diverse image datasets, demonstrating impressive performance in promptable interactive segmentation and zero-shot generalizability in natural images. In the medical imaging domain, despite being promoted as foundation models learned from vast amounts of (public) data across various imaging modalities, SAM-style segmentation models~\cite{ma2024segment,Zhao2023one,Ouyang2024towards,zhao2024foundation} with text prompts or even bounding box prompts (which is a much easier scenario), still evidently and quantitatively underperform the dataset-specific trained segmentation models by substantial margins when evaluated fairly and rigorously. For example, the latest prompt-required BiomedParse~\cite{zhao2024foundation} offers clearly inferior performance compared to the automated nnUNet~\cite{isensee2021nnu} model in organ segmentation using 3D CT scans (Table~\ref{tab:bmp_compare}), potentially making it far from clinically useful. The underwhelming performance observed for SAM-style models in the medical domain can be attributed to the following factors: 1) Unlike natural images, where targets often display distinct visual features, medical image segmentation is challenging due to the difficulty in delineating ambiguous local boundaries. Anatomical targets in medical scans frequently exhibit a subtle visual contrast with adjacent structures or the background. Consequently, Transformer architectures (adopted in SAM), known for their exceptional ability to model long-range spatial dependencies, offer limited advantages in capturing local subtle visual differences and, meanwhile, are difficult to optimize effectively. 2) The dimensional gap between 3D medical scans and 2D natural images significantly hinders SAM's transferability. Meanwhile, directly training SAM in 3D leads to a dramatic increase in model parameters, further exacerbating the optimization challenges. This raises the core question: \textit{how to build an effective and accurate universal medical segmentation model that is capable of segmenting all clinically relevant anatomies of the human body on CT scans (the most prevalent imaging modality used in radiology/oncology), and consistently achieving comparable or even better segmentation performance in accuracy and speed than the dataset-specific trained state-of-the-art nnUNet~\cite{isensee2021nnu} models}?

%our solution
To resolve all aforementioned problems, we propose to tackle universal segmentation in CT from a new perspective of continual learning and introduce a unified, scalable, and nonforgetting continuous learning method (CL-Net), capable of harnessing the synergies of a large number of partially labeled datasets to accurately segment the union of all anatomies present across all datasets. Continual learning is a sequential process that aims to dynamically adapt and extend the model to learn new targets or update existing targets without (re)access to previous training datasets~\cite{goodfellow2013empirical,li2017learning,Delang2021PAMI,van2022three}. Our \ac{clnet} is an architectural-based framework, which comprises a shared/general encoder, a set of sequentially optimized and pruned light-weighted decoders (one decoder per organ or per group of organs) and a body-part-guided output merging module. The key characteristics of \ac{clnet} include: 1) A trained then frozen general encoder coupled with sequentially added decoders for subsequent training datasets can extract sufficiently representative image features to segment new or update existing anatomies. The frozen encoder enables \ac{clnet} to dynamically adapt and segment new anatomies, while preserving all previously acquired knowledge, even in the absence of access to prior training data. 2) By assigning each organ (that is, major anatomy) or group of relevant organs (e.g., rib instances) to independent decoders, multiple datasets containing specific organ groups can be used sequentially or simultaneously and integrated together to optimize the associated decoders, ensuring that more data variations (from all healthy and pathological subjects) are observed/captured by \ac{clnet}. This decoder setup naturally handles the partial labeling in multi-dataset segmentation. %, harvesting universal information among multiple datasets. 
3) Optimizing each decoder with \ac{lth}-based pruning strategy~\cite{frankle2018lottery,frankle2019stabilizing}, \ac{clnet} only needs to maintain a very small fraction of effective parameters for each decoder (normally 1$\sim$3\% model size of the default full size decoder), ensuring its modeling scalability. 4) The body-part guided merging module, which combines outputs from all decoders, can effectively reduce distal false-positive segmentation from different decoders. A preliminary technical version~\cite{Ji_2023_ICCV} appeared in ICCV-2023, and this work introduces substantial methodological extensions, together with entirely new and significantly much larger-scale experiments and analysis. For example, replacing the dataset-wise decoder design with organ-wise decoder substantially increases the model's flexibility in managing partially labeled datasets and its scalability for learning from large-scale data and diverse anatomy types. Furthermore, the new LTH-based pruning strategy achieves a higher model compression rate compared to the knowledge distillation-based pruning in the original version~\cite{Ji_2023_ICCV}. Furthermore, the updated framework supports both continual segmentation (dataset sequentially accessible) and universal segmentation (dataset simultaneously accessible).
%that are often observed in segmenting whole-body anatomies from dataset specifically trained 
%\textcolor{red}{(with much lower parameter growth rate)}

We build the \ac{clnet} model using 13,952 CT scans from 20 public (4,855) and 16 private (9,097) high-quality partially labeled CT datasets of different vendors, different contrast phases and pathologies (based on a collection of our previous peer-reviewed work), which can automatically and accurately parse a set of \ac{orgn} whole-body human anatomies, encompassing 193 organs, 33 lymph node stations, and 9 \ac{gtv}s. Compared to existing work, more fine-grained vessels, muscle substructures and glands are covered, as well as lymph node stations in head \& neck and chest regions, to facilitate both oncologic and non-oncologic clinical diagnosis. In addition, our private datasets are primarily collected from pathological patient cohorts, which contain more disease-oriented organs as a valuable addition to the data diversity of public datasets. Through extensive comparison, \ac{clnet} consistently outperforms or matches the performance of an ensemble of \ac{dsn} specialist nnUNets~\cite{isensee2021nnu} trained on each dataset (Dice-Sørensen coefficient [DSC]: $86.1$\% vs $83.9$\%) and the nnUNet-extended universal model MultiTalent~\cite{ulrich2023multitalent} ($+2.7$\% DSC), and significantly outperforms the latest leading SAM-style medical foundation models (on a subset of anatomies that appeared in their training data) by evidently large margins, e.g.,
VISTA3D~\cite{he2024vista3d} ($+2.9$\% DSC), SAT-Pro~\cite{Zhao2023one} ($+9.9$\% DSC) and BiomedParse~\cite{zhao2024foundation} (\textcolor{black}{$+67.0$\% DSC in 3D}). We will make our \ac{clnet} network model publicly available (for non-commercial use), providing scientists and physicians with the latest comprehensive and precise \ac{orgn} whole-body CT anatomy segmentation tool as a useful routine clinical imaging semantic quantification and assessment tool.

\section{Results}
\label{sec:result}

%% reset results subsection counter
\newcounter{res_cnt}
% \setcounter{res_cnt}{0} 
% \arabic{res_cnt}
% \stepcounter{res_cnt}

%%------ Overview of Datasets --------

\subsection{Overview of Datasets}
\label{sec:dataset}

%%------ Fig. Dataset Stats ---------%%
\begin{figure*}[htpb]
\renewcommand{\figurename}{Fig.}
    \centering
    \includegraphics[width=\textwidth]{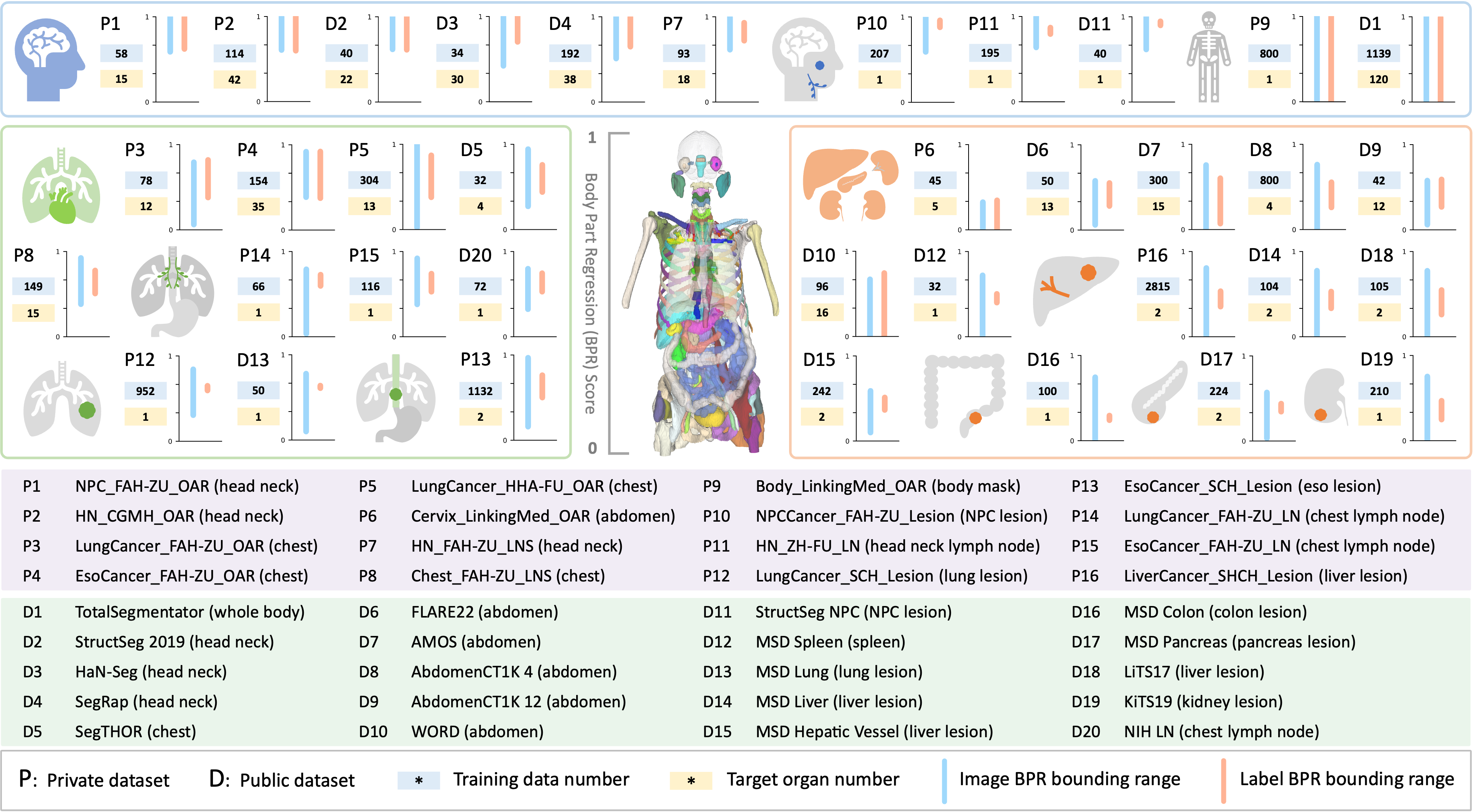}
    \captionsetup{belowskip=0pt} % Remove space below caption
    \caption{\textbf{\textbar~Dataset fingerprints.} Overview of \ac{dsn} datasets used for model development and internal validation, including 16 private (P) and 20 public (D) datasets. Datasets are categorized into head-neck (blue box), chest (green box), and abdomen (orange box) groups, based on the primary body region covered by their target organs, expect for \ac{totalseg} and \ac{body-linkmed-1}, which span the entire body. For each dataset, the number of training CT scans, number of target organs, and vertical range of \acfl{bpr} scores of CT scans and foreground labels are provided. The \acs{bpr} score represents the relative vertical position of each slice in a CT scan, normalized from the bottom of pelvis (0) to the top of head (1). Additional dataset fingerprint details are available in Supplementary Sec. \ref{sec:dataset_details}. }
    \label{fig:dataset}
    %\vspace{-2em}
\end{figure*}

%% Intro
The goal of \ac{clnet} is to train \textit{continuously or simultaneously} a unified model capable of segmenting a comprehensive set of fine-grained whole-body anatomies with high precision on 3D CT scans. To accomplish this, \ac{dsn} CT data sets are assembled for \textbf{model training and internal testing}, comprising 20 public and 16 private data sets in-house (a total of 13,952 CT scans, Fig. \ref{fig:dataset}) of different institutions with various vendors, phases, and pathologies. These datasets cover the entire human body region, including head \& neck, chest, abdomen, and pelvis, and contain a wide range of anatomies, such as major organs, bones, muscles, vessels, glands, lymph node stations, and \ac{gtv}s, etc. In total, there are \ac{orgn} anatomy types (Supplementary Sec. \ref{sec:dataset_details}, Table~\ref{tab:dataset_stats}). The 20 public datasets contain 4,855 CT scans ($35$\% of all data), which are divided to 3,904 for training and 951 for testing (detailed data split of each dataset in the Supplementary Table~\ref{tab:dataset_stats}). Together, these datasets include 162 anatomical structures or substructures (Table~\ref{tab:dataset_task_d}).  Our 16 private datasets, collected from multiple institutions, comprise a total of 9,097 CT scans (65\% of the entire data), including 7,278 for training and 1,819 for testing (detailed data split of each data set in the Supplementary Table~\ref{tab:dataset_stats}). These private datasets contain 125 anatomical structures (Table~\ref{tab:dataset_task_p}), including 52 overlapping and 73 additional fine-grained structures not present in the public data sets. 
%For instance, to facilitate both oncologic and non-oncologic clinical diagnosis, we include 33 lymph node stations in head \& neck and chest regions, and many fine-grained vessel and muscle substructures in chest and abdominal regions. Moreover, these private datasets are primarily collected from pathological patient cohorts, containing more disease-oriented organs as valuable addition to the data diversity of public datasets. 
Regarding the annotation process, anatomies in our private datasets have been carefully labeled by our clinical collaborators over the past several years and have been extensively used in our previous peer-reviewed work~\cite{jin2019accurate,jin2019deep,guo2020organ,ye2022comprehensive,guo2021deepstationing,liu2021same,guo2022thoracic,guo2024towards,jin2021deeptarget,jin2022towards,wang2023anatomy,zhang2023deep,yan2022sam,yan2023liver,yan2023anatomy,yu2025effective,yu2024slice,ye2022multi,li2023lvit,Ji_2023_ICCV,tian2023same++,zhu2024low,li2024leveraging,ye2024development}. 
%It is important to note that, unlike public datasets, which are accessible to external centers, the data privacy and accessibility of in-house datasets are strictly controlled; as a result, in real-world scenarios, \acs{css} model is required to be trained at different sites, making the final performance susceptible to the learning sequence of the multi-site datasets. %external datasets
We also collect seven datasets that did not appear in the model development for the \textbf{external testing} (Supplementary Table~\ref{tab:supp_ex_dataset}), which includes 1,979 CT scans of 43 anatomy types (39 organs and four lesions). Specifically, external testing covers 12 head \& neck, 14 chest and 13 abdominal organs, and four lesions of various regions of the body. Moreover, we ensure that an external dataset of a specific body-part/lesion-type comes from an institution that does not provide training data of this body-part/lesion-type. For example, the HN\_SMU\_OAR consists of 268 CT scans of head \& neck cancer patients obtained from a medical center in south China that is different from all sources of head \& neck organ datasets used in training (NPC\_FAH\_ZU\_OAR, HN\_CGMH\_OAR, StructSeg\_2019, HaN-Seg and SegRap).

%% Figure--Dataset: BPR Fingerprints
% As shown in Fig. \ref{fig:dataset}, the \ac{dsn} datasets are grouped into head and neck, chest, and abdomen, according to the mainly covered body part of each dataset's target organs (except for \ac{totalseg} and WholeBody\_LinkingMed\_1 which cover the entire body). It also illustrates some important fingerprints of all private (P) and public (D) datasets, including the training set size, target organ number, body part \& organ group, and \acll{bpr} (BPR) axial bounding range of both image CT scans and foreground labels. Specifically, BPR is a self-supervised body part regression model that outputs corresponding axial scores for each slice of a CT scan, mapping the start of the pelvis to \textbf{0} and the top of the head to \textbf{1}. To quantitatively assess the body axial ranges covered by the images and labels in each dataset, we compute the BPR scores for all slices within each dataset and report the score ranges for all images and labels as dataset fingerprints. 
% The dataset distribution of each group and other detailed fingerprints%, such as test set size, median image shape and voxel spacing, and numerical BPR scale, 
% are provided in Supplementary %Table~\ref{tab:dataset_stats}--\ref{tab:dataset_task_p} in 
% Sec. \ref{sec:dataset_details}. 

%%------ Overview of CL-Net Configuration --------

\subsection{Two \ac{clnet} Configurations}
\label{sec:overview_clnet}

\begin{figure*}[htpb]
\renewcommand{\figurename}{Fig.}
\centering
\includegraphics[width=0.9\textwidth]{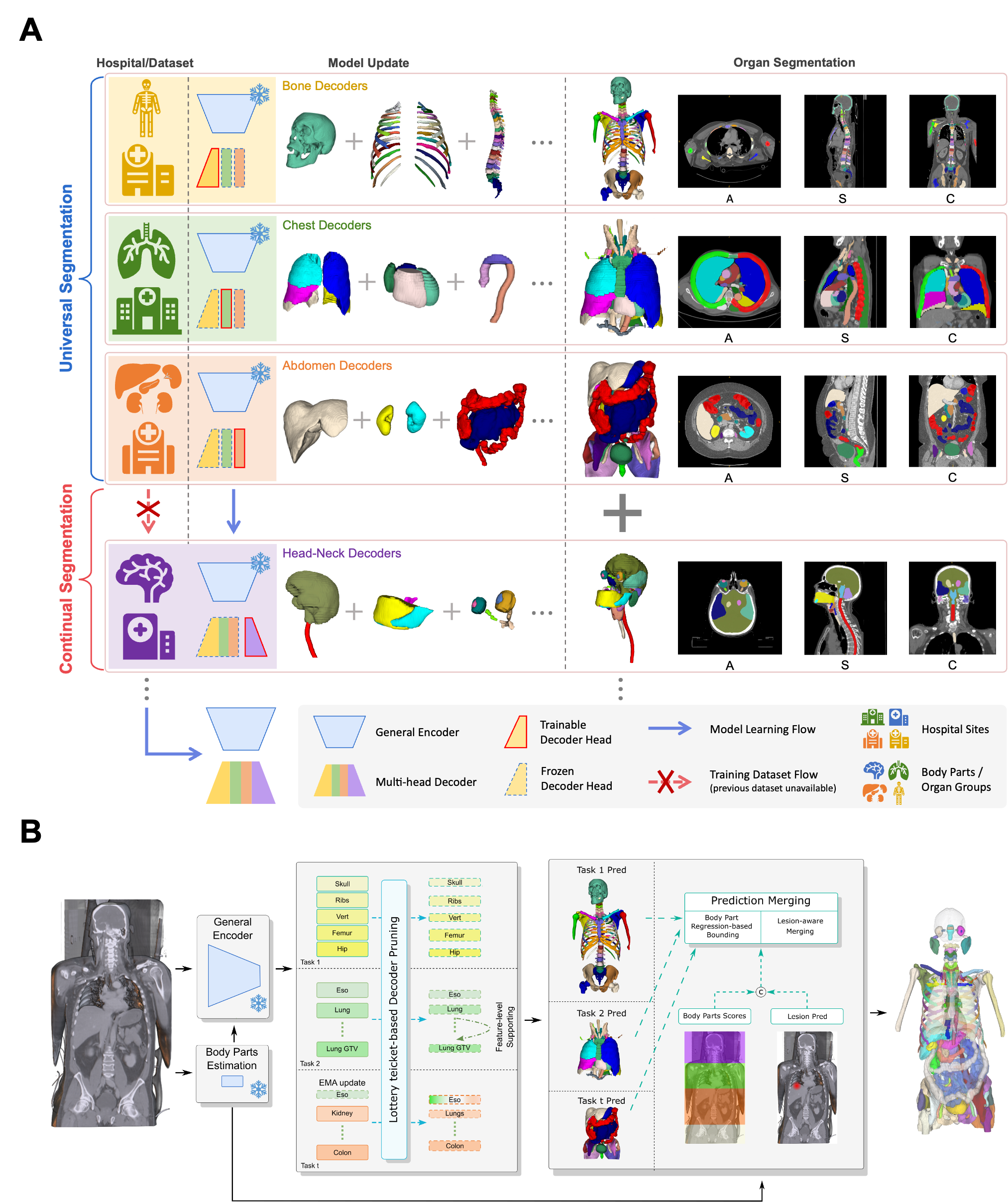}
\caption{\textbf{\textbar~Illustration of the learning process and architecture of CL-Net.} \textbf{A}. \ac{clnet} can be trained or updated from both \ac{pl} and \ac{cs} settings, with the pre-trained \ac{ge} being frozen. In the partial label segmentation setting, with simultaneous access to datasets of different body parts, the model directly learns different decoders to segment whole-body organs. In the continual segmentation setting, with sequentially available new datasets and no access to previous ones, lightweight decoders of corresponding anatomy structures are added or updated, enabling segmentation of all learned organs without forgetting. \textbf{B}. An overview of the \ac{clnet} framework: a GE for feature extraction, multiple decoders for organ-wise segmentation, a \ac{lth}-based decoder pruning module, and a prediction merging module. }
\label{fig:motivate}
\end{figure*}

\ac{clnet} is designed to support both \textit{\acl{pl} (partial label learning)} and \textit{\acl{cs}} of large-scale whole-body anatomies in CT scans (Fig.~\ref{fig:motivate}). Its learning process begins with training a general encoder using one or multiple datasets (e.g., \ac{totalseg}~\cite{wasserthal2022totalsegmentator}). Then, the general encoder is frozen in subsequent steps. 
% PL
In the \textbf{\ac{pl}} configuration, \ac{clnet} constructs individual decoders for each type of anatomy or group of relevant anatomy (e.g., rib instances) across the entire collection of datasets, as all datasets are available simultaneously for training. Each decoder is then sequentially trained using all data samples that contain the corresponding anatomy type(s). 
% CSS
In the \textbf{\ac{cs}} configuration, where only one dataset is accessible at a time, \ac{clnet} is sequentially trained on each dataset. Using the available dataset at each step, \ac{clnet} incrementally adds and trains new decoders for previously unseen anatomy types or fine-tunes existing decoders via \ac{ema} for anatomy types already appearing in previous datasets. 
% pros
In both configurations, \ac{clnet} facilitates accurate, efficient, and scalable segmentation of whole-body anatomies, accommodating sequential and simultaneous dataset availability, respectively. 
%Furthermore, \ac{clnet} provides exceptional generalizability and scalability, enabling effortless and efficient adaptation to new datasets and segmentation tasks across both settings. 

%Under \ac{css}, a specific organ labeled in multiple datasets has to be learned in several steps,  Although only one dataset is available at a learning step, the proposed CL-Net framework (frozen encoder and separate decoders with exponential moving average (EMA) update for existing organs) can preserve the old knowledge and meanwhile gradually learn the new knowledge. 

%%------ Overview of Experiment Setting --------

%---------- Fig. Overview ------------%

\subsection{2.1 Overview of Experimental Setting}
\label{sec:exp}
%% exp intro
%, and \textcolor{red}{the external results on seven datasets}
We perform extensive experimental evaluation in three dataset settings: 1) full internal evaluation using all 36 internal datasets, 2) comparative internal evaluation using five representative internal public datasets, and 3) external evaluation using seven external datasets. %Additionally, the effectiveness of key components of \ac{clnet} is demonstrated using the 5 representative internal public datasets (dataset setting-2).
First, the \textbf{full internal evaluation} is performed by developing the \ac{cln_u36} and \ac{cln_c36} using 13,952 CT scans of the 36 internal datasets under both partial label (simultaneous access to the dataset) and continual training configurations (sequential access to the dataset). The detailed training and testing split of \ac{dsn} datasets is summarized in the Supplementary Table~\ref{tab:dataset_stats}.
The complete internal evaluation provides comprehensive results for \ac{clnet} across the entire set of \ac{orgn} whole-body anatomies, and compares to the performance of fully optimized~\footnote{The TotalSegmentator nnUNet model was trained for 8,000 epochs, while the other models were trained for 1,000 epochs. All nnUNet models utilized the default ``3d\_fullres'' network setting with ``more data augmentation'' training scheme.}  \ac{dsn}-nnUNet~\cite{isensee2021nnu} specialists ensemble (\ac{nnu_e36}), which should be considered as the true state-of-the-art segmentation results. We also evaluate the impact of decoder pruning by comparing \ac{cln_u36} with its original unpruned counterpart model (\ac{cln_u36_unprn}). 

Next, the \textbf{internal comparative evaluation} is conducted on five representative public datasets (a total 1,471 CT scans and 137 anatomy types), including \ac{totalseg} (whole-body)~\cite{wasserthal2022totalsegmentator}, \ac{structseg} (head \& neck)~\cite{StructSeg2019}, \ac{segthor} (chest)~\cite{lambert2020segthor}, \ac{flare} (abdomen)~\cite{ma2023unleashing}, and \ac{kits} (\ac{gtv})~\cite{heller2020state}. We develop \ac{cln_u5} and \ac{cln_c5} using the training data of these five internal public datasets under partial label and continual training configurations, respectively.  We also train and compare to the leading multi-dataset approach Multi-Talent~\cite{ulrich2023multitalent} on these five internal public datasets. We further compare the performance of the latest SAM-style medical foundation models, SAT-Pro~\cite{Zhao2023one}, VISTA3D~\cite{he2024vista3d} and BiomedParse~\cite{zhao2024foundation}, as they can segment many organ types in these five public datasets.  Finally, the performance of \ac{cln_c5} (as a continual segmentation model) is compared to three of the most popular and leading continual segmentation methods, MiB~\cite{cermelli2020modeling}, PLOP~\cite{douillard2021plop}, and CSCLIP~\cite{zhang2023continual} under continual training configuration.  Detailed experimental settings are described in Supplementary Sec.~\ref{sec:exp_set_details}. Note that we did not compare Multi-Talent~\cite{ulrich2023multitalent} or SAM-style medical foundation models in the full evaluation of 36 internal datasets, because SAM-style foundation models~\cite{Zhao2023one,Zhao2024BiomedParse} are not capable of segmenting many fine-grained anatomies in our private datasets, and Multi-Talent, on the other hand, lacks the scalability to train on full-scale datasets (it takes over 1,000 GPU hours to converge when training on only five datasets of 1,471 CT scans). Finally, the \textbf{external evaluation} is carried out on seven external datasets (a total 1,979 CT scans and 43 anatomy types), including organs and lesions at head \& neck, chest, abdomen regions. We compare our CL-Net (\ac{cln_u36}) with the ensemble of \ac{nnu_e36}, Multi-Talent~\cite{ulrich2023multitalent}, SAT-Pro~\cite{Zhao2023one}, VISTA3D~\cite{he2024vista3d} and BiomedParse~\cite{zhao2024foundation} by directly performing inference on these seven external datasets.
%\textcolor{red}{The external testing are conducted on xxx datasets (xx public and xx private), which cover head \& neck, chest, abdomen (AbdomenAtlas1.1-Trauma~\cite{li2024abdomenatlas}), and several lesions.  Performance of \ac{clnet}, ensemble of \ac{nnu_e36}, Multi-Talent and SAT are directly evaluated and compared using the ground-truth mask on external datasets.}
In addition, using five representative internal public datasets, \textbf{ablation studies} are performed to evaluate the effectiveness of several key method components of CL-Net, such as the general encoder training scheme, decoder's \ac{fls} function, \ac{ema}-based decoder fine-tuning, etc.

\subsection{2.2 Full Internal Results on \ac{dsn} Datasets} 
\label{sec:css_full}

%% reset results subsection counter
% \newcounter{res_cnt}
\setcounter{res_cnt}{0} 
% \arabic{res_cnt}
% \stepcounter{res_cnt}

%% exp intro (U36 v.s. C36 v.s. nnUNet_E36)
We report the organ-wise segmentation performance of \ac{orgn} organs for 
%and report the organ-wise dice score statistics in Fig. \ref{fig:organ-dsc} and Fig. \ref{fig:organ-box}, which separately shows the organ-wise DSC box plots for three models: 
\ac{clnet} under both \ac{pl} (\acs{cln_u36}) and \ac{cs} settings (\acs{cln_c36}) and compare with the \ac{dsn}-nnUNets ensemble (\acs{nnu_e36}). The investigated anatomies (except `Body') are grouped as Head \& Neck, Chest, Abdomen, Bone, Lymph node station (LNS), and \Ac{gtv}, according to different body parts and anatomy types. 
%as explained in our experiment setting, 
%and the organ names of each group are colored separately with a bracket in the same color showing the range of its organ group on horizontal axis. 

%Detailed DSC and ASD scores of each anatomy are reported in \textcolor{red}{supplementary Sec.~\nameref{sec:supp_table}. }

%%------ Results ---------

% \usepackage{booktabs}
% \usepackage{multirow}
% \usepackage{graphicx}
\begin{table}[!t]
\centering
\caption{\textbar~\textbf{Quantitative segmentation results of \ac{nnu_e36}, \ac{cln_u36_unprn}, \ac{cln_c36} and \ac{cln_u36} on \ac{orgn} whole-body anatomies.} Results are grouped into different body regions and anatomy types. Model complexity is also reported. DSC: Dice-Sørensen coefficient (\%), ASD: average surface distance (mm), Param\#: network parameter size (MB). The best results are highlighted in bold. }
%\textbf{*Note}: \ac{nnu_e36} param\# includes both encoder and decoder, since the ensemble nnUNets have independent encoder for each dataset, which is a non-negligible model growth. For \ac{clnet}, the frozen encoder size is 15.65M, which has the same structure as nnUNet encoder. The whole model sizes of \ac{cln_c36} and \ac{cln_u36} are 55.5M and 55.6M, separately. }
\label{tab:main}
\resizebox{\textwidth}{!}{%
\begin{tabular}{@{}clccccccc|c@{}}
\toprule
\textbf{Metric} &
  \textbf{Method} &
  \textbf{Head \& Neck (49)} &
  \textbf{Chest (52)} &
  \textbf{Abdomen (28)} &
  \textbf{Bone (63)} &
  \textbf{LNS (33)} &
  \textbf{\Ac{gtv} (9)} &
  \textbf{Body (1)} &
  \textbf{All (\ac{orgn})} \\ \midrule
 &
  \ac{nnu_e36} &
  75.9 &
  85.5 &
  90.8 &
  94.3 &
  72.1 &
  69.1 &
  97.0 &
  83.9 \\
 &
  \ac{cln_u36_unprn} &
  \textbf{80.6} &
  \textbf{88.1} &
  \textbf{91.8} &
  \textbf{94.6} &
  \textbf{74.3} &
  69.6 &
  \textbf{97.2} &
  \textbf{86.1} \\
 &
  \ac{cln_c36} &
  79.6 &
  88.0 &
  91.5 &
  94.5 &
  74.1 &
  69.4 &
  97.1 &
  85.7 \\
\multirow{-4}{*}{DSC (\%)$\uparrow$} &
  \ac{cln_u36} &
  \textbf{80.6} &
  \textbf{88.1} &
  \textbf{91.8} &
  \textbf{94.6} &
  74.2 &
  \textbf{69.7} &
  97.1 &
  \textbf{86.1} \\ \midrule
 &
  \ac{nnu_e36} &
  1.18 &
  1.37 &
  1.14 &
  1.06 &
  1.35 &
  3.84 &
  \textbf{2.20} &
  1.32 \\
 &
  \ac{cln_u36_unprn} &
  0.96 &
  1.18 &
  0.98 &
  0.99 &
  1.21 &
  3.90 &
  2.29 &
  1.18 \\
 &
  \ac{cln_c36} &
  1.01 &
  1.19 &
  1.02 &
  0.99 &
  1.22 &
  3.92 &
  2.33 &
  1.20 \\
\multirow{-4}{*}{ASD (mm)$\downarrow$} &
  \ac{cln_u36} &
  \textbf{0.95} &
  \textbf{1.17} &
  \textbf{0.97} &
  \textbf{0.98} &
  \textbf{1.20} &
  \textbf{3.72} &
  2.50 &
  \textbf{1.17} \\ \midrule
 &
  \ac{nnu_e36} &
  --- &
  --- &
  --- &
  --- &
  --- &
  --- &
  --- &
  1,126.8 \\
 &
  \ac{cln_u36_unprn} &
  438.2 &
  485.2 &
  297.4 &
  172.2 &
  31.3 &
  156.5 &
  15.7 &
  1,612.1 \\
 &
  \ac{cln_c36} &
  12.2 &
  6.9 &
  3.5 &
  2.0 &
  3.8 &
  11.6 &
  0.1 &
  55.6 \\
\multirow{-4}{*}{Param\# (MB)$\downarrow$} &
  \ac{cln_u36} &
  12.1 &
  6.9 &
  3.5 &
  2.0 &
  3.8 &
  11.6 &
  0.1 &
  \textbf{55.5} \\ \bottomrule
\end{tabular}%
}
\end{table}

%% PL results
% \subsubsection{Partial Label (PL) \Ac{pl} Segmentation Results on \ac{dsn} Datasets.} 

\stepcounter{res_cnt}
\subsection{\arabic{res_cnt}) \ac{cln_u36}, simultaneously trained on \ac{dsn} datasets under \ac{pl} setting, demonstrates superior segmentation performance across \ac{orgn} anatomies compared to 36 nnUNet ensemble \ac{nnu_e36}.} 
\begin{figure*}[htpb]
\renewcommand{\figurename}{Fig.}
    \centering
    \includegraphics[width=0.9\textwidth]{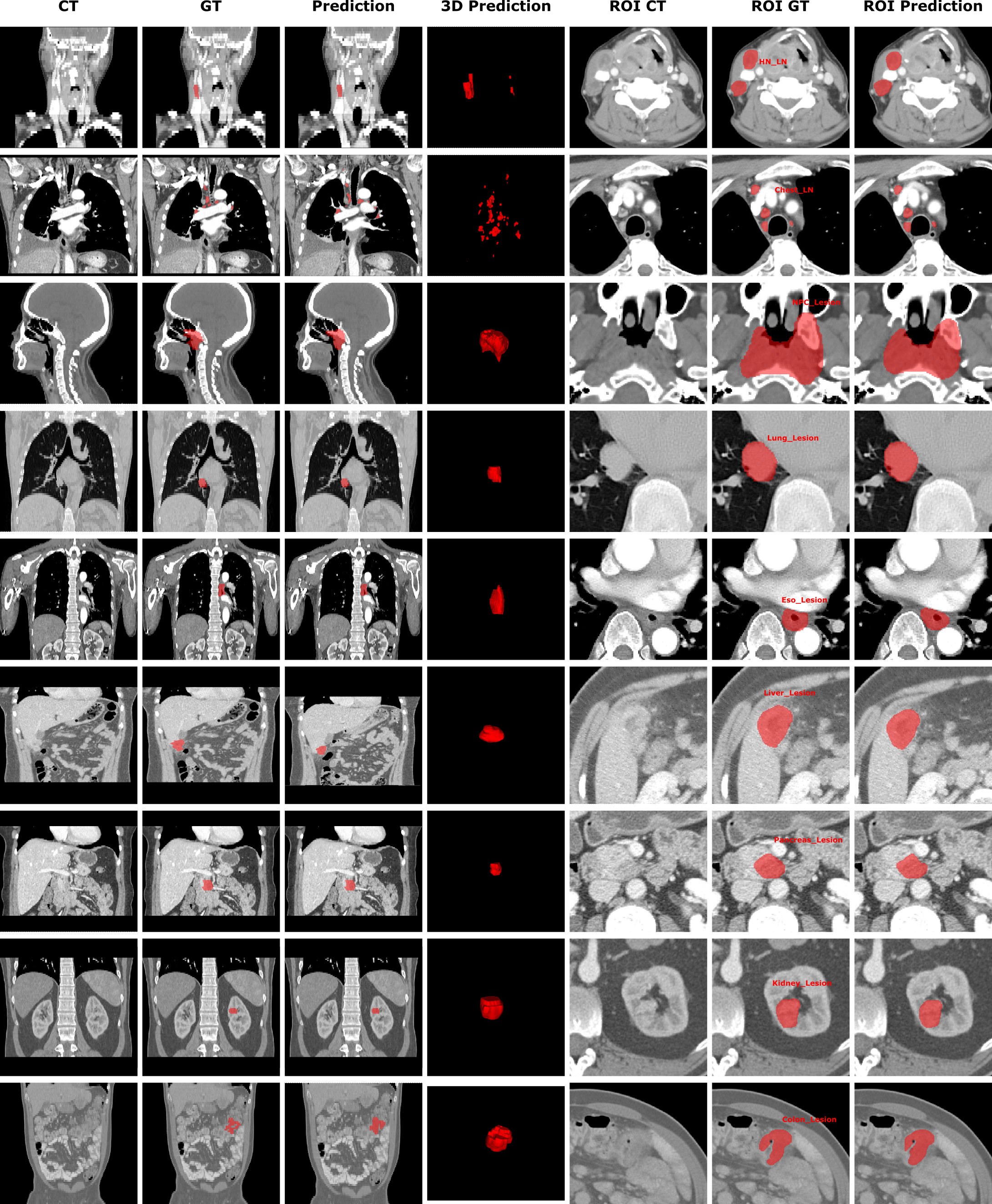}
    \caption{\textbf{\textbar~Qualitative visualization of \ac{clnet} lymph node and \ac{gtv} segmentation.} The first two rows depict the segmentation results of head \& neck and chest regions lymph nodes segmentation, respectively. For better illustration, lymph node stations are superimposed onto the CT images. The third to ninth rows show the \ac{gtv} segmentation at nasopharynx, lung, esophagus, liver, pancreas, kidney, and colon, respectively. }
    % 1) \ac{nnu_e36} ensemble induced inferior performance -- some dataset only contain partial IVC
    % 2) For better illustration, some sub-group organs (e.g., lung, heart) are merged and chest-wall is excluded.  
    \label{fig:quali-v2}
\end{figure*}

%% mean DSC/ASD over all organs, comp. w. nnUNet
The quantitative results are summarized in Table~\ref{tab:main} and Fig.~\ref{fig:organ-dsc}, and several observations are made. First, our \ac{cln_u36} achieves the organ-wise mean DSC of $86.1$\% and the mean ASD of $1.17$mm, markedly outperforming \ac{nnu_e36} by $2.2$\% DSC increase and $11.5$\% (0.15mm) ASD error reduction for \ac{orgn} anatomies. 
%% better/worse organs stats
Second, as illustrated in the detailed organ-wise DSC accuracy (Fig.~\ref{fig:organ-dsc}), \ac{cln_u36} outperforms \ac{nnu_e36} in 206 out of \ac{orgn} anatomies with an average improvement in DSC of $2.5$\%. For the remaining 29 anatomies where \ac{cln_u36} yields lower DSC, the mean difference in DSC is relatively small with an average DSC reduction of $0.4$\%. In terms of ASD, \ac{cln_u36} exhibits lower ASD errors in 214 out of \ac{orgn} anatomies, with an average ASD reduction of $0.19$mm. Third, \ac{cln_u36} exhibits significant DSC improvement over \ac{nnu_e36} on several organs in head \& neck and chest regions, such as pituitary ($10.1$\%), optical nerves ($11.8$\%) and chiasm ($11.9$\%), hypothalamus ($13.9$\%), pineal gland ($10.5$\%), vertebral arteries ($9.5$\%), and brachial plexus ($13.8$\%).  These organs are more difficult to segment, as they are small and have poor intensity contrast with adjacent anatomies on CT imaging. Due to the organ-specific decoder design and \ac{fls} (The detailed decoder-wise \ac{fls} information is reported in Supplementary Table~\ref{tab:dec2org}), each decoder can zoom in to focus on segmenting a particular organ, which leads to easier optimization compared to training a single nnUNet to segment all organs simultaneously. Lastly, in terms of model complexity, \ac{cln_u36} has $55.5$MB parameters, which are less than two nnUNet model sizes ($62.6$MB). In comparison, the ensemble of 36 nnUNet specialists contains $1,126.8$MB parameters, 20 times larger than \ac{cln_u36}. These results highlight the superior performance and scalability of \ac{cln_u36} for segmenting \ac{orgn} fine-grained whole-body anatomies on 3D CT scans (some qualitative segmentation examples are illustrated in Fig.~\ref{fig:quali-v1} and Fig.~\ref{fig:quali-v2}).
%and feature-level decoder supporting scheme
%The most significant improvements are observed for HypoThalamus, which shows a DSC increase of 36.90\% and an ASD reduction of 1.76mm compared to \ac{nnu_e36}. 
%% large improve on some organ groups v.s. nnUNet E36
%% some discuss
%These results highlight the exceptional segmentation performance of \ac{cln_u36}, particularly for small and complex structures such as head and neck substructures, nerves, blood vessels, and bronchi, showcasing its distinct advantages in handling intricate anatomical regions. %% conclusion
%The markedly enhanced segmentation performance of \ac{cln_u36} across \ac{orgn} diverse whole-body organs and individual organ groups, compared to the state-of-the-art nnUNet model, underscores the superior accuracy, generalizability, and robustness of our approach for large-scale whole-body organ segmentation in clinical CT scans. 

%% group wise stats
When stratified by different body regions and anatomy types, \ac{cln_u36} delivers high mean DSC scores for the head \& neck ($80.6$\%), chest ($88.1$\%), abdomen ($91.8$\%), bone ($94.6$\%) and body ($97.1$\%) subgroups, respectively. \ac{cln_u36} also yields small ASD errors ($\sim$1mm) in all subgroups except the \ac{gtv} group ($3.72$mm ASD). This is because lesions are usually more difficult to segment due to their ambiguous boundaries, small size, and irregular shapes~\cite{antonelli2022medical}. Compared to \ac{nnu_e36}, \ac{cln_u36} consistently shows higher mean DSC scores in all subgroups, with notable DSC improvements of $4.7$\%, $2.6$\%, $1.0$\% and $2.1$\% in head \& neck, chest, abdomen and lymph node station (LNS), respectively. Similar error reductions in mean ASD are observed across the subgroups. Note that LNSs are regions that contain different anatomical levels of lymph nodes and their segmentation is highly context-dependent~\cite{liu2016mediastinal,guo2021deepstationing}, i.e., LNS boundaries are restricted by anatomical organs. Hence, using the \ac{fls} function in \ac{cln_u36}, we can segment LNS more accurately by injecting decoder features of the adjacent organs of LNS, resulting in $2.1$\% improved DSC and $10.4$\% reduction of ASD error compared to \ac{nnu_e36}. 

%Organ-wise results of \ac{cln_u36} (Fig. \ref{fig:organ-box}) reveal that most organs from the head-neck, chest, abdomen and bone groups exhibit higher DSC scores with moderate variation. This consistency is likely attributed to the large size, stable shape and relative fixed anatomical position of these organs. In contrast, organs with smaller sizes, complex or irregular shapes, and diverse anatomical locations -- such as LNS and \ac{gtv}s, as well as some chest vessels -- tend to have lower DSC scores with wider variations. 

%Although the more challenging LNS and \ac{gtv} groups do not reach a mean DSC of 80\%, they still achieve respectable scores of 72.27\%, and 70.89\%, respectively. For ASD metric, all organ groups, except the \ac{gtv} group with a relatively larger mean ASD of 3.48mm, maintain mean ASD values within a narrow range of 0.82mm to 1.29mm. 

%When considering individual organ groups, \ac{cln_u36} delivers high mean DSC scores of 80.64\%, 88.15\%, 91.84\%, and 92.28\% for the head-neck, chest, abdomen, and bone groups, respectively. Although the more challenging LNS and lesion groups do not reach a mean DSC of 80\%, they still achieve respectable scores of 72.27\%, and 70.89\%, respectively. For ASD metric, all organ groups, except the lesion group with a relatively larger mean ASD of 3.48mm, maintain mean ASD values within a narrow range of 0.82mm to 1.29mm. 
%% some discuss

%% CSS results
% \subsubsection{Continual Semantic Segmentation \Ac{css} Results on \ac{dsn} Datasets.} 

\stepcounter{res_cnt}
\subsection{\arabic{res_cnt}) \ac{cln_c36}, continuously trained on \ac{dsn} datasets under \ac{cs} setting, achieves closely comparable performance from \ac{cln_u36}. } 
%% mean DSC/ASD over all organs, comp. w. cln_u36 & nnu_e36
As shown in Table~\ref{tab:main}, \ac{cln_c36} attains a mean DSC of $85.7$\% and a mean ASD of $1.20$mm across all \ac{orgn} organs. When compared to \ac{cln_u36} that simultaneously trained on \ac{dsn} datasets, \ac{cln_c36} shows very close performance with a minor performance gap of $0.4$\% DSC and $0.03$mm ASD. Although a slightly increased performance variation between \ac{cln_c36} and \ac{cln_u36} is observed in head \& neck ($1.0$\% DSC gap) and the lesion subgroups ($0.20$mm ASD gap), the results for most subgroups are very close in both metrics, with DSC ranging from $0.02$\% to $0.3$\% and ASD from $0.01$mm to $0.06$mm. These results demonstrate that \ac{cln_c36} exhibits superior universal segmentation performance even in the challenging \ac{cs} setting, where datasets are sequentially accessible to the model (only one dataset is available in a learning step).  This is primarily due to the design of CL-Net framework that allows to retain the existing anatomical segmentation capability while simultaneously adding and training new decoders for previously unseen anatomy types or updating existing decoders for anatomy types present in earlier datasets using \ac{ema}-based fine-tuning scheme. Note that \ac{cln_c36} also markedly outperforms the ensemble of 36 nnUNet specialists \ac{nnu_e36} by $1.8$\% DSC improvement and $9.2$\%($0.12$mm) ASD error reduction.

\stepcounter{res_cnt}
% \subsection{\arabic{res_cnt}) The decoder pruning of \ac{clnet} effectively reduces the parameter size of decoder heads and limits the parameter increasing rate on new tasks.} 
\subsection{\arabic{res_cnt}) \ac{cln_u36} significantly reduces the model size and parameter growth rate compared to \ac{cln_u36_unprn}, while maintaining similar segmentation accuracy.} 
\begin{figure*}[htpb]
\renewcommand{\figurename}{Fig.}
    \centering
    \includegraphics[width=\textwidth]{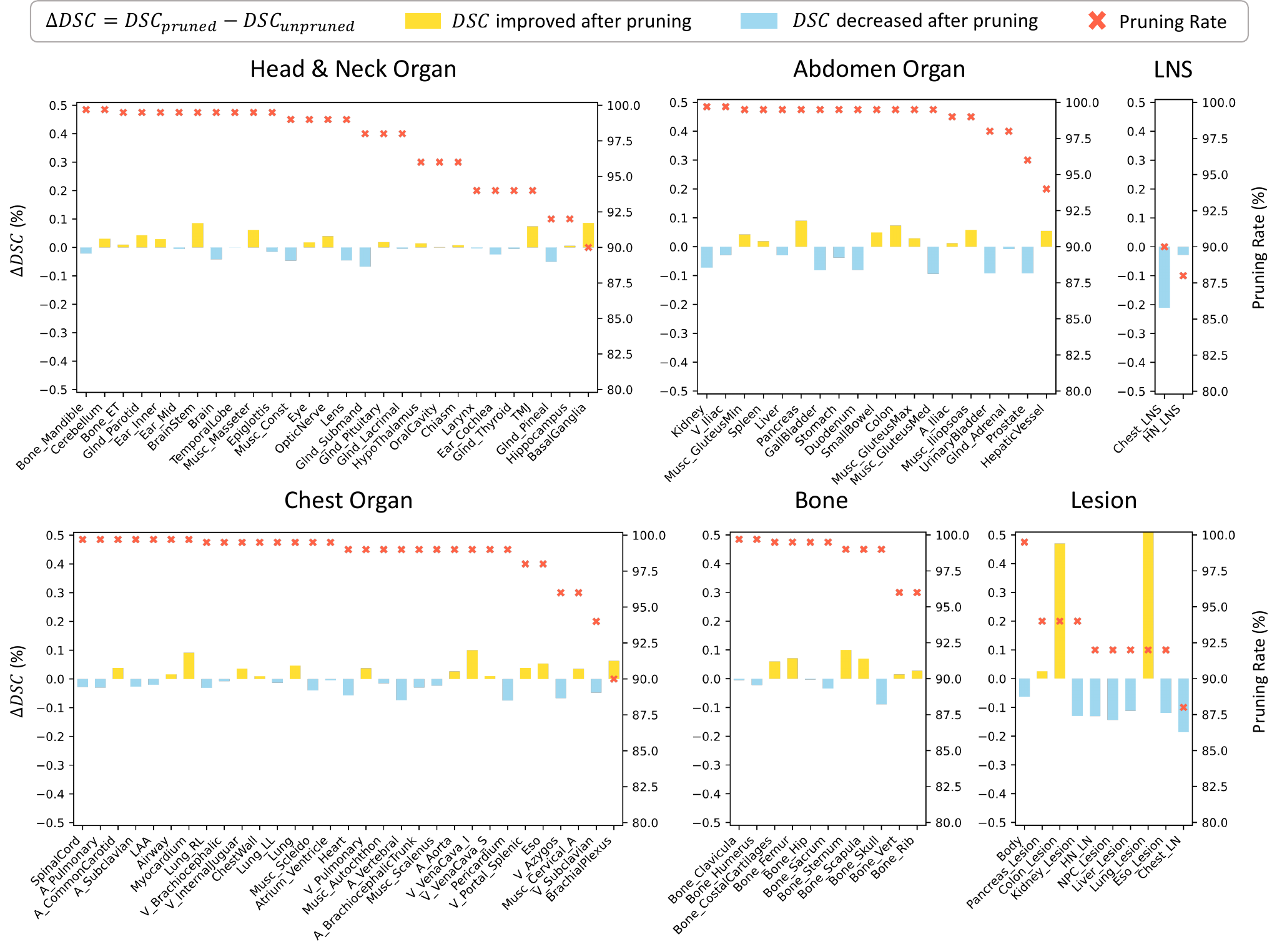}
    \caption{\textbar~\textbf{The decoder-wise pruning rates and DSC score differences of \ac{cln_u36} after pruning.} Evaluation of parameter pruning rates ($\mathcal{T}$, \%) and DSC score differences ($\Delta$DSC, \%) for \ac{decn} decoders between \ac{cln_u36_unprn} and \ac{cln_u36}. Decoders, except `Body', are grouped into Head \& Neck (28 decoders), Chest (31), Abdomen (19), Bone (11), LNS (2), and \Ac{gtv} (9). Yellow/blue bars (left axis) represent positive/negative $\Delta$DSC for each decoder post-pruning, while red cross markers (right axis) indicate decoder-wise $\mathcal{T}$. Results show minimal $\Delta$DSC (mostly within $\pm0.1\%$) and consistently high $\mathcal{T}$ (above 90\% for most decoders), highlighting the efficiency of pruning with negligible impact on segmentation performance. Detailed organ-wise metrics and decoder-wise pruning rates are provided in Supplementary Sec. \ref{sec:ablation_dec} and \ref{tab:ds36_hn}--\ref{tab:ds36_gtv}. }
    \label{fig:prune}
\end{figure*}

The performance and model parameters of CL-Net without decoder pruning (\ac{cln_u36_unprn}) are shown in Table~\ref{tab:main}. It can be seen that \ac{cln_u36_unprn} has a large model size of $1,612.1$MB, due to the incrementally expanded decoders that rapidly grow the model size. In comparison, with the optimized and pruned decoders, \ac{cln_u36} has only $55.5$M parameters, which is $3.4$\% of \ac{cln_u36_unprn} model size. As a result, the decoders in \ac{clnet} achieve a substantial mean pruning rate of $97.5$\%, indicating a very small parameter growth rate of $2.5$\% averaged across all decoders. The detailed pruning rates of each decoder are illustrated in Fig.~\ref{fig:prune}. The pruning rate for 99 out of 101 decoders in CL-Net is observed to exceed $90$\%, of which $80$ decoders' pruning rates further exceed $95$\% (some of the most pruned decoders even reach $99.7$\% pruning rates). Under significant pruning rates, \ac{cln_u36} still achieves approximately the same performance as the unpruned counterpart \ac{cln_u36_unprn}. For example, \ac{cln_u36_unprn} and \ac{cln_u36} exhibit the same \ac{orgn} organ-wise mean DSC of $86.1$\%. Specifically, $47.5$\% decoders in \ac{cln_u36} have improved DSC by an average of $0.07$\% after pruning, while $52.5$\% of its decoders have marginally decreased an average of $0.05$\% DSC.  These results demonstrate the effectiveness and robustness of our decoder pruning in CL-Net, which balances the model scale and performance accuracy.

\subsection{2.3 Comparative Internal Results on Five Representative Public Datasets} %Validation Results on Five Datasets
\label{sec:css_trial}
Trained and validated on five internal representative public datasets (total 137 anatomy types), we compare CL-Net to the ensemble of five nnUNet specialists (\ac{nnu_e5}) the leading multi-dataset approach, MultiTalent~\cite{ulrich2023multitalent}. The performance of the latest SAM-style medical foundation models~\cite{Zhao2023one,he2024vista3d,zhao2024foundation} is also compared. For comparison with SAM-style models, we only report the quantitative performance of organs that appeared in their training data (unseen organs are not tested in SAM-style models). For example, SAT-Pro~\cite{Zhao2023one} can segment all 137 anatomies in these five datasets, while VISTA3D~\cite{he2024vista3d} can segment 115 anatomies. In contrast, BiomedParse~\cite{zhao2024foundation} mainly segments abdomen organs; therefore, we only compare its performance on the \ac{flare}~\cite{ma2023unleashing} dataset. Finally, as a continual segmentation model, \ac{cln_c5} is further compared to three leading and popular continual segmentation methods, MiB~\cite{cermelli2020modeling}, PLOP~\cite{douillard2021plop}, and CSCLIP~\cite{zhang2023continual} under continual training configuration.

%has used all five public datasets (except StructSeg19~\cite{StructSeg2019}) in its model development, hence, it can segment all anatomy types in the five datasets, which are compare to the CL-Net. In contrast, for the latest foundation model BiomedParse~\cite{zhao2024foundation}, we only report its performance on \ac{flare}~\cite{ma2023unleashing}, because BiomedParse cannot semantically segment most of anatomy types in these five datasets. 

%% reset results subsection counter
% \newcounter{res_cnt}
\setcounter{res_cnt}{3} 
% \arabic{res_cnt}
% \stepcounter{res_cnt}

%% PL U5 dataset-wise statistics comparison (multitalent, sat-pro)
% \usepackage{booktabs}
% \usepackage{multirow}
% \usepackage{graphicx}
\begin{table}[!ht]
\centering
\caption{\textbar~\textbf{Performance comparison of \ac{pl} models on five representative public datasets.} The mean DSC (\%) and ASD (mm) of \ac{cln_u5}, \ac{nnu_e5}, MultiTalent~\cite{ulrich2023multitalent}, SAT-pro~\cite{Zhao2023one}, and VISTA3D~\cite{he2024vista3d} are evaluated on five representative public datasets. The best results are highlighted in bold. Note that, the officially released SAT-pro inference model removes spacing information from the original image headers, making ASD calculation infeasible. Additionally, VISTA3D does not support segmentation of head-neck organs or kidney tumors, precluding its evaluation on \ac{structseg} and \ac{kits}. } % Unavailable results are marked as gray. 
\label{tab:pl_res_avg}
\resizebox{1.0\textwidth}{!}{%
\begin{tabular}{@{}clccccc|c@{}}
\toprule
\textbf{Metric} &
  \textbf{Method} &
  \textbf{\ac{totalseg}} &
  \textbf{\ac{structseg}} &
  \textbf{\ac{flare}} &
  \textbf{\ac{segthor}} &
  \textbf{\ac{kits}} &
  \textbf{Mean} \\ \midrule
 &
  \ac{nnu_e5} &
  94.1 &
  86.4 &
  89.9 &
  92.6 &
  86.7 &
  92.5 \\
 &
  MultiTalent~\cite{ulrich2023multitalent} &
  92.5 &
  79.0 &
  89.3 &
  90.7 &
  79.6 &
  90.2 \\
 &
  SAT-Pro~\cite{Zhao2023one} &
  86.4 &
  62.9 &
  85.1 &
  90.3 &
  80.0 &
  83.0 \\
 &
  VISTA3D~\cite{he2024vista3d} & % \cellcolor[HTML]{D8D8D8}
  91.6 &
  --- &
  85.9 &
  90.4 &
  --- &
  --- \\
\multirow{-5}{*}{DSC (\%)$\uparrow$} &
  \ac{cln_u5} &
  \textbf{94.3} &
  \textbf{87.0} &
  \textbf{90.7} &
  \textbf{93.7} &
  \textbf{87.2} &
  \textbf{92.9} \\ \midrule
 &
  \ac{nnu_e5} &
  0.97 &
  0.29 &
  1.14 &
  0.34 &
  1.18 &
  0.87 \\
 &
  MultiTalent~\cite{ulrich2023multitalent} &
  \textbf{0.81} &
  1.95 &
  1.83 &
  1.44 &
  9.94 &
  1.13 \\
 &
  SAT-Pro~\cite{Zhao2023one} &
  --- &
  --- &
  --- &
  --- &
  --- &
  --- \\
 &
  VISTA3D~\cite{he2024vista3d} &
  3.41 &
  --- &
  1.95 &
  1.52 &
  --- &
  --- \\
\multirow{-5}{*}{ASD (mm)$\downarrow$} &
  \ac{cln_u5} &
  0.85 &
  \textbf{0.27} &
  \textbf{1.04} &
  \textbf{0.32} &
  \textbf{1.03} &
  \textbf{0.77} \\ \bottomrule
\end{tabular}%
}

\end{table}
%% PL U5 organ-wise comparison on FLARE (nnu_e5, cln_u5, biomedparse)
% \usepackage{booktabs}
% \usepackage{multirow}
% \usepackage{graphicx}
\begin{table}[!ht]
\centering
\caption{\textbar~\textbf{Performance comparison of \ac{cln_u5}, \ac{nnu_e5}~\cite{isensee2021nnu}, and BiomedParse~\cite{zhao2024foundation} on 13 organs of \ac{flare}.} \ac{cln_u5} shows the best DSC and ASD performance. 
$\dagger$Note: ``DSC* (2D organ-slice)'' is the mean DSC computed only on 2D CT slices that contain the foreground target organs (slices are pre-selected using ground truth annotations), which is used in BiomedParse for evaluation. In this table, values for this 2D DSC metric (shown in grey) are directly extracted from BiomedParse paper for reference only and are not intended for comparison. }
\label{tab:bmp_compare}
\resizebox{\textwidth}{!}{%
\begin{tabular}{@{}clccccccc@{}}
\toprule
\textbf{Metric} &
  \textbf{Method} &
  \textbf{Liver} &
  \textbf{Kidney\_R} &
  \textbf{Kidney\_L} &
  \textbf{Spleen} &
  \textbf{Pancreas} &
  \textbf{A\_Aorta} &
  \textbf{V\_VenaCava\_I} \\ \midrule
DSC* (2D organ-slice) &
  BiomedParse$\dagger$~\cite{Zhao2024BiomedParse} &
  \cellcolor[HTML]{E8E6E7}96.6 &
  \cellcolor[HTML]{E8E6E7}95.7 &
  \cellcolor[HTML]{E8E6E7}96.3 &
  \cellcolor[HTML]{E8E6E7}96.1 &
  \cellcolor[HTML]{E8E6E7}86.7 &
  \cellcolor[HTML]{E8E6E7}94.8 &
  \cellcolor[HTML]{E8E6E7}89.6 \\ \midrule
 &
  BiomedParse~\cite{Zhao2024BiomedParse} &
  24.1$\pm$19.8 &
  53.9$\pm$7.5 &
  64.9$\pm$3.6 &
  62.3$\pm$6.1 &
  2.4$\pm$0.4 &
  27.2$\pm$0.5 &
  11.6$\pm$1.5 \\
 &
  \ac{nnu_e5} &
  98.1$\pm$14.5 &
  92.4$\pm$16.9 &
  92.8$\pm$16.9 &
  97.9$\pm$15.6 &
  91.6$\pm$17.2 &
  96.7$\pm$16.5 &
  88.8$\pm$10.1 \\
\multirow{-3}{*}{DSC$\uparrow$} &
  \ac{cln_u5} &
  98.3$\pm$5.8 &
  94.3$\pm$7.1 &
  92.9$\pm$6.7 &
  98.0$\pm$6.6 &
  92.4$\pm$7.0 &
  97.1$\pm$16.5 &
  88.6$\pm$9.9 \\ \midrule
 &
  BiomedParse~\cite{Zhao2024BiomedParse} &
  48.34$\pm$26.44 &
  57.48$\pm$14.48 &
  69.38$\pm$10.18 &
  51.41$\pm$8.09 &
  60.92$\pm$0.61 &
  64.15$\pm$3.13 &
  70.89$\pm$8.90 \\
 &
  \ac{nnu_e5} &
  0.83$\pm$0.98 &
  1.81$\pm$0.74 &
  1.27$\pm$0.74 &
  0.30$\pm$0.37 &
  0.80$\pm$0.69 &
  0.21$\pm$0.95 &
  1.08$\pm$1.03 \\
\multirow{-3}{*}{ASD$\downarrow$} &
  \ac{cln_u5} &
  0.81$\pm$1.00 &
  1.44$\pm$0.76 &
  1.18$\pm$0.69 &
  0.19$\pm$0.37 &
  0.72$\pm$0.38 &
  0.18$\pm$0.83 &
  1.12$\pm$0.98 \\ \midrule
\textbf{Metric} &
  \textbf{Method} &
  \textbf{Glnd\_Adrenal\_R} &
  \textbf{Glnd\_Adrenal\_L} &
  \textbf{Gallbladder} &
  \textbf{Eso} &
  \textbf{Stomach} &
  \multicolumn{1}{c|}{\textbf{Duodenum}} &
  \textbf{Mean} \\ \midrule
DSC* (2D organ-slice) &
  BiomedParse$\dagger$~\cite{Zhao2024BiomedParse} &
  \cellcolor[HTML]{E8E6E7}76.3 &
  \cellcolor[HTML]{E8E6E7}79.0 &
  \cellcolor[HTML]{E8E6E7}86.8 &
  \cellcolor[HTML]{E8E6E7}85.0 &
  \cellcolor[HTML]{E8E6E7}92.8 &
  \multicolumn{1}{c|}{\cellcolor[HTML]{E8E6E7}81.1} &
  \cellcolor[HTML]{E8E6E7}89.0 \\ \midrule
 &
  BiomedParse~\cite{Zhao2024BiomedParse} &
  0.0$\pm$0.0 &
  0.0$\pm$0.0 &
  1.0$\pm$1.0 &
  59.4$\pm$13.7 &
  0.8$\pm$0.0 &
  \multicolumn{1}{c|}{1.0$\pm$1.0} &
  23.7 \\
 &
  \ac{nnu_e5} &
  85.9$\pm$12.3 &
  88.0$\pm$13.3 &
  80.2$\pm$24.7 &
  84.4$\pm$3.1 &
  91.7$\pm$16.4 &
  \multicolumn{1}{c|}{79.8$\pm$25.9} &
  89.9 \\
\multirow{-3}{*}{DSC$\uparrow$} &
  \ac{cln_u5} &
  87.9$\pm$13.2 &
  87.7$\pm$12.5 &
  83.1$\pm$8.7 &
  86.0$\pm$3.0 &
  92.4$\pm$6.5 &
  \multicolumn{1}{c|}{81.0$\pm$16.1} &
  \textbf{90.7} \\ \midrule
 &
  BiomedParse~\cite{Zhao2024BiomedParse} &
  79.70$\pm$2.33 &
  102.87$\pm$16.02 &
  134.81$\pm$19.78 &
  34.80$\pm$28.08 &
  59.16$\pm$2.60 &
  \multicolumn{1}{c|}{58.08$\pm$7.51} &
  68.62 \\
 &
  \ac{nnu_e5} &
  0.39$\pm$0.37 &
  0.26$\pm$1.16 &
  2.16$\pm$0.83 &
  1.70$\pm$0.15 &
  1.24$\pm$0.67 &
  \multicolumn{1}{c|}{2.71$\pm$1.29} &
  1.14 \\
\multirow{-3}{*}{ASD$\downarrow$} &
  \ac{cln_u5} &
  0.35$\pm$0.39 &
  0.21$\pm$0.38 &
  2.12$\pm$0.83 &
  1.61$\pm$0.14 &
  1.17$\pm$0.69 &
  \multicolumn{1}{c|}{2.39$\pm$1.18} &
  \textbf{1.04} \\ \bottomrule
\end{tabular}%
}

\end{table}

%% Compare with MultiTalent; SAM-based method 
%% compare cln_u5 w. MultiTalent; SAT (clip-based) & BiomedParse (SAM-based)
\stepcounter{res_cnt}
\subsection{\arabic{res_cnt}) \ac{clnet} substantially outperforms the leading multi-dataset segmentation approach Multi-Talent~\cite{ulrich2023multitalent}.}
%and the ensemble of five dataset-specific trained nnUNets~\cite{isensee2021nnu}
%% Multi-Talent
As shown in Table~\ref{tab:pl_res_avg}, averaged across the five datasets, \ac{cln_u5} achieves a DSC of $92.9$\% and an ASD of $0.77$mm, substantially outperforming MultiTalent~\cite{ulrich2023multitalent} by $2.7$\% absolute DSC improvement and $31.9$\% relative ASD error reduction. Among individual datasets, the multi-talent approach yields much inferior accuracy in segmenting head \& neck organs ($-8.0$\% DSC) and kidney lesions ($-7.6$\% DSC).  Because MultiTalent simply expands nnUNet's final convolutional layer (output) to generate output probabilities for existing anatomies in all datasets with the sigmoid activation function (relying only on the last convolutional layer); hence, encountering difficulty to segment small and hard anatomies (like lesions or many small organs in head \& neck region). 

%marginally outperforming \ac{nnu_e5} by +0.35\% in DSC and -0.10mm in ASD. On individual dataset, \ac{cln_u5} consistently delivers superior DSC scores compared to both methods, with the largest DSC improvement of 1.13\% over \ac{nnu_e5} on \ac{segthor} and 7.99\% over MultiTalent on \ac{structseg}.Two qualitative examples are shown in Fig.~\ref{fig:quali-v1}. 

\stepcounter{res_cnt}
\subsection{\arabic{res_cnt}) \ac{clnet} significantly outperforms SAM-style medical segmentation foundation models.} 
%under \ac{pl} setting.

%% SAT, BiomedParse
The comparison results on five public datasets between CL-Net and the latest 3D SAM-style text prompt foundation model SAT-Pro~\cite{Zhao2023one} are shown in Table~\ref{tab:pl_res_avg}. Although SAT-Pro, the strongest SAT model~\cite{Zhao2023one}, is trained using more than 28 CT datasets (including the training set of these five public datasets), it has a large performance gap with Multi-Talent, nnUNet ensemble and our CL-Net (with the largest margin of $-9.9$\% DSC). In individual dataset, SAT-Pro shows the largest performance decrease in \ac{structseg} (a significant drop in DSC $24.1$\% compared to our \ac{clnet}. Although the PDDCA~\cite{raudaschl2017evaluation} and SegRap2023~\cite{luo2023segrap2023} datasets already include the organs in \ac{structseg}, and have been used in SAT-Pro training data, SAT-Pro still lacks the generalizability to accurately segment head \& neck organs (normally harder than chest and abdomen) of a different cohort. Except for \ac{structseg}, the other four test datasets have all appeared in SAT-Pro’s training set. Even under this condition, SAT-Pro consistently and considerably produces inferior performance in other datasets compared to our CL-Net, e.g., $-7.9$\% DSC in \ac{totalseg}, $-5.6$\% DSC in \ac{flare}, $-3.4$\% DSC in SegThor and $-7.2$\% DSC in KiTS21. 
%Although claimed as the foundation model, the poor performance of SAT-Pro on head \& neck organs indicates its limited generalizability. This is because organ types in \ac{structseg} already exist in several training datasets of SAT-Pro, such as PDDCA~\cite{raudaschl2017evaluation} and SegRap2023~\cite{luo2023segrap2023}.

%To assess the advantages of our method over emerging trends in text-driven segmentation and interactive segmentation using large foundation models, 
% text-driven segmentation method based on multi-modality image-text pretrained model such as CLIP, and interactive segmentation method based on large foundation model such as SAM, 
%SAT~\cite{Zhao2023one} is a 3D SAM-type segmentation model and BiomedParse~\cite{Zhao2024BiomedParse} is a 2D SAM-type multi-modality large foundation model. 

%% SAT
%Averaged across five datasets, \ac{cln_u5} significantly outperforms SAT~\cite{Zhao2023one} (a 3D SAM-type model using text prompts), achieving a 9.88\% higher DSC score. 

%% BiomedParse
Compared to BiomedParse~\cite{zhao2024foundation}, the latest 2D SAM-type medical segmentation model, we restrict the comparison to 13 organs in \ac{flare} (since its officially released model can only segment a limited number of abdomen organs). BiomedParse reports 2D slice-level DSC scores on a different test set of \ac{flare} (copied in Table~\ref{tab:bmp_compare}, denoted as BiomedParse$\dagger$), which is different from the common 3D volume-level DSC. Because in the BiomedParse$\dagger$ evaluation, CT slices to calculate accuracy have been pre-selected using the ground-truth 3D mask of the specific organ (2D evaluation under unrealistic assumption). However, even under this substantially easier evaluation condition, BiomedParse$\dagger$ has a mean DSC of $89.6$\%, which is $0.9$\% and $1.7$\% lower than the nnUNet ensemble \ac{nnu_e5} and our CL-Net \ac{cln_u5} (evaluated in a strict 3D volumetric manner without pre-selected slices). To ensure a fair comparison, we also apply the official BiomedParse model to inference on all CT slices of \ac{flare} test data using text prompts and calculate its final 3D metrics. Surprisingly, the results reveal that BiomedParse performs poorly in 3D organ segmentation, with only $23.7$\% DSC (Table~\ref{tab:bmp_compare}) reported. Qualitative examples (Fig.~\ref{fig:quali-v4}) illustrate that the substantially low DSC scores and high ASD errors of BiomedParse are due to the large number of false positive predictions outside the target organ region. This is probably due to the intrinsic limitations of its 2D model architecture, which fails to associate the text prompts with the corresponding organs in the 3D space.
%% conclusion
%These comparison results highlight the strong and superior performance of our \ac{clnet} for whole-body anatomy segmentation. 

%However, this metric is not directly comparable to the common 3D volume-level DSC because CT slices, which only contain a specific organ, have been pre-selected using the organ's ground-truth 3D mask for evaluation.   

%that \ac{cln_u5} dramatically outperforms BiomedParse, with a 67.0\% improvement in DSC and a -67.58mm reduction in ASD, averaged on \ac{flare} (Table~\ref{tab:bmp_compare}). 

%As a continual learning model, CL-Net is also compared to three leading continual segmentation methods, MiB~\cite{cermelli2020modeling}, PLOP~\cite{douillard2021plop}, and CSCLIP~\cite{zhang2023continual} under CSS configuration. 

%% CSS learning, no forgetting

%---------- Fig. Forgeting Curve ------------%
\begin{figure*}[htpb]
\renewcommand{\figurename}{Fig.}
    \centering
    \includegraphics[width=\textwidth]{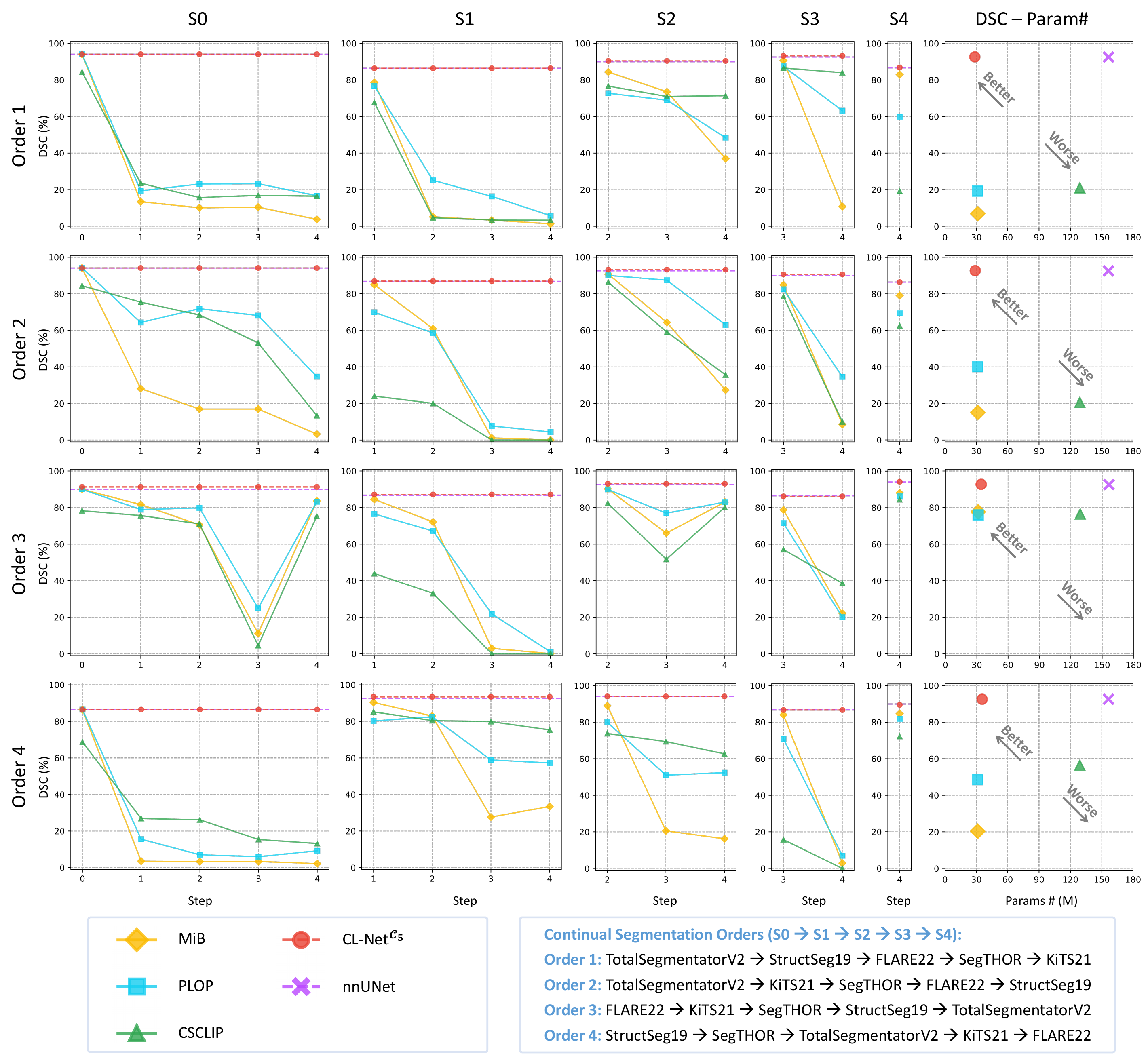}
    \caption{\textbar~\textbf{Comparison of \ac{cln_c5} with popular \ac{cs} methods.} Step-wise segmentation performance of \ac{cln_c5}, MiB, PLOP, CSCLIP, and nnUNet (upper bound) is evaluated across four \ac{cs} orders (each row for each order) on five public datasets covering various body parts. Columns S0-S4 show the forgetting curves for each method in each dataset and order. The column `DSC-Param\#' compares mean DSC scores against the parameter sizes of final models for all methods in each order. \Ac{cs} orders of the five datasets are detailed at the bottom. \ac{cln_c5} demonstrates no forgetting of previously learned tasks during continual learning across all orders and achieves the highest mean DSC scores with much smaller model size comparable to the ensemble of nnUNet. The detailed numerical metrics are provided in Table \ref{tab:css_res_avg}, \ref{tab:css_comp_res}, and \ref{tab:css_totalseg_organ}--\ref{tab:css_kits}. }
    \label{fig:css}
\end{figure*}

\stepcounter{res_cnt}
\subsection{\arabic{res_cnt}) As a \ac{cs} method, \ac{cln_c5} achieves high accuracy (no forgetting) and excellent scalability, significantly surpassing other leading continual segmentation approaches.} 
% results
The forgetting curves of \ac{cln_c5} and other three \ac{cs} methods are summarized in Fig.~\ref{fig:css}. It is observed that \ac{cln_c5}'s mean DSC scores at consecutive continual learning steps remain stable with high accuracy on every dataset across all four \ac{cs} orders. In contrast, the forgetting curves of other \ac{cs} methods (MiB~\cite{cermelli2020modeling}, PLOP~\cite{douillard2021plop}, and CSCLIP~\cite{zhang2023continual}) exhibit continuous declines or extensive fluctuations, leading to much lower \ac{dsc} scores in the final step. When evaluating the mean DSC of the final models in all datasets, \ac{cln_c5} achieves superior segmentation accuracy and model scalability, significantly outperforming other methods in all \ac{cs} orders (Fig. \ref{fig:css}, column ``DSC-Param\#''). Specifically, \ac{cln_c5} has a final mean DSC of $92.6$\% and a mean ASD of $0.85$mm, averaged over all \ac{cs} orders (Table~\ref{tab:css_res_avg} and Table~\ref{tab:css_comp_res}). In comparison, MiB, PLOP, and CSCLIP all produce catastrophic forgetting, which results in only $30.0$\%, $46.0$\%, and $43.6$\% mean DSC scores, respectively. Meanwhile, the model size of \ac{cln_c5} is comparable to MiB and PLOP and notable smaller than CSCLIP. %Detailed organ-wise performance of \ac{cln_c5} is provided in Supplementary Sec. \nameref{sec:supp_table}. 
% discussion
Model stability and plasticity, as two essential characteristics in continual learning, are often considered trade-off roles: a stable model resists catastrophic forgetting, while a plastic model adapts effectively to new tasks. However, the above results highlight that our \ac{clnet} achieves both remarkable model stability (nonforgetting) and model plasticity (scalability) in continual learning. \ac{clnet} avoids the knowledge forgetting by freezing the encoder and sequentially adding and training new decoders for new classes or fine-tuning existing decoders for previously existing classes using \ac{ema}-based decoder updating scheme. This framework demonstrates strong segmentation performance since \ac{clnet} matches the performance upper bound of \ac{nnu_e5} separately trained on each dataset. Meanwhile, utilizing the \ac{lth}-based pruning strategy~\cite{frankle2018lottery,frankle2019stabilizing} for the decoders, \ac{clnet} maintains an overall small model size. More detailed results of \ac{cln_c5} on four \ac{cs} orders are provided in the Supplementary Sec. \ref{sec:supp_val_res}.

\begin{table}[!t]
\centering
\caption{\textbar~\textbf{External Testing: Performance comparison of SAT-pro, VISTA3D, MultiTalent, \ac{nnu_e36}, \ac{cln_c36} and \ac{cln_u36}}. The mean DSC (\%) and ASD (mm) are evaluated on seven external testing datasets: NPC\_SMU (head neck), LungCancer\_HHA-FU 24 (chest), BTCV~\cite{landman2015miccai} (abdomen), NPC\_CGMH  (NPC lesion), EsoCancer\_CGMH (eso lesion), LiverCancer\_CGMH (liver lesion), and KidneyCancer\_SHCH (kidney lesion).  We evaluated the average segmentation performance of common chest structures (marked with an asterisk). Note that the SAT-pro inference framework removes image metadata headers during inference, making it impossible to calculate ASD. The best results are highlighted in bold. }
\label{tab:pl_ex_res_main}
\resizebox{\textwidth}{!}{%
\begin{tabular}{@{}clcrccccc|c@{}}
\toprule
\textbf{Metric} &
  \textbf{Method} &
  \textbf{Head \& Neck (12) } &
  \textbf{Chest (14 / 8*)} &
  \textbf{Abdomen (13)} &
  \textbf{NPC\_\Ac{gtv}} &
  \textbf{Eso\_\Ac{gtv}} &
  \textbf{Liver\_\Ac{gtv}} &
  \textbf{Kidney\_\Ac{gtv}} &
  \textbf{ Mean} \\ \midrule
 &
  SAT-Pro &
  46.1 &
  --- / 73.9* &
  --- &
  64.9 &
  --- &
  33.7 &
 --- & 
  --- \\
&
  Vista3D &
  --- &
  --- / 82.9* &
  85.2 &
  --- &
  --- &
  --- &
  --- &
  --- \\
 &
  MultiTalent &
  73.1 &
  --- / 80.2* &
  84.2 &
  --- &
  --- &
  --- &
  81.3 &
  --- \\
 &
  \ac{nnu_e36} &
  75.4 &
  85.5 / 88.5* &
  84.8 &
  70.4 &
   75.0 &
  69.9 &
  83.6 &
  81.5\\
 &
  \ac{cln_c36} &
  76.7 &
  86.8 / 89.7* &
  85.0 &
  \textbf{70.5} &
    74.9 &
  69.9 &
  86.6 &
  82.4 \\
\multirow{-6}{*}{DSC (\%)$\uparrow$} &
  \ac{cln_u36} &
  \textbf{77.1} &
  \textbf{87.4 / 90.2*} &
  \textbf{85.4} &
  70.3 &
    \textbf{75.1} &
  \textbf{70.4} &
  \textbf{86.7} &
  \textbf{82.8 } \\ \midrule
 &
 SAT-Pro &
  --- &
  --- / --- &
  --- &
  --- &
  --- &
  --- &
  --- & 
  --- \\ % \cellcolor[HTML]{D8D8D8}
 &
  Vista3D &
  --- &
  --- / 1.55* &
  0.85 &
  --- &
  --- &
  --- &
  --- & 
  ---  \\
 &
 MultiTalent &
  0.96 &
  --- / 37.19* &
  1.39 &
  --- &
  --- &
  --- &
  7.22 & 
  ---  \\
 &
  \ac{nnu_e36} &
  0.94 &
  1.38 / 1.21* &
  1.33 &
  2.02 &
  8.85 &
  8.41 &
  2.20 &
  1.64  \\
 &
  \ac{cln_c36} &
  0.76 &
  1.18 / 1.07* &
  0.85 &
  2.01 &
    8.87 &
  8.43 &
  1.01 &
  1.48 \\

\multirow{-6}{*}{ASD (mm)$\downarrow$} &
  \ac{cln_u36} &
  \textbf{0.70} &
  \textbf{1.08 / 0.99*} &
  \textbf{0.78} &
  \textbf{1.98} &
    \textbf{8.67} &
  \textbf{7.53} &
  \textbf{1.00} &
  \textbf{1.40} \\ \bottomrule
\end{tabular}%
}

\end{table}

%external results
\subsection{2.4 External Testing Results on Seven Datasets} 
\label{sec:external_results}
The external testing results of \ac{cln_u36} and other comparing methods are presented in Table~\ref{tab:pl_ex_res_main}. Several conclusions can be drawn. First, SAM-style medical foundation models either segment a limited number of anatomy types (e.g., VISTA3D~\cite{he2024vista3d} can only segment 21 chest and abdomen organs among the 43 external anatomies), or exhibit lower segmentation accuracy in external testing (evidently, SAT-Pro~\cite{Zhao2023one} has $46.1$\% and $73.9$\% DSC when segmenting organs in head \& neck and chest regions, and $64.9$\% and $33.7$\% DSC when segmenting head \& neck and liver lesions, showing larger performance gap with \ac{clnet} when compared to that in the internal testing (Table~\ref{tab:pl_res_avg}). This indicates that existing medical segmentation foundation models struggle to comprehensively and accurately segment common anatomies in CT imaging, which is the most widely used diagnostic modality. Second, Multi-Talent~\cite{ulrich2023multitalent} shows a notable performance drop in external tests compared to our CL-Net. Specifically, Multi-Talent~\cite{ulrich2023multitalent} has a mean DSC gap of $2.7$\% as compared to our \ac{cln_u5} in internal testing (Table~\ref{tab:pl_res_avg}), while the performance gap increases to $4.4$\% DSC in external testing. Finally, our CL-Net performs consistently on par or higher than the ensemble of \ac{dsn}-nnUNet specialists in external testing with a mean improvement of $1.3$\% DSC and a reduction of $14.6$\% ASD error. Notably, our \ac{cln_u36} outperforms \ac{nnu_e36} in kidney lesion segmentation by markedly increasing the DSC by $3.1$\% and reducing the error of half the ASD. Since our external kidney lesion dataset contains a large number of patients (978 testing cases) with various severity of the disease (both very small and large kidney lesions), it is observed that \ac{nnu_e36} sometimes has false positive segmentation outside the kidney region or misses small lesions. In contrast, CL-Net utilizes the features of the kidney organ decoder to support the segmentation of kidney lesion (\ac{fls}), reducing the difficulty in segmenting the kidney lesion. 
%(\textcolor{red}{Fig. xx})
%although Multi-Talent~\cite{ulrich2023multitalent} achieves relative high performance in internal testing with a mean DSC of 2.7\% gap as compared to our \ac{cln_u5} (Table~\ref{tab:pl_res_avg}), it shows notable performance drop in external testing when compared to our \ac{cln_c36} and \ac{cln_u36} that are trained using full \ac{dsn} datasets (with a mean DSC difference of 4.4\% DSC). This demonstrates the importance of dataset size and variation for universal segmentation model development, as Multi-Talent has relative limited scalability to train on a large number of data to segment a large number of anatomies.

%to be modified...
\subsection{2.5 Ablation Results on Five Representative Public Datasets} %Validation Results on Five Datasets
\label{sec:css_ablation}
The ablation results evaluating the effectiveness of CL-Net key components are described in Supplementary Sec~\ref{sec:ablation_ge} and Sec~\ref{sec:ablation_dec}, including the general encoder training scheme, general encoder impacts on downstream tasks, the decoder \ac{fls} function, the \ac{ema}-based decoder fine-tuning, etc.

% \input{results/val_supp_css}

%% ablation studies on 5 datasets
% \subsection{Ablation Studies on \ac{clnet}.} 
% \input{results/val_supp_abl}

%% Universal/PartialLabel results
% \subsection{Partial Label \ac{pl} Segmentation Results on Five Datasets.} 
% \input{results/val_supp_pl}

%%------ Discussion ---------%%
\section{Discussion}
\label{sec:discuss}
%what are the requirements in medical universal segmentaion model
In the medical imaging domain, we refer to a successful universal segmentation model if it meets the following criteria:
\begin{itemize}

\item {\bf Comprehensive target coverage:} Trained on a large and diverse dataset, the model should accurately segment a wide range of clinically relevant targets (human anatomical structures and/or even sub-structures if clinically desirable), with a preference for finer-grained structures.

\item {\bf Consistently high performance:} The model consistently achieves an accuracy at least comparable to and often higher than the dataset-specific trained state-of-the-art methods/models under rigorous quantitative evaluation.

\item {\bf Adaptability:} The model can effectively segment new targets through flexible, adaptive fine-tuning or be efficiently updated to segment new targets using additional datasets, with minimal performance degradation for previously learned targets.

\end{itemize}

%summary
Under these considerations, we build \ac{clnet}, a universal 3D whole-body anatomy segmentation model in CT imaging. Adopting a novel perspective on continuous learning, \ac{clnet} integrates information from 20 public and 16 private  partially labeled datasets to segment \ac{orgn} fine-grained and clinically comprehensive whole-body anatomies, including 193 organs, 33 lymph node stations and 9 \ac{gtv} types. \ac{clnet} proves its clinical utility by providing high segmentation precision, consistently outperforming or matching the performance of an ensemble of 36 dataset-specific trained nnUNet~\cite{isensee2021nnu} specialists (DSC: $86.1$\% vs. $83.9$\%) $-$ the current ``true'' state-of-the-art  benchmark~\cite{isensee2024nnu} under rigorous training and quantitative comparison. In terms of inference time, \ac{clnet} segments all \ac{orgn} anatomies in a typical whole-body CT scan in approximately 7 minutes on a single NVIDIA V100 GPU, 2 minutes on 4 GPUs, and just 1 minute on 8 GPUs. In contrast, the ensemble of \ac{dsn}-nnUNet requires approximately 25 minutes to process a single whole-body CT scan.  Furthermore, \ac{clnet} can be efficiently updated to segment new anatomies or refine existing anatomies using new datasets without compromising previously acquired knowledge/capacity, overcoming a major limitation of current deep segmentation approaches. Compared to recent universal SAM-style medical models~\cite{ma2024segment,wu2023medical,chen2024ma,Zhao2023one,Ouyang2024towards,zhao2024foundation,he2024vista3d} that can only segment a limited subset of fine-grained CT anatomies, \ac{clnet} achieves substantially higher accuracy in both internal and external tests. \textcolor{black}{For example, \ac{clnet} outperforms the recent 3D SAT-Pro~\cite{Zhao2023one} by $9.9$\% DSC in internal testing and $24.7$\% DSC in external testing, despite SAT-Pro being developed with a significantly larger dataset of 22,186 CT scans—nearly double the size of our training data. \ac{clnet} also evidently surpasses the latest 2D BiomedParse~\cite{zhao2024foundation} in abdominal organ segmentation by $67.0$\% DSC under rigorous 3D quantitative evaluation~\cite{isensee2024nnu}.} 

\begin{figure*}[htpb]
\renewcommand{\figurename}{Fig.}
    \centering
    \includegraphics[width=\textwidth]{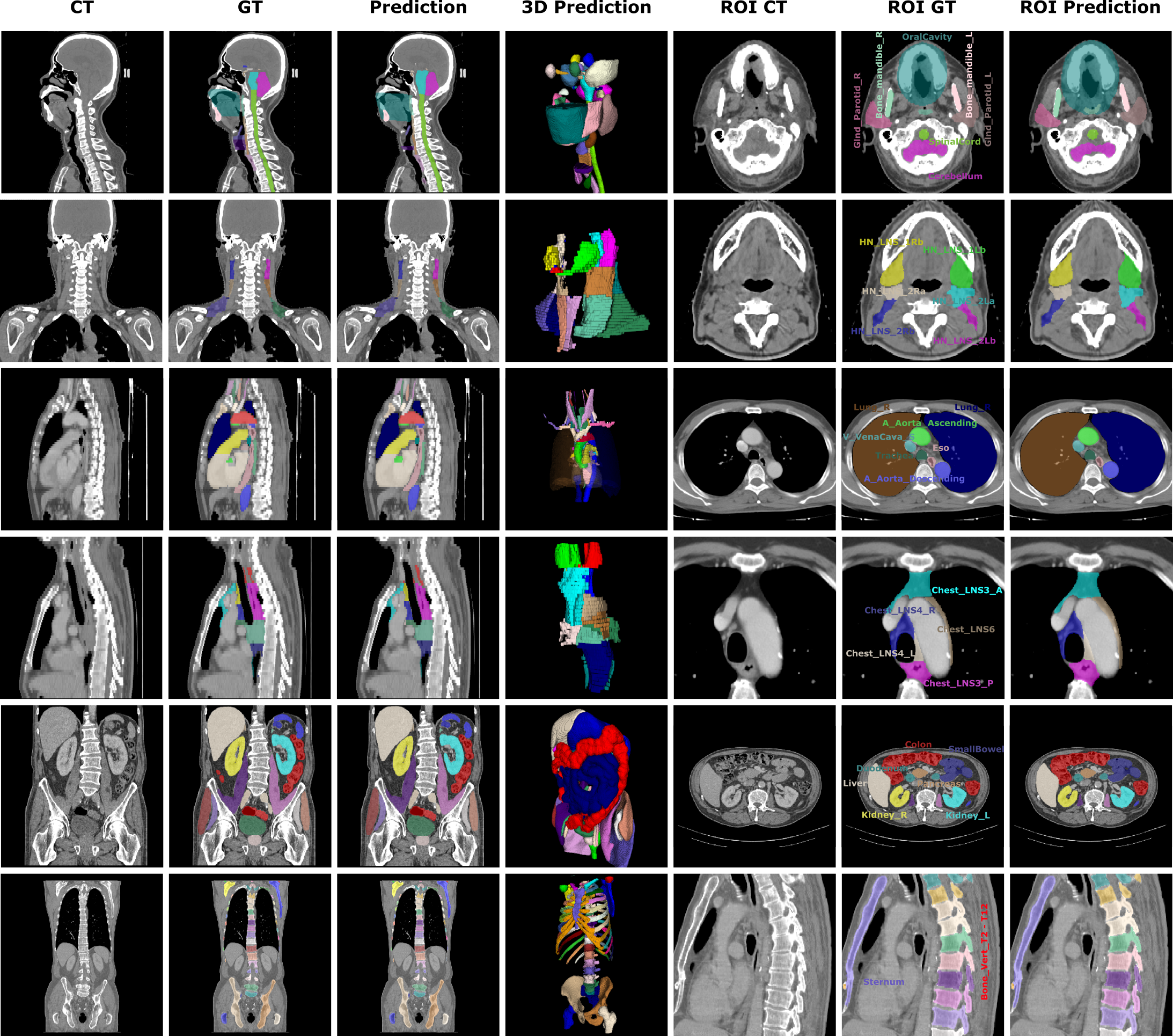}
    \caption{\textbf{\textbar~Qualitative visualization of \ac{clnet} whole-body anatomy structure segmentation.} The first two rows depict the segmentation results of head \& neck organs and corresponding lymph node stations, respectively. The third and fourth rows show the segmentation of chest organs and corresponding lymph node stations. The fifth row illustrates the segmentation of abdominal organs, while the final row shows bone segmentation. For better illustration, some organs (such as ChestWall) has been excluded, and certain sub-group organs, such as lung lobes and heart atria and ventricles, have been merged and rendered semi-transparent.}
    % 1) \ac{nnu_e36} ensemble induced inferior performance -- some dataset only contain partial IVC
    % 2) For better illustration, some sub-group organs (e.g., lung, heart) are merged and chest-wall is excluded.  
    \label{fig:quali-v1}
\end{figure*}

\begin{figure*}[htpb]
\renewcommand{\figurename}{Fig.}
    \centering
    \includegraphics[width=0.9\textwidth]{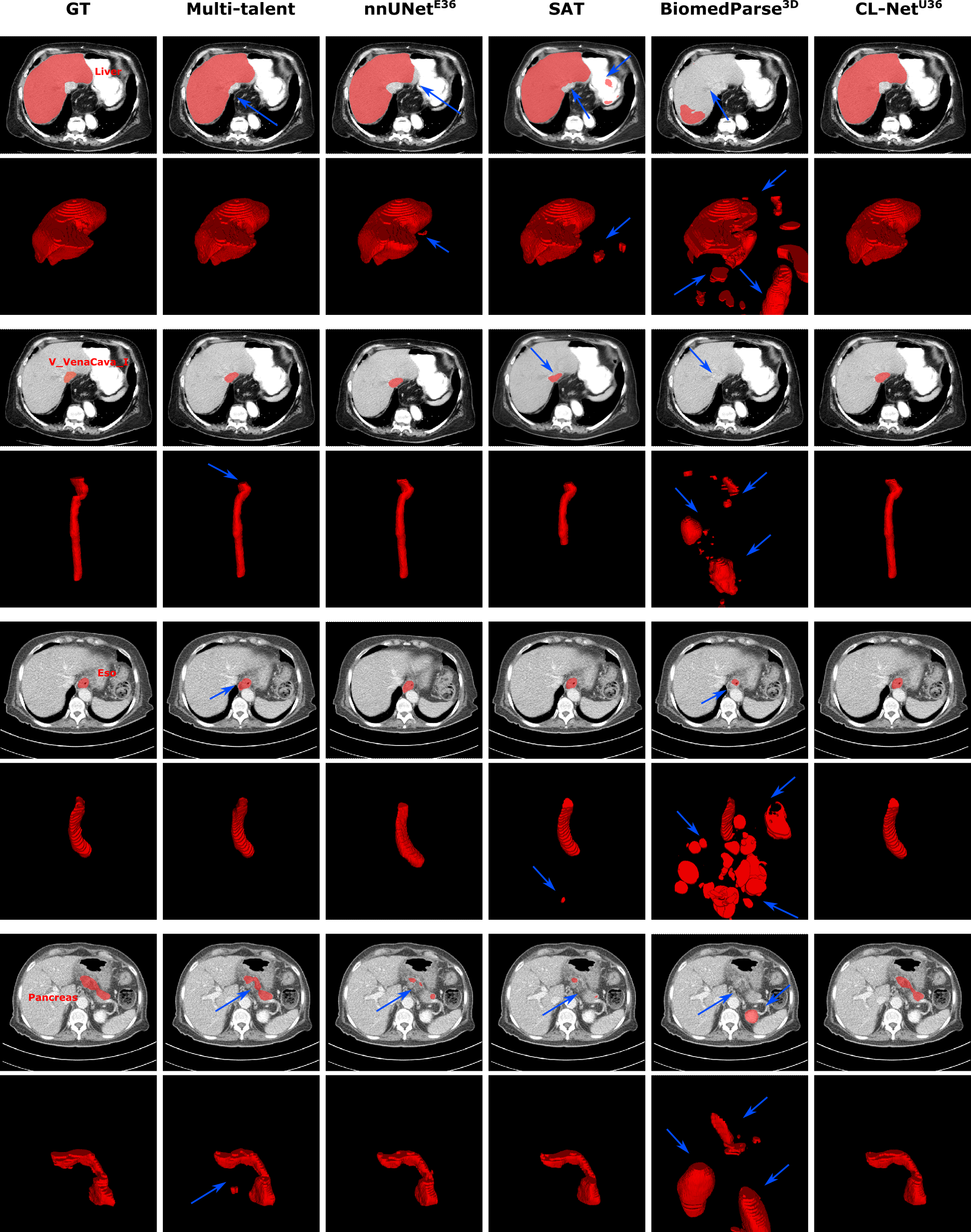}
    \caption{\textbf{\textbar~Examples of qualitative comparisons.} We compared the proposed \ac{clnet} to nnUNet ensemble~\cite{isensee2021nnu}, MultiTalent~\cite{ulrich2023multitalent}, SAT-Pro~\cite{Zhao2023one} and BiomedParse~\cite{zhao2024foundation} on predicting liver, inferior vena cava (IVC), esophagus, and pancreas.  SAT-Pro and BiomedParse are tested using their official published models. Blue arrow indicates the segmentation errors. }
    % 1) \ac{nnu_e36} ensemble induced inferior performance -- some dataset only contain partial IVC
    % 2) For better illustration, some sub-group organs (e.g., lung, heart) are merged and chest-wall is excluded.  
    \label{fig:quali-v4}
\end{figure*}

%Why SAM fail in medical segmentaiton, and what is the requirement in medical segmentation to be useful in clinical practice
Although SAM~\cite{kirillov2023segment} has excelled in natural image segmentation, demonstrating impressive performance in interactive (point or text promptable) and zero-shot segmentation, SAM-style models face significant challenges in the medical imaging domain. Our experiments show that the latest foundation models, SAT-Pro~\cite{Zhao2023one} and BiomedParse~\cite{zhao2024foundation}, perform substantially inferior to nnUNet~\cite{isensee2021nnu} and our \ac{clnet} in segmenting 3D anatomies in CT scans (Table~\ref{tab:bmp_compare}), rendering them impractical in clinical use. As a 2D SAM-based model, BiomedParse~\cite{zhao2024foundation} produces numerous false positives on 3D CT scans (Figure~\ref{fig:quali-v4}), highlighting its inability to effectively link text prompts to correct visual features under the 2D SAM architecture.  Although SAT-Pro~\cite{Zhao2023one} employs a 3D SAM-based network where the text prompt can roughly associate with 3D anatomical locations, its boundary accuracy is notably inferior to nnUNet or our \ac{clnet} (Figure~\ref{fig:quali-v1}). Unlike natural images in which the targets often have distinct visual features, medical imaging targets often exhibit low contrast with adjacent structures or the background. In such cases, Transformer architectures, known for their ability to model long-range spatial dependencies, offer limited advantages. Additionally, SAT-Pro's use of 3D Transformer dramatically increases the number of parameters, further complicating its network optimization. It is important to note that medical segmentation models are clinically useful only if they achieve high accuracy $-$ that ideally matches the variability between observers~\cite{ye2022comprehensive,ye2022multi} or higher, or at least comparable to (or better than) the nnUNet benchmark~\cite{isensee2021nnu}. Moreover, fully automated segmentation is preferred over interactive approaches, as inputting anatomy names or point prompts consumes physicians' valuable time (interrupting the regular clinical workflow), making such approaches less practical for routine clinical usage.

%Why our model is successful as unviersal CT segmentaion
The strong universal performance of our \ac{clnet} can be attributed to several key factors. 1) \textbf{Comprehensive and high-quality clinically curated datasets}: We compiled a large and diverse set of high-quality, partially annotated CT datasets, including 20 public and 16 private datasets. Public datasets were reviewed to ensure consistency in annotation standards and the absence of obvious errors. Private datasets, comprising primarily pathological patients, were carefully annotated and iterated by our clinical collaborators over the past decade (utilized in many previous peer-reviewed work~\cite{jin2019accurate,jin2019deep,guo2020organ,ye2022comprehensive,guo2021deepstationing,liu2021same,guo2022thoracic,guo2024towards,jin2021deeptarget,jin2022towards,wang2023anatomy,zhang2023deep,yan2022sam,yan2023liver,yan2023anatomy,yu2025effective,yu2024slice,ye2022multi,li2023lvit,Ji_2023_ICCV,tian2023same++,zhu2024low,li2024leveraging,ye2024development}). For instance, to parse the mediastinal lymph node stations~\cite{guo2021deepstationing}, 35 anatomies were labeled in the chest region, creating the most detailed chest organ dataset. In total, our dataset includes 13,952 CT scans (4,855 public and 9,097 private) covering \textit{a diverse range of vendors, different contrast phases and pathologies}, \textcolor{black}{which is also remarkedly larger than the developing dataset of the latest medical segmentation foundation models (e.g., 4,391 CT scans in BiomedParse~\cite{zhao2024foundation} and 4,633 CT scans in VISTA3D~\cite{he2024vista3d})}. 2) \textbf{A novel \ac{clnet} model network architecture}: the success of \ac{clnet} lies in its model design, where a trained then frozen general encoder paired with independently trainable decoders can extract sufficient features to accurately segment new or update existing organs while preserving existing segmentation capacity. This design naturally handles the issue of partial labeling and ensures that multiple datasets containing a specific organ can be used sequentially or simultaneously; and then integrated to optimize the associated decoder so that more data variations (from all healthy and pathological subjects) are captured by \ac{clnet}. Furthermore, by transforming the simultaneous segmentation of many anatomies from a large number of (often imbalanced) datasets into the sequential segmentation of one or a group of relevant organs, \ac{clnet} significantly reduces optimization complexity and accelerates convergence (e.g., compared to MultiTalent~\cite{ulrich2023multitalent} which takes more than 1,000 GPU hours to converge when training on 1,471 CT scans of 137 anatomies). 3) \textbf{Scalability through decoder pruning}: \ac{clnet} maintains only a small fraction of effective parameters for each decoder (normally 1$\sim$3\% model size of the default full-sized decoder) by optimizing each decoder with \ac{lth}-based pruning strategy~\cite{frankle2018lottery,frankle2019stabilizing}. These important factors collectively enable \ac{clnet} to deliver highly accurate, comprehensive, and scalable segmentation performance, which may serve as a universal model for clinical practice.
%Inspired by continual learning, we demonstrate that \ac{clnet} can serve as a 3D universal segmentation model in CT imaging. 

%\ac{clnet} as an efficient and effective multi-dataset method, a counterpart multi-dataset nNUNet method, support both sequential learning and simultanous learning
In a single dataset scenario, nnUNet~\cite{isensee2021nnu} achieves the state-of-the-art segmentation performance compared to the more recent Transformer-based~\cite{he2024vista3d,hatamizadeh2022unetr,cao2022swin} and Mamba-based approaches~\cite{ma2024u} when they are trained and evaluated rigorously without bias~\cite{isensee2024nnu,huang2023stu,zhu2024low}. Our \ac{clnet} inherits all the advantages of nnUNet (such as intensive data augmentation, auto-optimized patch spacing and size, empirically determined and auto-adaptive hyperparameters, etc.) and more importantly, has the ability to efficiently handle and fully utilize the synergies from a large number of partially labeled datasets no matter whether either sequential or simultaneous data accessibility is available. In contrast, other multi-dataset approaches~\cite{zhang2021dodnet,xie2023learning,fang2020multi,shi2021marginal,liu2023clip,ulrich2023multitalent} have limited modeling scalability or capacity when the datasets and organ types increase, e.g., Multi-Talent~\cite{ulrich2023multitalent} experiences extremely slow convergence and takes over 1,000 GPU hours when training on only five datasets with 1,471 CT scans (yet not fully optimized with $3$\%$\sim$$4$\% \ac{dsc} gap to our \ac{clnet}). Additionally, the \ac{clnet} architecture is flexible and can continually and dynamically expand its segmentation capability without affecting previous learned knowledge when new datasets (either new or existing anatomies) are added. \textcolor{black}{By open-sourcing our pretrained models, we empower peers to adaptively fine-tune specific decoders of \ac{clnet} by incorporating their own data in the continual learning manner and releasing the updated model to foster accelerated development within the research community.}  This architectural characteristic also ensures that the model size stays fixed when anatomy types in new datasets already exist, or that the model has a very slow and linear parameter growth rate when new anatomy classes are added. The high accuracy, scalability and extendability of \ac{clnet} make it a benchmark model in the multi-dataset segmentation scenario.
% (as compared to nnUNet in a single dataset)

%Finally, compared to other multi-datasets approaches, \ac{clnet} can inference on $16$ GiB GPU RAM, suitable for deployment in clinical setting where often limited computational resources are available (in contrast, other methods requires xxx
%model size easy deliver to clinical setting with small computational resource requirement

%Flexibility of \ac{clnet}:  1) new organs or new dataset of existing organs, model size stays stable or slowly growing; 2) \ac{clnet} every effective, requiring much less data as compared to other SAM foundation models; 3) fast trian and require less data to segment new organ or update existing organs. 
%\ac{clnet} is flexible and efficient when expanding its capability. (1) 

%discuss GE training; 
As one of the core components of \ac{clnet}, the trained then frozen general encoder is important to train subsequent decoders, which affect the accuracy of the final segmentation. Ideally, for whole-body anatomy segmentation, we expect to construct a sufficiently representative general encoder that extracts deep image features to capture and encode all visual information inside the full human body. Compared to image statistics of broad natural images, medical images appear in a much more confined semantic domain, that is, the human body is anatomically and intrinsically structured composing of distinct body parts. This makes it feasible to learn a strong universal general encoder competently capturing the holistic human body CT imaging statistics using large or not so-limited multi-organ datasets. We demonstrate that TotalSegmentator~\cite{wasserthal2022totalsegmentator} (1204 CT scans with 104 organ types) is a suitable dataset for this purpose. Adding more datasets or applying self-supervised pretraining~\cite{chen2020simple,chen2021exploring} for general encoder training does not bring obvious performance improvements or pruning benefits to the final universal model (as shown in Table~\ref{tab:impact_ge_totalseg}). 
%Alternatively, other relative small datasets can also be used to train the general encoder, especially under the help of self-supervised pretraining. When training the general encoder using the StructSeg19 dataset with much less training scans (292 vs. 1204 CT scans) and organ classes (22 vs. 103 anatomical structures), self-supervised SimSiam pretraining~\cite{chen2020simple} would increase the subsequent segmentation by an average 1.7\% DS improvement along with 10\% increased decoder pruning rate. Compared to using TotalSegmentator, a moderate performance drop (94.1\% to 92.8\% DS) is observed in the segmentation results. 

%\ac{clnet}'s design of frozen general encoder and independently trainable decoder not only handles partial labeling issue in multi-datasets segmentation, but also tr Using \ac{ema} can help stabilize the fine-tuning process and learn more robust features by reducing the impact of abrupt gradient changes. The \ac{ema} update works effectively when the input body part and output classes of the decoding path remain consistent. However, a potential drawback arises when different decoders with varying output classes are sequentially used for \ac{ge} updates. In such cases, \ac{ema} might not effectively `preserve' previously learned features but smooth out the distinctive features of each task on-the-fly. This smoothing effect can blur the specific characteristics unique to each task, potentially reducing the \ac{ge}'s ability to accurately capture the finer details necessary for distinguishing between different body parts and segmentation classes. 

With the advent of automated AI-enabled tools, there is now a rapidly growing appreciation of the potential clinical value added by CT-based segmentation of incidental tissues and organs in CT, often referred to as opportunistic screening~\cite{pickhardt2024harnessing,pickhardt2022value}. Systematic quantitative cardiometabolic assessment of CT scans can opportunistically yet conveniently produce robust risk stratification related to all-cause mortality, cardiovascular events, osteoporotic fractures, diabetes, metabolic syndrome, sarcopenia, and more~\cite{pickhardt2020automated,pickhardt2021utilizing,pickhardt2020automated_1,tallam2022fully,nachit2023ai}. For patients diagnosed with cancer, undergoing major surgery, or in need of organ transplant, CT biomarkers can also provide pre- and post-treatment frailty assessment, in addition to any specific radiomic tasks. Such opportunistic screening processes and assessments can be broadly applied to nearly any body CT scan, and an automated universal segmentation approach provides almost limitless clinical potential. Furthermore, no additional patient time, effort, or radiation exposure is required for this repurposed usage of data, which may not only be highly cost-effective, but may even save cost~\cite{pickhardt2023ai} in practice.

This work has several limitations. First, we do not include detailed brain anatomies or more \ac{gtv} types in current \ac{clnet}. However, given the demonstrated effectiveness of \ac{clnet} in segmenting numerous fine-grained anatomies and nine \ac{gtv} types, we are confident that incorporating more anatomies would not change observations and conclusions derived from the current edition. Second, although we demonstrate the effectiveness of current CNN-based encoder in 3D whole-body universal segmentation, other encoder architectures, e.g., Transformer-, Mamba- or hybrid-based general encoders, may be examined to explore the possibility of further improving the universal segmentation performance. Third, while \ac{lth}-based pruning has effectively reduced the number of parameters in \ac{clnet}, it does not alter the channel count or network topology, leaving the number of FLOPs unchanged. Exploring alternative pruning methods to address this limitation is part of our future work. Lastly, current \ac{clnet} includes lymph node stations (critical for cancer diagnosis and treatment planning) at only head \& neck and chest regions. To serve as a universal CT segmentation model in radiology and oncology, we plan to systemically curate the annotation of lymph node stations in the abdominal and pelvic regions (ongoing work). These will be efficiently integrated into \ac{clnet} in the near future without affecting previously learned knowledge, thanks to the expandably designed network architecture of \ac{clnet}.
\section{Method}
\label{sec:method}

\noindent\textbf{Problem Formulation}. We aim to simultaneously or continuously learn a whole-body organ segmentation model from $T$ partially labeled datasets $D=\{D_1, \dots, D_T\}$. In \ac{pl} setting, all datasets in $D$ are simultaneously accessible. While in the \ac{cs} setting, $\{D_1, \dots, D_T\}$ are sequentially available at each step, which means that all previous training data $\{D_k, k<t\}$ are not accessible when learning on the $t^{th}$ dataset $D_t$. 
% For $D_t=\{{X_i}^t, {Y_i}^t\}_{i=1}^{n_t}$ with $n_t$ samples, the organ class set $\Psi_t$ of $D_t$ can be further stratified into organ-specific sub-class groups $\psi_t$. 
Let $\Psi_t$ denotes the target organ set of $D_t$, the total output organ set is: 
\begin{equation}
    \Psi = \bigcup_{t=1}^{T} \Psi_t 
\end{equation}
When using $\psi$ to denote an anatomy type, $\Psi$ can also be represented as $\Psi=\{ \psi_1, \dots, \psi_M \}$, where $M$ is the total number of anatomy types. 
% We drop the task index $t$ for simplicity, as the pre-defined $Y_{\psi}$ can be shared across different datasets. 
% Let $X$ and $Y_{\psi}$ denote the input image and the corresponding organ label. 
Let $X$ be the input image. The prediction map $\hat{Y}_{\psi}$ for a particular anatomy $\psi \in \Psi$ is given by: 
\begin{align} 
    \hat{Y}_{\psi} &= f_{\psi}\left(f_e\left(X; W_e\right); W_{\psi}\right) \\
    \hat{\mathbf{Y}} &= \bigcup_{\psi \in \Psi} {\hat{Y}_{\psi}} 
    \label{eq:gs}
\end{align} 
where $f_e$, $f_{\psi}$, $W_e$, and $W_{\psi}$ denote the CNN functions and the corresponding parameters of the encoding and decoding paths, respectively. For simplicity, we drop explicit notation of voxel spatial locations. The final prediction, $\hat{\mathbf{Y}}$, is formed by taking the union of the predictions from all decoders, allowing each voxel to potentially have multiple labels.

% \noindent\textbf{Overall Framework}. 

\noindent\textbf{Overall Framework}. 
The proposed \ac{clnet} is designed to address both partial label segmentation and continual semantic segmentation tasks. In the partial label setting, the model utilizes all datasets' samples and labels simultaneously, whereas, in the continual setting, it is trained using one dataset at a time. The \ac{clnet} architecture comprises a shared encoder to extract general characteristics, as illustrated in Figure~\ref{fig:motivate}B, multiple stratified and pruned decoders (each corresponding to a pre-defined anatomy or a group of relevant anatomies, Supplementary Table~\ref{tab:dec2org}), a \ac{bpr} module, and a prediction-merging module. The process begins by training an encoding segmentation base network with a general encoder. We show that the well-trained encoder, referred to as \ac{ge}, can effectively extract universal features across various organs and datasets, thus enhancing the performance of subsequent downstream learning tasks. Once \ac{ge} is trained, it remains fixed, while additional trainable decoders are updated. This approach forms a unified, scalable, and nonforgetting architecture. At each decoder training stage, optimization and pruning are performed to control model complexity and prevent model escalation. Finally, predictions are merged based on body part regions (e.g., head and neck, chest, and abdomen) and potential lesion locations, resulting in a single unified model capable of segmenting all target organs. The \ac{clnet} supports both single and multi-GPU inference setups, optimizing deployment costs, improving prediction efficiency, and streamlining clinical workflows. The following subsections provide a detailed explanation of each component and the methods employed in \ac{clnet}'s implementation.

\acreset{bpr}
\subsection{4.1 \acf{bpr}}
The \ac{bpr} scores can be obtained based on axial CT slice scores predicted by an automated body part regression algorithm~\cite{yan2018unsupervised}. As the slice score monotonically correlates with the patient's anatomical height, each targeted anatomy can be located based on its relative position. By incorporating this additional task, the \ac{clnet} is trained to explicitly recognize the anatomical region (body part) associated with each target. This enhanced understanding of anatomy distribution is beneficial for learning better voxel representations of target organs and tissues, leading to more precise segmentation and reducing the likelihood of false positives from regions outside of the target anatomy. 
\acreset{ge}
\subsection{4.2 \acf{ge} Training}
In the segmentation of whole-body organs, our goal is to develop a \ac{ge} that is both representative and universally effective. This encoder should be capable of extracting deep image features that comprehensively capture and encode all visual information pertaining to the entire human body. Unlike the broad image statistics commonly seen in general natural image datasets, medical images, especially radiology scans such as CT and MRI, are confined to a more specific semantic domain: the human body, which is anatomically structured and often includes targets with ambiguous boundary and potential diseases. Based on this, our objective is to develop a robust universal \ac{ge} that effectively and adeptly captures the comprehensive statistical characteristics of the human body for segmenting whole-body anatomies in CT scans. To train \ac{ge} for whole-body organ segmentation, we propose starting with the publicly available \ac{totalseg} dataset~\cite{wasserthal2022totalsegmentator} using the proposed \ac{clnet} framework. In this setup, different types of anatomical structure (e.g., bones, muscles, organs) are assigned to separate decoders, while the \ac{ge} is updated alongside all decoding paths. In addition to utilizing comprehensive CT datasets for training, we also supplement the details of \ac{ge} training under the condition that only limited training data or a restricted variety of training classes is available, as described in the Supplementary.

\acreset{lth}
\acreset{ema}
\acreset{fls}
\subsection{4.3 Multi-path Decoding}
% The \ac{clnet} complexity may escalate when exploiting the full-size UNet decoder for each organ-specific decoding path. 
% Revisiting \ac{clnet} framework: 
Following completion and fixation of \ac{ge} training, trainable decoding heads targeting subsequent downstream tasks are learned. By independently training each decoding path, knowledge forgetting can be mitigated. However, several questions regarding the clinical applicability and generalizability of this approach remain unresolved: 1) As the segmentation target extends, the proposed model complexity may escalate. How can we effectively prune the network without compromising segmentation performance? 2) Automated segmentation of anatomical structures continues to present challenges, particularly for small and difficult-to-identify organs. How can we improve segmentation accuracy for these targets? 3) In a clinical setting, the continuous collection of new data is a routine practice. How can we efficiently and effectively update the decoders with minimal cost? To address these challenges, we propose the following strategies: 1) \textit{\acf{lth}-based Pruning},  2) \textit{\acf{fls}}, and 3) \textit{\acf{ema}-based Decoder Updating}. 

% To adequately address these questions, we propose: 1) \textit{Lottery Ticket Hypothesis-based Pruning}: We supplement the decoding head with a \ac{lth}-based pruning strategy to scale down the decoder complexity without compromising performance. 2) \textit{Feature-level Supporting}:  We introduce \ac{fls} by incorporating features from previously learned decoders into the current decoder being trained. This approach enhances segmentation accuracy and potentially achieves a better pruning rate, particularly for challenging targets. 3) \textit{Exponential Moving Average-based Decoder Updating}: We adopt \ac{ema}-based updating to continuously refine each stratified organ head. This approach leverages the previously pruned architecture and supporting features, ensuring efficient and cost-effective model updates. In the following subsections, we discuss each component in detail. 

\textbf{\ac{lth}-based Pruning}.
The \ac{lth}, introduced by Frankle and Carbin~\cite{frankle2018lottery}, defines a ``winning ticket'' as a sub-network that, when trained in isolation from its original initialization, achieves comparable or even better performance to the full network. This hypothesis implies that not all weights in a large network are essential for learning; instead, a critical subset of weights, if properly initialized~\cite{malach2020proving}, can effectively drive the learning process. In our implementation, we focus on pruning the decoding heads while leaving the \ac{ge} intact. To stabilize the pruning process, each decoding head is initialized using the default nnUNet decoder and trained to convergence without pruning~\cite{frankle2019stabilizing}. %e.g., 300 epochs. 
Then, the proposed \ac{lth}-based pruning process is repeated iteratively, with each round involving training, pruning, and rewinding, to progressively identify a smaller winning ticket. During each pruning iteration, we evaluate the performance using the \ac{dsc} on the validation set. If the performance drop exceeds a pre-defined threshold ($\delta>1\%$ for $5\% - 95\%$ percentile \ac{dsc}) compared to the validation performance before pruning, the pruning process is stopped and the network is rewound to the previous pruning stage. After pruning is complete, additional training epochs $20$ are applied to recover any potential performance drop. 

\textbf{\ac{fls}}.
To further enhance segmentation accuracy and robustness, we introduce a \ac{fls} mechanism. The \ac{fls} is intuitively simple yet effective. This involves incorporating features from previously learned decoders into the current decoder during training. Using these previously learned features, the current decoder can build on a richer and more informative set of features, leading to improved performance. This approach helps to capture finer details and ensures more consistent and accurate segmentation, particularly for challenging targets. For each decoding block, learnable projection convolutions $[1\times 1\times 1]$ are used to adapt relevant features from the output of the previously learned decoder while maintaining the same feature dimensions. The projected features are channel-wise concatenated to the current decoder's features, providing enriched information for improved segmentation accuracy. Note that the previously learned decoder head is fixed and is not pruned during the learning of the current decoder head. This ensures that the valuable features learned in the previous stages are preserved. The supporting projection convolutions and the current decoding head are targeted for pruning. 

\textbf{\ac{ema}-based Decoder Fine-tuning}.
We adopt \ac{ema} to ensure stability and preserve knowledge in continual segmentation setting when new data is included for the update of the decoder. The decay rate is set to $\alpha=0.999$. The \ac{ema}-based decoder fine-tuning is:
\begin{equation}
    f_{\psi}^{EMA} = \alpha f_{\psi}^{EMA} + (1-\alpha) f_{\psi}.
\end{equation}
% where $f_{\psi}$ is the CNN function of the decoder with target organs $\psi$. 
Leveraging on \ac{ema}, the decoder can be efficiently trained with a mitigated risk of overfitting. During model inference, by default, \ac{clnet} loads the weights of the EMA module to ensure stability and preserve the learned knowledge. 
Note that \ac{ema} updating is optional based on the user's design; however, we strongly recommend adopting \ac{ema} updating as the default. In the proposed \ac{clnet}, the \ac{ema} module is updated only when the decoder pruning is complete. Please note that the pruned mask of the decoder, which identifies the weights to be zeroed out based on their insignificance, is recommended to be \textbf{\textit{not}} updated during training with new data, ensuring consistent performance and stability. On the other hand, only for the decoding head that includes an \ac{ema} module, the corresponding pruned mask can be updated. Starting from the existing pruning rate, instead of applying the pruning to the non-smoothed counterpart, the pruning process is conducted to the \ac{ema} module, and only the pruned mask of the \ac{ema} module is saved.

% In a clinical setting, continuously collecting new data is a standard practice. However, training decoding head from scratch using both previously collected and newly included data can be inefficient. On the other hand, fine-tuning the decoding head with a limited volume of new data increases the risk of over-fitting. To address this, 

% To find the winning ticket, the vanilla \ac{lth} pruning process is followed: first, the original network is trained for a certain number of epochs. Then, a percentage of the smallest magnitude weights (e.g., L1-norm, the sum of the absolute values of the vector) are removed. The remaining weights are then reset to their initial values and the network is re-trained. 

% This challenge arises because the model may not be exposed to the diversity and breadth of data required to generalize effectively, potentially leading to degraded performance on unseen cases.
\subsection{4.4 Prediction Merging}
The prediction merging in the proposed \ac{clnet} is \textbf{\textit{NOT}} designed to combine all predictions into a single prediction map. This is because organ-wise predictions include both primary organs and their substructures, as well as potentially overlapping structures between predictions, such as the lungs vs. lung lobes and organs vs. lymph node stations. Forcefully merging all predictions together could lead to overlapping segmentation masks and inaccuracies. Instead, the merging process is carefully structured to consolidate these predictions while considering \ac{bpr}-based anatomical regions and any identified potential \acp{gtv}, ensuring an anatomical structured and contextually accurate anatomical representation. Since the \ac{gtv}s could be distributed throughout the human body, \ac{bpr} is not applied to the \ac{gtv} decoding heads. Consequently, simply overlaying \ac{gtv} predictions onto the predicted anatomy may cause false alarms. For example, this approach could mistakenly predict kidney cancer in healthy parotid regions, highlighting the need for more precise integration strategies. The proposed prediction merging includes: \textit{1) \ac{bpr}-based Bounding}, and \textit{2) \Ac{gtv}-aware Merging}. 

\textbf{\ac{bpr}-based Bounding}. For each decoding path, we pre-compute the \ac{bpr} scores $\text{BPR}_{\psi}$ of target organs of the decoder. Let $z$ denote the input CT image dimension in the axial direction and $\sigma_{\psi}$ denote the standard deviation of the depth of the anatomies of the $\psi$ for the decoder in the axial direction. The body part range of the decoder is determined using:
\begin{align}
    \text{BPR}^{U}_{\psi} &= \min\left(z, P_{95}\left(\text{BPR}_{\psi}\right) + 2\sigma_{\psi}\right), \\
    \text{BPR}^{L}_{\psi} &= \max\left(0, P_{5}\left(\text{BPR}_{\psi}\right) - 2\sigma_{\psi}\right), \\
    \mathcal{B}_{\psi} &= J_{[\text{BPR}^{L}_{\psi}, \text{BPR}^{U}_{\psi}]},
\end{align}
where $P_{*}(\cdot)$ denotes the $*$ percentile calculation, $\text{BPR}^{U}_{\psi}$ and $\text{BPR}^{L}_{\psi}$ denote the upper and lower body part bounds of the decoder, $J$ denote the matrix of ones, and $\mathcal{B}_{\psi}$ denote the foreground target bounding range of the decoder in the head-to-foot/axial direction, respectively. During model inference, for each decoding head, the prediction $\hat{Y}_{\psi}$ is updated as $\hat{Y}_{\psi}=\hat{Y}_{\psi} \odot \mathcal{B}_{\psi}$ before merging the decoder output, ensuring that predictions outside the bounding range are removed, where $\odot$ denote element-wise multiplication. This guarantees that the out-of-bound false positives of each decoder prediction are fully cleaned so that potential label conflicts during output merging are mitigated.

\textbf{\Ac{gtv}-aware Merging}. The proposed \Ac{gtv}-aware Merging aims to merge the predicted \ac{gtv} to the target structure considering the target's \ac{bpr} range. Let $\hat{Y}^{\epsilon}_{\psi}$ denote the \ac{gtv} prediction of the target structure $\psi$, and let $\text{BPR}_{\psi}$ denote the body-part bounding range of the target structure. The weighting map $M_{\psi}$ for the decoder $f_{\psi}$ is calculated such that only when $\hat{Y}^{\epsilon}_{\psi} \rightarrow 0$ and $\mathcal{B}_{\psi} \rightarrow 1$, s.t., the $M_{\psi} \rightarrow 1$; whereas $M_{\psi} \rightarrow 0.5$ for the rest states. 
\begin{align}
M_{\psi} &= J - \frac{1}{2} \left(J - \mathcal{B}_{\psi} + \hat{Y}^{\epsilon}_{\psi} \odot \mathcal{B}_{\psi}\right),  \\
H_{\psi} &= - \left(M_{\psi} \odot \hat{Y}_{\psi}\right) \log\left(M_{\psi} \odot \hat{Y}_{\psi}\right), \\
\mathbf{H}(j) &= \bigcup_{\forall \hat{Y}(j)_{\psi} \neq 0} H_{\psi}(j), \\
\hat{\mathbf{Y}}(j) &= \hat{Y}^{\text{argmin} (\mathbf{H}(j))}(j). \label{eq:final}
\end{align}
For each voxel $j$, we collect a set $\mathbf{H}(j)$, for all $\hat{Y}(j)_{\psi} \neq 0$. Depicted in Eq.~(\ref{eq:final}), the final output class $\hat{\mathbf{Y}}(j)$ is determined using the prediction $\hat{Y}_{\psi}(j)$, of which the smallest is $H_{\psi}(j)$. Note that if \ac{gtv} is not specific to any particular anatomy, such as metastasis cancers or lymph nodes, the bounding range $\mathcal{B}$ is set to encompass the entire image.

% To mitigate this issue, we propose using potentially diseased anatomy's $\text{BPR}_{\psi}$-based confidence scores while taking the target anatomy's \ac{gtv} prediction into account.
% % data preprocessing
% %  -- hyper-parameters
% % data postprocessing
% % training + inference time
\subsection{4.5 Implementation Overview}
We provide an overview of the \ac{clnet} implementation. More detailed implementation is reported in the Supplementary Sec.~\ref{sec:impl_details}. 

\textbf{Image Preprocessing}. \ac{clnet} adopts its general training framework from nnUNet~\footnote{\url{https://github.com/MIC-DKFZ/nnUNet}}. A CT windowing range of $[-1,024, 1,024]$ HU is applied to each CT image, which is subsequently pre-processed to the RAI (Right-Anterior-Inferior) orientation. Following this, the standard nnUNet pre-processing pipeline is applied. By default, all CT scans are resampled to a uniform resolution of $1.0\times 1.0\times 1.5$ mm and configured with a fixed input patch size of $128\times96\times112$ for all tasks. Additionally, similar to nnUNet, \ac{clnet} also supports automatic resample resolution and patch size selection based on the training dataset, provided that this option is enabled by the user. 

% Unlike nnUNet, which adapts the input patch size for each task, \ac{clnet} sets a fixed default input patch size of $160\times160\times64$ (or later determined by user) voxels for all tasks. 

\textbf{Network Architecture}. The architecture backbone is adapted on the basis of nnUNet's ``3D full-resolution'' setting. \ac{clnet} includes a 6-block encoding path and a 5-block decoding path, designed to provide robust performance and detailed feature extraction. Similarly to nnUNet, the base number of features for \ac{ge} is $32$ and the maximum number of features (toward the bottleneck block) is capped at $320$. Each \textit{encoding} convolutional block consists of two convolutional layers, with instance normalization and leaky ReLU activation applied. Downsampling is performed using strided convolution. The \textit{decoding} block consists of two convolutional layers, also featuring instance normalization and leaky ReLU activation, and is paired with an \ac{ema} module. The decoder's base feature number is configured according to a rule-based adaptation. Upsampling is achieved with transposed convolution, which upscales the output features of each decoding block. The data augmentation pipeline is implemented with the public available ``BatchGenerator'' framework~\footnote{\url{https://github.com/MIC-DKFZ/batchgeneratorsv2}}.

% \textbf{Hardware Requirement}. All experiments were trained on a server with $8$ Nvidia V100 GPUs, each with 16GiB of RAM, 330GiB of CPU memory, and an Intel(R) Xeon(R) Platinum \textit{8163} CPU, featuring $82$ threads. We have tested that all training tasks can be performed using a single GPU with a standard $12$GiB RAM and $64$GiB of CPU memory. The default number of threads for training and testing is set to $12$. It is preferred to use \ac{ddp} to leverage all available GPUs for efficient computation. 

% We have tested that all training tasks can be performed using a single GPU with a standard $12$GiB RAM and $64$GiB of CPU memory.
% During the prediction process, the average GPU memory consumption is approximately $6$ GiB of RAM.

\textbf{Training Efficiency}.
We have tested that all training tasks can be performed using a single GPU with a standard $12$ GiB RAM and $64$ GiB of CPU memory. The average training time for a two-class head (e.g., left and right counterparts) in \ac{clnet} is approximately 86 seconds per epoch on a single Nvidia V100 GPU, which can be accelerated to 32 seconds per epoch using a \ac{ddp} setup with four Nvidia V100 GPUs. The average GPU memory consumption is under $10$ GiB during the training of each decoding head. Each trained decoding head has a model size of approximately $80$ MiB, which can be reduced to just $18$ MiB when saved as a sparse model using PyTorch's sparse format~\footnote{\url{https://pytorch.org/docs/stable/sparse.html}}. Please note that sparse computation was not applied during model training. Hence, GPU RAM consumption remains unchanged during the training process. The network optimizer used is SGD with the Poly learning rate scheduler.

\textbf{Inference Efficiency}. The \ac{clnet} inference supports both single and multi-GPU setups, aiming to minimize deployment costs, improve prediction efficiency, and streamline clinical workflows. In a single GPU setup, segmentation is performed sequentially by processing each decoding head one at a time. In a multi-GPU setup, decoding heads are evenly distributed across GPUs to accelerate the process. The inference time on a single Nvidia V100 GPU is averagely less than $5$ seconds per stratified head. The total inference time for our 235-organ segmentation on a single Nvidia V100 GPU averages $430$ seconds per patient, comprising approximately $10$s for image pre-processing, $5$s for model loading (only once), $400$s for inference, $5$s for merging and saving, as well as additional system overhead. We strongly suggest using Solid-State Drive (SSD) to store the training data, as SSD improves input/output (BIOS) speed and facilitates faster inter-communication with CPU and GPU memories. When parallel predictions are performed using 4 or 8 GPUs, the total inference time can be reduced to approximately $118$s or $58$s per patient, respectively. The average consumption of GPU memory is approximately $6$ GiB of RAM throughout the entire prediction process.

\pagebreak

\section*{References}
\bibliographystyle{naturemag}
\bibliography{refs}

\section*{Acknowledgments} We'd like to cordially thank Prof. Gregory D. Hager from Department of Computer Science, the Johns Hopkins University who read our manuscript and provided invaluable suggestions, feedback and encouragement on its technical feasibility and possible clinical utility.

%%%%%%% Appendix/Supplementary Start Here %%%%%%%
\pagebreak
% \appendix
% \newpage
% -------------- CONVERT SECTION NOTATION --------------%
\setcounter{section}{0}
\setcounter{table}{0}
\setcounter{figure}{0}
\renewcommand\thesection{A\arabic{section}}%{\Alph{section}}
\renewcommand{\thetable}{S\arabic{table}}
\renewcommand{\thefigure}{S\arabic{figure}}
% -------------- CONVERT SECTION NOTATION --------------%

% \counterwithin{figure}{section}
% \counterwithin{table}{section}

% \newcounter{supp_sec_cnt}
% \setcounter{supp_sec_cnt}{1} 

%%%%%% start section here %%%%%%
\section{Details of Datasets}
\label{sec:dataset_details}
% \section*{A\arabic{supp_sec_cnt}. Dataset Details}
% \customlabel{sec:dataset_details}{A\arabic{supp_sec_cnt}}

%% training and internal validation datasets
An overview of the public and private datasets used in our experiments is provided in Fig. \ref{fig:dataset} and Tab. \ref{tab:dataset_stats}. The complete \ac{dsn} training dataset collection consists of 16 private datasets and 20 public datasets, spanning whole-body regions and various anatomical structures, including the head and neck, chest, abdomen, bone, lymph node station (LNS) and lesion. This comprehensive dataset collection contains a total of 7,278 private and 3,904 public CT scans, along with 37,827 private and 172,607 public annotations. Together, these annotations cover \ac{orgn} anatomical structures, 125 from private sources and 162 from public sources with 52 overlapping classes. 
Tab. \ref{tab:dataset_stats} provides a comprehensive summary of the fingerprints of all datasets. In addition to standard statistics, such as the name of the dataset, the number of training and testing samples, the number of target organs, the median image size, and the median voxel spacing, we also report two specialized statistics to better characterize the body part distribution of each dataset. These include the axial bounding ranges of body part regression (BPR)~\cite{yan2018unsupervised} for both the images and the foreground organs. Further details on the target organs for each private and public dataset are presented in Tab. \ref{tab:dataset_task_p} and Tab. \ref{tab:dataset_task_d}, respectively. 

As shown in Fig. \ref{fig:dataset}, the \ac{dsn} datasets are grouped by the corresponding main body part of the dataset's targets, i.e. head and neck, chest, and abdomen, except for TotalSegmentator and Body\_LinkingMed\_OAR which widely cover the entire body. The dataset distribution of each group is as follows: 
Head and neck group has 5 private datasets and 4 public datasets, including brain regions, visual organs, auditory organs, oral and larynx, glands, LNS, and \ac{npc} \ac{gtv}, where the image BPR scores mainly range from mid chest (0.6) to upper head (1.0); 
Chest group has 8 private datasets and 3 public datasets, covering respiratory organs, cardiovascular organs, LNS, as well as lung and esophagus \ac{gtv}, where the image BPR scores mainly range from mid-lower abdomen (0.3) to upper throat (0.8); 
Abdomen group has two private datasets and 12 public datasets, targeting digestive organs, urinary organs, vessels and glands, and common abdomen \ac{gtv}s, where the image BPR scores mainly range from lower pelvis (0.0) to upper chest (0.7); 
Whole-body group has one public dataset -- TotalSegmentator, which contributes bones and muscles such as ribs, vertebrae, and limbs, and one private dataset -- Body\_LinkingMed\_OAR, which provides whole-body mask to segment the foreground body area. The image BPR scores cover the entire body from 0 to 1.

%% External validation datasets
To further demonstrate the generalizability and robustness of our method, we collect 7 external datasets for validation, comprising 1 public dataset (BTCV) and 6 newly acquired private datasets. These external datasets span whole-body anatomical regions, covering 1,979 testing samples and 43 target anatomical structures in the head and neck, chest, abdomen and \acp{gtv}. Four types of \acp{gtv} are included: \ac{npc}, esophageal, liver, and kidney lesions. Comprehensive fingerprints of the external datasets, including dataset name, body part coverage, number of target classes, number of testing samples, median image size, median voxel spacing, and a detailed target organ list, are provided in Tab. \ref{tab:supp_ex_dataset}.

%% Hospital center abbr.

% \stepcounter{supp_sec_cnt}

%% Full-scale & Validation exp settings
\section{Details on Experimental Settings}
\label{sec:exp_set_details}

%% full-scale exp setting
\subsection{Internal Full Evaluation Experimental Setting}
\label{sec:exp_full}

%% motivation
The goal is to train a unified model that not only accurately and robustly segments a wide range of whole-body organs in CT but also has sufficient generalizability and scalability to be easily and efficiently extended to new datasets and segmentation tasks. Therefore, our internal full experiment results are evaluated on the full-scale \ac{clnet} trained across the entire \ac{dsn} dataset collection, covering a large amount of 11,182 training samples and \ac{orgn} whole-body anatomical structures, involving the following settings: 1) \ac{pl} setting, in which \ac{clnet} is fully trained across all datasets simultaneously and compared with the segmentation upper-bound nnUNet ensemble; 2) continual semantic segmentation setting, where we continuously train \ac{clnet} on \ac{dsn}-dataset sequence; 3) ablation study for decoder pruning module. 

% \begin{itemize}
%     %% PL exp
%     \item \ac{pl} experiment: We adapt our method to the \ac{pl} setting and directly train the unified model with simultaneous access to the entire dataset collection using partial label supervision. The organ-wise segmentation results are also compared with the \ac{css} model to investigate the performance difference between the two settings. 
%     %% CSS exp
%     \item \ac{css} experiment: We continuously train \ac{clnet} on the \ac{dsn} datasets and report the organ-wise performance. The \ac{dsn}-nnUNet ensemble is used as the upper bound for comparison. 
%     %% decoder pruning
%     \item Ablation study: We compare the \ac{clnet} without decoder pruning process with the pruned model and assess the impact of decoder pruning on segmentation accuracy and parameter size of each decoder. 
% \end{itemize}

%% dataset selection
\textit{\textbf{Dataset Setting:}}
As outlined in Sec. \nameref{sec:dataset}, all \ac{dsn} datasets are used to train the full-scale \ac{clnet} segmentation model, covering whole-body region. The dataset collection contains a total of 13,952 CT scans -- 11,182 for training and 2,770 for testing -- annotating \ac{orgn} target anatomical structures. 
To conveniently report organ-wise statistical results of our model, we divide the \ac{orgn} whole-body organs (excluding ``Body Mask'') into 6 groups according to their body parts and anatomical types: head and neck (head-neck, 49 organs), chest (52 organs), abdomen (28 organs), bone (63 organs), lymph node station (LNS, 33 organs), and \ac{gtv} (9 organs). 
% The result of the extra `Body' mask is also reported. 
Note that each dataset only labels a subset of our universal anatomy set and is therefore treated as a partially labeled dataset. 
The dataset details are introduced in Supplementary Sec. \ref{sec:dataset_details}.

%% pl setting
\textit{\textbf{Universal Segmentation Setting:}} 
Considering the ever-increasing amount of public medical imaging datasets, it is possible to directly get a large collection of partially labeled datasets for medical segmentation. Thus, in \ac{pl} setting, we directly train \ac{clnet} on \ac{dsn} datasets with simultaneous access to all training samples, resulting in a full-scale, unified model (\ac{cln_u36}) with comprehensive and integrated organ segmentation capabilities. \ac{cln_u36} follows the multi-decoder network architecture, and all decoders are pruned at the end of training. Note that only \ac{decn} decoders are included in our final model, fewer than the total number of anatomies, as symmetric anatomical instances and substructures are typically bundled and segmented by the same decoder head. To showcase the superior performance of our model, organ-wise segmentation results are evaluated for \ac{cln_u36} and compared with the \ac{dsn}-nnUNet ensemble (\ac{nnu_e36}) as the upper bound for multi-dataset multi-organ segmentation, with each nnUNet specialist trained on separate dataset.

%% css setting
\textit{\textbf{\ac{css} Setting:}}
Compared to \ac{pl} setting where all training data and labels are accessible, \ac{css} setting is more general and practical in real world situation. In \ac{css} setting, we continuously train and evaluate a \ac{css} version \ac{clnet} (\ac{cln_c36}) on \ac{dsn} dataset sequence. The continual learning dataset sequence follows the default order of public datasets (D1-D20) and private datasets (P1-P16) listed in Fig. \ref{fig:dataset}. We firstly pretrain and freeze the \acll{ge} on \ac{totalseg} (D1) at initial step, and then continually train \ac{clnet} on the subsequent datasets
% without access to previous training datasets
, incrementally adding or updating decoder heads for new anatomies and labels. At the end of each learning step, the newly added decoder heads are pruned. In accordance with the \ac{css} protocol, only the current dataset is available at each continual learning step, while all previous training datasets remain inaccessible. The segmentation performance of \ac{cln_c36} is also evaluated and compared with \ac{cln_u36} to confirm the effectiveness of our method’s \ac{css} capability.

%% ablation: decoder pruning, cln_u36_unprn
\textit{\textbf{Ablation Study:}}
To evaluate the impact of decoder pruning on segmentation accuracy and decoder parameter size, we compare the DSC difference and the pruning rate of each \ac{cln_u36} decoder against the unpruned model \ac{cln_u36_unprn}, as shown in Fig. \ref{fig:prune}. The DSC score for each decoder is computed as the mean DSC across its corresponding target anatomies. Consistent with the organ-wise results, the performance of all \ac{decn} decoders is also reported within 6 organ groups. 

%% Internal Validation Exp Setting on 5 representative datasets
\subsection{Internal Comparative Evaluation Experimental Setting}
\label{sec:exp_val}

%% motivation
%provide a comprehensive and detailed picture of 
We aim to comprehensively validate the effectiveness and robustness of \ac{clnet} in whole-body organ segmentation under both \ac{css} and \ac{pl} settings, compare it with other common continual segmentation methods and leading multi-dataset organ segmentation models, and explore the impact of each key module. Directly conducting all validation experiments on full-scale datasets would require excessive computational resources and training time, making it impractical and environmentally unfriendly. Moreover, some comparison methods are not scalable to a large number of datasets or organ classes, leading to unfair comparisons and rendering such experiments infeasible. 
Therefore, we conduct a small-scale internal evaluation experiment using five representative public datasets selected from the original \ac{dsn} datasets. These datasets include 1,471 training samples and 137 organs in various body parts and organ types, covering the head and neck, chest, abdomen, bone, and \ac{gtv}. This experimental design allows us to efficiently and comprehensively validate and evaluate our method. 
The evaluation experiments involve: 1) \ac{css} setting, in which we continuously train \ac{clnet} on the five datasets, evaluate the performance of different dataset orders, and compare our method with other continual segmentation methods; 2) \ac{pl} setting, where \ac{clnet} is simultaneously trained on the five datasets and compared its performance with other leading multi-dataset multi-organ segmentation models; 3) Ablation study for the key modules of \ac{clnet}. 

\textit{\textbf{Dataset Selection:}}
From the \ac{dsn} datasets collected in Sec. \ref{sec:dataset_details}, we carefully select five representative public datasets spanning different body regions and organ groups to ensure experimental reproducibility and a comprehensive evaluation of the segmentation capacity of our method across various body parts and anatomical structures. Brief descriptions of the five datasets are listed below: 
\begin{itemize}
    \item \textbf{\ac{totalseg}}: This large and diverse dataset contains totally 1,228 CT scans covering all regions of the body and are officially divided into 1,139 training data and 89 testing data. Totally 115 whole-body organs are annotated (left \& right kidney cysts are removed); 
    \item \textbf{\ac{structseg}}: The dataset focuses mainly on head and neck regions and includes 50 CT scans, divided into training and testing sets with 40 and 10 images, separately. The annotations cover 22 head and neck organs, in which 21 organs are not included in \ac{totalseg} and only spinal cord is overlapped; 
    \item \textbf{\ac{segthor}}: This dataset has 40 chest CT scans with only four chest organs, which overlap with \ac{totalseg}. We split the dataset into training and testing sets with a ratio of 4:1; 
    \item \textbf{\ac{flare}}: The dataset provides the annotations of 13 abdominal organs, which are also included in \ac{totalseg}. 50 training data and 20 testing data are officially released to public; 
    \item \textbf{\ac{kits}}: This is a large dataset specifically collected for kidney \ac{gtv} segmentation, where kidney and its \ac{gtv}s are carefully annotated. In this experiment, since \ac{kits} is simply used to demonstrate the performance of \ac{clnet} on \ac{gtv} segmentation, only kidney \ac{gtv} annotations are used for training while kidney labels are removed. The 300 CT scans are officially divided into 210 training data and 90 testing data. 
\end{itemize}

%% pl setting
\textit{\textbf{\Ac{pl} Setting:}}
% In contrast to the practical \ac{css} setting, where datasets with different target organs are trained sequentially and previous datasets are no longer accessible, \ac{pl} setting requires simultaneous access to all datasets during training. While this is not always feasible in real-world scenario, particularly when a large number of datasets is involved, it is possible to train a unified model with comprehensive and integrated medical pattern and knowledge and may yield better performance than \ac{css} setting. 
We denote the \ac{pl} version of \ac{clnet} trained on five representative public datasets as \ac{cln_u5}. 
Under the \ac{pl} setting, \ac{cln_u5} is trained with all decoder heads across the 5 datasets simultaneously to segment all 137 organs. For training samples from any specific partially labeled dataset, only the decoder heads corresponding to the target organ classes of the dataset are updated. At the final training stage, all decoder heads are pruned to reduce the model's scale. The performance of \ac{cln_u5} is evaluated on the test set of each dataset for its respective target organs, with an overall assessment conducted across all 137 organs. In particular, the performance of the model under the \ac{pl} setting is independent of the order of the dataset, since all datasets are available concurrently. 
Further implementation details of \ac{cln_u5} are provided in Supplementary Sec. \ref{sec:impl_details}. For comparison, we used nnUNet, the most widely adopted and leading method for medical segmentation, as a benchmark. we independently train nnUNet specialists on each of the five datasets and create a 5-nnUNets ensemble (\ac{nnu_e5}) to represent the upper-bound performance. Additionally, we re-implement and evaluate Multi-talent~\cite{ulrich2023multitalent}, a stat-of-the-art \ac{pl} approach for multi-organ segmentation, using the same experimental setup. Furthermore, considering recent advances in large foundation models, we assess the segmentation performance of CLIP-/SAM-based multi-organ segmentation models, including SAT-pro~\cite{Zhao2023one}, VISTA3D~\cite{he2024vista3d}, and BiomedParse~\cite{zhao2024foundation}, by testing their officially released models on the five datasets.

%% css setting
\textit{\textbf{\ac{css} Setting:}}
To comprehensively evaluate the performance of continual segmentation, the rate of forgetting, and the sensitivity of the order of \ac{clnet}, we design four distinct dataset orders for training in the \ac{css} setting. Each order consists of five sequential learning steps, S0 to S4, corresponding to incremental learning of the five datasets. The \ac{css} version of \ac{clnet} trained on the five datasets is denoted as \ac{cln_c5}. 
%% S0 initial step
During the initial training phase (S0), the \acll{ge} of \ac{cln_c5} is pretrained on the initial dataset of the current order to acquire general visual feature extraction capabilities. Concurrently, the decoder heads corresponding to the target organs in the S0 dataset are trained and subsequently pruned. 
%% S1-S4 continual learning steps
In each subsequent continual learning phases (S1-S4), the \acll{ge} and decoder heads for organs not present in the current dataset are frozen to mitigate catastrophic forgetting of previously acquired knowledge. \ac{cln_c5} is sequentially trained and extended on each dataset for new segmentation tasks. For previously unseen organs, new decoder heads are introduced, while existing decoders are updated for overlapping organs, if present. At the end of each step, the added and updated decoders are pruned to reduce the model's parameter size. Importantly, in accordance with the \ac{css} setting, only the current dataset is accessible during training and pruning at each continual learning step, while previously learned datasets remain unavailable. 

%% CSS orders O1-O4
The four CSS learning orders (O1-O4) of the five datasets are outlined at the bottom of Fig. \ref{fig:css} and are carefully designed to reflect various sequences of body parts or anatomical types. Both O1 and O2 begin with the same initial dataset (S0) as \ac{totalseg}, which covers a diverse range of whole-body organs. In O1, the sequence progresses through body parts in the order of head-neck, abdomen, chest, and \ac{gtv}, while O2 reverses this order to \ac{gtv}, chest, abdomen, and head-neck. In contrast, O3 starts with the abdomen dataset, followed by \ac{gtv}, chest, head-neck, and whole-body organs, whereas O4 rearranges the sequence to head-neck, chest, whole-body, \ac{gtv}, and abdomen. 
By training \ac{clnet} under \ac{css} settings with different dataset sequences, we explore the impact of body part and organ type order on segmentation performance. Specifically, the comparative design of O1 and O2 allows us to evaluate the order sensitivity of \ac{cln_c5} when using an identical initial step and shared \ac{ge}. 
For the technical implementation and hyperparameter settings of \ac{cln_c5}, please refer to Supplementary Sec. \ref{sec:impl_details}. 

%% MiB; PLOP; CSCLIP; \ac{nnu_e5} nnUNet (upper bound).
To highlight the advantages of \ac{clnet}, we reimplement three \ac{css} methods -- MiB~\cite{cermelli2020modeling}, PLOP~\cite{douillard2021plop}, and CSCLIP~\cite{zhang2023continual} -- and compare their performance with our model under the same experimental conditions. 
MiB and PLOP are widely used regularization-based \ac{css} benchmarks in both natural and medical imagery domains, employing knowledge distillation to alleviate catastrophic forgetting. CSCLIP, the latest \ac{css} method for continual multi-organ segmentation in the abdomen, mitigating knowledge forgetting through pseudo-labeling of previously learned organs and task prompts from CLIP-based text embeddings of organ names. 
For details on the re-implementation of these comparison methods, please refer to Supplementary Sec. \ref{sec:reimpl_details}.

%% ablation study
\textit{\textbf{Ablation Study:}}
Beyond the evaluation of the performance of the model, the key components of \ac{clnet} are further validated through comprehensive ablation studies. Given that model size and parameter growth rate are the major challenges for architecture-based \ac{css} methods, we evaluate the final performance and the model size of \ac{cln_c5} model without applying the decoder pruning process, referred to as \ac{cln_unprn}. Comparing \ac{cln_c5} with \ac{cln_unprn} demonstrates the effectiveness of decoder pruning in reducing the model size and its potential impact on segmentation accuracy. To investigate other essential components of \ac{clnet}, we pre-train \ac{ge} using different initial datasets and \acll{ssl} methods to explore how pre-training configurations influence the \ac{ge}'s ability to extract general visual semantic features. In addition, the contributions of \ac{fls} and \ac{ema} to improving hard organ segmentation and updating existing decoder heads are examined in detail. For further information on the experimental settings for these ablation studies, please refer to Supplementary Sec. \ref{sec:ablation_ge} and \ref{sec:ablation_dec}.

%% External Validation Exp Setting on 7 external datasets
\subsection{External Evaluation Experimental Setting}
\label{sec:exp_extest}
To further demonstrate the advanced segmentation accuracy, generalizability, and robustness of \ac{clnet}, an external evaluation is performed on one public dataset (BTCV) and six newly collected private datasets, covering whole-body regions, four \acp{gtv}, 43 anatomical classes, and a total of 1,979 testing samples (Tab. \ref{tab:supp_ex_dataset}). Models from the main experiments, including \ac{cln_u36}, \ac{cln_c36}, and \ac{nnu_e36}, are evaluated. Additionally, comparative methods from the \ac{pl} experimental setting in the internal validation, including Multi-talent~\cite{ulrich2023multitalent}, SAT-pro~\cite{Zhao2023one}, and VISTA3D~\cite{he2024vista3d}, are also evaluated on these external datasets. Consistent with internal comparative experiments, SAT-pro and VISTA3D are tested using their official released models, which are limited to segmenting only the organs covered within their predefined target classes. Therefore, we report their performance exclusively for these overlapping organ classes.

%% Evaluation Metrics
\subsection{Evaluation Metrics} 
To assess the segmentation performance of our method, we employ the \acf{dsc} and \acf{asd}, two widely used metrics in organ and \ac{gtv} segmentation studies. DSC measures the overlap between the predicted segmentation mask and ground-truth annotation, serving as a key indicator of volumetric accuracy. In contrast, ASD evaluates the accuracy of the boundary and the shape by quantifying the average distance between the predicted and ground truth surfaces, providing additional insights into structural fidelity. For comprehensive evaluations across entire datasets or organ groups, we report mean DSC (mDSC) and mean ASD (mASD), offering averaged measures of segmentation accuracy and surface agreement across whole-body anatomical structures. These metrics collectively provide a robust measurement for evaluating the precision of segmentation performance.

%% external testing results
%\section{External Evaluation Results of \ac{clnet} on 7 External Datasets}
%\label{sec:supp_extest_res}
%\input{supp/external_test_results}

%% 5-dataset CSS validation results
\section{\acl{cs} Results of \ac{clnet} on Five Representative Public Datasets}
\label{sec:supp_val_res}
%%----- CSS statistical tables here -----%%

%%----- CSS 5 datasets results analysis -----%%
%% 5 datasets CSS results

\subsection{\ac{clnet} outperforms other leading \ac{css} methods across all datasets under various orders.} Fig. \ref{fig:css} illustrates the forgetting curves of \ac{clnet} and other leading \ac{css} methods on each dataset under four orders. Notably, the dice scores of our method are consistently above MiB, PLOP and CSCLIP at every step of each dataset in all orders, with a particularly notable superiority in the final step of most datasets, e.g., \ac{clnet} keeps 94.1\% DSC at step 4 on Order1-S0 while the DSCs of MiB, PLOP and CSCLIP drop to 3.8\%, 16.7\% and 16.5\%, respectively; On Order3-S1 (\ac{kits}), \ac{clnet}'s DSC remains 87.1\% at the last step, in contrast, the other three methods entirely forget the \ac{gtv} segmentation task, with DSCs less than 1.0\%. 
The superior performance of \ac{clnet} can also be supported by the final mean dice scores after the entire \ac{css} training process illustrated ``DSC-Param\#'' plots, in which our \ac{cln_c5} achieves the top mean DSC and remarkably outperforms other comparison \ac{css} methods in all orders. Compared to MiB, PLOP and CSCLIP, \ac{clnet} demonstrates significant improvements in mean DSC by 85.7\%, 73.3\% and 71.6\% on order 1, 77.6\%, 52.5\% and 72.1\% on order 2, 15.1\%, 16.8\% and 16.3\% on order 3, and 72.2\%, 43.9\% and 36.1\% on order 4, respectively. Please refer to supplementary Table~\ref{tab:css_totalseg_organ}-\ref{tab:css_kits} for detailed organ-wise performance of \ac{cln_c5} on all datasets of each order.

\subsection{\ac{clnet} shows quantitative segmentation performance on par with or even surpassing the nnUNet upper bound performance.} As nnUNet performance can be treated as the single model upper bound performance trained on each dataset, the dice score curves of our method (red dashed lines) mostly overlaps with, and sometimes even be above, the nnUNet dice scores (purple dashed lines), which indicates that \ac{clnet} can reach or even exceed the single model upper bound even under \ac{css} setting. \ac{clnet} has very similar (but slightly higher) DSCs as nnUNet in Order1-S0 (\ac{totalseg}) (94.09\% vs. 94.05\%) and Order1-S1 (\ac{structseg}) (86.40\% vs. 86.39\%); in Order3-S0 (\ac{flare}), \ac{clnet} reaches a significantly higher score than nnUNet (91. 3\% vs. 89. 7\%).
The comparable final dice score of \ac{cln_c5} and the 5-nnUNets ensemble (\ac{nnu_e5}) of each order in ``DSC-Param\#'' column further highlights the stable and superior performance of our method. The final \ac{cln_c5} achieves similar or slightly higher mean DSCs as 92.6\%, 92.6\%, 92.7\% and 92.5\% over order 1-4 separately, in comparison to the 92.5\% upper bound dice score of \ac{nnu_e5}. The above results also reveal that our general encoder is able to guarantee top-level segmentation performance even without fine-tuning. 
Please refer to supplementary Table~\ref{tab:css_totalseg_organ}-\ref{tab:css_kits} for detailed organ-wise performance of nnUNet on each dataset.

%% discussion of model stability and plasticity
\subsection{\Ac{clnet} demonstrates performance advantages in both model stability and plasticity.} These two characteristics play significant roles and trade-off in continual learning and segmentation, i.e. a stable model resists catastrophic forgetting, while a plastic model learns better on new tasks. 
%% i) SUNSeg has better stability (non-forgetting)
The advanced model stability of \ac{clnet} is mainly supported by its non-forgetting ability. As illustrated in Fig.\ref{fig:css}, a flat trend is found in the forgetting curves of \ac{clnet} (blue and red) in every dataset under all orders (except the final dataset which is only validated once), which demonstrates that \ac{clnet} has no forgetting on all the previous learned datasets during continual learning. In comparison, the forgetting curves of other CSS leading methods mainly have a descending trend, which reveals a severe catastrophic forgetting drawback of these methods. The performance of MiB, PLOP and CSCLIP shows a general decrease of 86. 1\%, 73. 1\% and 67. 9\% in Order1-S0 and 76. 1\%, 69. 0\% and 55. 5\% in Order4-S0, respectively; severe dice drops are observed at step 3 on S1 in order 2 \& 3 (1-step DSC drop: MiB -- 59.7\% \& 69.1\%; PLOP -- 50.8\% \& 45.4\%; CSCLIP -- 20.1\% \& 33.1\%) and the last step in S3 in all orders (order 1-4, last-step DSC drop: MiB -- 79.8\%, 76.4\%, 56.6\% \& 81.2\%; PLOP -- 24.3\%, 47.9\%, 51.5\% \& 63.8\%). Two cases are found in which the dice scores of the comparison methods increase at some specific CSS steps. MiB, PLOP and CSCLIP resume DSC by 72.4\%, 58.2\% \& 70.8\% at step 4 of S0 (\ac{flare}) and 16.9\%, 6.1\% \& 28.4\% at step 4 of S2 (\ac{segthor}) in order 3, separately. This performance recovery phenomenon is mainly caused by the model learning in \ac{totalseg} at the last step, which contains the same overlapping organs as \ac{flare} at step 0 and \ac{segthor} at step 2, thus helping the model rehearsal and regains dice on these old tasks. 

%% ii) SUNSeg has better plasticity (reach nnunet upperbound, higher dsc in the first step of S1-S4) (generalizability, GE v.s. KD)
On the other hand, \ac{clnet} also has a superior plasticity, which is gained from the generalizability of \ac{ge} and continually incremented decoder heads. 
%Model plasticity improves the model's capability to be transferred to new tasks. 
Although \ac{clnet} freezes its encoder when it continually learns new datasets, the general and robust visual features extracted from \ac{ge} is able to cover most of the body contents in CT scans. Furthermore, the newly added decoder head provides additional parameters for knowledge updating. 
Since the performance of the first training step on each sequential dataset (step 1 on S1, step 2 on S2, step 3 on S3, and the last single step on S4) is %solely depended on model plasticity and
not affected by the forgetting issue, we inspect the model plasticity of \ac{clnet} at the first step of the forgetting curves on S1-S4 within Fig. \ref{fig:css}. As \ac{clnet} reaches the upper bound of nnUNet performance when it is adapted to sequential datasets of each order for the first time, our model demonstrates extraordinary plasticity on various datasets. In comparison, all the other \ac{css} methods remain some gaps to the upper bound at the first training step on S1-S4, specifically, PLOP and CSCLIP show poor plasticity, where large DSC gaps are observed at the first steps in several datasets. PLOP and CSCLIP only get DSCs 59.9\% and 19.2\% on Order1-S4 (\ac{kits}), result in large gaps as 27.3\% and 68.0\% compared to the nnUNet upper bound 87.2\%; Similarly, large gaps to the upper bound are also observed on \ac{kits} in other orders (Order2-S1, Oder3-S1 \& Order4-S3) and other datasets (Order3-S3 \ac{structseg} \& Order4-S2 \ac{totalseg}). 
The deficient plasticity of the comparison methods is mainly caused by their knowledge distillation strategy, which enhances stability, alleviates forgetting but severely reduces the plasticity and makes the model hard for new task adaptation.

%% 2) final dice and param#, SUNSeg has least param size but highest dice score
\subsection{\Ac{clnet} achieves the best overall organ segmentation performance while maintaining significantly tiny model scale and low parameter increasing rate.} Considering model deployment and upgrading issues in the real world scenario, both the prediction accuracy and parameter size should be of great importance for a comprehensive \ac{css} method evaluation. After continually learning the entire sequence of the dataset, the model with a smaller parameter size and higher dice score demonstrates superior performance. Thus, we report the mean dice score for all organs and the number of parameters of the final model of each \ac{css} method %as the vertical and horizontal axes 
in the ``DSC-Param\#'' plots of Fig. \ref{fig:css}, where a better method with a smaller size and higher precision appears in the upper left corner, while the larger model size with a lower dice score in the lower right corner indicates the worse method. As illustrated in the ``DSC-Param\#'' plots, \ac{cln_c5} (red) consistently locates on the most top-left corner over all orders, outperforms other regularization-based CSS methods (MiB, PLOP, CSCLIP) while keeps merely about a single nnUNet size (31.3M), which is much lower than 5-nnUNets ensemble (nnUNet$^E_{5}$) or Transformer-based CSCLIP. Compared with the 92.5\% upper bound dice score and 156.5M model size (five times the size of nnUNet) of nnUNet$^E_{5}$, the final \ac{cln_c5} achieves slightly higher mean DSCs as 92.6\%, 92.6\%, 92.7\% and 92.5\% while limits its parameter number to 28.5M, 28.9M, 34.2M and 35.5M in order 1-4, respectively. The segmentation performance stability of our final model on all CSS orders also demonstrates the robustness of \ac{clnet} to CSS order variations.

%% 3)* Discuss dataset orders and body parts/organ number/organ overlaps

%% 5 datasets pruning ablation

%% ablation study - decoder pruning
\subsection{The decoder pruning process effectively maintains \ac{clnet} performance while efficiently reducing the model size.} In practical model training, the trade-off between model scale and model accuracy, i.e., the scaling-law, must be deliberately designed so that the model avoids holding an extensive cost of physical storage and GPU memory. This is even more significant in the architecture-based \ac{css} method like \ac{clnet}, as our model incrementally expands the decoder head over each new task. Such a considerable parameter increasing rate would rapidly grow the model size during CSS and eventually cause model parameter explosion. Therefore, we develop the decoder pruning process in \ac{clnet} to effectively constrain the parameter increasing rate and the final model scale. 
Fig. \ref{fig:css} illustrates the dice scores of both pruned (\ac{cln_c5}, red) and unpruned (\ac{cln_unprn}, blue) settings to demonstrate the effectiveness of our decoder pruning. As shown in the forgetting curves of all datasets and orders, \ac{cln_c5} remains the similar DSC scores and overlaps the curve with \ac{cln_unprn}. This can be further supported by the overall segmentation performance of the final \ac{cln_c5} and \ac{cln_unprn} models, where \ac{cln_c5} consistently keep the same level of mean DSC and ASD values as \ac{cln_unprn}, with a marginal average increase in DSC of 0. 045\% and an ASD reduction of 0.002mm in all orders. Moreover, the final parameter size in ``DSC-Param\#'' plot underscores the robust and effective model compression capability of our decoder pruning, where more than 60\% parameter reductions are observed across all orders. Compared to 93.9M parameters of the unpruned \ac{clnet}, the modal size of each order after pruning is trimmed to 33.84$\pm$3.83\% of original size on average. 
%% ablation study - GE, EMA, FLS
%% \item briefly review the results of \ac{ge} (a large portion in method, however have less impact on main results - go to supp); in \ac{css}, \ac{ge}s are \textbf{trained according to the first dataset} in each \ac{css} order (the \ac{ge} are all trained in \ac{ssl} + fine-tuning fashion), the capacity of \ac{ge} v.s. final performance and pruning rate.
%% \ac{ema} impacts on organs that co-exist in multiple datasets, detailed -- supp
%% FL-support, improve dice score -- KiTS21 supp
To further validate the effectiveness of other main modules/processes of \ac{clnet} and explore their impacts on final performance, we conduct comprehensive ablation studies on \ac{ge}, \ac{ema} and \ac{fls}. The detailed experimental results and analyses are reported in the supplementary Sec. \ref{sec:ablation_ge} \& \ref{sec:ablation_dec}. 

%% Universal/PartialLabel results

%% clnet implementation
\section{Details on \ac{clnet} Implementation}
\label{sec:impl_details}
% % data preprocessing
% %  -- hyper-parameters
% % data postprocessing
% % training + inference time
\subsection{Image Preprocessing} \ac{clnet} adapts its general training framework from the powerful nnUNet~\footnote{\url{https://github.com/MIC-DKFZ/nnUNet}}, leveraging its robust preprocessing and training strategies. For input preparation, a CT windowing range of $[-1,024, 1,024]$ HU is applied uniformly to all CT images, followed by resampling to a consistent resolution of $1.0\times 1.0\times 1.5$ mm to ensure spatial uniformity. The images are pre-processed to the RAI (Right-Anterior-Inferior) orientation before undergoing the pre-processing pipeline, which includes intensity normalization, ROI cropping, and outlier removal. While nnUNet dynamically adjusts the input patch size for each task, \ac{clnet} enhances this approach by introducing a standardized default patch size of $128\times96\times112$ voxels (or a user-defined size), providing a balanced solution that combines consistency with task-specific flexibility.

\subsection{Network Architecture} The architecture backbone of \ac{clnet} is built on the \textit{3D full resolution} configuration of nnUNet, tailoring its design to provide robust performance and fine-grained feature extraction. It features a 6-block encoding path and a 5-block decoding path, ensuring a deep and hierarchical representation of the input data. The base number of features for the general encoder (\ac{ge}) starts at 32, following a systematic expansion of the features as the network deepens. Each \textit{encoding} block comprises two convolutional layers, both incorporating instance normalization and leaky ReLU activation for efficient feature transformation and gradient flow. Downsampling between encoding blocks is performed via stride convolution, which reduces spatial dimensions while preserving critical feature information. The \textit{decoding} blocks also contain two convolutional layers with instance normalization and leaky ReLU activation. In addition, each decoding block integrates an \ac{ema} (Exponential Moving Average) module to enhance feature refinement and temporal stability in training. Upsampling within the decoding path is handled by transposed convolution, which restores spatial resolution while enabling learnable upscaling of feature maps. The default \ac{ge} configuration of the $[[1, 3, 3], [3, 3, 3], [3, 3, 3], [3, 3, 3], [3, 3, 3], [3, 3, 3]]$ convolution kernels. For each decoding path, we recommend a mirroring of the \ac{ge} convolutional kernels. Additionally, \ac{clnet}, like nnUNet, supports optional automatic kernel selection for enhanced flexibility. 

% A rule-based adaptation governs the decoder's base feature configuration, balancing model capacity with computational efficiency.

\subsection{Evaluation Metrics} The \ac{dsc}, \ac{asd}, and pruning rate $\mathcal{T}$ are used for evaluation. To better understand the implications of the pruning rate $\mathcal{T}$, we review U-shaped CNN networks, focusing on the learned features of downstream decoding heads. Typically, better segmentation performance with fewer predicted targets correlates with a higher pruning rate, indicating a denser distribution of features in the latent space. In contrast, performing segmentation on difficult-to-predict organs or inferring a large number of output labels may result in a lower pruning rate.

\subsection{Hardware Requirement} All experiments were performed on a server with $8$ Nvidia V100 GPUs, each with $16$ GiB of RAM, $330$ GiB of CPU memory, and an Intel(R) Xeon(R) Platinum \textit{8163} CPU, featuring $82$ threads. We have tested that all training tasks can be performed using a single GPU with a standard $12$ GiB RAM and $64$ GiB CPU memory. The default number of threads for training and testing is set to $12$. It is preferred to use \ac{ddp} to leverage all available GPUs for efficient computation. 

\subsection{Training Efficiency} The average training time for a two-target head (e.g., left and right counterparts) in \ac{clnet} is approximately 86 seconds per epoch on a single Nvidia V100 GPU, which can be accelerated to $32$ seconds per epoch using a \ac{ddp} setup with four Nvidia V100 GPUs. The average GPU memory consumption is less than $10$ GiB during the training of each decoding head. Each trained decoding head has a model size of approximately $80$ MiB, which can be reduced to just $18$ MiB when saved as a sparse model using PyTorch's sparse format~\footnote{\url{https://pytorch.org/docs/stable/sparse.html}}. Please note that sparse computation was not applied during model training. Hence, GPU RAM consumption remains unchanged during the training process. The data augmentation pipeline is implemented with the public available `BatchGenerator' framework~\footnote{\url{https://github.com/MIC-DKFZ/batchgeneratorsv2}}. 

%  Yet, mature of the sparse package available, we can easily adapt \ac{clnet} to sparse computation in the future. 
\subsection{Inference Efficiency} The \ac{clnet} inference supports single- and multi-GPU setups, with the aim of minimizing deployment costs, improving prediction efficiency, and streamlining clinical workflows. In a single GPU setup, segmentation is performed sequentially, with each decoding head processed one at a time. In a multi-GPU setup, decoding heads are evenly distributed across GPUs to accelerate the process. The inference time on a single Nvidia V100 GPU is averagely less than $5$ seconds per stratified head. The total inference time for our 235-organ segmentation on a single V100 GPU averages 430 seconds per patient, comprising approximately 10s for image pre-processing, 5s for model loading (only once), 400s for inference, 5s for merging and saving, as well as additional system overhead. We strongly suggest using Solid-State Drive (SSD) to store the training data, as SSD improves basic input/output system (BIOS) speed and facilitates faster inter-communication with CPU and GPU memories. When performing parallel predictions using 4 or 8 GPUs, the total inference time can be reduced to approximately 118 or 58 seconds per patient, respectively. The average GPU memory consumption is approximately $6$ GiB of RAM throughout the entire prediction process.

% When equipped with high-end GPUs (e.g., sm80), with an average pruning rate of 97.48\%, TorchSparse~/cite{tang2023torchsparse++}.
\subsection{\ac{ge} Training Using Comprehensive Dataset}
The network architecture is based on the encoder of the nnUNet's default ``3D full resolution'' architecture. The base number of features of \ac{ge} is $32$, and each encoding convolutional block consists of $2$ convolutional layers. Similarly to nnUNet, the maximum number of features (toward the bottleneck block) is capped at $320$. Instance normalization and leaky ReLU are also applied. All decoding paths share the same architecture, consisting of two convolutional layers, with transposed convolution used to upscale the output features of each decoding block. To enhance training effectiveness, side supervisions are applied throughout the model training process. We follow nnUNet's training strategy and use the SGD optimizer for network training, starting with an initial learning rate of $0.01$. The learning rate is adjusted using a polynomial schedule. The batch size is set to $2$. This design ensures the adaptability and scalability of the network while maintaining an efficient memory footprint, requiring less than $16$ GiB of RAM for \ac{ge} and three decoding paths.

\subsection{\ac{ge} Training Using Constrained Datasets}
\label{sec:ge_train}
In addition to leveraging comprehensive CT datasets for training, we enhance the generalized encoder \ac{ge}'s training capability in scenarios where only limited training data (e.g., MRI or PET imaging) or a restricted variety of training classes (e.g., disease-specific imaging) are available. To achieve effective and efficient GE pre-training, we employ \ac{ssl}, which enables \ac{ge} to extract meaningful knowledge from constrained datasets. Furthermore, the domain gaps~\cite{zhang2023dive} between \ac{ssl} and semantic segmentation can be mitigated by fine-tuning the pre-trained \ac{ssl} network using semantic labels. Specifically, we begin by pre-training only the encoder component of the network using the SimSiam framework~\cite{chen2021exploring}. This approach employs two \ac{ge}s:  predictor encoder $f_e^p$ and target encoder $f_e$. These encoders iteratively refine data representations without explicit labels. The predictor generates augmented representations of the input data, while the target encoder learns robust features by ensuring consistency between different augmentations of the same input. For the pre-training process, we adapted the nnUNet data augmentation scheme for view augmentation. Let $\mathcal{A}$ denote the view augmentation function. We follow SimSiam and use negative cosine similarity loss $\mathcal{L}_{ncos}$ for training. The pseudo-code for proposed \ac{ge} pre-training is detailed in Algorithm~\ref{alg:simsiam}. Once pre-training with SimSiam is completed, the target encoder $f_e$ is then paired with a randomly initialized decoding path to minimize the domain gaps between \ac{ssl} and segmentation tasks via fine-tuning. To better preserve previously learned features for various body parts, motivated by MoCo~\cite{he2020momentum}, we employ a momentum queue to store a large number of feature representations of \ac{ge}. To improve efficiency, the momentum queue is detached from the GPU and set to a total size of $1,024$, significantly reducing CPU, GPU, and memory usage while maintaining training speed. This strategy ensures efficient and scalable \ac{ge} training, even in resource-constrained environments.

\begin{algorithm}
\caption{SimSiam-based General Encoder Pre-training}
\KwIn{Dataset $D=\{X_i, Y_i\}_{i=1}^n$, Target GE $f_e$, Predictor GE $f_e^p$}
\KwOut{Trained target GE $f_e$}
\For{$i \gets 1$ \KwTo $n$}{
    $X_i^1, X_i^2 \gets \mathcal{A}(X_i), \mathcal{A}(X_i)$\;
    $Z_i^1, Z_i^2 \gets f_e(X_i^1), f_e(X_i^2)$\;
    $P_i^1, P_i^2 \gets f_e^p(Z_i^1), f_e^p(Z_i^2)$\;
    $\mathcal{L} \gets \frac{1}{2} \mathcal{L}_{ncos}(P_i^1, Z_i^2) + \frac{1}{2} \mathcal{L}_{ncos}(P_i^2, Z_i^1)$\;
}
\label{alg:simsiam}
\end{algorithm}

\textbf{\ac{ssl} Pre-training Setup} 
We adopt SimSiam for \ac{ge} pre-training. The implementation of SimSiam is based on the publicly released code~\footnote{\url{https://github.com/facebookresearch/simsiam?tab=readme-ov-file}}. Unlike SimSiam for natural images, our pre-training augmented views do not include large resizing and cropping or head-to-foot flipping. These augmentations could eliminate small targets or confuse the model with respect to CT anatomical structures. We adapt the nnUNet data augmentation scheme for view augmentation:
\begin{itemize}
    \item Rotation and scaling: Scaling and rotation transformations are applied with a triggering probability of 50\% for scaling, 50\% for rotation, and 50\% for both scaling and rotation. The rotation angles are constrained within the range of $[-45, 45]$ degrees and applied sequentially to the directions from left to right, front to back, and head to foot. The PyTorch's~\footnote{\url{https://pytorch.org/}} tri-linear and nearest neighbor interpolations are applied to the image and label, respectively. The scaling factor is sampled in a random ratio between $[0.7, 1.4]$ for all patch types.
    \item Gaussian noise \& blur. For each voxel, the Gaussian noise with a mean $\mu=0$ and a random standard deviation $\sigma \in [0, 0.1]$ is added. The Gaussian blur with kernel $\sigma \in [0.5, 1.5]$ is applied to the input image. Blurring is triggered with a probability of $50\%$ per sample.
    \item Intensity re-scaling. The intensity of each voxel is multiplied by a factor between $[0.65, 1.5]$. The CT intensity values are clipped to a range of $[-1,024, 1,024]$. 
    \item Simulation of low resolution. The sampled 3D image patches are first down-sampled by a factor between $[1, 2]$ on the axial or xy plane and a random factor between $[2, 4]$ along the head-to-foot direction using nearest neighbor interpolation. Then, those patches are sampled back to their original size using tri-linear interpolation. This augmentation is triggered with a probability of $50\%$.
    \item Gamma augmentation. The Gamma augmentation is the same as the default nnUNet's Gamma augmentation. 
    \item Mirroring. All patches are only mirrored with a probability of $50\%$ along the axial and coronal axes. 
\end{itemize}
The initial learning rate is set to $1\times 10^{-3}$. The cosine annealing strategy is adopted to schedule the learning rate. The total training epoch is $5,000$ with $256$ mini-batches. 

\textbf{Momentum Multi-dataset Fine-tuning Setup}
The \ac{ge} momentum queue is stored independently of the GPU and is configured to a size of $1,024$. The default \ac{ge} momentum queue $\mathcal{Q}_{1,024}$ requires approximately 90 GiB of CPU memory. During fine-tuning on multiple datasets, the paired decoder \textit{ -- used exclusively for \ac{ge} fine-tuning and usually exhibiting inferior/less optimized performance -- } is re-initialized for each dataset, and the output convolutional layer's channel dimension is adjusted to match the number of output classes for each specific dataset. Each dataset is trained for $1,000$ with mini-batch sizes of $256$. The fine-tuning learning rate for \ac{ge} is set to $1\times 10^{-4}$, while the learning rate for the paired decoder is set to $1\times 10^{-3}$. Data augmentation follows the nnUNet ``more data augmentation'' scheme. 

% \subsubsection{\ac{ge} Training Conclusion} 
% When a comprehensive dataset is available, we recommend training the \ac{ge} directly using such datasets, e.g., TotalSegmentator. This approach is preferable because the additional resource consumption required for \ac{ssl} pre-training and momentum fine-tuning may outweigh the modest performance gains achieved. By directly utilizing a comprehensive dataset, the training process becomes more efficient without severely compromising the quality of the segmentation results. However, when only limited training data or a restricted variety of training classes are available, we recommend adopting \ac{ssl} pre-training and momentum fine-tuning for \ac{ge} training. These strategies enhance the \ac{ge}'s ability to generalize and improve performance, making it more robust and effective in scenarios with constrained data availability.

\subsection{Stratified Decoder Training Details}
\label{sec:stratified-decoder}
The decoding path, with or without \ac{fls}, is determined on the basis of the user's design. We recommend parsing target anatomies into stratified structure-based decoders, such as the Eyes decoder, Liver decoder, and Ribs decoder. Since training from scratch and fixing \ac{ge} can occasionally lead to large updates of the decoder in the early stages of training~\cite{gotmare2018closer}, we initially warm up the decoder and then train the decoder until convergence, followed by \ac{lth}-based weight pruning. Once the decoding path is pruned, the \ac{ema} component of the decoding path is updated. Inherited from nnUNet, the data augmentation is nnUNet ``more data augmentation'' scheme. We recommend training the decoding paths using \ac{ddp} implementations, as it significantly accelerates the training process.

% We default warm up the decoder for 5 epochs （with an initial learning rate of $1\times 10 ^{-3}$) and trained for total 500 epochs 
% The base number of features for the decoding path is configured on-the-fly by \ac{clnet} according to the decoder's target class of the task at hand. Rule-based dynamic adaptation is configured as follows:
% \begin{itemize}
%     \item If the target classes $\leq5$, decoder's base number of features is $16$.
%     \item If the target classes $\in (5,10]$, decoder's base number of features is $24$.
%     \item If the target classes $> 10$, decoder's base number of features is $32$.
% \end{itemize}
\textbf{Decoding Path Initial Training \& Pruning Setup}
By default, we warm up the decoder for five epochs with an initial learning rate of $1\times 10 ^{-3}$. Following this, the momentum is reset and the decoder's learning rate is adjusted to $1\times10^{-2}$ for training over 300 epochs. Then, the proposed \ac{lth}-based pruning process is applied iteratively, where each cycle of training, pruning, and rewinding progressively identifies a smaller winning ticket. Heuristically, most stratified decoders target only two organs, such as left and right counterparts, leading us to assume that the weights within these decoding heads are sparsely distributed. We suggest starting with the progressive overall pruning rates ranging from $80\%$ to $92\%$, with an interval of $4\%$. If the performance drop criteria ($\delta \le 1\%$) are still met, the pruning continues from $93\%$ to $99\%$, with an interval of $2\%$. If the criteria continue to hold, the pruning proceeds from $99.1\%$ to $99.7\%$, with intervals of $0.2\%$. The PyTorch built-in ``global-unstructured'' pruning with the ``L1Unstructured'' method is adopted throughout all pruning processes. The pruning process is applied to all convolution layers, targeting both weights and biases for pruning. To smooth the pruning process, we linearly increase the mini-batches per epoch, up to $512$, in proportion to the pruning rate. After pruning is complete, the number of minibatches per epoch is reset to 256. To mitigate any potential performance drop, additional training epochs $20$ are applied. 
 
\textbf{\ac{ema} Module Updating Configuration}
Most of \ac{ema} updating begins at the start of training. However, during the pruning process, the \ac{ema} updating might experience significant performance fluctuations. The assumed reason is that, with a decay rate of $\alpha=0.999$ and a mini-batch size for pruning of $256$, the \ac{ema} update process is effectively 1,000 times slower than its non-smoothed counterpart. This indicates that \ac{ema} could have only updated $25.6\%$ of its parameters compared to the unpruned state, making it difficult for the weights in the \ac{ema} module to timely adapt to the newly pruned state. Although enlarging the size of training mini-batches or using finer pruning steps could address this issue, it would lead to extended training times and increased resource consumption. To address this issue in \ac{clnet}, after pruning is complete, we copy the pruned counterpart to \ac{ema} and subsequently update only the unpruned weights in the \ac{ema} module.

\textbf{Continuous Decoder Updating}
Benefiting from the \ac{ema} updating, the decoder of the target organs can be continuously and smoothly updated. We recommend using a smaller learning rate (e.g., $1\times10^{-4}$) to update both the non-smooth decoder and the \ac{ema} counterpart. To maintain consistency, it is recommended that the pruned mask remain unchanged. However, the pruned mask can optionally be updated when using new data. Instead of applying the pruning process to the non-smoothed counterpart, we perform the pruning process directly on the \ac{ema} parameters using a refined pruning interval of $0.1\%$. During model inference, the weights are loaded from the \ac{ema} module to leverage the benefits of \ac{ema} smoothing.

% is shared across multiple datasets (e.g., the \textit{Liver} decoder is shared in both TotalSegmentator and FLARE22 datasets), 

% \ac{ema} updating is highly recommended but optional based on the user's design. We cannot conduct pruning directly on the \ac{ema} module. 

% Following this routine, most organ segmentation performance of the \ac{ema} module shows little impact when the pruned mask is directly applied to the \ac{ema} weights.

%% general encoder results and discussion
\section{Details on Ablation Studies of General Encoder (\ac{ge})}
\label{sec:ablation_ge}
The primary objective of \ac{ge} is to extract representative features that are applicable in a wide range of potential downstream tasks. To achieve this, it is essential for \ac{ge} to maintain a high degree of comprehensiveness, ensuring versatility and robustness in feature representation. Based on our experiments, we have identified two pivotal factors that profoundly influence the performance of downstream tasks: 1) the comprehensiveness of the training datasets and 2) the necessity of the popular self-supervised pre-training. To evaluate these factors, for \ac{ge} training, we conducted comprehensive validation experiments using the following dataset configurations: 1) A single comprehensive dataset -- \ac{totalseg}. 2) A single head \& neck region dataset -- \ac{structseg}. 3) The union of separate body part-wise datasets $\mathcal{U}_4$ -- \ac{structseg} (head \& neck), \ac{segthor} (thoracic), \ac{flare} (abdomen), and \ac{kits} (kidney \ac{gtv}). 4) The union of all five $\mathcal{U}_5$ datasets. Once \ac{ge} is fully trained using the aforementioned four \ac{ge} training setups, the trained \ac{ge}s are fixed and subsequently utilized for downstream validation experiments. Two downstream tasks were selected for evaluation:  1) \ac{totalseg} and 2) FLARE22 downstream tasks, which provided whole-body segmentation and local (abdomen) region segmentation, respectively. Consistent with the previously determined decoding path configuration,, the \ac{totalseg} downstream task includes 48 stratified decoders, and the FLARE22 downstream task includes 11 stratified decoders. We also calculated the cumulative sum of the complements, $\mathcal{T}^{\mathlarger{\prime}} = 1-\mathcal{T}$, of the pruning rates in the target evaluation dataset $\sum\mathcal{T}^{\mathlarger{\prime}}$. This sum represents the percentage of the total size of the decoder models after pruning compared to the default unpruned counterpart, $\sum\mathcal{T}^{\mathlarger{\prime}}=100\%$, with smaller values indicating more efficient pruning.

\subsection{Impact of Training Datasets}
The experiments designed to evaluate the impact of training datasets in two major respects:
\begin{itemize}
    \item \textit{Image-wise Body Part Coverage}. It determines the overall range of anatomical structures present in the images, impacting the \ac{ge}'s ability to recognize and generalize across different body regions. A dataset with a broad coverage of body parts ensures that \ac{ge} can identify a wide array of anatomical characteristics, facilitating robust learning in downstream tasks.
    
    \item \textit{Label Completeness}. It refers to the variety of anatomical structures annotated within the dataset. This diversity is crucial for training the \ac{ge} to distinguish and extract relevant features from different anatomical regions. A diverse set of labeled structures enhances \ac{ge}'s capacity to learn distinctions between various tissues and organs, thus improving its segmentation capabilities across a range of downstream tasks.

    % \item \textit{Label-wise body part coverage} focuses on the specific anatomical regions labeled within each image, influencing the \ac{ge}'s ability to extract features from local regions. 
    
    % \item \textit{Label-wise body part coverage} focuses on the specific anatomical regions labeled within each image, influencing the \ac{ge}'s ability to extract features from local regions. This aspect is essential for refining the performance of downstream tasks, particularly in detailed, focal segmentation tasks. Comprehensive label-wise coverage ensures that the downstream decoders can effectively segment specific areas, providing precise delineations of structures within those regions.
\end{itemize}
For completeness, we follow the same \ac{ge} training setup discussed in Section~\ref{sec:ge_train} and conducted experiments with both the \ac{ge} pre-trained by SimSiam and the \ac{ge} trained from scratch. 

\subsection{Image-wise Body Part Coverage}
The experiments are designed for two primary comparisons: 1) to assess how well a \ac{ge} trained in whole-body images generalizes to different body parts, and 2) to determine the effectiveness of a \ac{ge} trained in localized body parts when applied to other anatomical regions. When training \ac{ge} from scratch, as shown in Tab~\ref{tab:impact_ge_totalseg}, \ac{ge} trained using a whole-body dataset achieved \ac{dsc} of $94.1\%$, \ac{asd} of $0.89$mm, and $\sum\mathcal{T}^{\mathlarger{\prime}}$ of $38.6\%$. When evaluated on the abdominal downstream task, as shown in Tab~\ref{tab:impact_ge_flare}, where the whole-body dataset-trained \ac{ge} yielded the \ac{dsc} of $90.4\%$, \ac{asd} of $1.05$ mm, and $\sum\mathcal{T}^{\mathlarger{\prime}}$ of $15.2\%$. In contrast, training \ac{ge} with the head and neck region dataset led to a decrease in performance on the TotalSegmentator downstream task, with a \ac{dsc} of $91.1\%$. Furthermore, this model showed a higher \ac{asd} of $1.20$ mm, indicating less precise segmentation boundaries, and a significantly larger $\sum\mathcal{T}^{\mathlarger{\prime}}$ of $72.4\%$, suggesting the need for compensation in the extraction of features for the \ac{ge} compared to the model trained with the whole-body dataset. Similar performance trends were observed when evaluating experiments using \ac{ge} pre-trained using SimSiam. Using a whole-body dataset for training, it yielded a \ac{dsc} of $94.2\%$, \ac{asd} of $0.90$mm and$\sum\mathcal{T}^{\mathlarger{\prime}}$ of $38.2\%$. However, when only head \& neck region dataset is available, the performance is decreased to a \ac{dsc} of $92.8\%$, \ac{asd} of $1.14$mm, and the$\sum\mathcal{T}^{\mathlarger{\prime}}$ of $62.4\%$.

To assess how well a \ac{ge} trained in specific body parts can generalize to mostly unseen regions, as demonstrated in Tab~\ref{tab:impact_ge_flare}, we evaluated the head and neck dataset trained \ac{ge} on the abdominal downstream task. In particular, there is minimal overlap between the head and neck dataset and the abdomen FLARE22 dataset. In this setup, \ac{clnet} exhibited degraded performance, with a \ac{dsc} of $89.5\%$, an \ac{asd} of $1.28$ mm, and $\sum\mathcal{T}^{\mathlarger{\prime}}$ of $26.0\%$. Using SimSiam to pre-train \ac{ge} could help downstream decoders improve segmentation performance, achieving a \ac{dsc} of $90.1\%$, an \ac{asd} of $1.20$ mm, and $\sum\mathcal{T}^{\mathlarger{\prime}}$ of $20.3\%$. The inferior performance could suggest that the model struggled to accurately segment abdominal structures, likely because its previous exposure was limited to head and neck anatomies. The lack of overlap in anatomical structures between the training and evaluation datasets can lead to ineffective adaptation of learned characteristics to new and distinct abdominal anatomies.

\subsection{Label Completeness}
We assessed the impact of the diversity of the labels by training \ac{ge} from scratch using the whole-body dataset (restricted to only the bone label set). This approach limited the model's exposure to a variety of anatomical structures, focusing exclusively on bone-related organs. When evaluated in the downstream TotalSegmentator task, \ac{clnet} showed a noticeable decline in performance. It led to a \ac{dsc} of $92.9\%$, \ac{asd} of $1.13$mm, and $\sum\mathcal{T}^{\mathlarger{\prime}}$ of $58.8\%$. Most organ segmentation tasks showed less effective performance, whereas bone stratified decoders achieved minor yet significant improved segmentation performance ($+0.2$\% \ac{dsc} and $-0.06$mm \ac{asd}, $p<0.1$), suggesting that \ac{ge} trained using the bone label set was optimized biased towards bone anatomy feature extraction. Using multiple datasets from separate body parts, called $\mathcal{U}_4$ data sets, for \ac{ge} training can effectively enhance the coverage of body parts and increase the variety of anatomical labels. To fully exploit all available labels, the most straightforward approach is to complement incomplete label sets with pseudo-labels. This approach resulted in a \ac{dsc} of $94.0\%$, \ac{asd} of $1.09$ mm, and $\sum\mathcal{T}^{\mathlarger{\prime}}$ of $56.8\%$. Although this performance shows improvement over using the head and neck label trained \ac{ge}, it remains inferior to the \ac{ge} trained with the whole-body dataset. This observed sub-optimal performance may be attributed to the error propagation in pseudo-labels.

%\subsection{Impact of Self-supervised Pre-training}

\subsection{Impact of View Augmentations}
To exclude the effects introduced by domain adaptation from the analysis of view augmentation schemes, we compared two sets of experimental setups on the TotalSegmentator downstream task: 
\begin{itemize}
    \item Evaluating the impact of view augmentation schemes by directly applying the SimSiam pre-trained \ac{ge} to the training of downstream task decoding heads without any fine-tuning.
    \item Evaluating the impact of view augmentation schemes by applying the SimSiam pre-trained \ac{ge} with full label set fine-tuning to the training of downstream task decoding heads. 
\end{itemize}
The experimental results are shown in Tab~\ref{tab:impact_ge_totalseg}. In the \textit{first} setup. Using the default view augmentation, it produced the least effective segmentation results. The \ac{clnet} achieved a \ac{dsc} of $88.6\%$, a \ac{asd} of $1.73$mm, and $\sum\mathcal{T}^{\mathlarger{\prime}}=328.6\%$. In contrast, with the proposed view enhancement, the SimSiam pre-trained \ac{ge} produced improved results on the same task, achieving a \ac{dsc} of $89.6\%$, an \ac{asd} of $1.28$ mm, and $\sum\mathcal{T}^{\mathlarger{\prime}}=315.8\%$. In the \textit{second} setup. A similar performance trend was observed. Using the default view augmentation, \ac{clnet} achieved an inferior \ac{dsc} of $93.0\%$, an \ac{asd} of $1.17$ mm, and $\sum\mathcal{T}^{\mathlarger{\prime}}=52.4\%$. In contrast, with the proposed view augmentation, \ac{clnet} achieved an improved \ac{dsc} of $94.2\%$, an \ac{asd} of $0.90$mm, and $\sum\mathcal{T}^{\mathlarger{\prime}}=38.2\%$. After carefully examining the results, we observed that the degradation in performance was accompanied by an increase in false-positive predictions. This issue highlighted the model's tendency to misidentify large organs as smaller structures or to make incorrect predictions, such as classifying parotid glands as kidneys or making errors in head-to-foot direction identification.

\subsection{Impact of Supervised Domain Adaptations}
The domain gap between\ac{ssl} pre-training and supervised semantic segmentation remains a challenge in medical imaging segmentation tasks~\cite{zhang2023dive}. We begin by first analyzing the conventional ``pre-train then fine-tune'' workflow domain adaptation approach for domain gap minimization. As demonstrated in Tab~\ref{tab:impact_ge_totalseg}, when directly applying SimSiam pre-trained \ac{ge}, it predicted a \ac{dsc} of $89.6\%$, an \ac{asd} of $1.28$ mm, and $\sum\mathcal{T}^{\mathlarger{\prime}}$ of $315.8\%$. Notably, simply fine-tuning \ac{ge} boosted performance significantly to a \ac{dsc} of $94.2\%$, an \ac{asd} of $0.90$ mm. Notably, $\sum\mathcal{T}^{\mathlarger{\prime}}$ dropped dramatically, reaching as low as $38.2\%$, which is $8.5$ times smaller than the $\sum\mathcal{T}^{\mathlarger{\prime}}$ observed when the \ac{ge} was not fine-tuned. This substantial improvement indicates that fine-tuning helps the model better adapt to specific tasks by refining its understanding of the target anatomical features. 

% This substantial increase indicates a marked decrease in model pruning and suggests that the decoders, beyond just learning features for semantic segmentation, are also trained in a manner that attempts to bridge the domain gaps between \ac{ssl} and segmentation.  

% there is a significant performance gap between the fine-tuned \ac{ge} and the one without fine-tuning. W

Fine-tuning \ac{ge} is a straightforward and effective approach to minimize the domain gap between \ac{ssl} and supervised semantic segmentation. However, in clinical practice, it is common to collect multiple separate datasets that aim for various types of diagnosis. While SimSiam-based pre-training does not depend on semantic labels and using multiple datasets can help cover more body parts, issues may arise when sequentially fine-tuning \ac{ge} on multiple individual or partially labeled datasets. This approach might lead to knowledge forgetting or over-fitting, where the \ac{ge} fails to retain knowledge across diverse datasets or becomes overly specialized in the most recent dataset on which it was trained. Using \ac{ema} can stabilize fine-tuning and enhance feature robustness by mitigating abrupt gradient changes. However, it may also smooth out task-specific features, such as those that distinguish head organs from the abdomen, potentially leading to a loss of distinct features. To address the issues of knowledge forgetting and dataset over-fitting, the momentum queue~\cite{he2020momentum} is adopted. This technique involves maintaining a dynamic memory bank of feature representations, which allows the model to retain information from earlier datasets while learning from new ones. By continuously updating the queue with representations from the most recent data, the model can effectively balance the retention of diverse knowledge, ensuring that it remains robust and generalizable across various medical imaging tasks. 

To evaluate the effectiveness of the momentum updating strategy, we evaluate the segmentation performance using $\mathcal{U}_4$ with an order of StructSeg19 $\rightarrow$ SegTHOR $\rightarrow$ FLARE22 $\rightarrow$ KiTS21. We conducted experiments using \ac{ema} with $\alpha=0.999$ for \ac{ge} fine-tuning. Furthermore, to understand the impact of the momentum queue size, we conducted three separate experiments with queue sizes of 128, 1,024, and 4,096. For comparison, when sequentially fine-tuning \ac{ge} using $\mathcal{U}_4$ datasets, the segmentation results showed a \ac{dsc} of $93.0\%$, \ac{asd} of $1.14$ mm, and $\sum\mathcal{T}^{\mathlarger{\prime}}$ of $62.6\%$. When using \ac{ema} for updating and following the same fine-tuning dataset sequence, we observed only very minor gains or comparable performance compared to sequential fine-tuning of the full label set. Notably, when tested on the FLARE22 downstream task, the \ac{ema} approach yielded inferior segmentation performance compared to the sequential full label set fine-tuning. The assumed reason for this is that the KiTS dataset, being the largest and last dataset used for fine-tuning, could have overly influenced \ac{ge}, leading to a model that is more specialized for kidney-related tasks and less effective in generalizing to the wider range of anatomical features required for the FLARE22 task.

When adopting the proposed momentum updating, we find that the smaller queue size, $\mathcal{Q}_{128}$, is still prone to knowledge forgetting, as small sized queue may not store enough representations from previous datasets, leading to $93.1\%$ in \ac{dsc}, $1.12$ mm in \ac{asd}, and $58.6\%$ in$\sum\mathcal{T}^{\mathlarger{\prime}}$. Momentum updating \ac{ge} with larger queue sizes leads to better performance. Using momentum queue size of $1,024$, \ac{clnet} with $\mathcal{Q}_{1,024}$ achieved $94.2\%$ in \ac{dsc}, $0.86$ mm in \ac{asd}, and $37.2\%$ in$\sum\mathcal{T}^{\mathlarger{\prime}}$. However, larger queue sizes entail longer training times and increased memory consumption. Specifically, for momentum queue sizes of $\mathcal{Q}_{1,024}$ and $\mathcal{Q}_{4,096}$, the memory consumption is approximately 90 GB and 380 GB of CPU memory, respectively. The training time when using a momentum queue size of $\mathcal{Q}_{4,096}$ is approximately $6\times$. Nevertheless, the additional performance gains achieved with the momentum queue of $\mathcal{Q}_{4,096}$ are marginal, yielding an improvement of only $+0.03\%$ in \ac{dsc}, $-0.01$ mm in \ac{asd}, and $-0.3\%$ in $\sum\mathcal{T}^{\mathlarger{\prime}}$. A similar segmentation trend was also observed when evaluating the FLARE22 downstream task, with results resulting in a \ac{dsc} of $90.4\%$, \ac{asd} of $1.04$mm, and $\sum\mathcal{T}^{\mathlarger{\prime}}$ of $15.2\%$. To evaluate the added value of incorporating multiple separate datasets into a comprehensive dataset when training \ac{ge}, we conduct experiments using the five datasets $\mathcal{U}_5$. The momentum update with a queue size of $1,024$ is adopted. Unfortunately, the results showed only marginal improvements, despite achieving the best performance in both the TotalSegmentator and FLARE22 segmentation downstream tasks. The results included a \ac{dsc} of $94.3\%$ ($+0.1\%$), \ac{asd} of $0.85$ mm ($-0.05$ mm), and $\sum\mathcal{T}^{\mathlarger{\prime}}$ of $35.8\%$ ($-2.4\%$) on the TotalSegmentator downstream task, as well as a \ac{dsc} of $90.5\%$ ($+0.03\%$), \ac{asd} of $1.04$ mm ($-0.01$ mm), and $\sum\mathcal{T}^{\mathlarger{\prime}}$ of $14.9\%$ ($-0.03\%$) on the FLARE22 downstream task. 

\subsection{Discussion on \ac{ge} }
In conclusion, when a comprehensive dataset is available, we recommend training the \ac{ge} directly using such datasets, e.g., TotalSegmentator. This approach is preferable because the additional resource consumption required for \ac{ssl} pre-training and momentum fine-tuning may outweigh the modest performance gains achieved. Using a comprehensive dataset directly, the training process becomes more efficient without severely compromising the quality of the segmentation results. However, when only limited training data or a restricted variety of training classes are available, we recommend adopting \ac{ssl} pre-training and momentum fine-tuning for \ac{ge} training. Given the balance between segmentation performance and resource consumption, we heuristically recommend a queue size of $1,024$. These strategies enhance the \ac{ge}'s ability to generalize and improve performance, making it more robust and effective in scenarios with constrained data availability. If training \ac{ge} from scratch is necessary, we strongly suggest using multiple GPUs with substantial CPU memory, e.g., a server system setup that includes 8 Nvidia V100 GPUs of 16GB RAM each and 64GB of CPU memory. The learned decoding features can serve both feature extraction and feature decoding purposes~\cite{tian2019decoders}. However, these two types of characteristics often exhibit distinguished distributions in the latent space~\cite{zhang2023dive}. For simplicity, assuming a constant level of decoder optimization difficulty, we hypothesize that a better optimized \ac{ge} may lead decoders to place greater emphasis on feature decoding, thereby allowing the learned features to cluster more effectively with improved sparsity in the latent space. Therefore, when training the same stratified decoder using different \ac{ge}s, a less pruned decoding head might indicate that the decoding head compensates for sub-optimal feature extraction from a less optimized \ac{ge}.

%% Decoder head results and discussion
\section{Details on Ablation Studies of Multi-path Decoding Heads}
\label{sec:ablation_dec}

In the \ac{clnet} framework, trainable decoding heads are introduced for subsequent downstream tasks after completing and fixing the \ac{ge} training. By independently training each decoding path, this approach mitigates knowledge forgetting and ensures that each task-specific model retains its learned features. We conducted experiments to assess the impacts of \ac{ema}, \ac{fls}, and the fine-tuning of the \ac{ge} on the downstream tasks. These tasks include TotalSegmentation, StructSeg19, SegTHOR, FLARE22, and KiTS datasets. The training order for the tasks follows `Order 1': TotalSegmentator $\rightarrow$ StructSeg19 $\rightarrow$ FLARE22 $\rightarrow$ SegTHOR $\rightarrow$ KiTS21. Please note that the StructSeg19 downstream task does not share any decoders with the other tasks. The \ac{ge} is pre-trained using SimSiam and later momentum-updated using $\mathcal{U}_5$ datasets. The baseline stratified decoders of the \ac{clnet} is trained without \ac{ema} update or \ac{fls}. 

% For those shared stratified decoders across different tasks, decoder fine-tuning was performed with a learning rate of $1\times10^{-4}$. During the fine-tuning process, the pruning mask is updated according to the same pruning criteria if a higher pruning rate is observed in the fine-tuning task. 

% TotalSeg	StructSeg	FLARE22	SegTHOR	KiTS

% In consideration of clinical practices, during the stratified decoder training for downstream tasks, three factors are identified as having great impacts: continuously updating stratified decoders using 
% \ac{ema} \ac{fls}

\subsection{Impact of Fine-tuning \ac{ge}}
As with most universal models, it is common practice to fine-tune the entire network to adapt to the downstream tasks. An important question arises: Despite the risk of catastrophic forgetting previously learned knowledge, does fine-tuning the entire \ac{clnet} guarantee upper bound segmentation performance? This fine-tuning process involves adjusting the model's parameters based on the characteristics and requirements of the new task, allowing it to specialize and optimize performance.  We conducted ablation experiments by making apple-to-apple comparisons between the segmentation performance of models with fixed \ac{ge} parameters and those with fine-tuned \ac{ge} parameters. Specifically, both \ac{ge} and stratified decoders were individually fine-tuned on each downstream task. The learning rate for the fine-tuning of the entire network is $1\times10^{-4}$, the total fine-tuning epoch is set to 300. 

As shown in Tab~\ref{tab:impact_decoder}, fine-tuning \ac{ge} led to improved overall pruning rates for downstream tasks. However, when evaluating segmentation performance, the quantitative results were comparable to those achieved without fine-tuning the whole \ac{clnet}. This suggests that while fine-tuning enhances model efficiency by increasing pruning rates, it does not necessarily translate into significantly better segmentation accuracy. However, for some tasks, fine-tuning the whole framework even degrades performance. For example, in the downstream task of StructSeg19, the fine-tuning \ac{ge} resulted in a decrease of $0.6\%$ in \ac{dsc} and an increase of $0.03$ mm in \ac{asd}. Upon closer examination of the predictions, we found that the segmentation of optical chiasm and optical nerves showed the most significant degradation, with an increase in false negatives. This degradation might be attributed to the fine-tuning process causing overfitting to a relatively small dataset, where the \ac{ge} could have become too specialized to the training set.

\subsection{Impact of \ac{ema}}
In clinical practice, when new images are collected, it is desirable to promptly and efficiently update the decoding head to incorporate the latest data. The \ac{ema} updating is an easy and powerful tool to stabilize training and mitigate overfitting to newly collected data. To assess the impact of \ac{ema}, we evaluated each downstream task with and without the \ac{ema} updating component. When using \ac{ema} for updating, we followed the same training setup discussed in Section~\ref{sec:stratified-decoder} for updating \ac{ema} component and pruning mask. In contrast, when the \ac{ema} component is not used, the decoder is trained sequentially for each task. As demonstrated in Tab~\ref{tab:impact_decoder}, exploiting \ac{ema} module for stratified decoder training consistently yielded better performance compared to training without it. We compared the performance gaps of the five downstream tasks: 
\begin{itemize}
    \item TotalSegmentator: $+0.1\%$ in \ac{dsc}, $-0.1$ mm in \ac{asd}, and $-1.9\%$ in $\sum\mathcal{T}^{\mathlarger{\prime}}$,
    \item StructSeg:  $+0.2\%$ in \ac{dsc} and $-2.5\%$ in $\sum\mathcal{T}^{\mathlarger{\prime}}$,
    \item SegTHOR: $+0.4\%$ in \ac{dsc} and $-0.3$ mm in \ac{asd},
    \item FLARE22:  $+0.5\%$ in \ac{dsc}, $-0.11$ mm in \ac{asd}, and $-2.8\%$ in $\sum\mathcal{T}^{\mathlarger{\prime}}$,
    \item KiTS21: $+1.6\%$ in \ac{dsc}, $-0.07$ mm in \ac{asd}, and $-2.0\%$ in $\sum\mathcal{T}^{\mathlarger{\prime}}$.
\end{itemize}

It was observed that a substantial portion of the CT images in the FLARE22 dataset contain kidney \ac{gtv}. The \ac{ema} module helps stabilize the training process by averaging the model parameters over time, reducing the impact of sudden changes and fluctuations in the decoder weights. This stabilization was particularly noticeable in the performance gains achieved with \ac{ema} in the segmentation of the kidney \ac{gtv} using the KiTS21 dataset. This approach mitigates over-fitting and helps the decoders retain knowledge across multiple tasks, leading to improved generalization and accuracy.

\subsection{Impact of \ac{fls}}
For demonstration purposes, only the kidney \ac{gtv} decoder is provided with \ac{fls} using the learned features of the kidney decoder, as shown in Tab~\ref{tab:impact_decoder}. Leveraging supporting organs for difficult organ segmentation can reduce optimization difficulty and potentially provide spatial context, which helps to reduce false positives. This approach allows the model to utilize anatomical relationships and spatial information from nearby structures, improving the accuracy and consistency of segmentation, especially for challenging cases. The conventional approach to leveraging easy organ predictions for segmenting harder anatomies involves concatenating the predictions of the easier organs as additional input channels. However, this method requires adjusting the number of input channels, which inevitably necessitates fine-tuning both the \ac{ge} and the decoding paths. This process can be resource-intensive and may introduce complexities in maintaining the model's generalization capabilities across various tasks. Nevertheless, we conducted experiments using predicted kidney masks as an additional channel input for kidney \ac{gtv} segmentation. In contrast, the decoder with \ac{fls} was trained using the fixed \ac{ge}. 

When training directly the kidney \ac{gtv} decoder using the KiTS21 dataset, it showed a \ac{dsc} of $84.2\%$, \ac{asd} of $1.18$ mm, and $\sum\mathcal{T}^{\mathlarger{\prime}}$ of $6.0\%$. We observed an significant ($P<0.01$) improvement by channel-wisely concatenating predicted kidney masks for the kidney \ac{gtv} segmentation, demonstrating $+1.7\%$ in \ac{dsc}. In comparison, when using \ac{fls} for training, it demonstrated $+2.1\%$ in \ac{dsc}. A similar performance trend was observed, but a further improved prediction was observed when using \ac{ema} for updating the decoding path, leading to the \ac{dsc} of $86.4\%$ vs. $86.9\%$, \ac{asd} of $1.14$ mm vs. $1.12$ mm. The predictions of the diseased kidney were found to contain errors, particularly in cases with very enlarged \ac{gtv}s located on top (head-direction) of the kidneys. The assumed reasons for the inferior performance compared to the model using \ac{fls} could be: 1) including \ac{ge} for fine-tuning, which may lead the model to capture noisy distributions and potentially overfitting to these noise patterns, and 2) unlike binarized predictions, providing support at the feature level can offer a smoother representation of the data. Feature-level support allows the model to leverage more information, which may help to better capture subtle variations and complexities in the data, thus improving the segmentation performance.

To further evaluate the impact of \ac{fls}, we conducted an ablation study using LNS datasets, which include 18 head \& neck LNSs and 15 chest LNSs. The performance of \ac{cln_c36} and \ac{cln_u36} were compared to their counterparts without \ac{fls}, denoted as \ac{cln_c36_nofls} and \ac{cln_u36_nofls}, respectively (Tab. \ref{tab:abl_fls_lns}). For head and neck LNS segmentation, the supporting structures included Musc\_Cervial\_A, Musc\_Scalenus, Musc\_Scalenus\_A, and Musc\_Scleido, while for chest LNS segmentation, the supporting structures were A\_Aorta, Bronchus\_L, Bronchus\_R, Eso, and Heart~\cite{guo2021deepstationing}. The results revealed performance declines in both \ac{cln_c36} and \ac{cln_u36} when \ac{fls} was excluded, with a 1.9\% decrease in DSC and a 0.08 mm increase in ASD in 33 LNS classes. These findings demonstrate that \ac{fls} enhances the segmentation of small and challenging structures such as LNSs, potentially by leveraging additional localization information derived from supporting organ features.

\subsection{Discussion on Stratified Decoder}
When a comprehensively pre-trained \ac{ge} is available, we recommend directly utilizing it for the training of downstream tasks without fine-tuning \ac{ge}. For optimal results, it is recommended to train the decoding paths with \ac{ema} and \ac{fls} to enhance stability and performance. Under an 8 Nvidia V100 GPUs \ac{ddp} setup, the average training time for a two-class head (left and right counterparts) in \ac{ddp} nnUNet is approximately $192$ seconds per epoch with mini-batch sizes of $256$. Using the froth of \ac{ge} during decoding path training and dynamically adapting the decoder's base number of features, the average training time is significantly reduced to approximately $32$ seconds per epoch using the same system setup. Furthermore, for a 22 output class (e.g., StructSeg19) decoding head, the average training time of \ac{ddp} nnUNet is approximately $242$ seconds, and the average training time of \ac{clnet} is approximately $43$ seconds.

%% CSS related works & MiB, PLOP, CSCLIP, (SAT, BiomedParse) reimplementation
\section{Related Work and Reimplementation of Comparison Methods}
\label{sec:reimpl_details}
%%%%%% overall training %%%%%%%
For all comparison methods, we start with the same pre-trained nnUNet model on the TotalSegmentator dataset, which has been trained using the 3D nnUNet setting for 8000 epochs, with 250 iterations per epoch and initial learning rate 0.01. After that, the model is fine-tuned sequentially on continual segmentation tasks (ChestOrgan, HNOrgan and EsoOrgan), where each dataset is fine-tuned for 500 epochs, with 250 iterations per epoch and initial learning rate 0.005. All other nnUNet settings, such as data augmentation, remain the same as our implementation. Moreover, since our segmentation datasets/tasks are 3D CT scans (different from the previous continual segmentation works in natural images), adjustments to these comparison methods are required (extending 2D methods to 3D), as well as transferring their implementations to the nnUNet framework. We describe the detailed re-implementation of previous methods, especially our modifications, in the following subsections.

%in order to make sure the model has completely converged on all 103 organs and the encoder has enough ability to extract general features.

%%%%%%%%% MiB %%%%%%%%%%%
\subsection{MiB}
MiB~\cite{cermelli2020modeling} proposes two marginal losses, or unbiased losses to solve the background shift issue in continual segmentation in their original paper: unbiased cross-entropy (UNCE) loss, which merges the probabilities of old classes to the background label, and unbiased knowledge distillation (UNKD) loss, which merges the probabilities of all new classes (belonging unseen classes of the old model) to the background label. Notice that the original unbiased loss assumes that new classes from the current dataset are completely disjoint with all the old classes, however, this assumption is not holding in our datasets. For example, TotalSegmentator and ChestOrgan contain four overlapping organs: inferior vena cava, trachea, esophagus, and pulmonary artery. Therefore, in order to reimplement MiB losses in the nnUNet framework and make them compatible with our datasets, we slightly modifies and generalizes both unbiased losses to handle overlapping labels in the continual learning setting. The modified UNCE loss is as follows: 

\begin{equation}
    \ell^{\theta^t}_{ce}(x,y) = -\frac{1}{\lvert \mathcal{I} \rvert}\sum_{i\in \mathcal{I}}\log \tilde{q}_x^t (i,y_i)
\end{equation}

\noindent where:

\begin{equation}\label{eq:unce}
    \tilde{q}_x^t (i,c) = 
    \begin{cases}
        q_x^t(i,c) &\text{if $c \neq b$}\\
        \sum_{k\in Y^{t-1} \textcolor{red}{\setminus C^t_{p}}} q_x^t(i,k) &\text{if $c = b$}
    \end{cases}
\end{equation}

\noindent Here, the same notation as that in the original article is used, except $C^t_{p}=Y^{t-1} \cup C^t - b$, which indicates the overlapping classes (excluding the background label) between the current data set $C^t$ and all previous classes $Y^{t-1}$ at the learning step $t$. When calculating the UNCE loss, we merge all the old labels in the background except the overlapping classes. 

Similarly, we adapt UNKD loss as follows: 

\begin{equation}
    \ell^{\theta^t}_{kd}(x,y) = -\frac{1}{\lvert \mathcal{I} \rvert}\sum_{i\in \mathcal{I}} \sum_{c \in Y^{t-1} \textcolor{red}{\setminus C^t_p}} q_x^{t-1}(i,c) \log \hat{q}_x^t (i,y_i)
\end{equation}

\noindent where:

\begin{equation}\label{eq:unkd}
    \hat{q}_x^t (i,c) = 
    \begin{cases}
        q_x^t(i,c) &\text{if $c \neq b$}\\
        \sum_{k\in C^t} q_x^t(i,k) &\text{if $c = b$}
    \end{cases}
\end{equation}

\noindent In the above formula, overlapping organs from the old class set are excluded so that the knowledge distillation works on the real old classes that cannot be learned from the current dataset. 

Using two modified losses, we always train the model with the latest labels and ignore the previously learned overlapping labels when overlapping organs occur. Thus, overlapping labels are trained directly using the cross-entropy loss and merged to the background in the knowledge distillation loss. In addition, we use the same hyperparameters as the MiB setting: the weights of UNKD loss are set as 10 with balanced classifier initialization strategy.

%%%%%%%%% PLOP %%%%%%%%%%%
\subsection{PLOP}
PLOP~\cite{douillard2021plop} was originally implemented for 2D images, especially its local distillation loss on multiple scales based on local POD. Local POD is a multi-scale feature pooling strategy consisting of computing width- and height-pooled slices on multi-scale regions, which aims to better retain both global and local spatial knowledge from the old model. However, since our data are all 3D CT scans with an additional depth dimension, we specifically extend the local POD to higher dimensions when reimplementing the method. Two pooling strategies can be adopted for 3D cases: 1) Pooling 3D feature map along each single dimension and extracting three 2D projections along each axis: 

\begin{align}
    \begin{split}
    \Phi(\rvx)=
    &\left( \frac{1}{H}\sum_{h=1}^{H} \rvx\left[ h,:,:,: \right] \middle\| \frac{1}{W}\sum_{w=1}^{W} \rvx\left[ :,w,:,: \right] \right. \\ 
    &\left. \middle\| \frac{1}{D}\sum_{d=1}^{D} \rvx\left[ :,:,d,: \right] \right) \in R^{(WD + HD + HW) \times C}
    \end{split}
\end{align}

\noindent where notations follow the original PLOP paper. This pooling method can preserve enough spatial information while providing some level of plasticity to the model. 2) Pooling 3D feature map on two dimensions and only extract 1D projection along the remaining axis: 

\begin{align}
    \begin{split}
    \Phi(\rvx)=
    &\left( \frac{1}{HW}\sum_{h=1}^{H}\sum_{w=1}^{W} \rvx\left[ h,w,:,: \right] \middle\| \frac{1}{WD}\sum_{w=1}^{W}\sum_{d=1}^{D} \rvx\left[ :,w,d,: \right] \right. \\ 
    &\left. \middle\| \frac{1}{HD}\sum_{h=1}^{H}\sum_{d=1}^{D} \rvx\left[ h,:,d,: \right] \right) \in R^{(H+W+D) \times C}
    \end{split}
\end{align}

\noindent This pooling strategy has similar feature shape; however, when pooling on two axes together, most of the spatial information is lost and POD loss cannot retain the former knowledge. After comparing the performance using two strategies, we select the former, which better handles the trade-off between model rigidity and plasticity. For {\bf hyperparameters}, the original paper uses the pod weighting factor of 0.01, which is too large for the 3D pooling case. The L2 norm of 3D pooled features is more than 10 times larger than that of 2D pooled features. In our experiments, we set this pod factor to 0.001. Other hyperparameters are consistent with those used in the original paper.

%%%%%%%%% CSCLIP %%%%%%%%%%%
\subsection{CSCLIP}
The original CSCLIP~\cite{zhang2023continual} empowers a continual multi-organ segmentation framework with pseudo-labels, lightweight three-layer class-specific heads and Contrastive Language–Image Pretraining (CLIP)~\cite{radford2021learning} text embeddings to mitigate catastrophic forgetting. Since the authors have made their source code public available, we can directly implement and adapt their method to our datasets and tasks. We would like to mention several adjustments of the code: (1) The original paper contains only 38 target organs and their corresponding organ name text embeddings. In order to match our 5-dataset \ac{css} experiments including 137 target organs under different orders, we replace the original CSCLIP text embeddings with our order-specific 137 text embeddings generated by the CLIP text encoder. (2) Similarly, we increase the number of lightweight class-specific heads to 137 to fully cover our organ classes. (3) Inference of the original code is fully processed in GPU memory, which is limited to 80GB on NVIDIA A100. This may result in an ``out of memory'' error when the input 3D full-body CT image has a large size exceeding a certain threshold. To prevent such a memory overflow issue, we equally partition the input scan into three segments with a 20\% overlap along the axial direction, predict each separately and subsequently concatenate them to restore the original size.

\pagebreak

%% Supplementary Table Container
\section*{Supplementary Figures and Tables}
\label{sec:supp_table}

%---------- Fig. Organ-wise DSC (U36, C36, nnE36) ------------%
\begin{figure*}[htbp]
\renewcommand{\figurename}{Fig.}
    \centering
    \includegraphics[width=\textwidth]{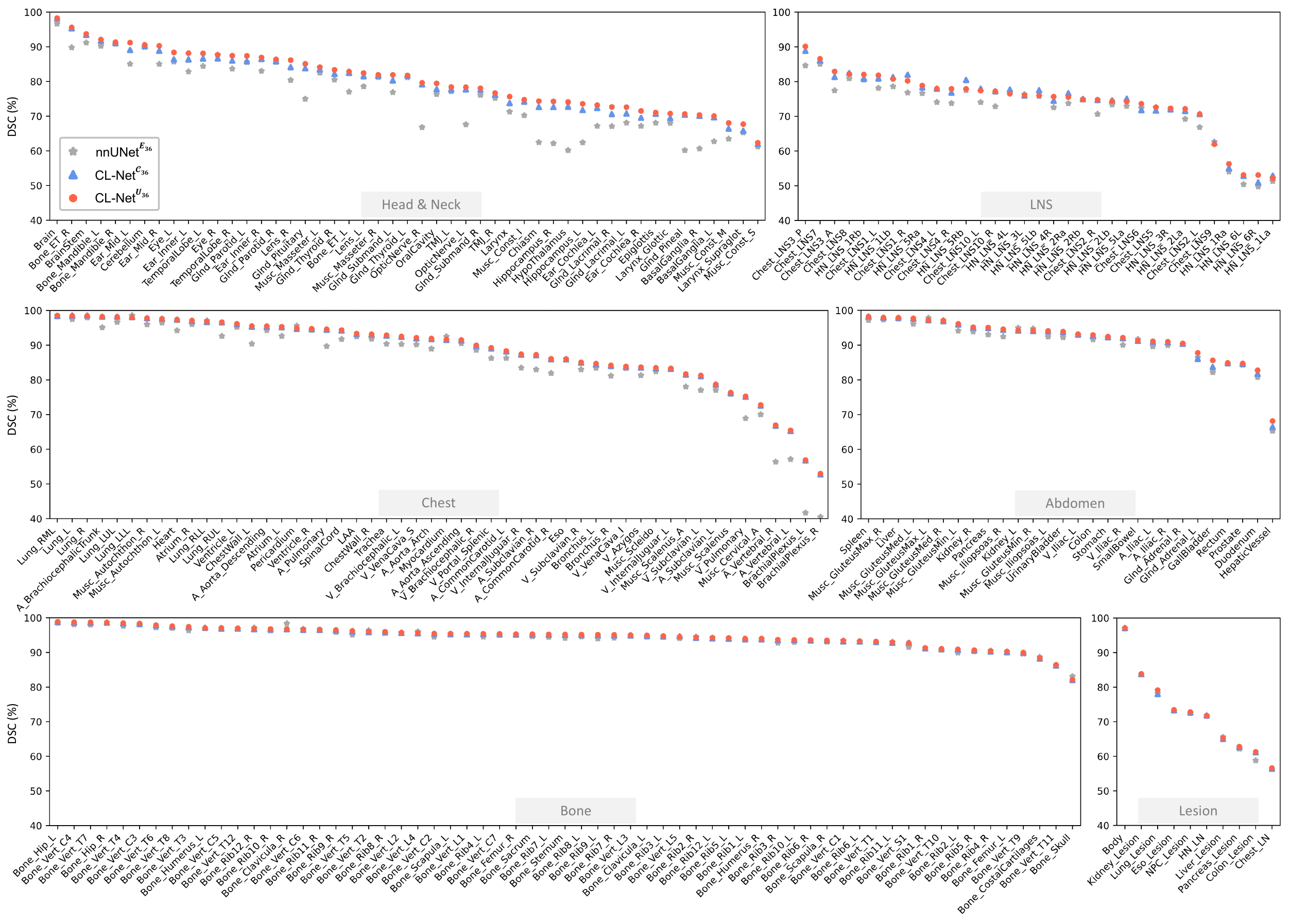}
    \caption{\textbar~\textbf{Organ-wise segmentation comparison of \ac{cln_u36}, \ac{cln_c36}, \ac{nnu_e36}.} DSC (\%) of \ac{orgn} individual anatomies are illustrated. Anatomies are grouped as Head \& Neck (49 organs), Chest (52), Abdomen (28), Bone (63), LSN (33), \Ac{gtv} (9), and Body (shown in \Ac{gtv} panel). }
    \label{fig:organ-dsc}
\end{figure*}

%---------- Fig S. Organ-wise Boxplot ------------%
\begin{figure*}[htbp]
\renewcommand{\figurename}{Fig.}
    \centering
    \includegraphics[width=0.9\textwidth]{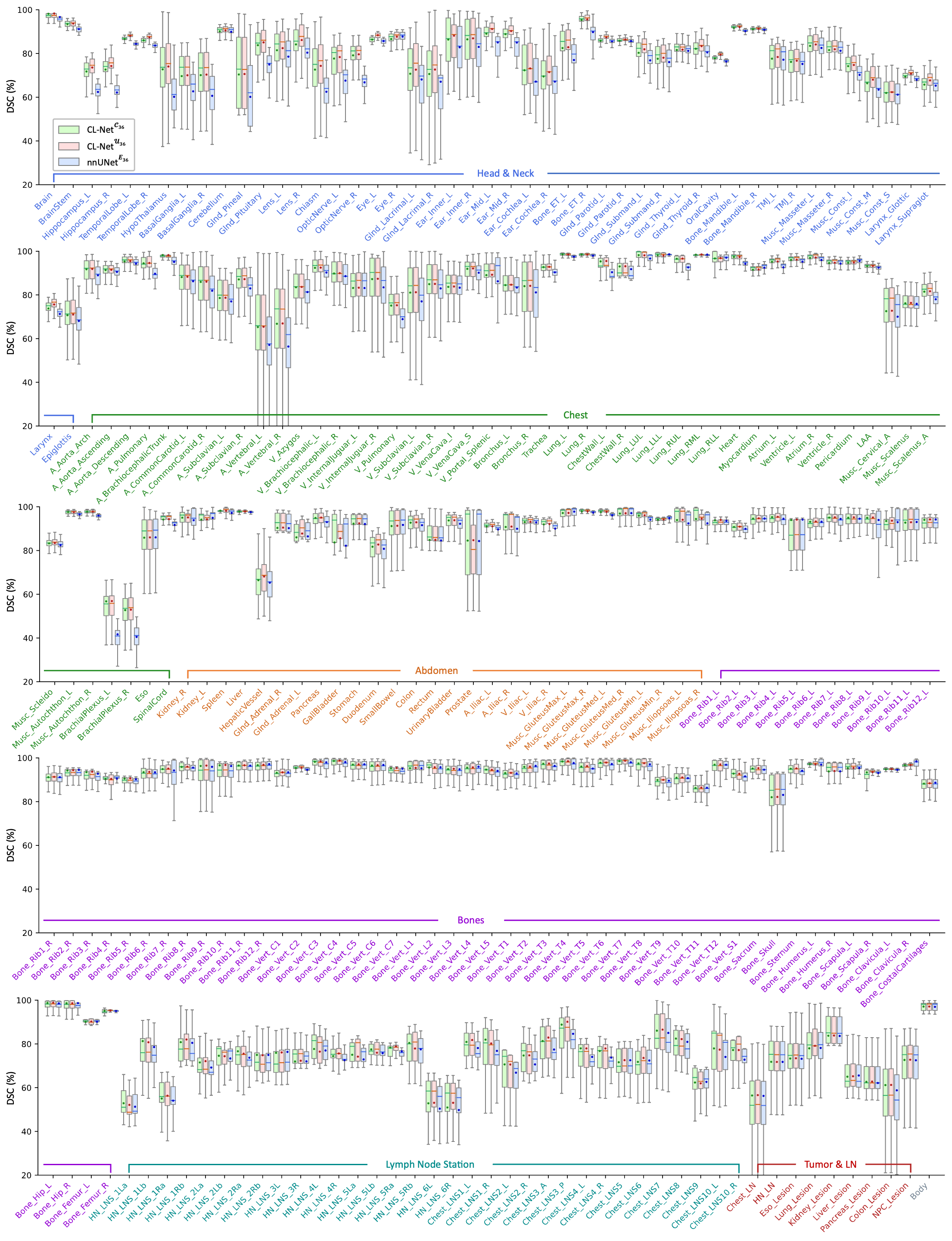}
    \caption{\textbar~The segmentation dice score (\%) of our method \ac{cln_u36} and \ac{cln_c36}, and the leading segmentation model \ac{nnu_e36} over all \ac{orgn} organs. Organs are grouped and colored by body parts or organ types. %The red, green and blue box plots demonstrate SUNSeg CSS, SUNSeg PL and nnUNet, separately.
    }
    \label{fig:organ-box}
\end{figure*}

%% Dataset Details
\begin{table}[htp]
\centering
\resizebox{\textwidth}{!}{%
\begin{tabular}{@{}|lccccccc|@{}}
\toprule

\multicolumn{1}{|l|}{{\textbf{Dataset Name}}} &
  \multicolumn{1}{c|}{{\textbf{Organ\#}}} &
  \multicolumn{1}{c|}{{\textbf{Train\#}}} &
  \multicolumn{1}{c|}{{\textbf{Test\#}}} &
  \multicolumn{1}{c|}{{\textbf{\begin{tabular}[c]{@{}c@{}}Median Size\\ (XYZ)\end{tabular}}}} &
  \multicolumn{1}{c|}{{\textbf{\begin{tabular}[c]{@{}c@{}}Median Spacing\\ (XYZ)\end{tabular}}}} &
  \multicolumn{1}{c|}{{\textbf{\begin{tabular}[c]{@{}c@{}}Image BPR Range\\ (Bottom-Top)\end{tabular}}}} &
  {\textbf{\begin{tabular}[c]{@{}c@{}}Label BPR Range\\ (Bottom-Top)\end{tabular}}} \\ \midrule\midrule

\multicolumn{8}{|c|}{\textbf{Private Datasets}} \\ \midrule

\multicolumn{1}{|l|}{{\acs{hn-zju-15}~\cite{guo2021comprehensive, zhu2024low}}} &
  \multicolumn{1}{c|}{{15}} &
  \multicolumn{1}{c|}{{58}} &
  \multicolumn{1}{c|}{{14}} &
  \multicolumn{1}{c|}{{512x512x128}} &
  \multicolumn{1}{c|}{{0.98x0.98x3.00}} &
  \multicolumn{1}{c|}{{0.59-1.00}} &
  {0.62-1.00} \\
  
\rowcolor[HTML]{F3F3F3} 
\multicolumn{1}{|l|}{{\acs{hn-cgmh-42}~\cite{guo2020organ, guo2021comprehensive}}} &
  \multicolumn{1}{c|}{{42}} &
  \multicolumn{1}{c|}{{114}} &
  \multicolumn{1}{c|}{{28}} &
  \multicolumn{1}{c|}{{478x312x284}} &
  \multicolumn{1}{c|}{{1.00x1.00x1.25}} &
  \multicolumn{1}{c|}{{0.61-1.00}} &
  {0.60-1.00} \\

\multicolumn{1}{|l|}{{\acs{ch-zju-lungcan-12}~\cite{guo2021deepstationing, liu2021same}}} &
  \multicolumn{1}{c|}{{12}} &
  \multicolumn{1}{c|}{{78}} &
  \multicolumn{1}{c|}{{20}} &
  \multicolumn{1}{c|}{{512x512x67}} &
  \multicolumn{1}{c|}{{0.68x0.68x5.00}} &
  \multicolumn{1}{c|}{{0.06-0.79}} &
  {0.38-0.82} \\
  
\rowcolor[HTML]{F3F3F3} 
\multicolumn{1}{|l|}{{\acs{ch-zju-esocan-35}~\cite{guo2021deepstationing,yan2022sam}}} &
  \multicolumn{1}{c|}{{35}} &
  \multicolumn{1}{c|}{{154}} &
  \multicolumn{1}{c|}{{38}} &
  \multicolumn{1}{c|}{{512x512x80}} &
  \multicolumn{1}{c|}{{0.68x0.68x5.00}} &
  \multicolumn{1}{c|}{{0.39-0.92}} &
  {0.37-0.92} \\

\multicolumn{1}{|l|}{{\acs{ch-zju-rtog-13}~\cite{guo2021deepstationing,zhu2024low}}} &
  \multicolumn{1}{c|}{{13}} &
  \multicolumn{1}{c|}{{304}} &
  \multicolumn{1}{c|}{{76}} &
  \multicolumn{1}{c|}{{512x512x80}} &
  \multicolumn{1}{c|}{{0.68x0.68x5.00}} &
  \multicolumn{1}{c|}{{0.02-1.00}} &
  {0.38-0.87} \\
  
\rowcolor[HTML]{F3F3F3} 
\multicolumn{1}{|l|}{{\acs{ab-td-5}}} &
  \multicolumn{1}{c|}{{5}} &
  \multicolumn{1}{c|}{{45}} &
  \multicolumn{1}{c|}{{11}} &
  \multicolumn{1}{c|}{{768x768x104}} &
  \multicolumn{1}{c|}{{0.65x0.65x3.00}} &
  \multicolumn{1}{c|}{{0.00-0.32}} &
  {0.00-0.35} \\

\multicolumn{1}{|l|}{{\acs{body-linkmed-1}}} &
  \multicolumn{1}{c|}{{1}} &
  \multicolumn{1}{c|}{{800}} &
  \multicolumn{1}{c|}{{201}} &
  \multicolumn{1}{c|}{{512x512x122}} &
  \multicolumn{1}{c|}{{0.98x0.98x3.00}} &
  \multicolumn{1}{c|}{{0.00-1.00}} &
  {0.00-1.00} \\
  
\rowcolor[HTML]{F3F3F3} 
\multicolumn{1}{|l|}{{\ac{hn-lns-18}~\cite{li2024semi}}} &
  \multicolumn{1}{c|}{{18}} &
  \multicolumn{1}{c|}{{93}} &
  \multicolumn{1}{c|}{{23}} &
  \multicolumn{1}{c|}{{512x512x129}} &
  \multicolumn{1}{c|}{{0.98x0.98x3.00}} &
  \multicolumn{1}{c|}{{0.61-1.00}} &
  {0.72-0.93} \\

\multicolumn{1}{|l|}{{\ac{ch-lns-15}~\cite{guo2021deepstationing, guo2022thoracic}}} &
  \multicolumn{1}{c|}{{15}} &
  \multicolumn{1}{c|}{{149}} &
  \multicolumn{1}{c|}{{37}} &
  \multicolumn{1}{c|}{{512x512x78}} &
  \multicolumn{1}{c|}{{0.68x0.68x5.00}} &
  \multicolumn{1}{c|}{{0.38-0.92}} &
  {0.51-0.78} \\
  
\rowcolor[HTML]{F3F3F3} 
\multicolumn{1}{|l|}{{\acs{hn-zju-npc-1}~\cite{zhang2023multi}}} &
  \multicolumn{1}{c|}{{1}} &
  \multicolumn{1}{c|}{{207}} &
  \multicolumn{1}{c|}{{52}} &
  \multicolumn{1}{c|}{{512x512x137}} &
  \multicolumn{1}{c|}{{0.98x0.98x3.00}} &
  \multicolumn{1}{c|}{{0.59-1.00}} &
  {0.88-0.96} \\

\multicolumn{1}{|l|}{{\acs{ch-wpy-lungcan-1}~\cite{wang2023anatomy,wang2024accurate}}} &
  \multicolumn{1}{c|}{{1}} &
  \multicolumn{1}{c|}{{952}} &
  \multicolumn{1}{c|}{{239}} &
  \multicolumn{1}{c|}{{512x512x63}} &
  \multicolumn{1}{c|}{{0.78x0.78x5.00}} &
  \multicolumn{1}{c|}{{0.31-0.84}} &
  {0.60-0.65} \\
  
\rowcolor[HTML]{F3F3F3} 
\multicolumn{1}{|l|}{{\acs{ch-wpy-esocan-2}~\cite{li2023lvit}}} &
  \multicolumn{1}{c|}{{2}} &
  \multicolumn{1}{c|}{{1132}} &
  \multicolumn{1}{c|}{{283}} &
  \multicolumn{1}{c|}{{512x512x94}} &
  \multicolumn{1}{c|}{{0.98x0.98x5.00}} &
  \multicolumn{1}{c|}{{0.16-0.97}} &
  {0.50-0.76} \\

\multicolumn{1}{|l|}{{\acs{ab-livercan-2}~\cite{yan2023liver}}} &
  \multicolumn{1}{c|}{{2}} &
  \multicolumn{1}{c|}{{2815}} &
  \multicolumn{1}{c|}{{704}} &
  \multicolumn{1}{c|}{{512x512x51}} &
  \multicolumn{1}{c|}{{0.74x0.74x5.00}} &
  \multicolumn{1}{c|}{{0.00-0.81}} &
  {0.35-0.53} \\
  
\rowcolor[HTML]{F3F3F3} 
\multicolumn{1}{|l|}{{\acs{hn-fd-ln-1}~\cite{yu2024slice, yu2025effective}}} &
  \multicolumn{1}{c|}{{1}} &
  \multicolumn{1}{c|}{{195}} &
  \multicolumn{1}{c|}{{47}} &
  \multicolumn{1}{c|}{{512x512x57}} &
  \multicolumn{1}{c|}{{0.49x0.49x5.00}} &
  \multicolumn{1}{c|}{{0.64-1.00}} &
  {0.79-0.87} \\

\multicolumn{1}{|l|}{{\acs{ch-zju-lungln-1}~\cite{yan2023anatomy,yu2025effective}}} &
  \multicolumn{1}{c|}{{1}} &
  \multicolumn{1}{c|}{{66}} &
  \multicolumn{1}{c|}{{16}} &
  \multicolumn{1}{c|}{{512x512x68}} &
  \multicolumn{1}{c|}{{0.69x0.69x5.00}} &
  \multicolumn{1}{c|}{{0.03-0.79}} &
  {0.60-0.73} \\
  
\rowcolor[HTML]{F3F3F3} 
\multicolumn{1}{|l|}{{\acs{ch-zju-esoln-1}~\cite{guo2021deepstationing, guo2022thoracic, yu2025effective}}} &
  \multicolumn{1}{c|}{{1}} &
  \multicolumn{1}{c|}{{116}} &
  \multicolumn{1}{c|}{{30}} &
  \multicolumn{1}{c|}{{512x512x80}} &
  \multicolumn{1}{c|}{{0.68x0.68x5.00}} &
  \multicolumn{1}{c|}{{0.38-0.92}} &
  {0.53-0.74} \\
 \midrule\midrule

\multicolumn{8}{|c|}{\textbf{Public Datasets}} \\ \midrule

\multicolumn{1}{|l|}{{TotalSegmentator~\cite{wasserthal2022totalsegmentator}}} &
  \multicolumn{1}{c|}{{120}} &
  \multicolumn{1}{c|}{{1139}} &
  \multicolumn{1}{c|}{{89}} &
  \multicolumn{1}{c|}{{240x229x228}} &
  \multicolumn{1}{c|}{{1.50x1.50x1.50}} &
  \multicolumn{1}{c|}{{0.00-1.00}} &
  {0.00-0.99} \\
  
\rowcolor[HTML]{F3F3F3} 
\multicolumn{1}{|l|}{{StructSeg 2019~\cite{StructSeg2019}}} &
  \multicolumn{1}{c|}{{22}} &
  \multicolumn{1}{c|}{{40}} &
  \multicolumn{1}{c|}{{10}} &
  \multicolumn{1}{c|}{{512x512x124}} &
  \multicolumn{1}{c|}{{1.14x1.14x3.00}} &
  \multicolumn{1}{c|}{{0.62-1.00}} &
  {0.61-1.00} \\

\multicolumn{1}{|l|}{{HaN-Seg~\cite{podobnik2023han}}} &
  \multicolumn{1}{c|}{{30}} &
  \multicolumn{1}{c|}{{34}} &
  \multicolumn{1}{c|}{{8}} &
  \multicolumn{1}{c|}{{1024x1024x185}} &
  \multicolumn{1}{c|}{{0.67x0.67x2.00}} &
  \multicolumn{1}{c|}{{0.42-1.00}} &
  {0.70-1.00} \\
  
\rowcolor[HTML]{F3F3F3} 
\multicolumn{1}{|l|}{{SegRap~\cite{luo2023segrap2023}}} &
  \multicolumn{1}{c|}{{38}} &
  \multicolumn{1}{c|}{{192}} &
  \multicolumn{1}{c|}{{48}} &
  \multicolumn{1}{c|}{{1024x1024x127}} &
  \multicolumn{1}{c|}{{0.54x0.54x3.00}} &
  \multicolumn{1}{c|}{{0.51-1.00}} &
  {0.65-1.00} \\

\multicolumn{1}{|l|}{{SegTHOR~\cite{lambert2020segthor}}} &
  \multicolumn{1}{c|}{{4}} &
  \multicolumn{1}{c|}{{32}} &
  \multicolumn{1}{c|}{{8}} &
  \multicolumn{1}{c|}{{512x512x171}} &
  \multicolumn{1}{c|}{{0.98x0.98x2.50}} &
  \multicolumn{1}{c|}{{0.27-0.94}} &
  {0.44-0.76} \\
  
\rowcolor[HTML]{F3F3F3} 
\multicolumn{1}{|l|}{{FLARE22~\cite{ma2023unleashing}}} &
  \multicolumn{1}{c|}{{13}} &
  \multicolumn{1}{c|}{{50}} &
  \multicolumn{1}{c|}{{20}} &
  \multicolumn{1}{c|}{{512x512x97}} &
  \multicolumn{1}{c|}{{0.80x0.80x2.50}} &
  \multicolumn{1}{c|}{{0.06-0.58}} &
  {0.28-0.55} \\

\multicolumn{1}{|l|}{{AMOS~\cite{ji2022amos}}} &
  \multicolumn{1}{c|}{{15}} &
  \multicolumn{1}{c|}{{300}} &
  \multicolumn{1}{c|}{{200}} &
  \multicolumn{1}{c|}{{512x512x104}} &
  \multicolumn{1}{c|}{{0.68x0.68x5.00}} &
  \multicolumn{1}{c|}{{0.00-0.76}} &
  {0.07-0.60} \\
  
\rowcolor[HTML]{F3F3F3} 
\multicolumn{1}{|l|}{{AbdomenCT1K 4~\cite{ma2021abdomenct}}} &
  \multicolumn{1}{c|}{{4}} &
  \multicolumn{1}{c|}{{800}} &
  \multicolumn{1}{c|}{{200}} &
  \multicolumn{1}{c|}{{512x512x102}} &
  \multicolumn{1}{c|}{{0.79x0.79x2.50}} &
  \multicolumn{1}{c|}{{0.00-0.76}} &
  {0.26-0.55} \\

\multicolumn{1}{|l|}{{AbdomenCT1K 12~\cite{ma2021abdomenct}}} &
  \multicolumn{1}{c|}{{12}} &
  \multicolumn{1}{c|}{{42}} &
  \multicolumn{1}{c|}{{8}} &
  \multicolumn{1}{c|}{{512x512x97}} &
  \multicolumn{1}{c|}{{0.80x0.80x2.50}} &
  \multicolumn{1}{c|}{{0.06-0.58}} &
  {0.27-0.59} \\
  
\rowcolor[HTML]{F3F3F3} 
\multicolumn{1}{|l|}{{WORD~\cite{luo2022word}}} &
  \multicolumn{1}{c|}{{16}} &
  \multicolumn{1}{c|}{{96}} &
  \multicolumn{1}{c|}{{24}} &
  \multicolumn{1}{c|}{{512x512x201}} &
  \multicolumn{1}{c|}{{0.98x0.98x3.00}} &
  \multicolumn{1}{c|}{{0.00-0.67}} &
  {0.00-0.74} \\

\multicolumn{1}{|l|}{{StructSeg \ac{npc}~\cite{StructSeg2019}}} &
  \multicolumn{1}{c|}{{1}} &
  \multicolumn{1}{c|}{{40}} &
  \multicolumn{1}{c|}{{10}} &
  \multicolumn{1}{c|}{{512x512x125}} &
  \multicolumn{1}{c|}{{1.14x1.14x3.00}} &
  \multicolumn{1}{c|}{{0.62-1.00}} &
  {0.90-0.95} \\
  
\rowcolor[HTML]{F3F3F3} 
\multicolumn{1}{|l|}{{MSD Spleen~\cite{antonelli2022medical}}} &
  \multicolumn{1}{c|}{{1}} &
  \multicolumn{1}{c|}{{32}} &
  \multicolumn{1}{c|}{{9}} &
  \multicolumn{1}{c|}{{512x512x90}} &
  \multicolumn{1}{c|}{{0.79x0.79x5.00}} &
  \multicolumn{1}{c|}{{0.01-0.72}} &
  {0.40-0.50} \\

\multicolumn{1}{|l|}{{MSD Lung~\cite{antonelli2022medical}}} &
  \multicolumn{1}{c|}{{1}} &
  \multicolumn{1}{c|}{{50}} &
  \multicolumn{1}{c|}{{13}} &
  \multicolumn{1}{c|}{{512x512x256}} &
  \multicolumn{1}{c|}{{0.78x0.78x1.25}} &
  \multicolumn{1}{c|}{{0.10-0.78}} &
  {0.61-0.64} \\
  
\rowcolor[HTML]{F3F3F3} 
\multicolumn{1}{|l|}{{MSD Liver~\cite{antonelli2022medical}}} &
  \multicolumn{1}{c|}{{2}} &
  \multicolumn{1}{c|}{{104}} &
  \multicolumn{1}{c|}{{27}} &
  \multicolumn{1}{c|}{{512x512x430}} &
  \multicolumn{1}{c|}{{0.77x0.77x1.00}} &
  \multicolumn{1}{c|}{{0.00-0.77}} &
  {0.33-0.53} \\

\multicolumn{1}{|l|}{{MSD Hepatic Vessel~\cite{antonelli2022medical}}} &
  \multicolumn{1}{c|}{{2}} &
  \multicolumn{1}{c|}{{242}} &
  \multicolumn{1}{c|}{{61}} &
  \multicolumn{1}{c|}{{512x512x49}} &
  \multicolumn{1}{c|}{{0.80x0.80x5.00}} &
  \multicolumn{1}{c|}{{0.10-0.59}} &
  {0.37-0.51} \\
  
\rowcolor[HTML]{F3F3F3} 
\multicolumn{1}{|l|}{{MSD Colon~\cite{antonelli2022medical}}} &
  \multicolumn{1}{c|}{{1}} &
  \multicolumn{1}{c|}{{100}} &
  \multicolumn{1}{c|}{{26}} &
  \multicolumn{1}{c|}{{512x512x95}} &
  \multicolumn{1}{c|}{{0.78x0.78x5.00}} &
  \multicolumn{1}{c|}{{0.00-0.74}} &
  {0.24-0.30} \\

\multicolumn{1}{|l|}{{MSD Pancreas~\cite{antonelli2022medical}}} &
  \multicolumn{1}{c|}{{2}} &
  \multicolumn{1}{c|}{{224}} &
  \multicolumn{1}{c|}{{57}} &
  \multicolumn{1}{c|}{{512x512x92}} &
  \multicolumn{1}{c|}{{0.80x0.80x2.50}} &
  \multicolumn{1}{c|}{{0.03-0.57}} &
  {0.35-0.44} \\
  
\rowcolor[HTML]{F3F3F3} 
\multicolumn{1}{|l|}{{LiTS17~\cite{bilic2023liver}}} &
  \multicolumn{1}{c|}{{2}} &
  \multicolumn{1}{c|}{{105}} &
  \multicolumn{1}{c|}{{26}} &
  \multicolumn{1}{c|}{{512x512x432}} &
  \multicolumn{1}{c|}{{0.77x0.77x1.00}} &
  \multicolumn{1}{c|}{{0.00-0.77}} &
  {0.26-0.55} \\

\multicolumn{1}{|l|}{{KiTS21~\cite{heller2019kits19,heller2020state}}} &
  \multicolumn{1}{c|}{{2}} &
  \multicolumn{1}{c|}{{210}} &
  \multicolumn{1}{c|}{{90}} &
  \multicolumn{1}{c|}{{512x512x110}} &
  \multicolumn{1}{c|}{{0.78x0.78x3.00}} &
  \multicolumn{1}{c|}{{0.00-0.75}} &
  {0.25-0.47} \\
  
\rowcolor[HTML]{F3F3F3} 
\multicolumn{1}{|l|}{{NIH LN~\cite{roth2015new,roth2014new}}} &
  \multicolumn{1}{c|}{{1}} &
  \multicolumn{1}{c|}{{72}} &
  \multicolumn{1}{c|}{{17}} &
  \multicolumn{1}{c|}{{512x512x154}} &
  \multicolumn{1}{c|}{{0.79x0.79x2.00}} &
  \multicolumn{1}{c|}{{0.32-0.80}} &
  {0.53-0.75} \\ \bottomrule
\end{tabular}%
}
\caption{\textbar~\textbf{Dataset fingerprints for \ac{clnet} training.} Overview of \ac{dsn} datasets used to train \ac{clnet}, including 16 private and 20 public datasets, spanning a wide range of segmentation tasks across various anatomical regions of the human body, including the Head \& Neck, Chest, Abdomen, Bones, lymph node stations (LNS), and Lesions. This comprehensive dataset collection comprises a total of 7,278 private and 3,904 public CT scans, along with 37,827 private and 172,607 public annotations, covering \ac{orgn} anatomical structures - 125 private and 162 public anatomical types with 52 organs overlapping between the two groups. In addition to standard statistics such as dataset name, training/testing sizes, number of target organ classes, median image size, and voxel spacing, we also report two specific metrics to better characterize the body part distribution within each dataset: the Body Part Regression (BPR, head -- 1, pelvic -- 0) axial bounding ranges for both the images and the foreground organs. }
\label{tab:dataset_stats}
\end{table}
\clearpage

% \usepackage{booktabs}
% \usepackage{graphicx}

% \newcolumntype{P}[1]{>{\raggedright\arraybackslash}p{#1}}
% \newcolumntype{M}[1]{>{\centering\arraybackslash}m{#1}}

% {
% \makeatletter
% \renewcommand\@cite[2]{{\tiny #1\if@tempswa , #2\fi}}
% \makeatother

\begin{table}[htp]
\renewcommand{\arraystretch}{1.2}
\centering
% \footnotesize
% \scriptsize
\resizebox{\textwidth}{!}{%
\begin{tabular}{@{}|l|M{1.2\textwidth}|@{}}
% \begin{tabular}{@{}|p{0.28\textwidth}|M{0.72\textwidth}|@{}}
\hline
{\textbf{Dataset Name}} &
  {\textbf{Target Organs}} \\ \hline
 
\acs{hn-zju-15}~\cite{guo2021comprehensive, zhu2024low} &
  BrainStem, Eye\_L, Eye\_R, Lens\_L, Lens\_R, Chiasm, OpticNerve\_L, OpticNerve\_R, Glnd\_Parotid\_L, Glnd\_Parotid\_R, TMJ\_L, TMJ\_R, SpinalCord, Musc\_Masseter\_L, Musc\_Masseter\_R \\
  
\rowcolor[HTML]{F3F3F3} 
\acs{hn-cgmh-42}~\cite{guo2020organ,guo2021comprehensive} &
  Bronchus\_L, Bronchus\_R, BasalGanglia\_L, BasalGanglia\_R, BrainStem, Cerebellum, Ear\_Cochlea\_L, Ear\_Cochlea\_R, Musc\_Const\_I, Musc\_Const\_M, Musc\_Const\_S, Ear\_Inner\_L, Ear\_Inner\_R, Eye\_L, Eye\_R, Epiglottis, Eso, Hippocampus\_L, Hippocampus\_R, HypoThalamus, Glnd\_Lacrimal\_L, Glnd\_Lacrimal\_R, Larynx, Bone\_Mandible\_L, Bone\_Mandible\_R, Chiasm, OpticNerve\_L, OpticNerve\_R, OralCavity, Glnd\_Parotid\_L, Glnd\_Parotid\_R, Glnd\_Pineal, Glnd\_Pituitary, SpinalCord, Glnd\_Submand\_L, Glnd\_Submand\_R, TemporalLobe\_L, TemporalLobe\_R, Glnd\_Thyroid\_L, Glnd\_Thyroid\_R, TMJ\_L, TMJ\_R \\
 
\acs{ch-zju-lungcan-12}~\cite{guo2021deepstationing, liu2021same} &
  Eso, Lung\_L, Lung\_R, Pericardium, SpinalCord, BrachialPlexus, ProximalBronchi, A\_Aorta, V\_VenaCava\_S, A\_Pulmonary, V\_Pulmonary, V\_VenaCava\_I, ChestWall\_L, ChestWall\_R \\
  
\rowcolor[HTML]{F3F3F3} 
\acs{ch-zju-esocan-35}~\cite{guo2021deepstationing,yan2022sam} &
  A\_Aorta\_Arch, A\_Aorta\_Ascending, A\_Aorta\_Descending, A\_Pulmonary, Bronchus\_L, Bronchus\_R, Heart, Lung\_L, Lung\_R, SpinalCord, Bone\_Sternum, Glnd\_Thyroid\_L, Glnd\_Thyroid\_R, Trachea, V\_VenaCava\_I, V\_VenaCava\_S, Musc\_Cervical\_A, Musc\_Scalenus, Musc\_Scalenus\_A, Musc\_Scleido, A\_CommonCarotid\_L, A\_CommonCarotid\_R, A\_Subclavian\_L, A\_Subclavian\_R, A\_Vertebral\_L, A\_Vertebral\_R, Eso, V\_Azygos, V\_Brachiocephalic\_L, V\_Brachiocephalic\_R, V\_InternalJuguar\_L, V\_InternalJuguar\_R, V\_Pulmonary, V\_Subclavian\_L, V\_Subclavian\_R, BrachialPlexus \\
 
\acs{ch-zju-rtog-13}~\cite{guo2021deepstationing,zhu2024low} &
  Eso, Lung\_L, Lung\_R, ChestWall\_L, ChestWall\_R, Pericardium, SpinalCord, BrachialPlexus, ProximalBronchi, A\_Aorta, A\_Pulmonary, V\_VenaCava\_S, V\_VenaCava\_I, V\_Pulmonary \\
  
\rowcolor[HTML]{F3F3F3} 
\acs{ab-td-5} &
  UrinaryBladder, Rectum, Colon, SmallBowel \\ % HR\_CTV
 
\acs{body-linkmed-1} &
  Body \\

\rowcolor[HTML]{F3F3F3} 
\acs{hn-lns-18}~\cite{li2024semi} &
  HN\_LNS\_1La, HN\_LNS\_1Lb, HN\_LNS\_1Ra, HN\_LNS\_1Rb, HN\_LNS\_2La, HN\_LNS\_2Lb, HN\_LNS\_2Ra, HN\_LNS\_2Rb, HN\_LNS\_3L, HN\_LNS\_3R, HN\_LNS\_4L, HN\_LNS\_4R, HN\_LNS\_5La, HN\_LNS\_5Lb, HN\_LNS\_5Ra, HN\_LNS\_5Rb, HN\_LNS\_6L, HN\_LNS\_6R \\
 
\acs{ch-lns-15}~\cite{guo2021deepstationing, guo2022thoracic} &
  Chest\_LNS1\_L, Chest\_LNS1\_R, Chest\_LNS2\_L, Chest\_LNS2\_R, Chest\_LNS3\_A, Chest\_LNS3\_P, Chest\_LNS4\_L, Chest\_LNS4\_R, Chest\_LNS5, Chest\_LNS6, Chest\_LNS7, Chest\_LNS8, Chest\_LNS9, Chest\_LNS10\_L, Chest\_LNS10\_R \\
  
\rowcolor[HTML]{F3F3F3} 
\acs{hn-zju-npc-1}~\cite{zhang2023multi,li2024leveraging} &
  \ac{npc}\_\Ac{gtv} \\

\acs{ch-wpy-lungcan-1}~\cite{wang2023anatomy,wang2024accurate} &
  Lung\_\Ac{gtv} \\

\rowcolor[HTML]{F3F3F3} 
\acs{ch-wpy-esocan-2}~\cite{li2023lvit} &
  Eso, Eso\_\Ac{gtv} \\

\acs{ab-livercan-2}~\cite{yan2023liver} &
  Liver, Liver\_\Ac{gtv} \\
  
\rowcolor[HTML]{F3F3F3} 
\acs{hn-fd-ln-1}~\cite{yu2024slice,yu2025effective} &
  HN\_LN \\

\acs{ch-zju-lungln-1}~\cite{yan2023anatomy, yu2025effective} &
  Chest\_LN \\
  
\rowcolor[HTML]{F3F3F3} 
\acs{ch-zju-esoln-1}~\cite{guo2021deepstationing, guo2022thoracic, yu2025effective} &
  Chest\_LN \\ \hline
\end{tabular}%
}
\caption{\textbar~\textbf{Detailed organ classes of 16 private datasets used for training \ac{clnet}.} The datasets cover 125 clinically relevant anatomical structures across whole-body regions, including 73 novel structures not available in public datasets. In addition to the common organ groups (Head \& Neck, Chest, Abdomen, Bones, and Lesions), lymph node stations (LNS) in the head-neck and chest regions are specifically included. All organ labels follow standardized naming conventions in radiation oncology~\cite{santanam2012standardizing}. }
\label{tab:dataset_task_p}
\end{table}
\clearpage

% }

% \newcolumntype{P}[1]{>{\raggedright\arraybackslash}p{#1}}
% \newcolumntype{M}[1]{>{\centering\arraybackslash}m{#1}}

\begin{table}[htp]
\renewcommand{\arraystretch}{1.2}
\centering
% \tiny
% \begingroup
% \scriptsize
\resizebox{\textwidth}{!}{%
\begin{tabular}{@{}|l|M{1.3\textwidth}|@{}}
% \begin{tabular}{@{}|M{0.18\textwidth}|M{0.82\textwidth}|@{}}
\hline
\cellcolor[HTML]{FFFFFF}{\color[HTML]{000000} \textbf{Dataset Name}} &
  {\color[HTML]{000000} \textbf{Target Organs}} \\ \hline
\rowcolor[HTML]{FFFFFF} 
{\color[HTML]{000000} TotalSegmentator~\cite{wasserthal2022totalsegmentator}} &
  {\color[HTML]{000000} Spleen, Kidney\_R, Kidney\_L, GallBladder, Liver, Stomach, Pancreas, Glnd\_Adrenal\_R, Glnd\_Adrenal\_L, Lung\_LUL, Lung\_LLL, Lung\_RUL, Lung\_RML, Lung\_RLL, Eso, Trachea, Glnd\_Thyroid, SmallBowel, Duodenum, Colon, UrinaryBladder, Prostate, Kidney\_Cyst\_L, Kidney\_Cyst\_R, Bone\_Sacrum, Bone\_Vert\_S1, Bone\_Vert\_L5, Bone\_Vert\_L4, Bone\_Vert\_L3, Bone\_Vert\_L2, Bone\_Vert\_L1, Bone\_Vert\_T12, Bone\_Vert\_T11, Bone\_Vert\_T10, Bone\_Vert\_T9, Bone\_Vert\_T8, Bone\_Vert\_T7, Bone\_Vert\_T6, Bone\_Vert\_T5, Bone\_Vert\_T4, Bone\_Vert\_T3, Bone\_Vert\_T2, Bone\_Vert\_T1, Bone\_Vert\_C7, Bone\_Vert\_C6, Bone\_Vert\_C5, Bone\_Vert\_C4, Bone\_Vert\_C3, Bone\_Vert\_C2, Bone\_Vert\_C1, Heart, A\_Aorta, V\_Pulmonary, A\_BrachiocephalicTrunk, A\_Subclavian\_R, A\_Subclavian\_L, A\_CommonCarotid\_R, A\_CommonCarotid\_L, V\_Brachiocephalic\_L, V\_Brachiocephalic\_R, LAA, V\_VenaCava\_S, V\_VenaCava\_I, V\_Portal\_Splenic, A\_Iliac\_L, A\_Iliac\_R, V\_Iliac\_L, V\_Iliac\_R, Bone\_Humerus\_L, Bone\_Humerus\_R, Bone\_Scapula\_L, Bone\_Scapula\_R, Bone\_Clavicula\_L, Bone\_Clavicula\_R, Bone\_Femur\_L, Bone\_Femur\_R, Bone\_Hip\_L, Bone\_Hip\_R, SpinalCord, Musc\_GluteusMax\_L, Musc\_GluteusMax\_R, Musc\_GluteusMed\_L, Musc\_GluteusMed\_R, Musc\_GluteusMin\_L, Musc\_GluteusMin\_R, Musc\_Autochthon\_L, Musc\_Autochthon\_R, Musc\_Iliopsoas\_L, Musc\_Iliopsoas\_R, Brain, Bone\_Skull, Bone\_Rib4\_R, Bone\_Rib3\_R, Bone\_Rib1\_L, Bone\_Rib2\_L, Bone\_Rib3\_L, Bone\_Rib4\_L, Bone\_Rib5\_L, Bone\_Rib6\_L, Bone\_Rib7\_L, Bone\_Rib8\_L, Bone\_Rib9\_L, Bone\_Rib10\_L, Bone\_Rib11\_L, Bone\_Rib12\_L, Bone\_Rib1\_R, Bone\_Rib2\_R, Bone\_Rib5\_R, Bone\_Rib6\_R, Bone\_Rib7\_R, Bone\_Rib8\_R, Bone\_Rib9\_R, Bone\_Rib10\_R, Bone\_Rib11\_R, Bone\_Rib12\_R, Bone\_Sternum, Bone\_CostalCartilages, Myocardium, Atrium\_L, Atrium\_R, Ventricle\_L, Ventricle\_R} \\
\rowcolor[HTML]{EFEFEF} 
{\color[HTML]{000000} StructSeg 2019~\cite{StructSeg2019}} &
  \cellcolor[HTML]{EFEFEF}{\color[HTML]{000000} BrainStem, Eye\_L, Eye\_R, Lens\_L, Lens\_R, OpticNerve\_L, OpticNerve\_R, Chiasm, TemporalLobe\_L, TemporalLobe\_R, Glnd\_Pituitary, Glnd\_Parotid\_L, Glnd\_Parotid\_R, Ear\_Inner\_L, Ear\_Inner\_R, Ear\_Mid\_L, Ear\_Mid\_R, TMJ\_L, TMJ\_R, SpinalCord, Bone\_Mandible\_L, Bone\_Mandible\_R} \\
\rowcolor[HTML]{FFFFFF} 
{\color[HTML]{000000} HaN-Seg~\cite{podobnik2023han}} &
  {\color[HTML]{000000} A\_CommonCarotid\_L, A\_CommonCarotid\_R, Arytenoid, Bone\_Mandible, BrainStem, BuccalMucosa, OralCavity, Ear\_Cochlea\_L, Ear\_Cochlea\_R, Cricopharyngeus, Eso, Eye\_AL, Eye\_AR, Eye\_PL, Eye\_PR, Glnd\_Lacrimal\_L, Glnd\_Lacrimal\_R, Glnd\_Submand\_L, Glnd\_Submand\_R, Glnd\_Thyroid, Larynx\_Glottic, Larynx\_Supraglot, Lips, Chiasm, OpticNerve\_L, OpticNerve\_R, Glnd\_Parotid\_L, Glnd\_Parotid\_R, Glnd\_Pituitary, SpinalCord} \\
\rowcolor[HTML]{EFEFEF} 
{\color[HTML]{000000} SegRap~\cite{luo2023segrap2023}} &
  \cellcolor[HTML]{EFEFEF}{\color[HTML]{000000} Brain, TemporalLobe\_L, TemporalLobe\_R, Hippocampus\_L, Hippocampus\_R, Glnd\_Pituitary, Chiasm, OpticNerve\_L, OpticNerve\_R, Eye\_L, Eye\_R, Lens\_L, Lens\_R, Ear\_Cochlea\_L, Ear\_Cochlea\_R, Mastoid\_L, Mastoid\_R, Ear\_Mid\_L, Ear\_Mid\_R, Bone\_ET\_L, Bone\_ET\_R, Bone\_Mandible\_L, Bone\_Mandible\_R, Glnd\_Submand\_L, Glnd\_Submand\_R, OralCavity, Glnd\_Parotid\_L, Glnd\_Parotid\_R, TMJ\_L, TMJ\_R, Larynx, Larynx\_Glottic, Larynx\_Supraglot, BrainStem, SpinalCord, Eso, Glnd\_Thyroid, Trachea} \\
\rowcolor[HTML]{FFFFFF} 
{\color[HTML]{000000} SegTHOR~\cite{lambert2020segthor}} &
  {\color[HTML]{000000} Eso, ProximalBronchi, Heart, A\_Aorta} \\
\rowcolor[HTML]{EFEFEF} 
{\color[HTML]{000000} FLARE22~\cite{ma2023unleashing}} &
  \cellcolor[HTML]{EFEFEF}{\color[HTML]{000000} Liver, Eso, Stomach, Duodenum, Kidney\_L, Kidney\_R, Spleen, Pancreas, A\_Aorta, V\_VenaCava\_I, Glnd\_Adrenal\_R, Glnd\_Adrenal\_L, GallBladder} \\
\rowcolor[HTML]{FFFFFF} 
{\color[HTML]{000000} AMOS~\cite{ji2022amos}} &
  {\color[HTML]{000000} Spleen, Kidney\_R, Kidney\_L, GallBladder, Eso, Liver, Stomach, Arota, V\_VenaCava\_I, Pancreas, Glnd\_Adrenal\_R, Glnd\_Adrenal\_L, Duodenum, UrinaryBladder, Prostate/Uterus} \\
\rowcolor[HTML]{EFEFEF} 
{\color[HTML]{000000} AbdomenCT1K 4~\cite{ma2021abdomenct}} &
  \cellcolor[HTML]{EFEFEF}{\color[HTML]{000000} Liver, Kidney, Spleen, Pancreas} \\
\rowcolor[HTML]{FFFFFF} 
{\color[HTML]{000000} AbdomenCT1K 12~\cite{ma2021abdomenct}} &
  {\color[HTML]{000000} Liver, Kidney, Spleen, Pancreas, A\_Aorta, V\_VenaCava\_I, Stomach, GallBladder, Eso, Glnd\_Adrenal\_R, Glnd\_Adrenal\_L, A\_CeliacTrunk} \\
\rowcolor[HTML]{EFEFEF} 
{\color[HTML]{000000} WORD~\cite{luo2022word}} &
  \cellcolor[HTML]{EFEFEF}{\color[HTML]{000000} Liver, Spleen, Kidney\_L, Kidney\_R, Stomach, GallBladder , Eso, Pancreas, Duodenum, Colon, SmallBowel, Glnd\_Adrenal, Rectum, UrinaryBladder, Bone\_FemurHead\_L, Bone\_FemurHead\_R} \\
\rowcolor[HTML]{FFFFFF} 
{\color[HTML]{000000} StructSeg \ac{npc}~\cite{StructSeg2019}} &
  {\color[HTML]{000000} \ac{npc}\_\Ac{gtv}} \\
\rowcolor[HTML]{EFEFEF} 
{\color[HTML]{000000} MSD Spleen~\cite{antonelli2022medical}} &
  \cellcolor[HTML]{EFEFEF}{\color[HTML]{000000} Spleen} \\
\rowcolor[HTML]{FFFFFF} 
{\color[HTML]{000000} MSD Lung~\cite{antonelli2022medical}} &
  {\color[HTML]{000000} Lung\_\Ac{gtv}} \\
\rowcolor[HTML]{EFEFEF} 
{\color[HTML]{000000} MSD Liver~\cite{antonelli2022medical}} &
  \cellcolor[HTML]{EFEFEF}{\color[HTML]{000000} Liver, Liver\_\Ac{gtv}} \\
\rowcolor[HTML]{FFFFFF} 
{\color[HTML]{000000} MSD Hepatic Vessel~\cite{antonelli2022medical}} &
  {\color[HTML]{000000} HepaticVessel, Liver\_\Ac{gtv}} \\
\rowcolor[HTML]{EFEFEF} 
{\color[HTML]{000000} MSD Colon~\cite{antonelli2022medical}} &
  \cellcolor[HTML]{EFEFEF}{\color[HTML]{000000} Colon\_\Ac{gtv}} \\
\rowcolor[HTML]{FFFFFF} 
{\color[HTML]{000000} MSD Pancreas~\cite{antonelli2022medical}} &
  {\color[HTML]{000000} Pancreas, Pancreas\_\Ac{gtv}} \\
\rowcolor[HTML]{EFEFEF} 
{\color[HTML]{000000} LiTS17~\cite{bilic2023liver}} &
  \cellcolor[HTML]{EFEFEF}{\color[HTML]{000000} Liver, Liver\_\Ac{gtv}} \\
\rowcolor[HTML]{FFFFFF} 
{\color[HTML]{000000} KiTS21~\cite{heller2019kits19,heller2020state}} &
  {\color[HTML]{000000} Kidney, Kidney\_\Ac{gtv}} \\
\rowcolor[HTML]{EFEFEF} 
{\color[HTML]{000000} NIH LN~\cite{roth2015new,roth2014new}} &
  \cellcolor[HTML]{EFEFEF}{\color[HTML]{000000} Chest\_LN} \\ \hline
\end{tabular}%
}
% \endgroup
\caption{\textbar~\textbf{Detailed organ classes of all 20 public datasets used for training \ac{clnet}.} The datasets cover 162 common anatomical structures across whole-body regions, including Head \& Neck, Chest, Abdomen, Bones, and Lesions. All organ labels follow standardized naming conventions in radiation oncology~\cite{santanam2012standardizing}. }
\label{tab:dataset_task_d}
\end{table}
\clearpage

\begin{table}[htp]
\renewcommand{\arraystretch}{1.2}
\centering
% \tiny
\resizebox{\textwidth}{!}{% 
\begin{tabular}{@{}|l|c|c|c|c|c|M{0.50\textwidth}|@{}}
% \begin{tabular}{@{}|M{0.20\textwidth}|C{0.07\textwidth}|C{0.03\textwidth}|C{0.03\textwidth}|C{0.07\textwidth}|C{0.08\textwidth}|M{0.30\textwidth}|@{}}
\hline
\cellcolor[HTML]{FFFFFF}\textbf{Dataset Name} &
  \textbf{Body Part} &
  \cellcolor[HTML]{FFFFFF}\textbf{Organ\#} &
  \cellcolor[HTML]{FFFFFF}\textbf{Test\#} &
  \cellcolor[HTML]{FFFFFF}\textbf{\begin{tabular}[c]{@{}c@{}}Med Shape\\ (XYZ)\end{tabular}} &
  \cellcolor[HTML]{FFFFFF}\textbf{\begin{tabular}[c]{@{}c@{}}Med Spacing\\ (XYZ)\end{tabular}} &
  \textbf{Target Organs} \\ \hline
\rowcolor[HTML]{F2F2F2} 
\acs{hn-gz-12} &
  Head \& Neck &
  12 &
  268 &
  406x227x95 &
  0.99x0.99x3.00 &
  BrainStem, Eye\_L, Eye\_R, Lens\_L, Lens\_R, Chiasm, OpticNerve\_L, OpticNerve\_R, Glnd\_Parotid\_L, Glnd\_Parotid\_R, TMJ\_L, TMJ\_R \\
\cellcolor[HTML]{FFFFFF}\acs{ch-zju-14} &
  Chest &
  14 &
  60 &
  512x512x93 &
  1.17x1.17x5.00 &
  Eso, Lung\_L, Lung\_R, Pericardium, SpinalCord, BrachialPlex, ProximalBronchi, A\_Aorta, V\_SVC, A\_Pulmonary, V\_Pulmonary, V\_IVC, ChestWall\_L, ChestWall\_R \\
\rowcolor[HTML]{F2F2F2} 
BTCV~\cite{landman2015miccai} &
  Abdomen &
  13 &
  30 &
  512x512x127 &
  0.76x0.76x3.00 &
  Spleen, Kidney\_R, Kidney\_L, GallBladder, Eso, Liver, Stomach, A\_Aorta, V\_VenaCava\_I, V\_Portal\_Splenic, Pancreas, Glnd\_Adrenal\_R, Glnd\_Adrenal\_L \\
\cellcolor[HTML]{FFFFFF}\acs{hn-cgmh-npc-1} &
  Head \& Neck &
  1 &
  319 &
  512x512x137 &
  0.98x0.98x3.00 &
  NPC\_\ac{gtv} \\
\rowcolor[HTML]{F2F2F2} 
\acs{ch-cgmh-esocan-1} &
  Chest &
  1 &
  148 &
  366x273x212 &
  1.00x1.00x2.50 &
  Eso\_\ac{gtv} \\
\cellcolor[HTML]{FFFFFF}\acs{ab-ch-kidneycan-1} &
  Abdomen &
  1 &
  978 &
  512x512x343 &
  0.74x0.74x1.00 &
  Kidney\_\ac{gtv} \\
\rowcolor[HTML]{F2F2F2} 
\acs{ab-cgmh-livercan-1} &
  Abdomen &
  1 &
  176 &
  512x512x224 &
  0.98x0.98x1.25 &
  Liver\_\ac{gtv} \\ \hline
\end{tabular}
}
\caption{\textbar~\textbf{Dataset fingerprints for \ac{clnet} external validation.} Overview of seven external datasets used for external validation, including one public dataset BTCV and six private datasets. The external datasets span whole-body anatomical regions and cover 1,979 testing samples and 43 targeting anatomical structures from head \& neck, chest, abdomen, and four \acp{gtv}. Basic dataset fingerprints are provided, including dataset name, body part coverage, number of target organ classes, testing number, median image size and voxel spacing, and detailed target organ list. }
\label{tab:supp_ex_dataset}
\end{table}
\clearpage

\begin{table}[htp]
\renewcommand{\arraystretch}{1.2}
\centering
\tiny
\resizebox{\textwidth}{!}{% 
\begin{tabular}{@{}|M{0.12\textwidth}|M{0.23\textwidth}|M{0.09\textwidth}||M{0.11\textwidth}|M{0.23\textwidth}|M{0.09\textwidth}|@{}}
\hline
\textbf{Decoder Name}        & \textbf{Structure List} & \textbf{\ac{fls}}                   & \textbf{Decoder Name}       & \textbf{Structure List}  & \textbf{\ac{fls}}        \\ \hline
\rowcolor[HTML]{F3F3F3} 
Brain & Brain &  & Pericardium & Pericardium & Heart \\
BrainStem & BrainStem &  & Airway & Bronchus\_L, Bronchus\_R, Trachea &  \\
\rowcolor[HTML]{F3F3F3} 
TemporalLobe & TemporalLobe\_L, TemporalLobe\_R &  & LAA & LAA &  \\
BasalGanglia & BasalGanglia\_L, BasalGanglia\_R & TemporalLobe & Eso & Eso &  \\
\rowcolor[HTML]{F3F3F3} 
Hippocampus & Hippocampus\_L, Hippocampus\_R & TemporalLobe & Musc\_Cervical\_A & Musc\_Cervical\_A &  \\
HypoThalamus & HypoThalamus & BrainStem & Musc\_Scalenus & Musc\_Scalenus, Musc\_Scalenus\_A &  \\
\rowcolor[HTML]{F3F3F3} 
Glnd\_Pineal & Glnd\_Pineal & BrainStem & Musc\_Scleido & Musc\_Scleido &  \\
Cerebellum & Cerebellum &  & Musc\_Autochthon & Musc\_Autochthon\_L, Musc\_Autochthon\_R &  \\
\rowcolor[HTML]{F3F3F3} 
Chiasm & Chiasm & OpticNerve, Glnd\_Pituitary & BrachialPlexus & BrachialPlexus\_L, BrachialPlexus\_R & A\_Subclavian, Musc\_Scalenus \\
Eye & Eye\_L, Eye\_R &  & SpinalCord & SpinalCord &  \\
\rowcolor[HTML]{F3F3F3} 
Lens & Lens\_L, Lens\_R & Eye & Liver & Liver &  \\
OpticNerve & OpticNerve\_L, OpticNerve\_R & Eye & Kidney & Kidney\_L, Kidney\_R &  \\
\rowcolor[HTML]{F3F3F3} 
Glnd\_Lacrimal & Glnd\_Lacrimal\_L, Glnd\_Lacrimal\_R & Eye & Pancreas & Pancreas &  \\
Glnd\_Pituitary & Glnd\_Pituitary &  & Spleen & Spleen &  \\
\rowcolor[HTML]{F3F3F3} 
Ear\_Mid & Ear\_Mid\_L, Ear\_Mid\_R &  & Prostate & Prostate &  \\
Ear\_Inner & Ear\_Inner\_L, Ear\_Inner\_R &  & Stomach & Stomach &  \\
\rowcolor[HTML]{F3F3F3} 
Ear\_Cochlea & Ear\_Cochlea\_L, Ear\_Cochlea\_R & Ear\_Inner & SmallBowel & SmallBowel &  \\
Bone\_ET & Bone\_ET\_L, Bone\_ET\_R &  & Colon & Colon, Rectum &  \\
\rowcolor[HTML]{F3F3F3} 
Bone\_Mandible & Bone\_Mandible\_L, Bone\_Mandible\_R &  & Duodenum & Duodenum & Stomach \\
TMJ & TMJ\_L, TMJ\_R &  & GallBladder & GallBladder &  \\
\rowcolor[HTML]{F3F3F3} 
Glnd\_Parotid & Glnd\_Parotid\_L, Glnd\_Parotid\_R &  & Glnd\_Adrenal & Glnd\_Adrenal\_L, Glnd\_Adrenal\_R & Kidney \\
Glnd\_Submand & Glnd\_Submand\_L, Glnd\_Submand\_R &  & HepaticVessel & HepaticVessel & Liver \\
\rowcolor[HTML]{F3F3F3} 
Glnd\_Thyroid & Glnd\_Thyroid\_L, Glnd\_Thyroid\_R &  & Musc\_GluteusMax & Musc\_GluteusMax\_L, Musc\_GluteusMax\_R &  \\
OralCavity & OralCavity &  & Musc\_GluteusMed & Musc\_GluteusMed\_L, Musc\_GluteusMed\_R &  \\
\rowcolor[HTML]{F3F3F3} 
Larynx & Larynx, Larynx\_Glottic, Larynx\_Supraglot &  & Musc\_GluteusMin & Musc\_GluteusMin\_L, Musc\_GluteusMin\_R &  \\
Epiglottis & Epiglottis &  & Musc\_Iliopsoas & Musc\_Iliopsoas\_L, Musc\_Iliopsoas\_R &  \\
\rowcolor[HTML]{F3F3F3} 
Musc\_Const & Musc\_Const\_I, Musc\_Const\_M, Musc\_Const\_S &  & UrinaryBladder & UrinaryBladder &  \\
Musc\_Masseter & Musc\_Masseter\_L, Musc\_Masseter\_R &  & A\_Iliac & A\_Iliac\_L, A\_Iliac\_R &  \\
\rowcolor[HTML]{F3F3F3} 
Lung & Lung\_L, Lung\_R &  & V\_Iliac & V\_Iliac\_L, V\_Iliac\_R &  \\
Lung\_LL & Lung\_LLL, Lung\_LUL &  & Bone\_Skull & Bone\_Skull &  \\
\rowcolor[HTML]{F3F3F3} 
Lung\_RL & Lung\_RLL, Lung\_RML, Lung\_RUL &  & Bone\_Clavicula & Bone\_Clavicula\_L, Bone\_Clavicula\_R &  \\
ChestWall & ChestWall\_L, ChestWall\_R & Lung & Bone\_Humerus & Bone\_Humerus\_L, Bone\_Humerus\_R &  \\
\rowcolor[HTML]{F3F3F3} 
A\_Aorta & A\_Aorta\_Arch, A\_Aorta\_Ascending, A\_Aorta\_Descending &  & Bone\_Scapula & Bone\_Scapula\_L, Bone\_Scapula\_R &  \\
A\_BrachiocephalicTrunk & A\_BrachiocephalicTrunk &  & Bone\_Sternum & Bone\_Sternum &  \\
\rowcolor[HTML]{F3F3F3} 
A\_CommonCarotid & A\_CommonCarotid\_L, A\_CommonCarotid\_R &  & Bone\_CostalCartilages & Bone\_CostalCartilages &  \\
A\_Pulmonary & A\_Pulmonary &  & Bone\_Femur & Bone\_Femur\_L, Bone\_Femur\_R &  \\
\rowcolor[HTML]{F3F3F3} 
A\_Subclavian & A\_Subclavian\_L, A\_Subclavian\_R &  & Bone\_Hip & Bone\_Hip\_L, Bone\_Hip\_R &  \\
A\_Vertebral & A\_Vertebral\_L, A\_Vertebral\_R & A\_CommonCarotid & Bone\_Sacrum & Bone\_Sacrum &  \\
\rowcolor[HTML]{F3F3F3} 
V\_Azygos & V\_Azygos &  & HN\_LN & HN\_LN & HN\_LNS \\
V\_Brachiocephalic & V\_Brachiocephalic\_L, V\_Brachiocephalic\_R &  & Chest\_LN & Chest\_LN & Chest\_LNS \\
\rowcolor[HTML]{F3F3F3} 
V\_InternalJuguar & V\_InternalJuguar\_L, V\_InternalJuguar\_R &  & NPC\_\Ac{gtv} & NPC\_\Ac{gtv} &  \\
V\_Portal\_Splenic & V\_Portal\_Splenic &  & Lung\_\Ac{gtv} & Lung\_\Ac{gtv} & Lung \\
\rowcolor[HTML]{F3F3F3} 
V\_Pulmonary & V\_Pulmonary & Heart & Eso\_\Ac{gtv} & Eso\_\Ac{gtv} & Eso \\
V\_Subclavian & V\_Subclavian\_L, V\_Subclavian\_R &  & Liver\_\Ac{gtv} & Liver\_\Ac{gtv} & Liver \\
\rowcolor[HTML]{F3F3F3} 
V\_VenaCava\_I & V\_VenaCava\_I &  & Kidney\_\Ac{gtv} & Kidney\_\Ac{gtv} & Kidney \\
V\_VenaCava\_S & V\_VenaCava\_S &  & Pancreas\_\Ac{gtv} & Pancreas\_\Ac{gtv} & Pancreas \\
\rowcolor[HTML]{F3F3F3} 
Heart & Heart &  & Colon\_\Ac{gtv} & Colon\_\Ac{gtv} & Colon \\
Atrium\_Ventricle & Atrium\_L, Atrium\_R, Ventricle\_L, Ventricle\_R &  & Body & Body &  \\
\rowcolor[HTML]{F3F3F3} 
Myocardium & Myocardium & Heart &  &  &  \\
HN\_LNS & HN\_LNS\_1La, HN\_LNS\_1Lb, HN\_LNS\_1Ra, HN\_LNS\_1Rb, HN\_LNS\_2La, HN\_LNS\_2Lb, HN\_LNS\_2Ra, HN\_LNS\_2Rb, HN\_LNS\_3L, HN\_LNS\_3R, HN\_LNS\_4L, HN\_LNS\_4R, HN\_LNS\_5La, HN\_LNS\_5Lb, HN\_LNS\_5Ra, HN\_LNS\_5Rb, HN\_LNS\_6L, HN\_LNS\_6R & Musc\_Cervical\_A, Musc\_Scalenus, Musc\_Scleido & Chest\_LNS & Chest\_LNS1\_L, Chest\_LNS1\_R, Chest\_LNS2\_L, Chest\_LNS2\_R, Chest\_LNS3\_A, Chest\_LNS3\_P, Chest\_LNS4\_L, Chest\_LNS4\_R, Chest\_LNS5, Chest\_LNS6, Chest\_LNS7, Chest\_LNS8, Chest\_LNS9, Chest\_LNS10\_L, Chest\_LNS10\_R & A\_Aorta, Airway, Eso, Heart \\
\rowcolor[HTML]{F3F3F3} 
Bone\_Rib & Bone\_Rib1\_L, Bone\_Rib2\_L, Bone\_Rib3\_L, Bone\_Rib4\_L, Bone\_Rib5\_L, Bone\_Rib6\_L, Bone\_Rib7\_L, Bone\_Rib8\_L, Bone\_Rib9\_L, Bone\_Rib10\_L, Bone\_Rib11\_L, Bone\_Rib12\_L, Bone\_Rib1\_R, Bone\_Rib2\_R, Bone\_Rib3\_R, Bone\_Rib4\_R, Bone\_Rib5\_R, Bone\_Rib6\_R, Bone\_Rib7\_R, Bone\_Rib8\_R, Bone\_Rib9\_R, Bone\_Rib10\_R, Bone\_Rib11\_R, Bone\_Rib12\_R &  & Bone\_Vert & Bone\_Vert\_C1, Bone\_Vert\_C2, Bone\_Vert\_C3, Bone\_Vert\_C4, Bone\_Vert\_C5, Bone\_Vert\_C6, Bone\_Vert\_C7, Bone\_Vert\_T1, Bone\_Vert\_T2, Bone\_Vert\_T3, Bone\_Vert\_T4, Bone\_Vert\_T5, Bone\_Vert\_T6, Bone\_Vert\_T7, Bone\_Vert\_T8, Bone\_Vert\_T9, Bone\_Vert\_T10, Bone\_Vert\_T11, Bone\_Vert\_T12, Bone\_Vert\_L1, Bone\_Vert\_L2, Bone\_Vert\_L3, Bone\_Vert\_L4, Bone\_Vert\_L5, Bone\_Vert\_S1 & 
\\ \hline
\end{tabular}
}
\caption{\textbar~\textbf{Anatomy structure mapping of all \ac{decn} decoders.} Anatomical structures of the same type are grouped and trained within a single decoder to streamline segmentation and improve efficiency. For more complex and challenging structures, the \ac{fls} is adopted, enhancing prediction accuracy and addressing spatial precision associated with these structures.}
\label{tab:dec2org}
\end{table}
\clearpage

%% CSS comparison methods statistics
% \usepackage{multirow}
% \usepackage{graphicx}
\begin{table}[htp]
\renewcommand{\arraystretch}{1.2}
\centering
\resizebox{\textwidth}{!}{%
\begin{tabular}{|l|cc|cc|cc|cc|cc|cc|c|}
\hline
 &
  \multicolumn{2}{c|}{\textbf{\ac{totalseg}}} &
  \multicolumn{2}{c|}{\textbf{\ac{structseg}}} &
  \multicolumn{2}{c|}{\textbf{\ac{flare}}} &
  \multicolumn{2}{c|}{\textbf{\ac{segthor}}} &
  \multicolumn{2}{c|}{\textbf{\ac{kits}}} &
  \multicolumn{2}{c|}{\textbf{All}} &
   \\ \cline{2-13}
 &
  DSC$\uparrow$ &
  ASD$\downarrow$ &
  DSC$\uparrow$ &
  ASD$\downarrow$ &
  DSC$\uparrow$ &
  ASD$\downarrow$ &
  DSC$\uparrow$ &
  ASD$\downarrow$ &
  DSC$\uparrow$ &
  ASD$\downarrow$ &
  DSC$\uparrow$ &
  ASD$\downarrow$ &
  \multirow{-2}{*}{\textbf{Param\# (MB)$\downarrow$}} \\ \hline
 \multicolumn{14}{|c|}{\textbf{\ac{cln_c5}}} \\ \hline
Order 1          & 94.1 & 0.89 & 86.4 & 0.28 & 90.4 & 1.05 & 93.2 & 0.33 & 86.9 & 1.32 & 92.6 & 0.81 & 28.5  \\
Order 2          & 94.1 & 0.89 & 86.4 & 0.28 & 90.6 & 1.06 & 93.2 & 0.33 & 86.9 & 1.32 & 92.6 & 0.81 & 28.9  \\
Order 3          & 94.2 & 0.97 & 86.1 & 0.29 & 91.3 & 1.10 & 93.1 & 0.48 & 87.1 & 0.92 & 92.7 & 0.87 & 34.2  \\
Order 4          & 94.1 & 0.99 & 86.4 & 0.29 & 89.5 & 1.28 & 93.5 & 0.32 & 86.6 & 1.13 & 92.5 & 0.90 & 35.5  \\
\cellcolor[HTML]{F3F3F3}Mean             & \cellcolor[HTML]{F3F3F3}94.1 & \cellcolor[HTML]{F3F3F3}0.94 & \cellcolor[HTML]{F3F3F3}86.3 & \cellcolor[HTML]{F3F3F3}0.28 & \cellcolor[HTML]{F3F3F3}90.5 & \cellcolor[HTML]{F3F3F3}1.12 & \cellcolor[HTML]{F3F3F3}93.2 & \cellcolor[HTML]{F3F3F3}0.36 & \cellcolor[HTML]{F3F3F3}86.9 & \cellcolor[HTML]{F3F3F3}1.17 & \cellcolor[HTML]{F3F3F3}92.6 & \cellcolor[HTML]{F3F3F3}0.85 & \cellcolor[HTML]{F3F3F3}31.8  \\ \hline
\multicolumn{14}{|c|}{\textbf{\ac{cln_unprn}}}  \\ \hline
Order 1          & 94.1 & 0.90 & 86.4 & 0.28 & 90.6 & 1.09 & 93.1 & 0.34 & 86.8 & 1.21 & 92.6 & 0.82 & 93.9  \\
Order 2          & 94.1 & 0.90 & 86.4 & 0.28 & 90.5 & 1.08 & 93.0 & 0.32 & 86.8 & 1.21 & 92.6 & 0.81 & 93.9  \\
Order 3          & 94.2 & 0.97 & 86.0 & 0.28 & 91.2 & 1.12 & 92.9 & 0.51 & 87.0 & 0.97 & 92.7 & 0.87 & 93.9  \\
Order 4          & 94.1 & 0.98 & 86.4 & 0.29 & 88.8 & 1.28 & 93.3 & 0.34 & 86.4 & 1.21 & 92.5 & 0.89 & 93.9  \\
\cellcolor[HTML]{F3F3F3} Mean             & \cellcolor[HTML]{F3F3F3}94.1 & \cellcolor[HTML]{F3F3F3}0.94 & \cellcolor[HTML]{F3F3F3}86.3 & \cellcolor[HTML]{F3F3F3}0.28 & \cellcolor[HTML]{F3F3F3}90.3 & \cellcolor[HTML]{F3F3F3}1.14 & \cellcolor[HTML]{F3F3F3}93.1 & \cellcolor[HTML]{F3F3F3}0.37 & \cellcolor[HTML]{F3F3F3}86.8 & \cellcolor[HTML]{F3F3F3}1.15 & \cellcolor[HTML]{F3F3F3}92.6 & \cellcolor[HTML]{F3F3F3}0.85 & \cellcolor[HTML]{F3F3F3}93.9  \\ \hline
\multicolumn{14}{|c|}{\textbf{\ac{nnu_e5}}}  \\ \hline
Mean & 94.1 & 0.97 & 86.4 & 0.29 & 90.0 & 1.14 & 92.6 & 0.34 & 86.7 & 1.18 & 92.5 & 0.87 & 156.5 \\ \hline
\end{tabular}%
}
\caption{\textbar~\textbf{\ac{css} order-wise and dataset-wise segmentation performance of \ac{cln_c5} and \ac{cln_unprn} on five representative public datasets.} The mean DSC (\%) and ASD (mm) are evaluated on four \ac{css} orders and five datasets separately for \ac{cln_c5} and \ac{cln_unprn}. The final parameter number of each order is also provided. \ac{nnu_e5} is used as segmentation upper bound of each dataset. }
\label{tab:css_res_avg}
\end{table}
\clearpage

% \usepackage{multirow}
% \usepackage{graphicx}
\begin{table}[htp]
\renewcommand{\arraystretch}{1.2}
\centering
\resizebox{\textwidth}{!}{%
\begin{tabular}{|l|cc|cc|cc|cc|cc|c|}
\hline
\multicolumn{1}{|c|}{} &
  \multicolumn{2}{c|}{\textbf{Order 1}} &
  \multicolumn{2}{c|}{\textbf{Order 2}} &
  \multicolumn{2}{c|}{\textbf{Order 3}} &
  \multicolumn{2}{c|}{\textbf{Order 4}} &
  \multicolumn{2}{c|}{\textbf{Mean}} &
  \multirow{2}{*}{\textbf{Param\# (MB)$\downarrow$}} \\ \cline{2-11}
\multicolumn{1}{|c|}{} &
  DSC$\uparrow$ &
  ASD$\downarrow$ &
  DSC$\uparrow$ &
  ASD$\downarrow$ &
  DSC$\uparrow$ &
  ASD$\downarrow$ &
  DSC$\uparrow$ &
  ASD$\downarrow$ &
  DSC$\uparrow$ &
  ASD$\downarrow$ &
   \\ \hline
MiB~\cite{cermelli2020modeling}         & 6.9  & 67.78 & 15.1 & 66.66 & 77.6 & 2.34 & 20.3 & 49.45 & 30.0 & 46.56 & 31.3 \\
PLOP~\cite{douillard2021plop}        & 19.3 & 68.66 & 40.1 & 8.33  & 75.9 & 2.24 & 48.6 & 14.52 & 46.0 & 23.44 & 31.3 \\
CSCLIP~\cite{zhang2023continual}      & 21.0 & 32.07 & 20.5 & 25.92 & 76.5 & 3.18 & 56.4 & 4.31  & 43.6 & 16.37 & 129  \\
\ac{cln_c5} &
  \textbf{92.6} &
  \textbf{0.81} &
  \textbf{92.6} &
  \textbf{0.81} &
  \textbf{92.7} &
  \textbf{0.87} &
  \textbf{92.5} &
  \textbf{0.90} &
  \textbf{92.6} &
  \textbf{0.85} &
  31.8 \\ \hline
\end{tabular}%
}
\caption{\textbar~\textbf{\ac{css} performance comparison of \ac{cln_c5}, MiB~\cite{cermelli2020modeling}, PLOP~\cite{douillard2021plop}, and CSCLIP~\cite{zhang2023continual} across four orders of five representative public datasets}. The order-wise mean DSC (\%) and ASD (mm) are evaluated on four \ac{css} orders of five representative public datasets. The final parameter number averaged over 4 orders are also provided. \ac{cln_c5} significantly outperforms other comparison methods, with similar small parameter size as MiB and PLOP. }
\label{tab:css_comp_res}
\end{table}
\clearpage

%% Details on Ablation Studies of General Encoder
\begin{table}[htp]
\centering
\resizebox{\textwidth}{!}{
\begin{tabular}{|l|c|c|c|cc|c|}
\hline
Pretrain Dataset & Pretrain Method & Supervision & View Augmentation & DSC$\uparrow$ & ASD$\downarrow$ & \multicolumn{1}{l|}{$\sum \mathcal{T}^{\mathlarger{\prime}}\downarrow$} \\ \hline
 & \cellcolor[HTML]{F3F3F3}Train from scratch & \cellcolor[HTML]{F3F3F3}Bone label set train & \cellcolor[HTML]{F3F3F3}---  & \cellcolor[HTML]{F3F3F3}92.9 & \cellcolor[HTML]{F3F3F3}1.13 & \cellcolor[HTML]{F3F3F3}58.8 \\
 
 & Train from scratch & Full label set train & --- & 94.1 & 0.89 & 38.6 \\

 & \cellcolor[HTML]{F3F3F3}SimSiam & \cellcolor[HTML]{F3F3F3}--- & \cellcolor[HTML]{F3F3F3}Default & \cellcolor[HTML]{F3F3F3}88.6 & \cellcolor[HTML]{F3F3F3}1.73 & \cellcolor[HTML]{F3F3F3}328.6 \\
 
 & SimSiam & --- & Proposed & 89.6 & 1.28 & 315.8 \\
 
 & \cellcolor[HTML]{F3F3F3}SimSiam & \cellcolor[HTML]{F3F3F3}Full label set fine-tune & \cellcolor[HTML]{F3F3F3}Default  & \cellcolor[HTML]{F3F3F3}93.0 & \cellcolor[HTML]{F3F3F3}1.17 & \cellcolor[HTML]{F3F3F3}52.4 \\
 
 & SimSiam & Bone label set fine-tune & Proposed & 93.2 & 1.02 & 51.6 \\
 
\multirow{-8}{*}{TotalSegmentator} & \cellcolor[HTML]{F3F3F3}SimSiam  & \cellcolor[HTML]{F3F3F3}Full label set fine-tune & \cellcolor[HTML]{F3F3F3}Proposed & \cellcolor[HTML]{F3F3F3}94.2 & \cellcolor[HTML]{F3F3F3}0.90 & \cellcolor[HTML]{F3F3F3}38.2 \\ \hline

 & Train from scratch & Head label set train & --- & 91.1 & 1.20 & 72.4 \\
 
\multirow{-2}{*}{StructSeg19} & \cellcolor[HTML]{F3F3F3}SimSiam & \cellcolor[HTML]{F3F3F3}Head label set fine-tune & \cellcolor[HTML]{F3F3F3}Proposed & \cellcolor[HTML]{F3F3F3}92.8 & \cellcolor[HTML]{F3F3F3}1.14 & \cellcolor[HTML]{F3F3F3}62.4 \\ \hline

  & Train from scratch & Pseudo label train & ---  & 94.0 & 1.09 & 56.8 \\ 

 & \cellcolor[HTML]{F3F3F3}SimSiam  & \cellcolor[HTML]{F3F3F3}Sequential full label set fine-tune & \cellcolor[HTML]{F3F3F3}Proposed & \cellcolor[HTML]{F3F3F3}93.0 & \cellcolor[HTML]{F3F3F3}1.14 & \cellcolor[HTML]{F3F3F3}62.6 \\ 

  & SimSiam  & \ac{ema} update & Proposed & 93.0 & 1.13 & 60.6 \\ 
  
 & \cellcolor[HTML]{F3F3F3}SimSiam  & \cellcolor[HTML]{F3F3F3}Full label set momentum update $\mathcal{Q}_{128}$ & \cellcolor[HTML]{F3F3F3}Proposed & \cellcolor[HTML]{F3F3F3}93.1 & \cellcolor[HTML]{F3F3F3}1.12 & \cellcolor[HTML]{F3F3F3}58.6 \\ 
 
 & SimSiam  & Full label set momentum update $\mathcal{Q}_{1,024}$ & Proposed & 94.2 & 0.86 & 37.2   \\ 
 
\multirow{-6}{*}{$\mathcal{U}_4$ Datasets}  & \cellcolor[HTML]{F3F3F3}SimSiam  & \cellcolor[HTML]{F3F3F3}Full label set momentum update $\mathcal{Q}_{4,096}$ & \cellcolor[HTML]{F3F3F3}Proposed & \cellcolor[HTML]{F3F3F3}94.2 & \cellcolor[HTML]{F3F3F3}0.85 & \cellcolor[HTML]{F3F3F3}36.9\\ \hline

$\mathcal{U}_5$ Datasets & SimSiam & Full label set momentum update $\mathcal{Q}_{1,024}$ & Proposed & 94.3 & 0.85 & 35.8 \\ \hline

\end{tabular}
}\caption{\textbar~\textbf{Impact of \ac{ge} pre-training on TotalSegmentator downstream task evaluation}:  The `$\mathcal{U}_4$ datasets' represent the union of body part-wise StructSeg19, SegTHOR, FLARE22, and KiTS21 datasets. The `$\mathcal{U}_5$ datasets' represent the combined of TotalSegmentator and $\mathcal{U}_4$ datasets. }
\label{tab:impact_ge_totalseg}
\end{table}
\clearpage

\begin{table}[htp]
\centering
\resizebox{\textwidth}{!}{
\begin{tabular}{|l|c|c|c|cc|c|}
\hline
Pretrain Dataset & Pretrain Method & Supervision & View Augmentation & \multicolumn{1}{l}{DSC$\uparrow$} & \multicolumn{1}{l|}{ASD$\downarrow$} & $\sum \mathcal{T}^{\mathlarger{\prime}}\downarrow$ \\ \hline
 % & \cellcolor[HTML]{F3F3F3}Train from scratch & \cellcolor[HTML]{F3F3F3}- & \cellcolor[HTML]{F3F3F3} Thoracic label set train & \cellcolor[HTML]{F3F3F3}89.8 & \cellcolor[HTML]{F3F3F3}1.12 & \cellcolor[HTML]{F3F3F3}18.0 \\
 
 & Train from scratch & Full label set train & ---  & 90.4 & 1.05 & 15.2 \\
 
\multirow{-2}{*}{TotalSegmentator} & \cellcolor[HTML]{F3F3F3}SimSiam & \cellcolor[HTML]{F3F3F3}Full label set fine-tune & \cellcolor[HTML]{F3F3F3}Proposed  & \cellcolor[HTML]{F3F3F3}90.5 & \cellcolor[HTML]{F3F3F3}1.05 & \cellcolor[HTML]{F3F3F3}14.9 \\ \hline

 & Train from scratch & Head label set train & ---  & 89.5 & 1.28 & 26.0 \\
 
\multirow{-2}{*}{StructSeg19} & \cellcolor[HTML]{F3F3F3}SimSiam & \cellcolor[HTML]{F3F3F3}Head label set fine-tune & \cellcolor[HTML]{F3F3F3}Proposed & \cellcolor[HTML]{F3F3F3}90.1 & \cellcolor[HTML]{F3F3F3}1.20 & \cellcolor[HTML]{F3F3F3}20.3 \\ \hline

& SimSiam & \ac{ema} update & Proposed & 89.9 & 1.14 & 18.2 \\

\multirow{-2}{*}{$\mathcal{U}_4$ Datasets} & \cellcolor[HTML]{F3F3F3}SimSiam & \cellcolor[HTML]{F3F3F3}Full label set momentum update $\mathcal{Q}_{1,024}$ & \cellcolor[HTML]{F3F3F3}Proposed & \cellcolor[HTML]{F3F3F3}90.4 & \cellcolor[HTML]{F3F3F3}1.04 & \cellcolor[HTML]{F3F3F3}15.2 \\ \hline

$\mathcal{U}_5$ Datasets & SimSiam & Full label set momentum update $\mathcal{Q}_{1,024}$ & Proposed & 90.5 & 1.04 & 14.9 \\ \hline

\end{tabular}
}\caption{\textbar~\textbf{Impact of \ac{ge} pre-training on FLARE22 downstream task evaluation}:  The `$\mathcal{U}_4$ datasets' represent the union of body part-wise StructSeg19, SegTHOR, FLARE22, and KiTS21 datasets. The `$\mathcal{U}_5$ datasets' represent the combined of TotalSegmentator and $\mathcal{U}_4$ datasets. }
\label{tab:impact_ge_flare}
\end{table}
\clearpage

%% Details on Ablation Studies of Multi-path Decoding Heads
\begin{table}[htp]
\centering
\resizebox{0.8\textwidth}{!}{
\begin{tabular}{|l|ccc|rr|c|}
\hline
Testing Dataset & \ac{ge} Params & Updating Scheme & Supporting  & \multicolumn{1}{l}{\ac{dsc}$\uparrow$} & \multicolumn{1}{l|}{\ac{asd}$\downarrow$} & $\sum \mathcal{T}^{\mathlarger{\prime}}\downarrow$ \\ \hline
 & \cellcolor[HTML]{F3F3F3}Fixed & \cellcolor[HTML]{F3F3F3}Fine-tune & \cellcolor[HTML]{F3F3F3}---  & \cellcolor[HTML]{F3F3F3}94.0 & \cellcolor[HTML]{F3F3F3}0.99 & \cellcolor[HTML]{F3F3F3}40.1 \\
 & Fine-tune & Fine-tune & --- & 94.0 & 0.97 & 38.2 \\
\multirow{-3}{*}{TotalSegmentator}& \cellcolor[HTML]{F3F3F3}Fixed  & \cellcolor[HTML]{F3F3F3}\ac{ema} update & \cellcolor[HTML]{F3F3F3}--- & \cellcolor[HTML]{F3F3F3}94.1 & \cellcolor[HTML]{F3F3F3}0.89 & \cellcolor[HTML]{F3F3F3}38.2 \\ \hline

 & Fixed & Fine-tune & --- & 86.2 & 0.28 & 27.3 \\
 & \cellcolor[HTML]{F3F3F3}Fine-tune & \cellcolor[HTML]{F3F3F3}Fine-tune & \cellcolor[HTML]{F3F3F3}--- & \cellcolor[HTML]{F3F3F3}85.6 & \cellcolor[HTML]{F3F3F3}0.31 & \cellcolor[HTML]{F3F3F3}24.2 \\
\multirow{-3}{*}{StructSeg19} & Fixed & \ac{ema} update & --- & 86.4 & 0.28 & 24.8 \\ \hline

 & \cellcolor[HTML]{F3F3F3}Fixed & \cellcolor[HTML]{F3F3F3}Fine-tune & \cellcolor[HTML]{F3F3F3}--- & \cellcolor[HTML]{F3F3F3}92.9 & \cellcolor[HTML]{F3F3F3}0.34 & \cellcolor[HTML]{F3F3F3}1.6 \\
 & Fine-tune & Fine-tune & --- & 92.6 & 0.33 & 1.6 \\
\multirow{-3}{*}{SegTHOR} & \cellcolor[HTML]{F3F3F3}Fixed & \cellcolor[HTML]{F3F3F3}\ac{ema} update & \cellcolor[HTML]{F3F3F3}--- & \cellcolor[HTML]{F3F3F3}93.3 & \cellcolor[HTML]{F3F3F3}0.31 & \cellcolor[HTML]{F3F3F3}1.6 \\ \hline

 & Fixed & Fine-tune & --- & 89.9 & 1.16 & 18.0 \\
 & \cellcolor[HTML]{F3F3F3}Fine-tune & \cellcolor[HTML]{F3F3F3}Fine-tune & \cellcolor[HTML]{F3F3F3}--- & \cellcolor[HTML]{F3F3F3}90.0 & \cellcolor[HTML]{F3F3F3}1.14 & \cellcolor[HTML]{F3F3F3}15.2 \\
\multirow{-3}{*}{FLARE22} & Fixed & \ac{ema} update & --- & 90.4 & 1.05 & 15.2 \\ \hline

 & \cellcolor[HTML]{F3F3F3}Fixed & \cellcolor[HTML]{F3F3F3}Fine-tune & \cellcolor[HTML]{F3F3F3}--- & \cellcolor[HTML]{F3F3F3}84.2 & \cellcolor[HTML]{F3F3F3}1.18 & \cellcolor[HTML]{F3F3F3}6.0 \\
 & Fine-tune & Fine-tune & --- & 84.6 & 1.20 & 2.0 \\
 & \cellcolor[HTML]{F3F3F3}Fixed & \cellcolor[HTML]{F3F3F3}\ac{ema} update & \cellcolor[HTML]{F3F3F3}--- & \cellcolor[HTML]{F3F3F3}85.0 & \cellcolor[HTML]{F3F3F3}1.18 & \cellcolor[HTML]{F3F3F3}4.0 \\
 & Fine-tune & Fine-tune & Channel-wise & 85.9 & 1.17 & 4.0 \\
 & \cellcolor[HTML]{F3F3F3}Fixed & \cellcolor[HTML]{F3F3F3}Fine-tune & \cellcolor[HTML]{F3F3F3}Feature-level & \cellcolor[HTML]{F3F3F3}86.3 & \cellcolor[HTML]{F3F3F3}1.14 & \cellcolor[HTML]{F3F3F3}2.0 \\
 & Fine-tune & \ac{ema} update & Channel-wise & 86.4 & 1.14 & 2.0 \\
\multirow{-7}{*}{KiTS21} & \cellcolor[HTML]{F3F3F3}Fixed & \cellcolor[HTML]{F3F3F3}\ac{ema} update & \cellcolor[HTML]{F3F3F3}Feature-level & \cellcolor[HTML]{F3F3F3}86.9 & \cellcolor[HTML]{F3F3F3}1.12 & \cellcolor[HTML]{F3F3F3}2.0 \\ \hline

\end{tabular}
}\caption{\textbar~\textbf{Impact of fine-tuning \ac{ge}, \ac{ema} update, and feature-level supporting}: The \ac{ge} is pre-trained using SimSiam and later momentum-updated using $\mathcal{U}_5$ datasets. The training order of the stratified decoders follows `Order 1': TotalSegmentator $\rightarrow$ StructSeg19 $\rightarrow$ FLARE22 $\rightarrow$ SegTHOR $\rightarrow$ KiTS21}
\label{tab:impact_decoder}
\end{table}
\clearpage

\begin{table}[htp]
\centering
\resizebox{\textwidth}{!}{
\begin{tabular}{|l|ll|r|ll|r|ll|r|ll|r|}
\hline
 & \multicolumn{3}{c|}{\ac{cln_c36_nofls}} & \multicolumn{3}{c|}{\ac{cln_c36}} & \multicolumn{3}{c|}{\ac{cln_u36_nofls}} & \multicolumn{3}{c|}{\ac{cln_u36}} \\ \cline{2-13} 
 
\multirow{-2}{*}{} & DSC & {ASD} & $\mathcal{T}$ & DSC & {ASD} & $\mathcal{T}$ & DSC & {ASD} & $\mathcal{T}$ & DSC & {ASD} & $\mathcal{T}$ \\ \hline
\rowcolor[HTML]{F3F3F3} 
HN\_LNS\_1La & 50.8$\pm$6.7 & {1.12$\pm$0.38} & 86.0 & 52.9$\pm$6.7 & {1.06$\pm$0.38} & 86.0 & 49.9$\pm$6.8 & {1.10$\pm$0.40} & 86.0 & 52.2$\pm$5.9 & {1.02$\pm$0.41} & 88.0 \\
\rowcolor[HTML]{F3F3F3} 
HN\_LNS\_1Lb & 78.4$\pm$6.9 & {0.25$\pm$0.74} &  & 81.3$\pm$6.8 & {0.20$\pm$0.74} &  & 78.9$\pm$7.0 & {0.28$\pm$0.75} &  & 80.8$\pm$7.7 & {0.21$\pm$0.76} &  \\
\rowcolor[HTML]{F3F3F3} 
HN\_LNS\_1Ra & 54.7$\pm$6.6 & {1.26$\pm$0.69} &  & 55.0$\pm$6.6 & {1.21$\pm$0.69} &  & 54.5$\pm$6.6 & {1.38$\pm$0.68} &  & 56.3$\pm$5.8 & {1.31$\pm$0.68} &  \\
\rowcolor[HTML]{F3F3F3} 
HN\_LNS\_1Rb & 79.9$\pm$8.7 & {1.06$\pm$0.83} &  & 80.8$\pm$8.1 & {1.00$\pm$0.82} &  & 80.1$\pm$8.6 & {1.10$\pm$0.82} &  & 82.0$\pm$8.0 & {1.04$\pm$0.83} &  \\
\rowcolor[HTML]{F3F3F3} 
HN\_LNS\_2La & 70.1$\pm$5.6 & {1.07$\pm$0.98} &  & 71.5$\pm$5.0 & {0.99$\pm$0.99} &  & 70.4$\pm$5.8 & {0.97$\pm$0.91} &  & 72.1$\pm$6.1 & {0.92$\pm$0.92} &  \\
\rowcolor[HTML]{F3F3F3} 
HN\_LNS\_2Lb & 72.7$\pm$6.5 & {1.77$\pm$0.69} &  & 74.6$\pm$5.8 & {1.68$\pm$0.70} &  & 71.9$\pm$6.2 & {1.68$\pm$0.66} &  & 74.4$\pm$7.2 & {1.58$\pm$0.67} &  \\
\rowcolor[HTML]{F3F3F3} 
HN\_LNS\_2Ra & 74.0$\pm$7.1 & {1.96$\pm$0.39} &  & 76.6$\pm$6.7 & {1.91$\pm$0.38} &  & 74.3$\pm$7.0 & {1.87$\pm$0.38} &  & 75.5$\pm$6.7 & {1.78$\pm$0.37} &  \\
\rowcolor[HTML]{F3F3F3} 
HN\_LNS\_2Rb & 74.9$\pm$13.3 & {1.62$\pm$0.57} &  & 75.0$\pm$14.1 & {1.53$\pm$0.56} &  & 74.0$\pm$13.5 & {1.72$\pm$0.60} &  & 74.9$\pm$13.9 & {1.64$\pm$0.60} &  \\
\rowcolor[HTML]{F3F3F3} 
HN\_LNS\_3L & 75.6$\pm$12.3 & {1.57$\pm$0.76} &  & 75.9$\pm$12.3 & {1.52$\pm$0.75} &  & 75.7$\pm$12.4 & {1.50$\pm$0.78} &  & 76.1$\pm$13.0 & {1.40$\pm$0.78} &  \\
\rowcolor[HTML]{F3F3F3} 
HN\_LNS\_3R & 72.0$\pm$6.9 & {1.76$\pm$0.64} &  & 72.0$\pm$7.9 & {1.68$\pm$0.63} &  & 72.1$\pm$6.7 & {1.61$\pm$0.62} &  & 72.3$\pm$5.7 & {1.56$\pm$0.62} &  \\
\rowcolor[HTML]{F3F3F3} 
HN\_LNS\_4L & 76.4$\pm$6.4 & {1.41$\pm$0.50} &  & 77.7$\pm$5.6 & {1.31$\pm$0.50} &  & 76.3$\pm$6.1 & {1.41$\pm$0.47} &  & 76.5$\pm$5.3 & {1.36$\pm$0.47} &  \\
\rowcolor[HTML]{F3F3F3} 
HN\_LNS\_4R & 73.4$\pm$6.4 & {1.51$\pm$0.47} &  & 74.5$\pm$6.3 & {1.43$\pm$0.47} &  & 72.7$\pm$6.7 & {1.37$\pm$0.51} &  & 75.7$\pm$5.7 & {1.30$\pm$0.50} &  \\
\rowcolor[HTML]{F3F3F3} 
HN\_LNS\_5La & 73.0$\pm$3.3 & {0.99$\pm$0.56} &  & 75.1$\pm$3.5 & {0.90$\pm$0.55} &  & 73.8$\pm$3.4 & {0.94$\pm$0.57} &  & 74.2$\pm$3.0 & {0.88$\pm$0.57} &  \\
\rowcolor[HTML]{F3F3F3} 
HN\_LNS\_5Lb & 75.4$\pm$3.7 & {0.81$\pm$0.20} &  & 77.5$\pm$3.3 & {0.75$\pm$0.19} &  & 75.2$\pm$3.7 & {0.88$\pm$0.20} &  & 76.0$\pm$2.8 & {0.80$\pm$0.19} &  \\
\rowcolor[HTML]{F3F3F3} 
HN\_LNS\_5Ra & 77.0$\pm$3.1 & {0.97$\pm$0.15} &  & 78.3$\pm$3.3 & {0.92$\pm$0.14} &  & 76.5$\pm$2.9 & {0.94$\pm$0.13} &  & 78.8$\pm$3.9 & {0.87$\pm$0.14} &  \\
\rowcolor[HTML]{F3F3F3} 
HN\_LNS\_5Rb & 77.7$\pm$5.8 & {1.48$\pm$0.09} &  & 80.5$\pm$6.4 & {1.45$\pm$0.08} &  & 77.5$\pm$5.8 & {1.59$\pm$0.09} &  & 77.9$\pm$5.2 & {1.52$\pm$0.09} &  \\
\rowcolor[HTML]{F3F3F3} 
HN\_LNS\_6L & 51.2$\pm$14.5 & {1.27$\pm$0.97} &  & 52.9$\pm$15.0 & {1.24$\pm$0.97} &  & 51.6$\pm$14.8 & {1.35$\pm$0.99} &  & 53.1$\pm$14.6 & {1.29$\pm$0.99} &  \\
\rowcolor[HTML]{F3F3F3} 
HN\_LNS\_6R & 50.2$\pm$14.5 & {1.46$\pm$0.96} &  & 50.9$\pm$14.7 & {1.39$\pm$0.96} &  & 50.9$\pm$14.5 & {1.59$\pm$0.96} &  & 53.1$\pm$15.3 & {1.55$\pm$0.95} &  \\ \hline
Chest\_LNS1\_L & 77.7$\pm$8.1 & {1.56$\pm$0.43} & 90.0 & 80.9$\pm$8.0 & {1.41$\pm$0.43} & 90.0 & 78.7$\pm$7.9 & {1.43$\pm$0.41} & 90.0 & 81.8$\pm$8.5 & {1.32$\pm$0.42} & 90.0 \\
Chest\_LNS1\_R & 77.7$\pm$10.1 & {1.78$\pm$0.60} &  & 82.0$\pm$10.1 & {1.68$\pm$0.60} &  & 77.1$\pm$10.2 & {1.85$\pm$0.61} &  & 80.2$\pm$10.2 & {1.69$\pm$0.61} &  \\
Chest\_LNS2.L & 67.2$\pm$10.7 & {1.40$\pm$0.48} &  & 70.6$\pm$10.7 & {1.22$\pm$0.48} &  & 66.7$\pm$10.7 & {1.36$\pm$0.49} &  & 70.6$\pm$10.7 & {1.25$\pm$0.49} &  \\
Chest\_LNS2\_R & 70.5$\pm$15.5 & {1.48$\pm$2.41} &  & 74.7$\pm$15.5 & {1.31$\pm$2.41} &  & 69.9$\pm$15.5 & {1.47$\pm$2.27} &  & 74.8$\pm$15.5 & {1.34$\pm$2.27} &  \\
Chest\_LNS3\_A & 77.7$\pm$19.7 & {0.96$\pm$3.46} &  & 81.3$\pm$19.7 & {0.78$\pm$3.46} &  & 78.3$\pm$19.4 & {0.95$\pm$3.61} &  & 82.9$\pm$19.4 & {0.77$\pm$3.61} &  \\
Chest\_LNS3\_P & 84.7$\pm$8.4 & {0.81$\pm$1.14} &  & 88.8$\pm$8.4 & {0.62$\pm$1.14} &  & 85.1$\pm$8.3 & {0.72$\pm$1.09} &  & 90.1$\pm$8.3 & {0.60$\pm$1.09} &  \\
Chest\_LNS4\_L & 74.3$\pm$10.3 & {0.96$\pm$0.91} &  & 78.0$\pm$10.3 & {0.85$\pm$0.91} &  & 74.7$\pm$10.3 & {1.02$\pm$0.92} &  & 78.0$\pm$10.3 & {0.86$\pm$0.92} &  \\
Chest\_LNS4\_R & 73.7$\pm$13.4 & {1.49$\pm$1.00} &  & 76.8$\pm$13.4 & {1.36$\pm$1.00} &  & 73.6$\pm$13.3 & {1.58$\pm$0.99} &  & 77.9$\pm$13.3 & {1.48$\pm$0.99} &  \\
Chest\_LNS5 & 72.4$\pm$12.1 & {1.28$\pm$0.98} &  & 71.6$\pm$12.5 & {1.27$\pm$0.98} &  & 73.3$\pm$11.9 & {1.31$\pm$0.97} &  & 72.6$\pm$12.7 & {1.31$\pm$0.97} &  \\
Chest\_LNS6 & 72.5$\pm$12.1 & {0.79$\pm$0.95} &  & 71.8$\pm$11.2 & {0.79$\pm$0.95} &  & 72.8$\pm$12.2 & {0.75$\pm$0.97} &  & 73.6$\pm$12.9 & {0.75$\pm$0.97} &  \\
Chest\_LNS7 & 85.7$\pm$16.3 & {0.90$\pm$1.09} &  & 86.1$\pm$15.7 & {0.90$\pm$1.10} &  & 86.2$\pm$16.2 & {0.96$\pm$1.04} &  & 86.6$\pm$15.5 & {0.97$\pm$1.04} &  \\
Chest\_LNS8 & 81.6$\pm$14.2 & {1.67$\pm$0.40} &  & 82.5$\pm$13.7 & {1.68$\pm$0.41} &  & 82.5$\pm$14.4 & {1.64$\pm$0.42} &  & 82.2$\pm$14.9 & {1.64$\pm$0.42} &  \\
Chest\_LNS9 & 62.8$\pm$17.2 & {1.38$\pm$1.43} &  & 62.4$\pm$16.2 & {1.37$\pm$1.44} &  & 62.5$\pm$17.2 & {1.24$\pm$1.51} &  & 62.0$\pm$17.2 & {1.25$\pm$1.50} &  \\
Chest\_LNS10\_L & 74.6$\pm$19.1 & {1.53$\pm$1.18} &  & 78.0$\pm$20.0 & {1.37$\pm$1.18} &  & 73.6$\pm$19.1 & {1.45$\pm$1.16} &  & 77.4$\pm$19.4 & {1.32$\pm$1.16} &  \\
Chest\_LNS10\_R & 73.3$\pm$13.6 & {1.51$\pm$1.26} &  & 77.2$\pm$12.6 & {1.39$\pm$1.26} &  & 73.7$\pm$13.4 & {1.40$\pm$1.28} &  & 77.2$\pm$13.2 & {1.20$\pm$1.27} &  \\ \hline

Mean & \multicolumn{1}{c}{72.2} & \multicolumn{1}{c|}{1.30} & \multicolumn{1}{c|}{88.0} & \multicolumn{1}{c}{74.1} & \multicolumn{1}{c|}{1.22} & \multicolumn{1}{c|}{88.0} & \multicolumn{1}{c}{72.3} & \multicolumn{1}{c|}{1.29} & \multicolumn{1}{c|}{88.0} & \multicolumn{1}{c}{74.2} & \multicolumn{1}{c|}{1.21} & \multicolumn{1}{c|}{89.0} \\ \hline
\end{tabular}
}
\caption{\textbar~\textbf{Impact of \ac{fls} on head \& neck and chest LNS downstream task evaluation}:  Comparison of DSC (\%) and ASD (mm) among \ac{cln_c36}, \ac{cln_c36_nofls}, \ac{cln_u36}, and \ac{cln_u36_nofls} for 18 head \& neck LNSs and 15 chest LNSs, respectively. Decoder-wise pruning rate ($\mathcal{T}$, \%) are also provided. The supporting structures for the head \& neck LNS segmentation are Musc\_Cervial\_A, Musc\_Scalenus, Musc\_Scalenus\_A, and Musc\_Scleido. The supporting structures for the chest LNS segmentation are A\_Aorta, Bronchus\_L, Bronchus\_R, Eso, and Heart~\cite{guo2021deepstationing}. \ac{cln_c36} and \ac{cln_u36} achieve the best performance, demonstrating their superior segmentation capabilities when using \ac{fls}.}

\label{tab:abl_fls_lns}
\end{table}
\clearpage

%% 235-organ-wise results of 6 organ groups (Table 1 details)
%% C36, U36, nnUNet_E36 236-organ-wise results
\begin{table}[htp]
\centering
\resizebox{0.9\textwidth}{!}{
\begin{tabular}{|l|ll|ll|ll|r|ll|r|}
\hline
\multicolumn{1}{|l|}{} & \multicolumn{2}{c|}{\cellcolor[HTML]{FFFFFF}{\color[HTML]{1F1F1F} \ac{nnu_e36}}} & \multicolumn{2}{c|}{\ac{cln_u36_unprn}} & \multicolumn{3}{c|}{\ac{cln_c36}} & \multicolumn{3}{c|}{\ac{cln_u36}} \\ \cline{2-11} 
\multicolumn{1}{|l|}{\multirow{-2}{*}{}} & DSC$\uparrow$ & \multicolumn{1}{l|}{ASD$\downarrow$} & DSC$\uparrow$ & \multicolumn{1}{l|}{ASD$\downarrow$} & DSC$\uparrow$ & ASD$\downarrow$ & \multicolumn{1}{r|}{$\mathcal{T}$$\uparrow$ } & DSC$\uparrow$ & ASD$\downarrow$ & \multicolumn{1}{r|}{$\mathcal{T}$$\uparrow$ } \\ \hline
\rowcolor[HTML]{F3F3F3} 
Hippocampus\_L & 62.4$\pm$6.1 & 1.99$\pm$0.25 & 73.5$\pm$5.9 & 1.27$\pm$0.23 & 71.8$\pm$5.9 & 1.44$\pm$0.22 & 92.0 & 73.6$\pm$5.8 & 1.25$\pm$0.21 & 92.0 \\
\rowcolor[HTML]{F3F3F3} 
Hippocampus\_R & 62.2$\pm$6.4 & 1.90$\pm$0.15 & 74.3$\pm$6.6 & 1.25$\pm$0.17 & 72.7$\pm$6.4 & 1.42$\pm$0.15 &  & 74.3$\pm$6.5 & 1.24$\pm$0.16 &  \\
TemporalLobe\_L & 84.4$\pm$1.0 & 2.18$\pm$0.06 & 88.1$\pm$1.4 & 1.53$\pm$0.10 & 86.6$\pm$1.4 & 1.70$\pm$0.10 & \cellcolor[HTML]{F3F3F3}99.0 & 88.1$\pm$1.3 & 1.51$\pm$0.08 & 99.5 \\
TemporalLobe\_R & 83.7$\pm$1.3 & 2.26$\pm$0.08 & 87.4$\pm$1.6 & 1.65$\pm$0.11 & 86.0$\pm$1.3 & 1.85$\pm$0.08 &  & 87.4$\pm$1.5 & 1.64$\pm$0.10 & {} \\
\rowcolor[HTML]{F3F3F3} 
HypoThalamus & 60.2$\pm$25.7 & 2.54$\pm$2.62 & 74.1$\pm$25.7 & 0.80$\pm$2.56 & 72.7$\pm$25.7 & 0.76$\pm$2.79 & 96.0 & 74.1$\pm$25.9 & 0.79$\pm$2.65 & 96.0 \\
BasalGanglia\_L & 62.7$\pm$11.7 & 2.66$\pm$1.28 & 70.0$\pm$11.9 & 2.24$\pm$1.14 & 69.7$\pm$12.1 & 2.37$\pm$1.27 & \cellcolor[HTML]{F3F3F3}92.0 & 70.1$\pm$11.7 & 2.24$\pm$1.26 & 90.0 \\
BasalGanglia\_R & 60.7$\pm$12.2 & 2.78$\pm$1.31 & 70.3$\pm$12.1 & 2.08$\pm$1.24 & 70.1$\pm$12.1 & 2.29$\pm$1.29 &  & 70.4$\pm$12.0 & 2.07$\pm$1.34 & {} \\
\rowcolor[HTML]{F3F3F3} 
BrainStem & 91.2$\pm$1.4 & 0.80$\pm$0.08 & 93.6$\pm$1.5 & 0.69$\pm$0.10 & 93.4$\pm$1.6 & 0.70$\pm$0.11 & 99.5 & 93.7$\pm$1.5 & 0.69$\pm$0.10 & 99.5 \\
Cerebellum & 90.0$\pm$3.1 & 1.20$\pm$2.61 & 90.5$\pm$3.0 & 0.91$\pm$2.44 & 90.2$\pm$2.9 & 1.00$\pm$2.59 & \cellcolor[HTML]{F3F3F3}99.5 & 90.6$\pm$2.8 & 0.91$\pm$2.49 & 99.7 \\
\rowcolor[HTML]{F3F3F3} 
Glnd\_Pineal & 60.2$\pm$19.3 & 1.06$\pm$0.50 & 70.7$\pm$19.0 & 0.60$\pm$0.48 & 70.4$\pm$19.5 & 0.60$\pm$0.53 & 90.0 & 70.6$\pm$19.1 & 0.59$\pm$0.49 & 92.0 \\
Brain & 96.7$\pm$2.8 & 0.59$\pm$2.02 & 98.3$\pm$2.9 & 0.27$\pm$2.04 & 98.2$\pm$2.9 & 0.29$\pm$1.93 & \cellcolor[HTML]{F3F3F3}99.5 & 98.3$\pm$2.8 & 0.26$\pm$2.13 & 99.5 \\
\rowcolor[HTML]{F3F3F3} 
Glnd\_Pituitary & 75.0$\pm$6.3 & 0.81$\pm$0.19 & 85.1$\pm$6.7 & 0.50$\pm$0.24 & 83.8$\pm$6.6 & 0.52$\pm$0.22 & 98.0 & 85.1$\pm$6.6 & 0.49$\pm$0.22 & 98.0 \\
Lens\_L & 78.6$\pm$9.7 & 0.57$\pm$0.16 & 82.4$\pm$9.7 & 0.50$\pm$0.17 & 81.5$\pm$9.7 & 0.53$\pm$0.17 & \cellcolor[HTML]{F3F3F3}99.0 & 82.5$\pm$9.7 & 0.50$\pm$0.16 & 99.0 \\
Lens\_R & 80.4$\pm$9.4 & 0.48$\pm$0.44 & 86.3$\pm$9.1 & 0.43$\pm$0.41 & 84.2$\pm$9.1 & 0.40$\pm$0.41 &  & 86.2$\pm$9.2 & 0.42$\pm$0.42 & {} \\
\rowcolor[HTML]{F3F3F3} 
Chiasm & 62.5$\pm$12.9 & 1.42$\pm$0.13 & 74.3$\pm$13.0 & 0.69$\pm$0.14 & 72.7$\pm$13.1 & 0.82$\pm$0.14 & 94.0 & 74.3$\pm$13.0 & 0.68$\pm$0.14 & 96.0 \\
OpticNerve\_L & 67.6$\pm$8.2 & 0.78$\pm$0.30 & 78.3$\pm$7.9 & 0.57$\pm$0.27 & 77.7$\pm$7.8 & 0.56$\pm$0.27 & \cellcolor[HTML]{F3F3F3}98.0 & 78.4$\pm$8.0 & 0.56$\pm$0.28 & 99.0 \\
OpticNerve\_R & 66.8$\pm$6.4 & 0.81$\pm$0.21 & 79.6$\pm$6.4 & 0.56$\pm$0.21 & 79.2$\pm$6.6 & 0.59$\pm$0.23 &  & 79.6$\pm$6.3 & 0.55$\pm$0.20 & {} \\
\rowcolor[HTML]{F3F3F3} 
Eye\_L & 85.7$\pm$1.3 & 0.82$\pm$0.08 & 88.4$\pm$1.3 & 0.61$\pm$0.08 & 86.4$\pm$1.3 & 0.63$\pm$0.08 & 99.5 & 88.4$\pm$1.2 & 0.61$\pm$0.07 & 99.0 \\
\rowcolor[HTML]{F3F3F3} 
Eye\_R & 87.7$\pm$4.0 & 0.55$\pm$0.09 & 87.6$\pm$4.2 & 0.56$\pm$0.11 & 86.6$\pm$4.4 & 0.60$\pm$0.13 &  & 87.7$\pm$4.1 & 0.56$\pm$0.10 &  \\
Glnd\_Lacrimal\_L & 68.1$\pm$12.6 & 1.05$\pm$0.74 & 72.6$\pm$12.7 & 0.60$\pm$0.75 & 70.8$\pm$12.5 & 0.59$\pm$0.72 & \cellcolor[HTML]{F3F3F3}98.0 & 72.6$\pm$12.6 & 0.58$\pm$0.74 & 98.0 \\
Glnd\_Lacrimal\_R & 67.1$\pm$15.5 & 1.37$\pm$0.88 & 72.7$\pm$15.1 & 0.65$\pm$0.85 & 70.6$\pm$15.3 & 0.61$\pm$0.86 &  & 72.7$\pm$15.2 & 0.65$\pm$0.86 & {} \\
\rowcolor[HTML]{F3F3F3} 
Ear\_Inner\_L & 82.8$\pm$9.4 & 0.57$\pm$0.36 & 88.1$\pm$9.2 & 0.59$\pm$0.35 & 86.4$\pm$9.3 & 0.58$\pm$0.35 & 99.5 & 88.2$\pm$9.4 & 0.58$\pm$0.36 & 99.5 \\
\rowcolor[HTML]{F3F3F3} 
Ear\_Inner\_R & 83.0$\pm$9.1 & 0.58$\pm$0.28 & 86.9$\pm$9.0 & 0.59$\pm$0.28 & 86.5$\pm$9.3 & 0.61$\pm$0.30 &  & 86.9$\pm$9.2 & 0.58$\pm$0.30 &  \\
Ear\_Mid\_L & 85.1$\pm$3.8 & 0.97$\pm$0.05 & 91.2$\pm$3.6 & 0.65$\pm$0.03 & 89.1$\pm$3.9 & 0.67$\pm$0.05 & \cellcolor[HTML]{F3F3F3}99.5 & 91.2$\pm$3.6 & 0.63$\pm$0.03 & 99.5 \\
Ear\_Mid\_R & 85.0$\pm$3.4 & 0.94$\pm$0.06 & 90.3$\pm$3.2 & 0.62$\pm$0.04 & 88.9$\pm$3.2 & 0.60$\pm$0.04 &  & 90.3$\pm$3.2 & 0.62$\pm$0.04 & {} \\
\rowcolor[HTML]{F3F3F3} 
Ear\_Cochlea\_L & 67.2$\pm$10.4 & 0.47$\pm$0.23 & 73.2$\pm$10.2 & 0.35$\pm$0.21 & 72.4$\pm$10.6 & 0.36$\pm$0.24 & 96.0 & 73.2$\pm$10.4 & 0.34$\pm$0.22 & 94.0 \\
\rowcolor[HTML]{F3F3F3} 
Ear\_Cochlea\_R & 67.2$\pm$11.4 & 0.47$\pm$0.26 & 71.5$\pm$11.2 & 0.48$\pm$0.24 & 69.5$\pm$11.3 & 0.45$\pm$0.25 &  & 71.5$\pm$11.2 & 0.46$\pm$0.24 &  \\
Glnd\_Parotid\_L & 85.6$\pm$1.9 & 1.05$\pm$0.16 & 87.4$\pm$1.9 & 0.97$\pm$0.16 & 85.8$\pm$1.9 & 0.99$\pm$0.16 & \cellcolor[HTML]{F3F3F3}99.0 & 87.4$\pm$2.1 & 0.97$\pm$0.18 & 99.5 \\
Glnd\_Parotid\_R & 85.7$\pm$3.3 & 1.09$\pm$0.51 & 86.3$\pm$3.4 & 1.11$\pm$0.52 & 85.8$\pm$3.6 & 1.05$\pm$0.54 &  & 86.4$\pm$3.4 & 1.09$\pm$0.52 & {} \\
\rowcolor[HTML]{F3F3F3} 
Glnd\_Submand\_L & 76.9$\pm$7.5 & 1.24$\pm$0.98 & 82.0$\pm$7.5 & 0.90$\pm$0.99 & 80.3$\pm$7.4 & 0.91$\pm$0.99 & 98.0 & 81.9$\pm$7.5 & 0.88$\pm$0.99 & 98.0 \\
\rowcolor[HTML]{F3F3F3} 
Glnd\_Submand\_R & 76.1$\pm$8.4 & 1.20$\pm$0.44 & 78.2$\pm$8.4 & 1.01$\pm$0.44 & 77.6$\pm$8.4 & 1.18$\pm$0.44 &  & 78.1$\pm$8.6 & 1.01$\pm$0.46 &  \\
Glnd\_Thyroid\_L & 81.2$\pm$4.4 & 0.85$\pm$0.26 & 81.8$\pm$4.4 & 0.78$\pm$0.25 & 81.6$\pm$4.8 & 0.74$\pm$0.29 & \cellcolor[HTML]{F3F3F3}94.0 & 81.8$\pm$4.6 & 0.77$\pm$0.27 & 94.0 \\
Glnd\_Thyroid\_R & 80.5$\pm$4.4 & 0.97$\pm$0.23 & 83.3$\pm$4.2 & 0.77$\pm$0.21 & 82.2$\pm$4.0 & 0.75$\pm$0.19 &  & 83.4$\pm$4.3 & 0.75$\pm$0.21 & {} \\
\rowcolor[HTML]{F3F3F3} 
OralCavity & 76.3$\pm$1.9 & 2.40$\pm$0.15 & 79.5$\pm$1.8 & 4.35$\pm$0.13 & 77.8$\pm$1.9 & 4.47$\pm$0.15 & 96.0 & 79.5$\pm$1.8 & 4.33$\pm$0.14 & 96.0 \\
TMJ\_L & 77.2$\pm$9.5 & 0.65$\pm$0.24 & 78.3$\pm$9.5 & 0.61$\pm$0.23 & 77.6$\pm$9.6 & 0.67$\pm$0.24 & \cellcolor[HTML]{F3F3F3}94.0 & 78.4$\pm$9.6 & 0.61$\pm$0.24 & 94.0 \\
TMJ\_R & 75.2$\pm$7.7 & 0.75$\pm$0.21 & 76.6$\pm$7.4 & 0.71$\pm$0.18 & 76.2$\pm$7.5 & 0.70$\pm$0.19 & {} & 76.7$\pm$7.4 & 0.70$\pm$0.18 & {} \\
\rowcolor[HTML]{F3F3F3} 
Bone\_ET\_L & 77.0$\pm$12.7 & 0.43$\pm$0.15 & 82.8$\pm$13.1 & 0.41$\pm$0.19 & 82.5$\pm$12.8 & 0.42$\pm$0.16 & 99.5 & 82.9$\pm$13.0 & 0.40$\pm$0.18 & 99.5 \\
\rowcolor[HTML]{F3F3F3} 
Bone\_ET\_R & 89.8$\pm$10.0 & 0.37$\pm$0.14 & 95.6$\pm$10.1 & 0.31$\pm$0.15 & 95.3$\pm$10.1 & 0.31$\pm$0.15 &  & 95.6$\pm$10.1 & 0.31$\pm$0.15 & \\
Bone\_Mandible\_L & 90.2$\pm$2.0 & 0.50$\pm$0.07 & 92.1$\pm$2.3 & 0.43$\pm$0.09 & 91.7$\pm$2.3 & 0.40$\pm$0.09 & {99.7} & 92.1$\pm$2.3 & 0.41$\pm$0.09 & {99.7} \\
Bone\_Mandible\_R & 90.8$\pm$2.3 & 0.45$\pm$0.06 & 91.4$\pm$2.5 & 0.45$\pm$0.08 & 91.1$\pm$2.4 & 0.43$\pm$0.07 &  & 91.3$\pm$2.4 & 0.44$\pm$0.07 &  \\

\rowcolor[HTML]{F3F3F3} 
Musc\_Masseter\_L & 82.5$\pm$6.0 & 1.15$\pm$1.11 & 84.1$\pm$5.9 & 0.91$\pm$1.06 & 83.5$\pm$6.0 & 1.05$\pm$1.19 & 99.5 & 84.1$\pm$5.8 & 0.91$\pm$1.14 & 99.5 \\
\rowcolor[HTML]{F3F3F3} 
Musc\_Masster\_R & 81.2$\pm$6.4 & 1.28$\pm$1.07 & 81.9$\pm$6.4 & 1.19$\pm$0.97 & 81.6$\pm$6.4 & 1.25$\pm$1.12 &  & 81.9$\pm$6.3 & 1.19$\pm$1.04 &  \\
Epiglottis & 68.1$\pm$8.9 & 1.33$\pm$6.01 & 71.1$\pm$9.2 & 1.04$\pm$5.85 & 70.7$\pm$9.0 & 1.25$\pm$5.79 & 99.5 & 71.0$\pm$9.0 & 1.02$\pm$6.17 & 99.5 \\
\rowcolor[HTML]{F3F3F3} 
Musc\_Const\_I & 70.2$\pm$5.9 & 1.10$\pm$0.49 & 74.8$\pm$6.1 & 0.98$\pm$0.50 & 74.2$\pm$5.9 & 1.03$\pm$0.49 & 99.0 & 74.7$\pm$5.9 & 0.98$\pm$0.49 & 99.0 \\
\rowcolor[HTML]{F3F3F3} 
Musc\_Const\_M & 63.5$\pm$8.4 & 1.71$\pm$0.41 & 68.1$\pm$8.6 & 1.60$\pm$0.43 & 66.4$\pm$8.3 & 1.63$\pm$0.40 &  & 68.0$\pm$8.7 & 1.59$\pm$0.43 &  \\
\rowcolor[HTML]{F3F3F3} 
Musc\_Const\_S & 61.2$\pm$7.5 & 1.86$\pm$1.22 & 62.4$\pm$7.7 & 1.63$\pm$1.03 & 62.0$\pm$7.6 & 1.85$\pm$1.15 &  & 62.4$\pm$7.5 & 1.63$\pm$1.12 &  \\
Larynex\_Glottic & 68.0$\pm$3.4 & 1.67$\pm$0.18 & 70.7$\pm$3.3 & 1.52$\pm$0.17 & 69.4$\pm$3.3 & 1.68$\pm$0.17 & 94.0 & 70.8$\pm$3.3 & 1.50$\pm$0.17 & 94.0 \\
Larynx\_Supraglot & 65.3$\pm$7.9 & 1.62$\pm$0.39 & 67.7$\pm$8.0 & 1.44$\pm$0.40 & 65.8$\pm$8.0 & 1.63$\pm$0.40 & {} & 67.7$\pm$8.1 & 1.43$\pm$0.42 & {} \\
Larynx & 71.3$\pm$5.5 & 1.46$\pm$0.18 & 75.7$\pm$5.6 & 1.45$\pm$0.20 & 73.8$\pm$5.7 & 1.44$\pm$0.21 & {} & 75.6$\pm$5.5 & 1.45$\pm$0.18 & {} \\
\rowcolor[HTML]{F3F3F3} 
Bone\_Skull & 83.1$\pm$35.4 & 3.91$\pm$0.20 & 82.2$\pm$35.5 & 1.44$\pm$0.22 & 82.0$\pm$35.3 & 1.34$\pm$0.20 & {\cellcolor[HTML]{F3F3F3}99.0} & 82.5$\pm$35.4 & 3.66$\pm$0.21 & {\cellcolor[HTML]{F3F3F3}99.0} \\
Bone\_Clavicula\_L & 94.5$\pm$0.8 & 0.6$\pm$0.10 & 94.9$\pm$0.7 & 0.30$\pm$0.09 & 94.7$\pm$0.8 & 0.30$\pm$0.10 & {99.7} & 95.2$\pm$0.9 & 0.45$\pm$0.11 & {99.5} \\
Bone\_Clavicula\_R & 98.4$\pm$16.1 & 0.53$\pm$0.21 & 96.8$\pm$16.1 & 0.48$\pm$0.21 & 96.7$\pm$15.8 & 0.51$\pm$0.18 &  & 98.3$\pm$16.0 & 1.37$\pm$0.20 &  
\\ \hline

Mean & \multicolumn{1}{c}{ 76.8 } & \multicolumn{1}{c|}{ 1.15 } & \multicolumn{1}{c}{ 81.3 } & \multicolumn{1}{c|}{ 0.94 } & \multicolumn{1}{c}{ 80.3 } & \multicolumn{1}{c|}{ 0.99 } & \multicolumn{1}{c|}{ 97.3 } & \multicolumn{1}{c}{ 81.3 } & \multicolumn{1}{c|}{ 0.94 }  & \multicolumn{1}{c|}{ 97.4} \\ \hline

\end{tabular}
}\caption{\textbar~\textbf{Organ-wise segmentation performance comparison on head \& neck region.} Comparison of DSC (\%) and ASD (mm) among \ac{nnu_e36}, \ac{cln_u36_unprn}, \ac{cln_c36}, and \ac{cln_u36} for 52 head \& neck structures. Decoder-wise pruning rate ($\mathcal{T}$, \%) of \ac{cln_c36} and \ac{cln_u36} are also provided. \ac{cln_u36} achieves high mean DSC, with only 0.1\% gap to top, and the lowest mean ASD, demonstrating its superior performance in the head \& neck region. }
\label{tab:ds36_hn}
\end{table}
\clearpage

\begin{table}[htp]
\centering
\resizebox{\textwidth}{!}{
\begin{tabular}{|l|ll|ll|ll|r|ll|r|}
\hline
\multicolumn{1}{|l|}{} & \multicolumn{2}{c|}{\cellcolor[HTML]{FFFFFF}{\color[HTML]{1F1F1F} \ac{nnu_e36}}} & \multicolumn{2}{c|}{\ac{cln_u36_unprn}} & \multicolumn{3}{c|}{\ac{cln_c36}} & \multicolumn{3}{c|}{\ac{cln_u36}} \\ \cline{2-11} 
\multicolumn{1}{|l|}{\multirow{-2}{*}{}} & DSC$\uparrow$ & \multicolumn{1}{l|}{ASD$\downarrow$} & DSC$\uparrow$ & \multicolumn{1}{l|}{ASD$\downarrow$} & DSC$\uparrow$ & ASD$\downarrow$ & \multicolumn{1}{r|}{$\mathcal{T}$$\uparrow$ } & DSC$\uparrow$ & ASD$\downarrow$ & \multicolumn{1}{r|}{$\mathcal{T}$$\uparrow$ } \\ \hline
\rowcolor[HTML]{F3F3F3} 
BrachialPlexus\_L & 41.7$\pm$9.0 & 3.20$\pm$0.71 & 56.8$\pm$8.9 & 2.83$\pm$0.73 & 56.8$\pm$8.9 & 2.87$\pm$0.70 & 90.0 & 56.9$\pm$8.9 & 2.81$\pm$0.74 & 90.0 \\
\rowcolor[HTML]{F3F3F3} 
BrachialPlexus\_R & 40.5$\pm$8.3 & 3.20$\pm$0.64 & 52.9$\pm$8.5 & 2.18$\pm$0.68 & 52.8$\pm$8.3 & 2.27$\pm$0.63 & {\cellcolor[HTML]{F3F3F3}} & 53.0$\pm$8.4 & 2.17$\pm$0.66 & {\cellcolor[HTML]{F3F3F3}} \\

Bone\_Humerus\_L & 96.9$\pm$17.5 & 0.54$\pm$0.20 & 97.1$\pm$17.9 & 0.53$\pm$0.24 & 97.1$\pm$17.9 & 0.50$\pm$0.24 & {99.7} & 97.1$\pm$17.9 & 0.52$\pm$0.24 & {99.7} \\
Bone\_Humerus\_R & 93.9$\pm$17.2 & 0.34$\pm$0.19 & 94.0$\pm$17.2 & 0.33$\pm$0.19 & 93.8$\pm$17.2 & 0.33$\pm$0.19 &  & 94.0$\pm$17.2 & 0.32$\pm$0.20 &  \\
\rowcolor[HTML]{F3F3F3} 
Bone\_Scapula\_L & 95.3$\pm$21.4 & 0.19$\pm$0.18 & 95.3$\pm$21.2 & 0.17$\pm$0.16 & 95.3$\pm$21.2 & 0.15$\pm$0.16 & {\cellcolor[HTML]{F3F3F3}99.0} & 95.4$\pm$21.1 & 0.16$\pm$0.14 & {\cellcolor[HTML]{F3F3F3}99.0} \\
\rowcolor[HTML]{F3F3F3} 
Bone\_Scapula\_R & 93.1$\pm$1.8 & 0.32$\pm$0.07 & 93.4$\pm$1.9 & 0.28$\pm$0.08 & 93.4$\pm$1.9 & 0.27$\pm$0.08 &  & 93.5$\pm$1.8 & 0.28$\pm$0.07 &  \\
Eso & 86.0$\pm$11.9 & 0.66$\pm$4.95 & 85.9$\pm$11.8 & 0.55$\pm$5.07 & 85.9$\pm$11.7 & 0.53$\pm$4.73 & 96.0 & 86.0$\pm$11.6 & 0.55$\pm$5.21 & 98.0 \\
\rowcolor[HTML]{F3F3F3} 
A\_Aorta\_Arch & 89.0$\pm$5.1 & 2.13$\pm$0.18 & 91.8$\pm$4.7 & 2.15$\pm$0.15 & 91.8$\pm$5.0 & 2.01$\pm$0.17 & 99.0 & 91.9$\pm$4.9 & 2.14$\pm$0.17 & 99.0 \\
\rowcolor[HTML]{F3F3F3} 
A\_Aorta\_Ascending & 90.5$\pm$2.7 & 1.80$\pm$0.29 & 91.4$\pm$2.5 & 1.81$\pm$0.28 & 91.4$\pm$2.8 & 1.66$\pm$0.30 & {\cellcolor[HTML]{F3F3F3}} & 91.5$\pm$2.7 & 1.79$\pm$0.29 & {\cellcolor[HTML]{F3F3F3}} \\
\rowcolor[HTML]{F3F3F3} 
A\_Aorta\_Descending & 94.3$\pm$2.9 & 0.90$\pm$0.54 & 95.6$\pm$2.9 & 0.76$\pm$0.55 & 95.3$\pm$3.1 & 0.75$\pm$0.55 & {\cellcolor[HTML]{F3F3F3}} & 95.5$\pm$2.9 & 0.74$\pm$0.55 & {\cellcolor[HTML]{F3F3F3}} \\
A\_Pulmonary & 89.7$\pm$2.9 & 0.94$\pm$0.19 & 94.6$\pm$3.4 & 0.81$\pm$0.21 & 94.5$\pm$3.0 & 0.84$\pm$0.20 & 99.7 & 94.6$\pm$3.3 & 0.80$\pm$0.20 & 99.7 \\
\rowcolor[HTML]{F3F3F3} 
A\_CommonCarotid\_L & 86.3$\pm$7.5 & 1.00$\pm$3.88 & 88.3$\pm$7.7 & 0.90$\pm$3.58 & 88.2$\pm$7.8 & 0.88$\pm$3.74 & 99.7 & 88.3$\pm$7.6 & 0.89$\pm$3.83 & 99.7 \\
\rowcolor[HTML]{F3F3F3} 
A\_CommonCarotid\_R & 81.9$\pm$9.7 & 0.69$\pm$3.56 & 86.0$\pm$9.3 & 0.57$\pm$3.42 & 86.0$\pm$9.6 & 0.55$\pm$3.47 & {\cellcolor[HTML]{F3F3F3}} & 86.1$\pm$9.4 & 0.55$\pm$3.72 & {\cellcolor[HTML]{F3F3F3}} \\
A\_Subclavian\_L & 77.0$\pm$8.6 & 2.75$\pm$28.19 & 78.8$\pm$8.7 & 2.29$\pm$24.81 & 78.6$\pm$8.9 & 2.63$\pm$28.08 & 99.7 & 78.7$\pm$8.7 & 2.27$\pm$26.56 & 99.7 \\
A\_Subclavian\_R & 83.0$\pm$5.8 & 1.35$\pm$6.93 & 87.2$\pm$5.7 & 1.27$\pm$6.16 & 87.1$\pm$5.6 & 1.20$\pm$6.92 & {} & 87.2$\pm$5.8 & 1.26$\pm$6.68 & {} \\
\rowcolor[HTML]{F3F3F3} 
A\_Vertebral\_L & 57.1$\pm$16.7 & 2.51$\pm$2.84 & 65.5$\pm$16.7 & 1.86$\pm$2.68 & 65.3$\pm$16.8 & 2.11$\pm$2.94 & 99.0 & 65.4$\pm$16.5 & 1.86$\pm$2.93 & 99.0 \\
\rowcolor[HTML]{F3F3F3} 
A\_Vertebral\_R & 56.4$\pm$16.0 & 2.98$\pm$2.35 & 67.0$\pm$16.1 & 2.30$\pm$2.11 & 66.8$\pm$16.3 & 2.27$\pm$2.31 & {\cellcolor[HTML]{F3F3F3}} & 66.9$\pm$16.1 & 2.29$\pm$2.15 & {\cellcolor[HTML]{F3F3F3}} \\
A\_BrachiocephalicTrunk & 89.8$\pm$16.2 & 0.32$\pm$0.43 & 98.3$\pm$16.3 & 0.64$\pm$0.44 & 98.2$\pm$16.2 & 0.61$\pm$0.44 & 99.0 & 90.8$\pm$16.3 & 0.22$\pm$0.44 & 99.0 \\
\rowcolor[HTML]{F3F3F3} 
V\_Azygos & 81.3$\pm$6.6 & 1.24$\pm$0.49 & 83.7$\pm$6.7 & 1.25$\pm$0.53 & 83.5$\pm$6.8 & 1.10$\pm$0.51 & 96.0 & 83.7$\pm$6.7 & 1.23$\pm$0.53 & 96.0 \\
V\_Brachiocephalic\_L & 90.3$\pm$5.0 & 0.45$\pm$0.12 & 92.5$\pm$5.2 & 0.37$\pm$0.14 & 92.4$\pm$5.1 & 0.30$\pm$0.13 & 99.5 & 92.5$\pm$5.2 & 0.36$\pm$0.14 & 99.5 \\
V\_Brachiocephalic\_R & 88.6$\pm$5.9 & 0.62$\pm$0.28 & 90.0$\pm$5.8 & 0.46$\pm$0.25 & 89.8$\pm$5.8 & 0.50$\pm$0.27 & {} & 90.0$\pm$5.8 & 0.45$\pm$0.25 & {} \\
\rowcolor[HTML]{F3F3F3} 
V\_InternalJuguar\_L & 83.2$\pm$13.9 & 0.75$\pm$1.56 & 83.3$\pm$14.1 & 0.60$\pm$1.63 & 83.2$\pm$13.7 & 0.61$\pm$1.64 & 99.5 & 83.3$\pm$13.9 & 0.58$\pm$1.69 & 99.5 \\
\rowcolor[HTML]{F3F3F3} 
V\_InternalJuguar\_R & 83.5$\pm$12.4 & 2.83$\pm$1.76 & 87.3$\pm$12.1 & 2.87$\pm$1.62 & 87.2$\pm$12.2 & 2.70$\pm$1.77 & {\cellcolor[HTML]{F3F3F3}} & 87.4$\pm$12.0 & 2.86$\pm$1.69 & {\cellcolor[HTML]{F3F3F3}} \\
V\_Pulmonary & 68.9$\pm$6.3 & 2.24$\pm$1.16 & 75.2$\pm$6.3 & 1.37$\pm$1.10 & 75.1$\pm$6.3 & 1.31$\pm$1.20 & 98.0 & 75.3$\pm$6.3 & 1.36$\pm$1.18 & 99.0 \\
\rowcolor[HTML]{F3F3F3} 
V\_Subclavian\_L & 77.0$\pm$14.2 & 2.75$\pm$8.88 & 81.2$\pm$14.2 & 2.49$\pm$8.61 & 81.1$\pm$14.2 & 2.63$\pm$8.93 & 94.0 & 81.2$\pm$14.2 & 2.47$\pm$8.62 & 94.0 \\
\rowcolor[HTML]{F3F3F3} 
V\_Subclavian\_R & 83.0$\pm$11.9 & 1.35$\pm$8.56 & 85.1$\pm$12.0 & 1.59$\pm$8.51 & 84.9$\pm$12.2 & 1.25$\pm$8.80 & {\cellcolor[HTML]{F3F3F3}} & 85.1$\pm$12.1 & 1.58$\pm$8.92 & {\cellcolor[HTML]{F3F3F3}} \\
V\_VenaCava\_I & 83.3$\pm$7.5 & 1.64$\pm$0.47 & 83.8$\pm$7.4 & 1.46$\pm$0.49 & 83.7$\pm$7.5 & 1.53$\pm$0.47 & 99.0 & 83.9$\pm$7.4 & 1.45$\pm$0.48 & 99.0 \\
\rowcolor[HTML]{F3F3F3} 
V\_VenaCava\_S & 90.2$\pm$4.5 & 0.88$\pm$0.26 & 92.1$\pm$4.2 & 0.76$\pm$0.26 & 91.9$\pm$4.4 & 0.78$\pm$0.25 & 99.0 & 92.1$\pm$4.1 & 0.75$\pm$0.25 & 99.0 \\
V\_Portal\_Splenic & 86.3$\pm$29.0 & 0.25$\pm$0.62 & 89.2$\pm$29.3 & 0.14$\pm$0.63 & 89.1$\pm$29.2 & 0.14$\pm$0.63 & 98.0 & 89.2$\pm$29.2 & 0.13$\pm$0.62 & 98.0 \\
\rowcolor[HTML]{F3F3F3} 
Trachea & 90.4$\pm$2.6 & 0.78$\pm$2.10 & 92.9$\pm$2.3 & 0.69$\pm$2.05 & 92.7$\pm$2.5 & 0.68$\pm$2.02 & 99.7 & 92.9$\pm$2.5 & 0.67$\pm$2.08 & 99.7 \\
\rowcolor[HTML]{F3F3F3} 
Bronchus\_L & 83.4$\pm$7.3 & 0.72$\pm$1.13 & 84.6$\pm$6.8 & 0.61$\pm$1.16 & 84.5$\pm$7.1 & 0.57$\pm$1.15 & {\cellcolor[HTML]{F3F3F3}} & 84.7$\pm$7.0 & 0.59$\pm$1.19 & {\cellcolor[HTML]{F3F3F3}} \\
\rowcolor[HTML]{F3F3F3} 
Bronchus\_R & 81.2$\pm$11.3 & 0.98$\pm$1.44 & 84.1$\pm$11.4 & 0.93$\pm$1.34 & 84.0$\pm$11.5 & 0.87$\pm$1.36 & {\cellcolor[HTML]{F3F3F3}} & 84.2$\pm$11.5 & 0.92$\pm$1.42 & {\cellcolor[HTML]{F3F3F3}} \\
Lung\_L & 97.5$\pm$0.7 & 0.75$\pm$0.33 & 98.4$\pm$0.6 & 0.63$\pm$0.31 & 98.5$\pm$0.7 & 0.61$\pm$0.32 & 99.5 & 98.5$\pm$0.6 & 0.61$\pm$0.31 & 99.5 \\
Lung\_R & 97.9$\pm$0.7 & 0.60$\pm$0.12 & 98.5$\pm$0.8 & 0.51$\pm$0.13 & 98.5$\pm$0.4 & 0.47$\pm$0.08 & {} & 98.5$\pm$0.6 & 0.50$\pm$0.11 & {} \\
\rowcolor[HTML]{F3F3F3} 
ChestWall\_L & 90.4$\pm$2.5 & 1.40$\pm$1.19 & 95.6$\pm$2.7 & 1.35$\pm$1.24 & 95.3$\pm$2.5 & 1.29$\pm$1.25 & 99.5 & 95.5$\pm$2.6 & 1.35$\pm$1.30 & 99.5 \\
\rowcolor[HTML]{F3F3F3} 
ChestWall\_R & 91.8$\pm$2.5 & 0.80$\pm$1.05 & 93.1$\pm$2.5 & 0.63$\pm$0.99 & 93.0$\pm$2.5 & 0.69$\pm$1.06 & {\cellcolor[HTML]{F3F3F3}} & 93.1$\pm$2.5 & 0.63$\pm$1.09 & {\cellcolor[HTML]{F3F3F3}} \\
Lung\_LUL & 97.8$\pm$3.6 & 0.31$\pm$0.93 & 98.2$\pm$3.8 & 0.57$\pm$0.61 & 98.2$\pm$3.9 & 0.62$\pm$0.64 & 99.5 & 97.9$\pm$3.7 & 0.17$\pm$0.62 & 99.7 \\
Lung\_LLL & 96.5$\pm$24.0 & 0.21$\pm$0.98 & 97.9$\pm$2.0 & 0.46$\pm$0.47 & 98.1$\pm$2.0 & 0.45$\pm$0.46 & {} & 96.5$\pm$2.1 & 0.21$\pm$0.47 & {} \\
\rowcolor[HTML]{F3F3F3} 
Lung\_RUL & 97.5$\pm$20.5 & 0.79$\pm$0.95 & 96.6$\pm$2.4 & 0.70$\pm$0.97 & 96.6$\pm$2.4 & 0.67$\pm$0.98 & 99.5 & 97.7$\pm$2.4 & 0.69$\pm$0.98 & 99.5 \\
\rowcolor[HTML]{F3F3F3} 
Lung\_RML & 92.0$\pm$6.2 & 0.47$\pm$1.41 & 98.6$\pm$3.2 & 0.60$\pm$1.41 & 98.4$\pm$3.3 & 0.61$\pm$1.48 & {\cellcolor[HTML]{F3F3F3}} & 92.3$\pm$3.3 & 0.22$\pm$1.50 & {\cellcolor[HTML]{F3F3F3}} \\
\rowcolor[HTML]{F3F3F3} 
Lung\_RLL & 97.4$\pm$26.8 & 0.16$\pm$2.45 & 96.9$\pm$3.5 & 0.47$\pm$1.07 & 96.7$\pm$3.5 & 0.45$\pm$1.20 & {\cellcolor[HTML]{F3F3F3}} & 97.2$\pm$3.6 & 0.21$\pm$1.19 & {\cellcolor[HTML]{F3F3F3}} \\
Heart & 94.2$\pm$1.8 & 1.62$\pm$0.22 & 97.4$\pm$1.8 & 1.35$\pm$0.20 & 97.4$\pm$1.7 & 1.48$\pm$0.21 & 99.0 & 97.4$\pm$1.7 & 1.34$\pm$0.20 & 99.0 \\
\rowcolor[HTML]{F3F3F3} 
Myocardium & 92.5$\pm$1.7 & 1.62$\pm$0.21 & 91.6$\pm$1.4 & 1.59$\pm$0.18 & 91.5$\pm$1.6 & 1.51$\pm$0.21 & 99.7 & 91.7$\pm$1.6 & 1.58$\pm$0.19 & 99.7 \\
Atrium\_L & 92.6$\pm$1.8 & 1.70$\pm$0.22 & 95.3$\pm$1.9 & 1.68$\pm$0.22 & 95.2$\pm$1.5 & 1.56$\pm$0.19 & 99.5 & 95.3$\pm$1.8 & 1.68$\pm$0.20 & 99.5 \\
Ventricle\_L & 95.2$\pm$1.5 & 1.70$\pm$0.20 & 96.1$\pm$1.6 & 1.47$\pm$0.21 & 96.0$\pm$1.6 & 1.58$\pm$0.20 & {} & 96.2$\pm$1.6 & 1.45$\pm$0.21 & {} \\
Atrium\_R & 96.0$\pm$1.7 & 1.65$\pm$0.21 & 97.1$\pm$1.9 & 1.29$\pm$0.24 & 96.9$\pm$1.6 & 1.51$\pm$0.20 & {} & 97.1$\pm$1.8 & 1.27$\pm$0.23 & {} \\
Ventricle\_R & 94.4$\pm$1.7 & 1.69$\pm$0.22 & 94.8$\pm$1.9 & 1.64$\pm$0.24 & 94.6$\pm$1.6 & 1.55$\pm$0.20 & {} & 94.7$\pm$1.8 & 1.62$\pm$0.23 & {} \\
\rowcolor[HTML]{F3F3F3} 
Pericardium & 95.6$\pm$1.7 & 1.52$\pm$0.21 & 94.9$\pm$1.8 & 1.49$\pm$0.22 & 94.7$\pm$1.9 & 1.40$\pm$0.22 & 99.0 & 94.9$\pm$1.8 & 1.48$\pm$0.22 & 99.0 \\
LAA & 92.6$\pm$3.4 & 0.46$\pm$0.15 & 93.3$\pm$3.4 & 0.79$\pm$0.14 & 93.2$\pm$3.6 & 0.78$\pm$0.18 & 99.7 & 91.7$\pm$3.3 & 0.16$\pm$0.15 & 99.5 \\
\rowcolor[HTML]{F3F3F3} 
Musc\_Cervical\_A & 70.0$\pm$15.0 & 1.15$\pm$2.70 & 72.7$\pm$15.1 & 0.99$\pm$2.55 & 72.6$\pm$15.0 & 1.03$\pm$2.82 & 96.0 & 72.8$\pm$15.0 & 0.97$\pm$2.76 & 96.0 \\
Musc\_Scalenus & 75.9$\pm$4.7 & 1.09$\pm$0.28 & 76.3$\pm$4.3 & 1.02$\pm$0.27 & 76.2$\pm$4.8 & 0.98$\pm$0.29 & 99.0 & 76.3$\pm$4.5 & 1.01$\pm$0.28 & 99.0 \\
Musc\_Scalenus\_A & 78.0$\pm$5.7 & 0.95$\pm$0.34 & 81.8$\pm$5.9 & 0.73$\pm$0.36 & 81.5$\pm$5.7 & 0.84$\pm$0.34 & {} & 81.7$\pm$5.9 & 0.72$\pm$0.36 & {} \\
\rowcolor[HTML]{F3F3F3} 
Musc\_Scleido & 82.4$\pm$2.9 & 0.99$\pm$0.21 & 83.5$\pm$2.9 & 0.79$\pm$0.20 & 83.3$\pm$2.6 & 0.88$\pm$0.17 & 99.5 & 83.5$\pm$2.8 & 0.79$\pm$0.19 & 99.5 \\
Musc\_Autochthon\_L & 94.2$\pm$0.8 & 0.37$\pm$0.11 & 97.7$\pm$0.9 & 1.84$\pm$0.10 & 97.6$\pm$0.8 & 1.99$\pm$0.11 & 99.0 & 93.9$\pm$0.9 & 0.32$\pm$0.12 & 99.0 \\
Musc\_Autochthon\_R & 94.6$\pm$0.8 & 0.41$\pm$0.12 & 97.7$\pm$0.8 & 1.79$\pm$0.12 & 97.8$\pm$0.8 & 1.93$\pm$0.10 & {} & 94.8$\pm$0.8 & 0.41$\pm$0.11 & {} \\
\rowcolor[HTML]{F3F3F3} 
SpinalCord & 91.7$\pm$3.0 & 0.86$\pm$0.25 & 94.4$\pm$3.2 & 0.68$\pm$0.24 & 94.2$\pm$3.3 & 0.72$\pm$0.27 & 99.7 & 94.4$\pm$3.2 & 0.67$\pm$0.25 & 99.7 
\\ \hline

Mean & \multicolumn{1}{c}{ 86.1 } & \multicolumn{1}{c|}{ 1.29 } & \multicolumn{1}{c}{ 88.6 } & \multicolumn{1}{c|}{ 1.12 } & \multicolumn{1}{c}{ 88.5 } & \multicolumn{1}{c|}{ 1.13 } & \multicolumn{1}{c|}{ 98.6 } & \multicolumn{1}{c}{ 88.6 } & \multicolumn{1}{c|}{ 1.11 }  & \multicolumn{1}{c|}{ 98.6 } \\ \hline

\end{tabular}
}\caption{\textbar~\textbf{Organ-wise segmentation performance comparison on thoracic region.} Comparison of DSC (\%) and ASD (mm) among \ac{nnu_e36}, \ac{cln_u36_unprn}, \ac{cln_c36}, and \ac{cln_u36} for 56 thoracic structures. Decoder-wise pruning rate ($\mathcal{T}$, \%) of \ac{cln_c36} and \ac{cln_u36} are also provided. \ac{cln_u36} achieves the highest mean DSC and the lowest mean ASD, demonstrating its superior performance in the thoracic region. }
\label{tab:ds36_ch}
\end{table}
\clearpage

\begin{table}[htp]
\centering
\resizebox{\textwidth}{!}{
\begin{tabular}{|l|ll|ll|ll|r|ll|r|}
\hline
\multicolumn{1}{|l|}{} & \multicolumn{2}{c|}{\cellcolor[HTML]{FFFFFF}{\color[HTML]{1F1F1F} \ac{nnu_e36}}} & \multicolumn{2}{c|}{\ac{cln_u36_unprn}} & \multicolumn{3}{c|}{\ac{cln_c36}} & \multicolumn{3}{c|}{\ac{cln_u36}} \\ \cline{2-11} 
\multicolumn{1}{|l|}{\multirow{-2}{*}{}} & DSC$\uparrow$ & \multicolumn{1}{l|}{ASD$\downarrow$} & DSC$\uparrow$ & \multicolumn{1}{l|}{ASD$\downarrow$} & DSC$\uparrow$ & ASD$\downarrow$ & \multicolumn{1}{r|}{$\mathcal{T}$$\uparrow$ } & DSC$\uparrow$ & ASD$\downarrow$ & \multicolumn{1}{r|}{$\mathcal{T}$$\uparrow$ } \\ \hline
\rowcolor[HTML]{F3F3F3} 
Kidney\_R & 93.9$\pm$6.8 & 1.20$\pm$0.73 & 95.2$\pm$6.9 & 0.98$\pm$0.74 & 95.0$\pm$7.1 & 1.06$\pm$0.76 & 99.3 & 95.2$\pm$7.0 & 0.97$\pm$0.75 & 99.7 \\
\rowcolor[HTML]{F3F3F3} 
Kidney\_L & 95.0$\pm$6.6 & 1.45$\pm$0.69 & 94.3$\pm$6.9 & 1.30$\pm$0.72 & 94.2$\pm$6.7 & 1.35$\pm$0.70 & {\cellcolor[HTML]{F3F3F3}} & 94.3$\pm$6.8 & 1.30$\pm$0.71 & {\cellcolor[HTML]{F3F3F3}} \\
Spleen & 97.2$\pm$6.8 & 0.20$\pm$0.39 & 98.2$\pm$6.7 & 0.10$\pm$0.38 & 98.3$\pm$6.9 & 0.10$\pm$0.40 & 99.5 & 98.2$\pm$6.6 & 0.10$\pm$0.37 & 99.5 \\
\rowcolor[HTML]{F3F3F3} 
Liver & 97.6$\pm$5.8 & 0.82$\pm$1.00 & 97.9$\pm$6.1 & 0.69$\pm$0.93 & 97.8$\pm$5.7 & 0.72$\pm$0.99 & 99.3 & 97.9$\pm$5.9 & 0.67$\pm$0.91 & 99.5 \\
HepaticVessel & 65.3$\pm$9.4 & 2.40$\pm$3.79 & 68.1$\pm$9.5 & 2.24$\pm$3.30 & 66.5$\pm$9.3 & 2.25$\pm$3.63 & 94.0 & 68.1$\pm$9.5 & 2.22$\pm$3.46 & 94.0 \\
\rowcolor[HTML]{F3F3F3} 
Glnd\_Adrenal\_R & 90.2$\pm$13.2 & 0.22$\pm$0.39 & 90.5$\pm$13.1 & 0.10$\pm$0.38 & 90.4$\pm$13.5 & 0.09$\pm$0.42 & 96.0 & 90.5$\pm$13.3 & 0.08$\pm$0.39 & 98.0 \\
\rowcolor[HTML]{F3F3F3} 
Glnd\_Adrenal\_L & 86.5$\pm$12.6 & 0.36$\pm$0.39 & 87.8$\pm$12.7 & 0.21$\pm$0.41 & 86.0$\pm$12.3 & 0.21$\pm$0.36 & {\cellcolor[HTML]{F3F3F3}} & 87.8$\pm$12.6 & 0.20$\pm$0.39 & {\cellcolor[HTML]{F3F3F3}} \\
Pancreas & 93.0$\pm$7.1 & 0.74$\pm$0.39 & 95.0$\pm$7.0 & 0.62$\pm$0.38 & 94.9$\pm$6.8 & 0.62$\pm$0.36 & 99.3 & 95.1$\pm$7.1 & 0.61$\pm$0.39 & 99.5 \\
\rowcolor[HTML]{F3F3F3} 
GallBladder & 82.2$\pm$8.6 & 2.13$\pm$0.82 & 85.7$\pm$8.6 & 1.87$\pm$0.82 & 83.7$\pm$8.7 & 2.02$\pm$0.83 & 99.3 & 85.6$\pm$8.6 & 1.85$\pm$0.82 & 99.5 \\
Stomach & 92.1$\pm$6.5 & 1.18$\pm$0.68 & 92.5$\pm$6.1 & 0.99$\pm$0.65 & 92.3$\pm$6.3 & 1.07$\pm$0.67 & 99.3 & 92.4$\pm$6.3 & 0.98$\pm$0.66 & 99.5 \\
\rowcolor[HTML]{F3F3F3} 
Duodenum & 80.8$\pm$15.9 & 2.40$\pm$1.30 & 82.8$\pm$16.0 & 2.10$\pm$1.25 & 81.8$\pm$15.8 & 2.30$\pm$1.29 & 99.5 & 82.8$\pm$16.2 & 2.09$\pm$1.28 & 99.5 \\
SmallBowel & 91.7$\pm$16.3 & 2.32$\pm$2.60 & 91.3$\pm$16.5 & 2.08$\pm$2.36 & 91.2$\pm$16.4 & 2.17$\pm$2.65 & 98.0 & 91.3$\pm$16.5 & 2.06$\pm$2.62 & 99.5 \\
\rowcolor[HTML]{F3F3F3} 
Colon & 91.6$\pm$15.8 & 3.01$\pm$1.22 & 92.9$\pm$15.9 & 2.65$\pm$1.15 & 92.7$\pm$15.8 & 2.87$\pm$1.21 & 99.0 & 92.9$\pm$16.0 & 2.64$\pm$1.27 & 99.5 \\
\rowcolor[HTML]{F3F3F3} 
Rectum & 84.6$\pm$6.8 & 3.25$\pm$1.48 & 84.8$\pm$6.7 & 3.03$\pm$1.56 & 84.8$\pm$7.0 & 3.11$\pm$1.55 & {\cellcolor[HTML]{F3F3F3}} & 84.9$\pm$6.7 & 3.02$\pm$1.57 & {\cellcolor[HTML]{F3F3F3}} \\
UrinaryBladder & 92.2$\pm$9.7 & 1.87$\pm$1.06 & 94.0$\pm$9.5 & 1.77$\pm$0.97 & 93.8$\pm$9.5 & 1.75$\pm$0.99 & 98.0 & 93.9$\pm$9.4 & 1.75$\pm$1.04 & 98.0 \\
\rowcolor[HTML]{F3F3F3} 
Prostate & 84.3$\pm$18.0 & 3.65$\pm$1.41 & 84.8$\pm$17.8 & 3.42$\pm$1.42 & 84.6$\pm$17.7 & 3.52$\pm$1.43 & 96.0 & 84.7$\pm$18.0 & 3.42$\pm$1.43 & 96.0 \\
A\_Iliac\_L & 94.4$\pm$3.1 & 0.61$\pm$0.19 & 91.1$\pm$3.3 & 0.29$\pm$0.20 & 90.9$\pm$3.2 & 0.30$\pm$0.19 & 99.0 & 95.1$\pm$3.3 & 0.50$\pm$0.20 & 99.5 \\
A\_Iliac\_R & 91.9$\pm$19.5 & 0.45$\pm$0.15 & 90.9$\pm$19.4 & 0.20$\pm$0.16 & 90.8$\pm$19.5 & 0.20$\pm$0.17 & {} & 91.9$\pm$19.6 & 0.39$\pm$0.18 & {} \\
\rowcolor[HTML]{F3F3F3} 
V\_Iliac\_L & 92.2$\pm$3.0 & 0.48$\pm$0.13 & 93.1$\pm$3.2 & 0.26$\pm$0.14 & 93.0$\pm$3.0 & 0.27$\pm$0.12 & 99.7 & 92.9$\pm$3.3 & 0.43$\pm$0.15 & 99.0 \\
\rowcolor[HTML]{F3F3F3} 
V\_Iliac\_R & 94.2$\pm$4.9 & 0.44$\pm$0.31 & 92.2$\pm$5.0 & 0.23$\pm$0.29 & 92.0$\pm$4.9 & 0.23$\pm$0.28 & {\cellcolor[HTML]{F3F3F3}} & 93.6$\pm$4.8 & 0.45$\pm$0.27 & {\cellcolor[HTML]{F3F3F3}} \\
Musc\_GluteusMax\_L & 98.9$\pm$5.2 & 0.84$\pm$1.41 & 97.2$\pm$5.5 & 0.40$\pm$1.33 & 97.2$\pm$5.3 & 0.44$\pm$1.48 & 99.5 & 98.7$\pm$5.4 & 0.76$\pm$1.50 & 99.5 \\
Musc\_GluteusMax\_R & 98.1$\pm$2.0 & 0.93$\pm$0.34 & 97.8$\pm$1.8 & 0.38$\pm$0.31 & 97.9$\pm$2.1 & 0.39$\pm$0.33 & {} & 97.9$\pm$2.2 & 0.71$\pm$0.34 & {} \\
\rowcolor[HTML]{F3F3F3} 
Musc\_GluteusMed\_L & 98.2$\pm$1.5 & 0.43$\pm$0.22 & 97.8$\pm$1.3 & 0.29$\pm$0.21 & 97.6$\pm$1.6 & 0.31$\pm$0.24 & 99.5 & 98.4$\pm$1.5 & 0.45$\pm$0.23 & 99.7 \\
\rowcolor[HTML]{F3F3F3} 
Musc\_GluteusMed\_R & 97.6$\pm$7.3 & 0.74$\pm$0.22 & 97.1$\pm$7.5 & 0.32$\pm$0.21 & 96.9$\pm$7.5 & 0.32$\pm$0.21 & {\cellcolor[HTML]{F3F3F3}} & 97.3$\pm$7.5 & 0.47$\pm$0.21 & {\cellcolor[HTML]{F3F3F3}} \\
Musc\_GluteusMin\_L & 97.1$\pm$2.3 & 0.56$\pm$0.16 & 96.1$\pm$2.5 & 0.27$\pm$0.18 & 96.0$\pm$2.3 & 0.28$\pm$0.16 & 99.5 & 97.3$\pm$2.2 & 0.42$\pm$0.16 & 99.7 \\
Musc\_GluteusMin\_R & 98.0$\pm$1.1 & 0.39$\pm$0.08 & 94.1$\pm$1.2 & 0.19$\pm$0.11 & 94.0$\pm$1.3 & 0.19$\pm$0.11 & {} & 98.2$\pm$1.1 & 0.36$\pm$0.10 & {} \\
\rowcolor[HTML]{F3F3F3} 
Musc\_Iliopsoas\_L & 94.3$\pm$19.7 & 0.37$\pm$0.18 & 94.0$\pm$19.5 & 0.21$\pm$0.18 & 94.0$\pm$19.5 & 0.20$\pm$0.18 & 99.0 & 93.5$\pm$19.5 & 0.36$\pm$0.19 & 99.5 \\
\rowcolor[HTML]{F3F3F3} 
Musc\_Iliopsoas\_R & 95.0$\pm$9.1 & 0.39$\pm$0.27 & 94.5$\pm$9.3 & 0.18$\pm$0.26 & 94.4$\pm$9.1 & 0.18$\pm$0.24 & {\cellcolor[HTML]{F3F3F3}} & 94.9$\pm$9.1 & 0.21$\pm$0.25 & {\cellcolor[HTML]{F3F3F3}} \\
Bone\_Hip\_L & 96.6$\pm$1.6 & 0.32$\pm$0.13 & 99.0$\pm$1.9 & 0.25$\pm$0.14 & 98.7$\pm$1.8 & 0.26$\pm$0.12 & {99.5} & 97.0$\pm$1.9 & 0.29$\pm$0.14 & {99.5} \\
Bone\_Hip\_R & 95.9$\pm$31.6 & 0.34$\pm$0.10 & 98.5$\pm$31.7 & 0.28$\pm$0.09 & 98.6$\pm$31.6 & 0.26$\pm$0.08 &  & 97.0$\pm$31.9 & 0.27$\pm$0.11 &  \\
\rowcolor[HTML]{F3F3F3} 
Bone\_Femur\_L & 97.4$\pm$15.4 & 3.51$\pm$0.29 & 90.2$\pm$0.7 & 5.51$\pm$0.11 & 90.1$\pm$0.6 & 5.66$\pm$0.10 & {\cellcolor[HTML]{F3F3F3}99.5} & 97.5$\pm$0.7 & 3.28$\pm$0.11 & {\cellcolor[HTML]{F3F3F3}99.7} \\
\rowcolor[HTML]{F3F3F3} 
Bone\_Femur\_R & 95.8$\pm$0.5 & 6.85$\pm$0.08 & 95.2$\pm$0.2 & 6.32$\pm$0.06 & 95.2$\pm$0.5 & 5.77$\pm$0.09 &  & 95.3$\pm$0.4 & 6.26$\pm$0.06 & \\

Body Mask & 97.0$\pm$2.9 & 2.20$\pm$1.31 & 97.2$\pm$2.8 & 2.29$\pm$1.11 & 97.1$\pm$2.7 & 2.33$\pm$1.27 & 99.5 & 97.1$\pm$2.8 & 2.50$\pm$1.24 & 99.5
\\ \hline

Mean & \multicolumn{1}{c}{ 91.6 } & \multicolumn{1}{c|}{ 1.40 } & \multicolumn{1}{c}{ 92.5 } & \multicolumn{1}{c|}{ 1.27 } & \multicolumn{1}{c}{ 92.2 } & \multicolumn{1}{c|}{ 1.30 } & \multicolumn{1}{c|}{ 98.6 } & \multicolumn{1}{c}{ 92.5 } & \multicolumn{1}{c|}{ 1.27 }  & \multicolumn{1}{c|}{ 98.9 } \\ \hline

\end{tabular}
}\caption{\textbar~\textbf{Organ-wise segmentation performance comparison on abdomen region.} Comparison of DSC (\%) and ASD (mm) among \ac{nnu_e36}, \ac{cln_u36_unprn}, \ac{cln_c36}, and \ac{cln_u36} for 32 abdomen structures. Decoder-wise pruning rate ($\mathcal{T}$, \%) of \ac{cln_c36} and \ac{cln_u36} are also provided. \ac{cln_u36} achieves the highest mean DSC and the lowest mean ASD, demonstrating its superior performance in the abdomen region. }
\label{tab:ds36_abd}
\end{table}
\clearpage

\begin{table}[htp]
\centering
\resizebox{\textwidth}{!}{
\begin{tabular}{|l|ll|ll|ll|r|ll|r|}
\hline
\multicolumn{1}{|l|}{} & \multicolumn{2}{c|}{\cellcolor[HTML]{FFFFFF}{\color[HTML]{1F1F1F} \ac{nnu_e36}}} & \multicolumn{2}{c|}{\ac{cln_u36_unprn}} & \multicolumn{3}{c|}{\ac{cln_c36}} & \multicolumn{3}{c|}{\ac{cln_u36}} \\ \cline{2-11} 
\multicolumn{1}{|l|}{\multirow{-2}{*}{}} & DSC$\uparrow$ & \multicolumn{1}{l|}{ASD$\downarrow$} & DSC$\uparrow$ & \multicolumn{1}{l|}{ASD$\downarrow$} & DSC$\uparrow$ & ASD$\downarrow$ & \multicolumn{1}{r|}{$\mathcal{T}$$\uparrow$ } & DSC$\uparrow$ & ASD$\downarrow$ & \multicolumn{1}{r|}{$\mathcal{T}$$\uparrow$ } \\ \hline
\rowcolor[HTML]{F3F3F3} 
Bone\_Rib1\_L & 93.6$\pm$3.5 & 0.16$\pm$0.23 & 94.0$\pm$3.1 & 0.15$\pm$0.19 & 93.8$\pm$3.1 & 0.15$\pm$0.19 & \multicolumn{1}{r|}{\cellcolor[HTML]{F3F3F3}96.0} & 94.0$\pm$3.1 & 0.15$\pm$0.19 & \multicolumn{1}{r|}{\cellcolor[HTML]{F3F3F3}96.0} \\
\rowcolor[HTML]{F3F3F3} 
Bone\_Rib2\_L & 89.9$\pm$1.6 & 0.95$\pm$0.97 & 90.8$\pm$1.6 & 0.85$\pm$0.90 & 90.8$\pm$1.6 & 0.91$\pm$0.94 &  & 90.9$\pm$1.7 & 0.85$\pm$0.92 &  \\
\rowcolor[HTML]{F3F3F3} 
Bone\_Rib3\_L & 94.7$\pm$17.2 & 0.91$\pm$1.04 & 94.8$\pm$17.3 & 0.84$\pm$1.00 & 94.6$\pm$17.0 & 0.87$\pm$1.09 &  & 94.7$\pm$17.1 & 0.84$\pm$1.05 &  \\
\rowcolor[HTML]{F3F3F3} 
Bone\_Rib4\_L & 94.5$\pm$18.7 & 0.76$\pm$1.74 & 95.4$\pm$19.0 & 0.74$\pm$1.75 & 95.3$\pm$19.0 & 0.72$\pm$1.79 &  & 95.4$\pm$18.9 & 0.73$\pm$1.79 &  \\
\rowcolor[HTML]{F3F3F3} 
Bone\_Rib5\_L & 94.2$\pm$29.3 & 1.52$\pm$4.82 & 94.1$\pm$29.4 & 1.44$\pm$4.49 & 94.0$\pm$29.3 & 1.48$\pm$4.82 &  & 94.2$\pm$29.3 & 1.42$\pm$4.79 &  \\
\rowcolor[HTML]{F3F3F3} 
Bone\_Rib6\_L & 93.2$\pm$24.2 & 1.64$\pm$3.35 & 93.3$\pm$24.1 & 1.47$\pm$3.10 & 93.1$\pm$23.9 & 1.59$\pm$3.49 &  & 93.3$\pm$24.2 & 1.45$\pm$3.20 &  \\
\rowcolor[HTML]{F3F3F3} 
Bone\_Rib7\_L & 94.5$\pm$20.5 & 2.71$\pm$4.19 & 95.2$\pm$20.9 & 2.66$\pm$3.81 & 95.1$\pm$20.7 & 2.66$\pm$3.96 &  & 95.2$\pm$20.8 & 2.64$\pm$4.18 &  \\
\rowcolor[HTML]{F3F3F3} 
Bone\_Rib8\_L & 94.6$\pm$20.8 & 4.29$\pm$4.04 & 95.1$\pm$21.0 & 4.34$\pm$4.16 & 95.0$\pm$21.0 & 4.27$\pm$4.05 &  & 95.2$\pm$21.0 & 4.34$\pm$4.20 &  \\
\rowcolor[HTML]{F3F3F3} 
Bone\_Rib9\_L & 94.0$\pm$19.3 & 2.83$\pm$4.83 & 95.1$\pm$19.5 & 2.95$\pm$4.70 & 95.0$\pm$19.4 & 2.81$\pm$4.84 &  & 95.1$\pm$19.5 & 2.94$\pm$5.17 &  \\
\rowcolor[HTML]{F3F3F3} 
Bone\_Rib10\_L & 93.0$\pm$13.1 & 1.51$\pm$6.56 & 93.6$\pm$13.2 & 1.37$\pm$6.44 & 93.5$\pm$13.2 & 1.48$\pm$6.60 &  & 93.6$\pm$13.2 & 1.35$\pm$6.80 &  \\
\rowcolor[HTML]{F3F3F3} 
Bone\_Rib11\_L & 93.0$\pm$12.1 & 0.36$\pm$8.12 & 93.0$\pm$12.3 & 0.41$\pm$7.70 & 92.8$\pm$12.1 & 0.34$\pm$8.73 &  & 92.9$\pm$12.3 & 0.41$\pm$8.26 &  \\
\rowcolor[HTML]{F3F3F3} 
Bone\_Rib12\_L & 94.2$\pm$16.3 & 0.54$\pm$0.40 & 94.2$\pm$16.3 & 0.52$\pm$0.41 & 94.1$\pm$16.4 & 0.51$\pm$0.41 &  & 94.2$\pm$16.4 & 0.50$\pm$0.42 &  \\
\rowcolor[HTML]{F3F3F3} 
Bone\_Rib1\_R & 91.2$\pm$5.7 & 0.11$\pm$0.16 & 91.3$\pm$5.9 & 0.10$\pm$0.18 & 91.2$\pm$5.5 & 0.08$\pm$0.15 &  & 91.3$\pm$5.8 & 0.08$\pm$0.17 &  \\
\rowcolor[HTML]{F3F3F3} 
Bone\_Rib2\_R & 94.5$\pm$3.4 & 0.23$\pm$0.92 & 94.3$\pm$3.6 & 0.20$\pm$0.93 & 94.2$\pm$3.6 & 0.19$\pm$0.93 &  & 94.4$\pm$3.6 & 0.19$\pm$0.93 &  \\
\rowcolor[HTML]{F3F3F3} 
Bone\_Rib3\_R & 92.8$\pm$9.0 & 0.48$\pm$1.13 & 93.6$\pm$9.2 & 0.48$\pm$1.10 & 93.5$\pm$8.9 & 0.45$\pm$1.07 &  & 93.7$\pm$9.1 & 0.47$\pm$1.13 &  \\
\rowcolor[HTML]{F3F3F3} 
Bone\_Rib4\_R & 90.5$\pm$18.7 & 0.83$\pm$1.86 & 90.3$\pm$18.7 & 0.77$\pm$1.82 & 90.3$\pm$18.6 & 0.80$\pm$1.89 &  & 90.4$\pm$18.8 & 0.77$\pm$1.94 &  \\
\rowcolor[HTML]{F3F3F3} 
Bone\_Rib5\_R & 90.3$\pm$16.4 & 1.70$\pm$2.58 & 90.7$\pm$16.4 & 1.67$\pm$2.38 & 90.5$\pm$16.5 & 1.69$\pm$2.46 &  & 90.7$\pm$16.3 & 1.66$\pm$2.56 &  \\
\rowcolor[HTML]{F3F3F3} 
Bone\_Rib6\_R & 93.4$\pm$13.1 & 1.52$\pm$2.43 & 93.6$\pm$13.1 & 1.38$\pm$2.35 & 93.5$\pm$13.1 & 1.47$\pm$2.54 &  & 93.6$\pm$13.0 & 1.37$\pm$2.52 &  \\
\rowcolor[HTML]{F3F3F3} 
Bone\_Rib7\_R & 94.2$\pm$21.2 & 1.56$\pm$3.51 & 95.0$\pm$21.1 & 1.42$\pm$3.27 & 94.9$\pm$21.2 & 1.53$\pm$3.33 &  & 95.1$\pm$21.3 & 1.41$\pm$3.60 &  \\
\rowcolor[HTML]{F3F3F3} 
Bone\_Rib8\_R & 95.5$\pm$18.1 & 1.88$\pm$4.26 & 95.9$\pm$17.9 & 1.78$\pm$4.08 & 95.9$\pm$18.1 & 1.83$\pm$4.23 &  & 96.0$\pm$18.1 & 1.76$\pm$4.15 &  \\
\rowcolor[HTML]{F3F3F3} 
Bone\_Rib9\_R & 95.9$\pm$21.4 & 2.48$\pm$3.21 & 96.4$\pm$21.4 & 2.46$\pm$3.16 & 96.3$\pm$21.3 & 2.47$\pm$3.42 &  & 96.5$\pm$21.3 & 2.46$\pm$3.48 &  \\
\rowcolor[HTML]{F3F3F3} 
Bone\_Rib10\_R & 96.3$\pm$23.3 & 2.50$\pm$7.32 & 96.8$\pm$23.4 & 2.41$\pm$7.03 & 96.6$\pm$23.3 & 2.46$\pm$7.19 &  & 96.8$\pm$23.2 & 2.41$\pm$7.18 &  \\
\rowcolor[HTML]{F3F3F3} 
Bone\_Rib11\_R & 96.4$\pm$22.9 & 1.85$\pm$6.92 & 96.5$\pm$23.0 & 1.71$\pm$6.64 & 96.5$\pm$22.8 & 1.81$\pm$6.79 &  & 96.6$\pm$22.9 & 1.70$\pm$6.82 &  \\
\rowcolor[HTML]{F3F3F3} 
Bone\_Rib12\_R & 97.1$\pm$27.2 & 0.63$\pm$0.08 & 96.8$\pm$27.2 & 0.63$\pm$0.08 & 96.7$\pm$27.1 & 0.61$\pm$0.07 &  & 96.8$\pm$27.3 & 0.62$\pm$0.09 &  \\
Bone\_Sternum & 94.1$\pm$5.7 & 0.63$\pm$0.55 & 95.1$\pm$5.7 & 0.59$\pm$0.55 & 95.0$\pm$5.8 & 0.62$\pm$0.56 & {99.0} & 95.2$\pm$5.8 & 0.59$\pm$0.56 & {99.0} \\
\rowcolor[HTML]{F3F3F3} 
Bone\_CostalCartilages & 88.7$\pm$5.0 & 0.18$\pm$0.74 & 88.3$\pm$5.0 & 0.16$\pm$0.74 & 88.2$\pm$5.2 & 0.14$\pm$0.76 & {\cellcolor[HTML]{F3F3F3}99.5} & 88.4$\pm$5.0 & 0.05$\pm$0.75 & {\cellcolor[HTML]{F3F3F3}99.5} \\
Bone\_Vert\_C1 & 93.4$\pm$14.4 & 0.25$\pm$0.24 & 93.3$\pm$14.3 & 0.23$\pm$0.24 & 93.2$\pm$14.5 & 0.21$\pm$0.25 & {96.0} & 93.4$\pm$14.2 & 0.22$\pm$0.22 & {96.0} \\
Bone\_Vert\_C2 & 94.5$\pm$12.8 & 0.25$\pm$0.14 & 95.5$\pm$13.0 & 0.25$\pm$0.16 & 95.3$\pm$12.8 & 0.24$\pm$0.15 &  & 95.5$\pm$12.9 & 0.25$\pm$0.16 &  \\
Bone\_Vert\_C3 & 98.0$\pm$0.8 & 0.31$\pm$0.04 & 98.4$\pm$0.9 & 0.26$\pm$0.05 & 98.2$\pm$1.0 & 0.27$\pm$0.06 &  & 98.4$\pm$0.8 & 0.26$\pm$0.04 &  \\
Bone\_Vert\_C4 & 98.1$\pm$28.0 & 0.16$\pm$0.04 & 98.6$\pm$28.0 & 0.08$\pm$0.05 & 98.6$\pm$28.0 & 0.11$\pm$0.05 &  & 98.7$\pm$28.0 & 0.08$\pm$0.05 &  \\
Bone\_Vert\_C5 & 97.0$\pm$22.8 & 0.34$\pm$0.37 & 97.0$\pm$22.8 & 0.31$\pm$0.37 & 96.9$\pm$23.0 & 0.29$\pm$0.39 &  & 97.0$\pm$22.9 & 0.29$\pm$0.38 &  \\
Bone\_Vert\_C6 & 96.8$\pm$8.9 & 0.22$\pm$0.20 & 96.6$\pm$8.9 & 0.21$\pm$0.20 & 96.5$\pm$9.1 & 0.21$\pm$0.22 &  & 96.6$\pm$8.9 & 0.20$\pm$0.20 &  \\
Bone\_Vert\_C7 & 95.0$\pm$2.4 & 0.60$\pm$0.19 & 95.4$\pm$2.4 & 0.58$\pm$0.19 & 95.2$\pm$2.7 & 0.56$\pm$0.22 &  & 95.4$\pm$2.4 & 0.58$\pm$0.19 &  \\
Bone\_Vert\_L1 & 95.2$\pm$10.5 & 0.41$\pm$0.94 & 95.4$\pm$10.6 & 0.27$\pm$0.94 & 95.3$\pm$10.3 & 0.37$\pm$0.92 &  & 95.4$\pm$10.6 & 0.27$\pm$0.94 &  \\
Bone\_Vert\_L2 & 95.7$\pm$17.5 & 1.27$\pm$1.71 & 95.8$\pm$17.5 & 1.23$\pm$1.53 & 95.6$\pm$17.2 & 1.22$\pm$1.69 &  & 95.7$\pm$17.4 & 1.22$\pm$1.64 &  \\
Bone\_Vert\_L3 & 94.8$\pm$19.4 & 0.29$\pm$3.44 & 95.1$\pm$19.9 & 0.26$\pm$3.13 & 94.9$\pm$19.6 & 0.25$\pm$3.61 &  & 95.0$\pm$19.7 & 0.26$\pm$3.33 &  \\
Bone\_Vert\_L4 & 96.0$\pm$15.6 & 0.24$\pm$3.49 & 95.7$\pm$15.6 & 0.21$\pm$3.14 & 95.5$\pm$15.3 & 0.21$\pm$3.38 &  & 95.6$\pm$15.5 & 0.21$\pm$3.19 &  \\
Bone\_Vert\_L5 & 94.1$\pm$10.7 & 0.43$\pm$0.48 & 94.6$\pm$10.6 & 0.42$\pm$0.48 & 94.6$\pm$10.8 & 0.40$\pm$0.49 &  & 94.5$\pm$10.7 & 0.40$\pm$0.49 &  \\
Bone\_Vert\_T1 & 92.9$\pm$5.2 & 0.21$\pm$0.74 & 93.1$\pm$5.3 & 0.20$\pm$0.75 & 93.1$\pm$5.2 & 0.20$\pm$0.74 &  & 93.2$\pm$5.4 & 0.19$\pm$0.76 &  \\
Bone\_Vert\_T2 & 96.3$\pm$15.9 & 0.20$\pm$1.37 & 96.0$\pm$16.4 & 0.19$\pm$1.31 & 95.9$\pm$16.0 & 0.17$\pm$1.35 &  & 96.0$\pm$16.3 & 0.18$\pm$1.39 &  \\
Bone\_Vert\_T3 & 96.3$\pm$12.2 & 0.15$\pm$1.69 & 97.3$\pm$12.2 & 0.09$\pm$1.52 & 97.3$\pm$12.4 & 0.10$\pm$1.62 &  & 97.4$\pm$12.4 & 0.08$\pm$1.58 &  \\
Bone\_Vert\_T4 & 97.6$\pm$12.9 & 0.36$\pm$1.27 & 98.4$\pm$12.9 & 0.34$\pm$1.24 & 98.4$\pm$13.1 & 0.32$\pm$1.26 &  & 98.5$\pm$13.0 & 0.34$\pm$1.29 &  \\
Bone\_Vert\_T5 & 95.1$\pm$13.3 & 0.25$\pm$1.36 & 96.1$\pm$13.3 & 0.18$\pm$1.29 & 96.0$\pm$13.4 & 0.23$\pm$1.33 &  & 96.2$\pm$13.4 & 0.17$\pm$1.32 &  \\
Bone\_Vert\_T6 & 97.2$\pm$19.0 & 0.32$\pm$1.23 & 97.9$\pm$19.1 & 0.32$\pm$1.06 & 97.8$\pm$19.1 & 0.31$\pm$1.22 &  & 97.9$\pm$19.1 & 0.31$\pm$1.14 &  \\
Bone\_Vert\_T7 & 98.0$\pm$17.3 & 0.48$\pm$1.51 & 98.6$\pm$17.5 & 0.39$\pm$1.36 & 98.6$\pm$17.4 & 0.44$\pm$1.50 &  & 98.7$\pm$17.5 & 0.38$\pm$1.50 &  \\
Bone\_Vert\_T8 & 97.0$\pm$14.4 & 0.48$\pm$0.43 & 97.6$\pm$14.1 & 0.41$\pm$0.40 & 97.5$\pm$14.4 & 0.44$\pm$0.42 &  & 97.6$\pm$14.2 & 0.40$\pm$0.41 &  \\
Bone\_Vert\_T9 & 89.6$\pm$16.3 & 1.12$\pm$1.05 & 90.0$\pm$16.1 & 1.14$\pm$1.01 & 89.9$\pm$16.4 & 1.11$\pm$1.06 &  & 90.0$\pm$16.2 & 1.13$\pm$1.05 &  \\
Bone\_Vert\_T10 & 90.7$\pm$12.3 & 0.92$\pm$0.91 & 91.1$\pm$12.2 & 0.92$\pm$0.91 & 90.9$\pm$12.3 & 0.90$\pm$0.97 &  & 91.0$\pm$12.1 & 0.92$\pm$0.90 &  \\
Bone\_Vert\_T11 & 86.3$\pm$12.2 & 1.15$\pm$1.00 & 86.5$\pm$12.0 & 1.13$\pm$0.98 & 86.2$\pm$12.2 & 1.12$\pm$0.99 &  & 86.4$\pm$11.9 & 1.13$\pm$0.97 &  \\
Bone\_Vert\_T12 & 96.8$\pm$13.3 & 1.58$\pm$1.06 & 97.0$\pm$13.4 & 1.57$\pm$0.98 & 96.9$\pm$13.2 & 1.55$\pm$0.97 &  & 97.0$\pm$13.4 & 1.56$\pm$0.98 &  \\
Bone\_Vert\_S1 & 91.5$\pm$10.2 & 1.08$\pm$0.51 & 92.6$\pm$10.3 & 1.19$\pm$0.52 & 92.8$\pm$10.3 & 1.23$\pm$0.51 &  & 92.7$\pm$10.5 & 1.19$\pm$0.54 &  \\
\rowcolor[HTML]{F3F3F3} 
Bone\_Sacrum & 94.6$\pm$2.0 & 0.53$\pm$0.15 & 95.3$\pm$2.3 & 0.49$\pm$0.18 & 95.1$\pm$2.1 & 0.48$\pm$0.16 & {\cellcolor[HTML]{F3F3F3}99.5} & 95.3$\pm$2.1 & 0.49$\pm$0.16 & {\cellcolor[HTML]{F3F3F3}99.5}\\ \hline

Mean & \multicolumn{1}{c}{ 94.4 } & \multicolumn{1}{c|}{ 1.08 } & \multicolumn{1}{c}{ 94.7 } & \multicolumn{1}{c|}{ 1.05 } & \multicolumn{1}{c}{ 94.6 } & \multicolumn{1}{c|}{ 1.05 } & \multicolumn{1}{c|}{ 98.0 } & \multicolumn{1}{c}{ 94.8 } & \multicolumn{1}{c|}{ 1.04 }  & \multicolumn{1}{c|}{ 98.0 } \\ \hline

\end{tabular}
}
\caption{\textbar~\textbf{Organ-wise segmentation performance comparison on costovertebral bones.} Comparison of DSC (\%) and ASD (mm) among \ac{nnu_e36}, \ac{cln_u36_unprn}, \ac{cln_c36}, and \ac{cln_u36} for 52 costovertebral bones. Decoder-wise pruning rate ($\mathcal{T}$, \%) of \ac{cln_c36} and \ac{cln_u36} are also provided. \ac{cln_u36} achieves the highest mean DSC and the lowest mean ASD, demonstrating its superior performance in costovertebral bones. }
\label{tab:ds36_bone}
\end{table}
\clearpage

\begin{table}[htp]
\centering
\resizebox{\textwidth}{!}{
\begin{tabular}{|l|ll|ll|ll|r|ll|r|}
\hline
\multicolumn{1}{|l|}{} & \multicolumn{2}{c|}{\cellcolor[HTML]{FFFFFF}{\color[HTML]{1F1F1F} \ac{nnu_e36}}} & \multicolumn{2}{c|}{\ac{cln_u36_unprn}} & \multicolumn{3}{c|}{\ac{cln_c36}} & \multicolumn{3}{c|}{\ac{cln_u36}} \\ \cline{2-11} 
\multicolumn{1}{|l|}{\multirow{-2}{*}{}} & DSC$\uparrow$ & \multicolumn{1}{l|}{ASD$\downarrow$} & DSC$\uparrow$ & \multicolumn{1}{l|}{ASD$\downarrow$} & DSC$\uparrow$ & ASD$\downarrow$ & \multicolumn{1}{r|}{$\mathcal{T}$$\uparrow$ } & DSC$\uparrow$ & ASD$\downarrow$ & \multicolumn{1}{r|}{$\mathcal{T}$$\uparrow$ } \\ \hline
\rowcolor[HTML]{F3F3F3} 
HN\_LNS\_1La & 51.3$\pm$6.7 & 1.06$\pm$0.38 & 52.6$\pm$6.2 & 1.05$\pm$0.39 & 52.9$\pm$6.7 & 1.06$\pm$0.38 & {\cellcolor[HTML]{F3F3F3}86.0} & 52.2$\pm$5.9 & 1.02$\pm$0.41 & {\cellcolor[HTML]{F3F3F3}88.0} \\
\rowcolor[HTML]{F3F3F3} 
HN\_LNS\_1Lb & 78.6$\pm$7.1 & 0.28$\pm$0.76 & 80.2$\pm$6.1 & 0.18$\pm$0.77 & 81.3$\pm$6.8 & 0.20$\pm$0.74 &  & 80.8$\pm$7.7 & 0.21$\pm$0.76 &  \\
\rowcolor[HTML]{F3F3F3} 
HN\_LNS\_1Ra & 54.1$\pm$6.6 & 1.39$\pm$0.69 & 57.0$\pm$6.1 & 1.36$\pm$0.66 & 55.0$\pm$6.6 & 1.21$\pm$0.69 &  & 56.3$\pm$5.8 & 1.31$\pm$0.68 &  \\
\rowcolor[HTML]{F3F3F3} 
HN\_LNS\_1Rb & 80.5$\pm$8.3 & 1.04$\pm$0.80 & 81.3$\pm$9.7 & 1.02$\pm$0.85 & 80.8$\pm$8.1 & 1.00$\pm$0.82 &  & 82.0$\pm$8.0 & 1.04$\pm$0.83 &  \\
\rowcolor[HTML]{F3F3F3} 
HN\_LNS\_2La & 69.2$\pm$5.7 & 1.00$\pm$0.99 & 72.8$\pm$5.2 & 0.89$\pm$0.93 & 71.5$\pm$5.0 & 0.99$\pm$0.99 &  & 72.1$\pm$6.1 & 0.92$\pm$0.92 &  \\
\rowcolor[HTML]{F3F3F3} 
HN\_LNS\_2Lb & 73.4$\pm$6.5 & 1.93$\pm$0.69 & 74.2$\pm$5.9 & 1.60$\pm$0.66 & 74.6$\pm$5.8 & 1.68$\pm$0.70 &  & 74.4$\pm$7.2 & 1.58$\pm$0.67 &  \\
\rowcolor[HTML]{F3F3F3} 
HN\_LNS\_2Ra & 73.7$\pm$7.1 & 2.11$\pm$0.39 & 75.7$\pm$8.1 & 1.77$\pm$0.40 & 76.6$\pm$6.7 & 1.91$\pm$0.38 &  & 75.5$\pm$6.7 & 1.78$\pm$0.37 &  \\
\rowcolor[HTML]{F3F3F3} 
HN\_LNS\_2Rb & 75.0$\pm$13.4 & 1.75$\pm$0.58 & 75.2$\pm$14.1 & 1.68$\pm$0.61 & 75.0$\pm$14.1 & 1.53$\pm$0.56 &  & 74.9$\pm$13.9 & 1.64$\pm$0.60 &  \\
\rowcolor[HTML]{F3F3F3} 
HN\_LNS\_3L & 76.3$\pm$12.6 & 1.51$\pm$0.79 & 76.5$\pm$12.3 & 1.40$\pm$0.77 & 75.9$\pm$12.3 & 1.52$\pm$0.75 &  & 76.1$\pm$13.0 & 1.40$\pm$0.78 &  \\
\rowcolor[HTML]{F3F3F3} 
HN\_LNS\_3R & 72.0$\pm$6.8 & 1.83$\pm$0.63 & 72.4$\pm$7.1 & 1.57$\pm$0.60 & 72.0$\pm$7.9 & 1.68$\pm$0.63 &  & 72.3$\pm$5.7 & 1.56$\pm$0.62 &  \\
\rowcolor[HTML]{F3F3F3} 
HN\_LNS\_4L & 77.0$\pm$6.3 & 1.56$\pm$0.49 & 76.8$\pm$6.4 & 1.35$\pm$0.47 & 77.7$\pm$5.6 & 1.31$\pm$0.50 &  & 76.5$\pm$5.3 & 1.36$\pm$0.47 &  \\
\rowcolor[HTML]{F3F3F3} 
HN\_LNS\_4R & 72.6$\pm$6.5 & 1.56$\pm$0.49 & 73.9$\pm$5.8 & 1.31$\pm$0.51 & 74.5$\pm$6.3 & 1.43$\pm$0.47 &  & 75.7$\pm$5.7 & 1.30$\pm$0.50 &  \\
\rowcolor[HTML]{F3F3F3} 
HN\_LNS\_5La & 72.9$\pm$3.2 & 0.90$\pm$0.54 & 74.3$\pm$3.0 & 0.91$\pm$0.59 & 75.1$\pm$3.5 & 0.90$\pm$0.55 &  & 74.2$\pm$3.0 & 0.88$\pm$0.57 &  \\
\rowcolor[HTML]{F3F3F3} 
HN\_LNS\_5Lb & 75.8$\pm$3.5 & 0.92$\pm$0.18 & 76.2$\pm$4.5 & 0.85$\pm$0.18 & 77.5$\pm$3.3 & 0.75$\pm$0.19 &  & 76.0$\pm$2.8 & 0.80$\pm$0.19 &  \\
\rowcolor[HTML]{F3F3F3} 
HN\_LNS\_5Ra & 76.6$\pm$2.8 & 1.00$\pm$0.12 & 78.7$\pm$3.5 & 0.91$\pm$0.12 & 78.3$\pm$3.3 & 0.92$\pm$0.14 &  & 78.8$\pm$3.9 & 0.87$\pm$0.14 &  \\
\rowcolor[HTML]{F3F3F3} 
HN\_LNS\_5Rb & 77.5$\pm$5.8 & 1.51$\pm$0.09 & 79.3$\pm$6.2 & 1.50$\pm$0.08 & 80.5$\pm$6.4 & 1.45$\pm$0.08 &  & 77.9$\pm$5.2 & 1.52$\pm$0.09 &  \\
\rowcolor[HTML]{F3F3F3} 
HN\_LNS\_6L & 50.4$\pm$14.5 & 1.38$\pm$0.96 & 51.7$\pm$14.0 & 1.27$\pm$0.97 & 52.9$\pm$15.0 & 1.24$\pm$0.97 &  & 53.1$\pm$14.6 & 1.29$\pm$0.99 &  \\
\rowcolor[HTML]{F3F3F3} 
HN\_LNS\_6R & 49.8$\pm$14.7 & 1.42$\pm$0.98 & 53.6$\pm$14.1 & 1.55$\pm$0.94 & 50.9$\pm$14.7 & 1.39$\pm$0.96 &  & 53.1$\pm$15.3 & 1.55$\pm$0.95 &  \\
Chest\_LNS1\_L & 78.1$\pm$8.1 & 1.48$\pm$0.43 & 82.9$\pm$7.7 & 1.23$\pm$0.42 & 80.9$\pm$8.0 & 1.41$\pm$0.43 & {90.0} & 81.8$\pm$8.5 & 1.32$\pm$0.42 & {90.0} \\
Chest\_LNS1\_R & 76.8$\pm$10.4 & 1.66$\pm$0.63 & 81.9$\pm$10.0 & 1.71$\pm$0.59 & 82.0$\pm$10.1 & 1.68$\pm$0.60 &  & 80.2$\pm$10.2 & 1.69$\pm$0.61 &  \\
Chest\_LNS2.L & 66.9$\pm$10.5 & 1.55$\pm$0.47 & 71.1$\pm$10.9 & 1.23$\pm$0.50 & 70.6$\pm$10.7 & 1.22$\pm$0.48 &  & 70.6$\pm$10.7 & 1.25$\pm$0.49 &  \\
Chest\_LNS2\_R & 70.7$\pm$15.5 & 1.67$\pm$2.44 & 74.1$\pm$15.5 & 1.35$\pm$2.17 & 74.7$\pm$15.5 & 1.31$\pm$2.41 &  & 74.8$\pm$15.5 & 1.34$\pm$2.27 &  \\
Chest\_LNS3\_A & 77.4$\pm$19.7 & 1.09$\pm$3.26 & 82.1$\pm$19.6 & 0.82$\pm$3.37 & 81.3$\pm$19.7 & 0.78$\pm$3.46 &  & 82.9$\pm$19.4 & 0.77$\pm$3.61 &  \\
Chest\_LNS3\_P & 84.6$\pm$8.2 & 0.86$\pm$1.19 & 90.0$\pm$8.3 & 0.63$\pm$1.07 & 88.8$\pm$8.4 & 0.62$\pm$1.14 &  & 90.1$\pm$8.3 & 0.60$\pm$1.09 &  \\
Chest\_LNS4\_L & 74.1$\pm$10.4 & 0.96$\pm$0.92 & 79.4$\pm$10.4 & 0.88$\pm$0.92 & 78.0$\pm$10.3 & 0.85$\pm$0.91 &  & 78.0$\pm$10.3 & 0.86$\pm$0.92 &  \\
Chest\_LNS4\_R & 73.8$\pm$13.4 & 1.65$\pm$1.06 & 77.0$\pm$13.4 & 1.40$\pm$0.98 & 76.8$\pm$13.4 & 1.36$\pm$1.00 &  & 77.9$\pm$13.3 & 1.48$\pm$0.99 &  \\
Chest\_LNS5 & 72.6$\pm$12.1 & 1.28$\pm$0.98 & 72.6$\pm$11.9 & 1.30$\pm$0.96 & 71.6$\pm$12.5 & 1.27$\pm$0.98 &  & 72.6$\pm$12.7 & 1.31$\pm$0.97 &  \\
Chest\_LNS6 & 72.4$\pm$12.2 & 0.88$\pm$0.91 & 75.1$\pm$13.2 & 0.76$\pm$0.96 & 71.8$\pm$11.2 & 0.79$\pm$0.95 &  & 73.6$\pm$12.9 & 0.75$\pm$0.97 &  \\
Chest\_LNS7 & 85.0$\pm$16.0 & 1.00$\pm$1.06 & 86.7$\pm$15.3 & 0.95$\pm$0.99 & 86.1$\pm$15.7 & 0.90$\pm$1.10 &  & 86.6$\pm$15.5 & 0.97$\pm$1.04 &  \\
Chest\_LNS8 & 80.9$\pm$14.1 & 1.66$\pm$0.40 & 81.8$\pm$13.9 & 1.65$\pm$0.42 & 82.5$\pm$13.7 & 1.68$\pm$0.41 &  & 82.2$\pm$14.9 & 1.64$\pm$0.42 &  \\
Chest\_LNS9 & 62.7$\pm$17.6 & 1.39$\pm$1.39 & 62.1$\pm$17.8 & 1.26$\pm$1.48 & 62.4$\pm$16.2 & 1.37$\pm$1.44 &  & 62.0$\pm$17.2 & 1.25$\pm$1.50 &  \\
Chest\_LNS10\_L & 74.1$\pm$19.1 & 1.69$\pm$1.19 & 77.4$\pm$19.4 & 1.30$\pm$1.11 & 78.0$\pm$20.0 & 1.37$\pm$1.18 &  & 77.4$\pm$19.4 & 1.32$\pm$1.16 &  \\
Chest\_LNS10\_R & 72.8$\pm$13.5 & 1.52$\pm$1.37 & 76.9$\pm$13.5 & 1.21$\pm$1.15 & 77.2$\pm$12.6 & 1.39$\pm$1.26 &  & 77.2$\pm$13.2 & 1.20$\pm$1.27 &  
\\ \hline

Mean & \multicolumn{1}{c}{ 72.1 } & \multicolumn{1}{c|}{ 1.35 } & \multicolumn{1}{c}{ 74.4 } & \multicolumn{1}{c|}{ 1.21 } & \multicolumn{1}{c}{ 74.1 } & \multicolumn{1}{c|}{ 1.22 } & \multicolumn{1}{c|}{ 88.0 } & \multicolumn{1}{c}{ 74.2 } & \multicolumn{1}{c|}{ 1.21 }  & \multicolumn{1}{c|}{ 89.0 } \\ \hline

\end{tabular}
}\caption{\textbar~\textbf{Organ-wise segmentation performance comparison on lymph node stations (LNS).} Comparison of DSC (\%) and ASD (mm) among \ac{nnu_e36}, \ac{cln_u36_unprn}, \ac{cln_c36}, and \ac{cln_u36} for 33 lymph node stations. Decoder-wise pruning rate ($\mathcal{T}$, \%) of \ac{cln_c36} and \ac{cln_u36} are also provided. \ac{cln_u36} achieves the highest mean DSC and the lowest mean ASD, demonstrating its superior performance in the LNS group. }
\label{tab:ds36_lns}
\end{table}
\clearpage

\begin{table}[htp]
\centering
\resizebox{\textwidth}{!}{
\begin{tabular}{|l|ll|ll|ll|r|ll|r|}
\hline
\multicolumn{1}{|l|}{} & \multicolumn{2}{c|}{\cellcolor[HTML]{FFFFFF}{\color[HTML]{1F1F1F} \ac{nnu_e36}}} & \multicolumn{2}{c|}{\ac{cln_u36_unprn}} & \multicolumn{3}{c|}{\ac{cln_c36}} & \multicolumn{3}{c|}{\ac{cln_u36}} \\ \cline{2-11} 
\multicolumn{1}{|l|}{\multirow{-2}{*}{}} & DSC$\uparrow$ & \multicolumn{1}{l|}{ASD$\downarrow$} & DSC$\uparrow$ & \multicolumn{1}{l|}{ASD$\downarrow$} & DSC$\uparrow$ & ASD$\downarrow$ & \multicolumn{1}{r|}{$\mathcal{T}$$\uparrow$ } & DSC$\uparrow$ & ASD$\downarrow$ & \multicolumn{1}{r|}{$\mathcal{T}$$\uparrow$ } \\ \hline
\rowcolor[HTML]{F3F3F3} 
Eso\_\Ac{gtv} & 73.2$\pm$10.9 & 3.70$\pm$2.23 & 73.6$\pm$10.9 & 3.73$\pm$2.09 & 73.3$\pm$11.0 & 3.93$\pm$2.03 & 92.0 & 73.4$\pm$11.0 & 3.56$\pm$2.19 & 92.0 \\
Lung\_\Ac{gtv} & 78.2$\pm$10.9 & 1.07$\pm$2.17 & 78.2$\pm$11.0 & 1.16$\pm$2.12 & 78.0$\pm$11.0 & 1.17$\pm$2.20 & 92.0 & 79.1$\pm$11.0 & 1.07$\pm$2.21 & 92.0 \\
\rowcolor[HTML]{F3F3F3} 
Liver\_\Ac{gtv} & 63.4$\pm$8.4 & 3.80$\pm$2.15 & 63.9$\pm$8.4 & 3.75$\pm$2.09 & 63.6$\pm$8.3 & 3.60$\pm$2.15 & 92.0 & 63.8$\pm$8.5 & 3.48$\pm$2.10 & 92.0 \\
Kidney\_\Ac{gtv} & 83.7$\pm$9.2 & 1.10$\pm$1.08 & 84.0$\pm$9.1 & 1.23$\pm$1.00 & 83.8$\pm$9.3 & 1.10$\pm$1.02 & 94.0 & 83.9$\pm$9.0 & 1.18$\pm$1.04 & 94.0 \\
\rowcolor[HTML]{F3F3F3} 
Colon\_\Ac{gtv} & 58.8$\pm$8.9 & 3.62$\pm$3.60 & 60.8$\pm$8.8 & 3.50$\pm$3.57 & 61.2$\pm$8.8 & 3.81$\pm$3.28 & 94.0 & 61.3$\pm$8.9 & 3.40$\pm$3.59 & 94.0 \\
% Hepatic\_Vessel\_GTV & 83.1$\pm$11.0 & 1.23$\pm$2.21 & 83.2$\pm$11.0 & 1.15$\pm$2.11 & 83.3$\pm$10.9 & 1.24$\pm$2.08 & 96.0 & 83.4$\pm$10.8 & 1.15$\pm$2.01 & 96.0 \\
Pancreas\_\Ac{gtv} & 62.2$\pm$8.5 & 3.75$\pm$2.02 & 62.8$\pm$8.4 & 3.74$\pm$2.09 & 62.7$\pm$8.5 & 3.75$\pm$2.10 & 94.0 & 62.8$\pm$8.3 & 3.62$\pm$1.89 & 94.0 \\
\rowcolor[HTML]{F3F3F3} 

\ac{npc}\_\Ac{gtv} & 72.5$\pm$12.4 & 2.40$\pm$2.09 & 72.9$\pm$12.4 & 2.06$\pm$1.95 & 72.7$\pm$12.4 & 2.36$\pm$1.92 & 92.0 & 72.8$\pm$12.6 & 2.23$\pm$2.02 & 92.0 \\

HN\_LN & 71.8$\pm$13.2 & 2.58$\pm$2.19 & 71.9$\pm$13.6 & 2.51$\pm$2.05 & 71.8$\pm$13.3 & 2.72$\pm$2.15 & 92.0 & 71.8$\pm$13.2 & 2.57$\pm$2.10 & 92.0 \\
\rowcolor[HTML]{F3F3F3} 
Chest\_LN & 56.2$\pm$18.8 & 12.70$\pm$4.91 & 56.8$\pm$19.1 & 13.62$\pm$4.59 & 56.4$\pm$18.7 & 12.97$\pm$4.50 & 88.0 & 56.6$\pm$18.9 & 12.58$\pm$4.61 & 88.0 
\\ \hline

Mean & \multicolumn{1}{c}{ 70.3 } & \multicolumn{1}{c|}{ 3.60 } & \multicolumn{1}{c}{ 70.8 } & \multicolumn{1}{c|}{ 3.65 } & \multicolumn{1}{c}{ 70.7 } & \multicolumn{1}{c|}{ 3.67 } & \multicolumn{1}{c|}{ 90.6 } & \multicolumn{1}{c}{ 70.9 } & \multicolumn{1}{c|}{ 3.48 }  & \multicolumn{1}{c|}{ 92.6 } \\ \hline

\end{tabular}
}\caption{\textbar~\textbf{Organ-wise segmentation performance comparison on lesions.} Comparison of DSC (\%) and ASD (mm) among \ac{nnu_e36}, \ac{cln_u36_unprn}, \ac{cln_c36}, and \ac{cln_u36} for 10 lesions. Decoder-wise pruning rate ($\mathcal{T}$, \%) of \ac{cln_c36} and \ac{cln_u36} are also provided. \ac{cln_u36} achieves the highest mean DSC and the lowest mean ASD, demonstrating its superior performance in the lesion group. }
\label{tab:ds36_gtv}
\end{table}
\clearpage

%% dataset-wise results of U5/E5/MT on 5 datasets
% \input{Tabs/tabs_ds5_pl/tab_pl_totalseg}
% \input{Tabs/tabs_ds5_pl/tab_pl_structseg}
% \input{Tabs/tabs_ds5_pl/tab_pl_segthor}
% \input{Tabs/tabs_ds5_pl/tab_pl_flare}
% \input{Tabs/tabs_ds5_pl/tab_pl_kits}

\begin{table}[htp]
\centering
\fontsize{10pt}{12pt}\selectfont
% \resizebox{\textwidth}{!}{
\begin{tabular}{|l|ll|ll|ll|r|}
\hline
 & \multicolumn{2}{c|}{\ac{nnu_e5}} & \multicolumn{2}{c|}{MultiTalent} & \multicolumn{3}{c|}{\ac{cln_u5}} \\ \cline{2-8} 
\multirow{-2}{*}{} & DSC$\uparrow$ & ASD$\downarrow$ & DSC$\uparrow$ & ASD$\downarrow$ & DSC$\uparrow$ & ASD$\downarrow$ & $\mathcal{T}\uparrow$ \\ \hline
\rowcolor[HTML]{F3F3F3} 
Spleen & 97.6$\pm$15.6 & 1.95$\pm$0.40 & 96.6$\pm$3.2 & 0.36$\pm$0.44 & 97.1$\pm$6.8 & 1.65$\pm$0.39 & 99.7 \\
Kidney\_R & 95.9$\pm$16.9 & 0.19$\pm$0.73 & 95.7$\pm$4.7 & 0.29$\pm$0.30 & 96.1$\pm$7.1 & 0.21$\pm$0.76 & 99.7 \\
Kidney\_L & 96.1$\pm$26.7 & 1.15$\pm$6.45 & 91.9$\pm$17.5 & 1.72$\pm$8.64 & 97.2$\pm$6.9 & 1.00$\pm$0.72 & {} \\
\rowcolor[HTML]{F3F3F3} 
GallBladder & 92.2$\pm$24.4 & 1.14$\pm$0.82 & 86.0$\pm$19.3 & 1.42$\pm$2.67 & 93.2$\pm$8.7 & 0.95$\pm$0.83 & 99.7 \\
Liver & 97.3$\pm$14.9 & 0.57$\pm$0.93 & 97.8$\pm$1.5 & 0.52$\pm$0.65 & 97.5$\pm$5.8 & 0.54$\pm$0.93 & 99.5 \\
\rowcolor[HTML]{F3F3F3} 
Stomach & 95.9$\pm$16.2 & 1.18$\pm$0.66 & 94.2$\pm$8.2 & 0.66$\pm$0.68 & 96.5$\pm$6.2 & 1.11$\pm$0.66 & 99.5 \\
Pancreas & 91.8$\pm$17.0 & 1.06$\pm$0.67 & 90.3$\pm$6.8 & 0.79$\pm$1.53 & 92.0$\pm$6.9 & 0.84$\pm$0.37 & 99.5 \\
\rowcolor[HTML]{F3F3F3} 
Glnd\_Adrenal\_R & 83.0$\pm$13.5 & 0.55$\pm$1.17 & 87.8$\pm$11.2 & 0.50$\pm$1.60 & 84.2$\pm$13.3 & 0.43$\pm$0.39 & 94.0 \\
\rowcolor[HTML]{F3F3F3} 
Glnd\_Adrenal\_L & 84.2$\pm$12.2 & 1.03$\pm$0.36 & 85.4$\pm$13.5 & 0.31$\pm$0.26 & 84.8$\pm$12.5 & 0.87$\pm$0.39 & {\cellcolor[HTML]{F3F3F3}} \\
Lung\_LUL & 97.8$\pm$3.6 & 0.31$\pm$0.93 & 97.3$\pm$4.5 & 1.15$\pm$4.41 & 97.9$\pm$3.7 & 0.17$\pm$0.62 & 99.7 \\
Lung\_LLL & 96.5$\pm$24.0 & 0.21$\pm$0.98 & 95.7$\pm$8.5 & 0.72$\pm$1.68 & 96.5$\pm$2.1 & 0.21$\pm$0.47 & {} \\
\rowcolor[HTML]{F3F3F3} 
Lung\_RUL & 97.5$\pm$20.5 & 0.30$\pm$1.48 & 92.0$\pm$19.5 & 0.77$\pm$1.37 & 97.7$\pm$2.5 & 0.20$\pm$0.98 & 99.5 \\
\rowcolor[HTML]{F3F3F3} 
Lung\_RML & 92.0$\pm$6.2 & 0.47$\pm$1.41 & 94.8$\pm$6.5 & 0.83$\pm$1.31 & 92.3$\pm$3.3 & 0.22$\pm$1.50 & {\cellcolor[HTML]{F3F3F3}} \\
\rowcolor[HTML]{F3F3F3} 
Lung\_RLL & 97.4$\pm$26.8 & 0.16$\pm$2.45 & 97.2$\pm$5.9 & 0.43$\pm$0.83 & 97.2$\pm$3.6 & 0.21$\pm$1.19 & {\cellcolor[HTML]{F3F3F3}} \\
Eso & 94.8$\pm$2.9 & 0.61$\pm$0.15 & 94.1$\pm$2.1 & 0.24$\pm$0.24 & 94.4$\pm$3.1 & 0.45$\pm$0.16 & 99.7 \\
\rowcolor[HTML]{F3F3F3} 
Trachea & 96.7$\pm$5.8 & 0.36$\pm$0.12 & 96.0$\pm$5.0 & 0.27$\pm$0.47 & 97.1$\pm$6.1 & 0.27$\pm$0.12 & 99.5 \\
Glnd\_Thyroid & 91.7$\pm$14.7 & 0.41$\pm$0.96 & 90.2$\pm$9.2 & 0.37$\pm$0.45 & 92.6$\pm$14.5 & 0.28$\pm$0.96 & 99.7 \\
\rowcolor[HTML]{F3F3F3} 
SmallBowel & 91.2$\pm$26.4 & 1.63$\pm$2.53 & 87.4$\pm$14.0 & 1.56$\pm$1.77 & 91.2$\pm$16.6 & 1.55$\pm$2.62 & 99.7 \\
Duodenum & 87.3$\pm$26.0 & 1.87$\pm$1.26 & 82.9$\pm$18.3 & 1.02$\pm$1.08 & 88.8$\pm$16.1 & 1.81$\pm$1.26 & 99.5 \\
\rowcolor[HTML]{F3F3F3} 
Colon & 90.2$\pm$16.0 & 2.54$\pm$1.21 & 89.9$\pm$7.9 & 1.55$\pm$1.90 & 90.4$\pm$15.8 & 2.36$\pm$1.23 & 99.5 \\
UrinaryBladder & 89.1$\pm$9.5 & 2.98$\pm$1.08 & 92.5$\pm$7.6 & 1.27$\pm$1.97 & 89.2$\pm$9.6 & 2.85$\pm$1.03 & 99.5 \\
\rowcolor[HTML]{F3F3F3} 
Prostate & 81.7$\pm$38.0 & 0.52$\pm$2.06 & 72.8$\pm$33.5 & 1.45$\pm$1.72 & 83.3$\pm$17.9 & 0.35$\pm$1.47 & 99.5 \\ \hline

Mean & \multicolumn{1}{c}{ 92.6 } & \multicolumn{1}{c|}{ 0.96 } & \multicolumn{1}{c}{ 91.3 } & \multicolumn{1}{c|}{ 0.83 } & \multicolumn{1}{c}{ 93.1 } & \multicolumn{1}{c|}{ 0.84 } & \multicolumn{1}{c|}{ 99.3 } \\ \hline
\end{tabular}
% }
\caption{\textbar~\textbf{Organ-wise performance comparison of \ac{nnu_e5}, MultiTalent, and \ac{cln_u5} on \ac{totalseg} `Organ' subgroup.} Organ-wise DSC (\%) and ASD (mm) of \ac{nnu_e5}, MultiTalent and \ac{cln_u5} are evaluated on \ac{totalseg} `Organ' subgroup. \ac{cln_u5} achieves the best mean DSC.}
\label{tab:pl_totalseg_organ_ds5}
\end{table}
\clearpage

\begin{table}[htp]
\centering
\fontsize{10pt}{12pt}\selectfont
% \resizebox{\textwidth}{!}{
\begin{tabular}{|l|ll|ll|ll|r|}
\hline
 & \multicolumn{2}{c|}{\ac{nnu_e5}} & \multicolumn{2}{c|}{MultiTalent} & \multicolumn{3}{c|}{\ac{cln_u5}} \\ \cline{2-8} 
\multirow{-2}{*}{} & DSC$\uparrow$ & ASD$\downarrow$ & DSC$\uparrow$ & ASD$\downarrow$ & DSC$\uparrow$ & ASD$\downarrow$ & $\mathcal{T}\uparrow$ \\ \hline
\rowcolor[HTML]{F3F3F3} 
Heart & 95.2$\pm$14.7 & 0.42$\pm$0.20 & 95.9$\pm$1.4 & 1.10$\pm$3.24 & 94.9$\pm$14.7 & 0.43$\pm$0.18 & \multicolumn{1}{r|}{\cellcolor[HTML]{F3F3F3}99.7} \\
A\_Aorta & 96.0$\pm$16.7 & 0.44$\pm$0.95 & 96.7$\pm$4.7 & 0.30$\pm$0.57 & 96.5$\pm$16.7 & 0.41$\pm$0.85 & \multicolumn{1}{r|}{99.7} \\
\rowcolor[HTML]{F3F3F3} 
V\_Pulmonary & 89.8$\pm$15.5 & 0.47$\pm$0.14 & 93.1$\pm$3.1 & 0.23$\pm$0.13 & 91.3$\pm$15.5 & 0.31$\pm$0.16 & \multicolumn{1}{r|}{\cellcolor[HTML]{F3F3F3}99.0} \\
A\_BrachiocephalicTrunk & 89.8$\pm$16.2 & 0.32$\pm$0.43 & 93.8$\pm$3.0 & 0.22$\pm$0.18 & 90.8$\pm$16.3 & 0.22$\pm$0.44 & \multicolumn{1}{r|}{99.0} \\
\rowcolor[HTML]{F3F3F3} 
A\_Subclavian\_L & 94.6$\pm$3.5 & 0.44$\pm$0.28 & 91.6$\pm$3.5 & 0.33$\pm$0.47 & 94.9$\pm$3.6 & 0.27$\pm$0.29 & 99.7 \\
\rowcolor[HTML]{F3F3F3} 
A\_Subclavian\_R & 94.1$\pm$3.3 & 0.23$\pm$0.17 & 91.5$\pm$3.4 & 0.22$\pm$0.15 & 93.3$\pm$3.4 & 0.17$\pm$0.18 & {\cellcolor[HTML]{F3F3F3}} \\
A\_CommonCarotid\_L & 91.2$\pm$5.4 & 0.35$\pm$0.28 & 90.3$\pm$4.8 & 0.23$\pm$0.22 & 91.1$\pm$5.4 & 0.24$\pm$0.28 & 99.5 \\
A\_CommonCarotid\_R & 96.6$\pm$3.3 & 0.28$\pm$0.99 & 93.2$\pm$2.4 & 0.15$\pm$0.13 & 97.4$\pm$3.3 & 0.26$\pm$1.02 & {} \\
\rowcolor[HTML]{F3F3F3} 
V\_Brachiocephalic\_L & 90.4$\pm$26.2 & 0.37$\pm$1.97 & 91.5$\pm$15.7 & 0.44$\pm$1.59 & 91.2$\pm$16.1 & 0.31$\pm$2.11 & 99.5 \\
\rowcolor[HTML]{F3F3F3} 
V\_Brachiocephalic\_R & 88.7$\pm$15.0 & 0.32$\pm$0.81 & 89.8$\pm$11.4 & 0.37$\pm$0.57 & 89.1$\pm$15.3 & 0.24$\pm$0.84 & {\cellcolor[HTML]{F3F3F3}} \\
LAA & 92.6$\pm$3.4 & 0.46$\pm$0.15 & 93.2$\pm$2.7 & 0.20$\pm$0.11 & 91.7$\pm$3.3 & 0.16$\pm$0.15 & 99.5 \\
\rowcolor[HTML]{F3F3F3} 
V\_VenaCava\_S & 93.8$\pm$21.4 & 0.23$\pm$2.35 & 88.4$\pm$24.2 & 0.42$\pm$0.83 & 93.7$\pm$21.4 & 0.18$\pm$2.44 & 99.5 \\
V\_VenaCava\_I & 93.9$\pm$9.8 & 1.03$\pm$1.05 & 92.3$\pm$6.3 & 0.47$\pm$0.38 & 95.2$\pm$9.8 & 0.77$\pm$1.00 & 99.5 \\
\rowcolor[HTML]{F3F3F3} 
V\_Portal\_Splenic & 85.5$\pm$29.3 & 1.29$\pm$0.62 & 85.1$\pm$12.1 & 0.67$\pm$0.72 & 86.7$\pm$29.1 & 1.12$\pm$0.63 & 99.0 \\
A\_Iliac\_L & 94.4$\pm$3.1 & 0.61$\pm$0.19 & 93.2$\pm$3.0 & 0.25$\pm$0.38 & 95.1$\pm$3.3 & 0.50$\pm$0.20 & 99.5 \\
A\_Iliac\_R & 91.9$\pm$19.5 & 0.45$\pm$0.15 & 93.5$\pm$3.8 & 0.19$\pm$0.26 & 91.9$\pm$19.6 & 0.39$\pm$0.18 & {} \\
\rowcolor[HTML]{F3F3F3} 
V\_Iliac\_L & 92.2$\pm$3.0 & 0.48$\pm$0.13 & 94.1$\pm$2.8 & 0.18$\pm$0.10 & 92.9$\pm$3.3 & 0.43$\pm$0.15 & 99.0 \\
\rowcolor[HTML]{F3F3F3} 
V\_Iliac\_R & 94.2$\pm$4.9 & 0.44$\pm$0.31 & 92.9$\pm$5.6 & 0.22$\pm$0.20 & 93.6$\pm$4.8 & 0.45$\pm$0.27 & {\cellcolor[HTML]{F3F3F3}}
\\ \hline

Mean & \multicolumn{1}{c}{ 92.5 } & \multicolumn{1}{c|}{ 0.48 } & \multicolumn{1}{c}{ 92.2 } & \multicolumn{1}{c|}{ 0.34 } & \multicolumn{1}{c}{ 92.9 } & \multicolumn{1}{c|}{ 0.38 } & \multicolumn{1}{c|}{ 99.4 } \\ \hline

\end{tabular}
% }
\caption{\textbar~\textbf{Organ-wise performance comparison of \ac{nnu_e5}, MultiTalent, and \ac{cln_u5} on \ac{totalseg} `Cardiac' subgroup.} Organ-wise DSC (\%) and ASD (mm) of \ac{nnu_e5}, MultiTalent and \ac{cln_u5} are evaluated on \ac{totalseg} `Cardiac' subgroup. \ac{cln_u5} achieves the best mean DSC.}
\label{tab:pl_totalseg_cardi_ds5}
\end{table}
\clearpage

\begin{table}[htp]
\centering
\fontsize{10pt}{12pt}\selectfont
% \resizebox{\textwidth}{!}{
\begin{tabular}{|l|ll|ll|ll|r|}
\hline
 & \multicolumn{2}{c|}{\ac{nnu_e5}} & \multicolumn{2}{c|}{MultiTalent} & \multicolumn{3}{c|}{\ac{cln_u5}} \\ \cline{2-8} 
\multirow{-2}{*}{} & DSC$\uparrow$ & ASD$\downarrow$ & DSC$\uparrow$ & ASD$\downarrow$ & DSC$\uparrow$ & ASD$\downarrow$ & $\mathcal{T}\uparrow$ \\ \hline
\rowcolor[HTML]{F3F3F3} 
Bone\_Humerus\_L & 96.9$\pm$17.5 & 2.84$\pm$0.20 & 93.5$\pm$7.1 & 0.57$\pm$0.97 & 97.1$\pm$17.6 & 2.77$\pm$0.21 & \multicolumn{1}{r|}{\cellcolor[HTML]{F3F3F3}99.5} \\
\rowcolor[HTML]{F3F3F3} 
Bone\_Humerus\_R & 96.6$\pm$17.3 & 4.31$\pm$0.17 & 96.5$\pm$2.0 & 5.20$\pm$12.27 & 97.1$\pm$17.0 & 3.88$\pm$0.17 & \multicolumn{1}{l|}{\cellcolor[HTML]{F3F3F3}} \\
Bone\_Scapula\_L & 94.1$\pm$21.1 & 1.16$\pm$0.18 & 94.8$\pm$1.9 & 0.49$\pm$1.16 & 93.8$\pm$21.4 & 0.97$\pm$0.17 & 99.5 \\
Bone\_Scapula\_R & 95.1$\pm$1.7 & 1.95$\pm$0.07 & 95.2$\pm$2.1 & 0.35$\pm$0.59 & 94.4$\pm$1.9 & 1.63$\pm$0.08 & {} \\
\rowcolor[HTML]{F3F3F3} 
Bone\_Clavicula\_L & 94.5$\pm$0.8 & 0.60$\pm$0.10 & 96.6$\pm$1.8 & 1.76$\pm$7.40 & 95.2$\pm$0.9 & 0.45$\pm$0.11 & 99.5 \\
\rowcolor[HTML]{F3F3F3} 
Bone\_Clavicula\_R & 98.4$\pm$16.1 & 1.52$\pm$0.19 & 95.5$\pm$3.2 & 1.72$\pm$9.22 & 98.3$\pm$16.0 & 1.37$\pm$0.20 & {\cellcolor[HTML]{F3F3F3}} \\
Bone\_Femur\_L & 97.4$\pm$15.4 & 3.51$\pm$0.29 & 94.6$\pm$14.3 & 9.92$\pm$40.53 & 97.5$\pm$0.7 & 3.28$\pm$0.11 & 99.7 \\
Bone\_Femur\_R & 95.8$\pm$0.5 & 6.85$\pm$0.08 & 98.4$\pm$0.4 & 0.52$\pm$1.08 & 96.3$\pm$0.2 & 6.26$\pm$0.06 & {} \\
\rowcolor[HTML]{F3F3F3} 
Bone\_Hip\_L & 96.6$\pm$1.6 & 0.32$\pm$0.13 & 97.3$\pm$1.5 & 0.44$\pm$0.89 & 97.0$\pm$1.9 & 0.29$\pm$0.14 & 99.5 \\
\rowcolor[HTML]{F3F3F3} 
Bone\_Hip\_R & 95.9$\pm$31.6 & 0.34$\pm$0.10 & 98.0$\pm$0.3 & 0.20$\pm$0.45 & 97.0$\pm$31.9 & 0.27$\pm$0.11 & {\cellcolor[HTML]{F3F3F3}} \\
SpinalCord & 92.7$\pm$16.8 & 0.27$\pm$0.24 & 94.5$\pm$3.7 & 0.20$\pm$0.13 & 92.7$\pm$16.5 & 0.26$\pm$0.24 & 99.5 \\
\rowcolor[HTML]{F3F3F3} 
Musc\_GluteusMax\_L & 98.9$\pm$5.2 & 0.84$\pm$1.41 & 93.8$\pm$7.7 & 3.56$\pm$10.31 & 98.7$\pm$5.4 & 0.76$\pm$1.50 & 99.5 \\
\rowcolor[HTML]{F3F3F3} 
Musc\_GluteusMax\_R & 98.1$\pm$2.0 & 0.93$\pm$0.34 & 95.2$\pm$4.0 & 2.29$\pm$4.92 & 97.9$\pm$2.2 & 0.71$\pm$0.34 & {\cellcolor[HTML]{F3F3F3}} \\
Musc\_GluteusMed\_L & 98.2$\pm$1.5 & 0.43$\pm$0.22 & 94.0$\pm$3.0 & 0.93$\pm$1.57 & 98.4$\pm$1.5 & 0.45$\pm$0.23 & 99.7 \\
Musc\_GluteusMed\_R & 97.6$\pm$7.3 & 0.74$\pm$0.22 & 93.7$\pm$3.8 & 0.76$\pm$0.73 & 97.3$\pm$7.5 & 0.47$\pm$0.21 & {} \\
\rowcolor[HTML]{F3F3F3} 
Musc\_GluteusMin\_L & 97.1$\pm$2.3 & 0.56$\pm$0.16 & 91.9$\pm$3.0 & 0.93$\pm$1.75 & 97.3$\pm$2.2 & 0.42$\pm$0.16 & 99.7 \\
\rowcolor[HTML]{F3F3F3} 
Musc\_GluteusMin\_R & 98.0$\pm$1.1 & 0.39$\pm$0.08 & 93.0$\pm$1.7 & 0.41$\pm$0.16 & 98.2$\pm$1.1 & 0.36$\pm$0.10 & {\cellcolor[HTML]{F3F3F3}} \\
Musc\_Autochthon\_L & 94.2$\pm$0.8 & 0.37$\pm$0.11 & 95.8$\pm$1.3 & 0.38$\pm$0.15 & 93.9$\pm$0.9 & 0.32$\pm$0.12 & 99.7 \\
Musc\_Autochthon\_R & 94.6$\pm$0.8 & 0.41$\pm$0.12 & 95.8$\pm$1.2 & 0.37$\pm$0.12 & 94.8$\pm$0.8 & 0.41$\pm$0.11 & {} \\
\rowcolor[HTML]{F3F3F3} 
Musc\_Iliopsoas\_L & 94.3$\pm$19.7 & 0.37$\pm$0.18 & 90.3$\pm$9.0 & 0.50$\pm$0.86 & 93.5$\pm$19.6 & 0.36$\pm$0.19 & 99.5 \\
\rowcolor[HTML]{F3F3F3} 
Musc\_Iliopsoas\_R & 95.0$\pm$9.1 & 0.39$\pm$0.27 & 90.0$\pm$9.5 & 0.44$\pm$0.57 & 94.9$\pm$9.1 & 0.32$\pm$0.25 & {\cellcolor[HTML]{F3F3F3}} \\
Brain & 97.9$\pm$36.4 & 0.56$\pm$1.01 & 88.3$\pm$28.0 & 7.85$\pm$24.00 & 98.2$\pm$16.3 & 0.51$\pm$1.07 & 99.5 \\
\rowcolor[HTML]{F3F3F3} 
Skull & 83.1$\pm$35.4 & 3.91$\pm$0.20 & 79.5$\pm$21.0 & 1.07$\pm$2.79 & 82.5$\pm$35.4 & 3.66$\pm$0.21 & 99.7
\\ \hline

Mean & \multicolumn{1}{c}{ 96.1 } & \multicolumn{1}{c|}{ 1.46 } & \multicolumn{1}{c}{ 94.2 } & \multicolumn{1}{c|}{ 1.78 } & \multicolumn{1}{c}{ 96.2 } & \multicolumn{1}{c|}{ 1.31 } & \multicolumn{1}{c|}{ 99.6 } \\ \hline
\end{tabular}
% }
\caption{\textbar~\textbf{Organ-wise performance comparison of \ac{nnu_e5}, MultiTalent, and \ac{cln_u5} on \ac{totalseg} `Muscle' subgroup.} Organ-wise DSC (\%) and ASD (mm) of \ac{nnu_e5}, MultiTalent and \ac{cln_u5} are evaluated on \ac{totalseg} `Muscle' subgroup. \ac{cln_u5} achieves the best mean DSC and ASD.}
\label{tab:pl_totalseg_musc_ds5}
\end{table}
\clearpage

\begin{table}[htp]
\centering
\fontsize{10pt}{12pt}\selectfont
% \resizebox{\textwidth}{!}{
\begin{tabular}{|l|ll|ll|ll|r|}
\hline
 & \multicolumn{2}{c|}{\ac{nnu_e5}} & \multicolumn{2}{c|}{MultiTalent} & \multicolumn{3}{c|}{\ac{cln_u5}} \\ \cline{2-8} 
\multirow{-2}{*}{} & DSC$\uparrow$ & ASD$\downarrow$ & DSC$\uparrow$ & ASD$\downarrow$ & DSC$\uparrow$ & ASD$\downarrow$ & $\mathcal{T}\uparrow$ \\ \hline
\rowcolor[HTML]{F3F3F3} 
Bone\_Rib1\_L & 93.6$\pm$3.3 & 0.16$\pm$0.23 & 92.4$\pm$8.9 & 0.15$\pm$0.13 & 93.3$\pm$3.3 & 0.14$\pm$0.21 & {\cellcolor[HTML]{F3F3F3}96.0} \\
\rowcolor[HTML]{F3F3F3} 
Bone\_Rib2\_L & 89.9$\pm$1.7 & 0.95$\pm$0.95 & 95.4$\pm$1.5 & 0.19$\pm$0.76 & 90.2$\pm$1.5 & 0.74$\pm$1.02 &  \\
\rowcolor[HTML]{F3F3F3} 
Bone\_Rib3\_L & 94.7$\pm$17.0 & 0.91$\pm$1.06 & 94.9$\pm$3.0 & 0.13$\pm$0.20 & 94.9$\pm$17.0 & 0.85$\pm$1.08 &  \\
\rowcolor[HTML]{F3F3F3} 
Bone\_Rib4\_L & 94.5$\pm$18.9 & 0.76$\pm$1.82 & 94.9$\pm$3.8 & 0.20$\pm$0.48 & 93.9$\pm$18.7 & 0.63$\pm$1.77 &  \\
\rowcolor[HTML]{F3F3F3} 
Bone\_Rib5\_L & 94.2$\pm$29.7 & 1.52$\pm$4.61 & 91.5$\pm$15.6 & 0.42$\pm$1.23 & 94.3$\pm$29.6 & 1.06$\pm$4.53 &  \\
\rowcolor[HTML]{F3F3F3} 
Bone\_Rib6\_L & 93.2$\pm$23.9 & 1.64$\pm$3.32 & 93.0$\pm$11.4 & 0.37$\pm$0.97 & 92.8$\pm$24.1 & 1.51$\pm$3.35 &  \\
\rowcolor[HTML]{F3F3F3} 
Bone\_Rib7\_L & 94.5$\pm$20.7 & 2.71$\pm$4.08 & 93.9$\pm$5.4 & 0.33$\pm$0.50 & 95.8$\pm$20.8 & 2.36$\pm$4.01 &  \\
\rowcolor[HTML]{F3F3F3} 
Bone\_Rib8\_L & 94.6$\pm$21.1 & 4.29$\pm$4.00 & 90.4$\pm$15.0 & 0.89$\pm$2.41 & 95.2$\pm$21.1 & 4.15$\pm$4.04 &  \\
\rowcolor[HTML]{F3F3F3} 
Bone\_Rib9\_L & 94.0$\pm$19.4 & 2.83$\pm$5.21 & 89.9$\pm$16.2 & 1.32$\pm$3.61 & 94.5$\pm$19.3 & 2.74$\pm$4.83 &  \\
\rowcolor[HTML]{F3F3F3} 
Bone\_Rib10\_L & 93.0$\pm$22.9 & 1.51$\pm$6.63 & 87.5$\pm$23.7 & 2.24$\pm$5.96 & 91.9$\pm$13.2 & 1.48$\pm$6.45 &  \\
\rowcolor[HTML]{F3F3F3} 
Bone\_Rib11\_L & 93.0$\pm$22.4 & 0.36$\pm$8.01 & 87.5$\pm$23.4 & 2.55$\pm$7.91 & 92.1$\pm$12.1 & 0.34$\pm$8.12 &  \\
\rowcolor[HTML]{F3F3F3} 
Bone\_Rib12\_L & 94.2$\pm$26.2 & 0.54$\pm$0.41 & 91.5$\pm$8.3 & 1.02$\pm$4.36 & 94.7$\pm$16.3 & 0.33$\pm$0.41 &  \\
\rowcolor[HTML]{F3F3F3} 
Bone\_Rib1\_R & 91.2$\pm$5.6 & 0.11$\pm$0.17 & 94.0$\pm$2.6 & 0.15$\pm$0.17 & 91.8$\pm$5.6 & 0.14$\pm$0.15 &  \\
\rowcolor[HTML]{F3F3F3} 
Bone\_Rib2\_R & 94.5$\pm$3.5 & 0.23$\pm$0.94 & 95.1$\pm$3.4 & 0.14$\pm$0.41 & 94.6$\pm$3.4 & 0.09$\pm$0.91 &  \\
\rowcolor[HTML]{F3F3F3} 
Bone\_Rib3\_R & 92.8$\pm$9.1 & 0.48$\pm$0.75 & 93.7$\pm$5.8 & 0.20$\pm$0.38 & 93.7$\pm$9.1 & 0.34$\pm$1.14 &  \\
\rowcolor[HTML]{F3F3F3} 
Bone\_Rib4\_R & 90.5$\pm$18.8 & 0.83$\pm$2.02 & 94.4$\pm$7.4 & 0.10$\pm$0.10 & 90.4$\pm$18.6 & 0.73$\pm$1.92 &  \\
\rowcolor[HTML]{F3F3F3} 
Bone\_Rib5\_R & 90.3$\pm$26.5 & 1.70$\pm$2.42 & 89.7$\pm$20.6 & 0.72$\pm$2.36 & 90.7$\pm$16.4 & 1.49$\pm$2.51 &  \\
\rowcolor[HTML]{F3F3F3} 
Bone\_Rib6\_R & 93.4$\pm$18.9 & 1.52$\pm$2.45 & 89.6$\pm$18.0 & 1.05$\pm$3.01 & 93.4$\pm$13.0 & 1.38$\pm$2.54 &  \\
\rowcolor[HTML]{F3F3F3} 
Bone\_Rib7\_R & 94.2$\pm$21.0 & 1.56$\pm$3.51 & 90.7$\pm$16.5 & 0.77$\pm$1.61 & 95.5$\pm$21.3 & 1.57$\pm$3.36 &  \\
\rowcolor[HTML]{F3F3F3} 
Bone\_Rib8\_R & 95.5$\pm$18.3 & 1.88$\pm$4.08 & 91.8$\pm$10.0 & 0.81$\pm$1.83 & 95.0$\pm$18.0 & 1.78$\pm$4.22 &  \\
\rowcolor[HTML]{F3F3F3} 
Bone\_Rib9\_R & 95.9$\pm$21.4 & 2.48$\pm$5.16 & 91.7$\pm$11.6 & 1.21$\pm$2.97 & 96.3$\pm$21.6 & 2.25$\pm$3.18 &  \\
\rowcolor[HTML]{F3F3F3} 
Bone\_Rib10\_R & 96.3$\pm$23.3 & 2.50$\pm$7.23 & 90.6$\pm$18.4 & 1.33$\pm$4.17 & 95.9$\pm$23.3 & 2.30$\pm$7.41 &  \\
\rowcolor[HTML]{F3F3F3} 
Bone\_Rib11\_R & 96.4$\pm$22.8 & 1.85$\pm$6.95 & 90.4$\pm$18.2 & 1.50$\pm$5.23 & 97.5$\pm$22.7 & 1.71$\pm$6.70 &  \\
\rowcolor[HTML]{F3F3F3} 
Bone\_Rib12\_R & 97.1$\pm$27.5 & 0.63$\pm$0.07 & 89.5$\pm$13.9 & 1.32$\pm$3.03 & 97.4$\pm$27.5 & 0.46$\pm$0.10 &  \\
Bone\_Sternum & 94.1$\pm$5.8 & 0.63$\pm$0.55 & 94.9$\pm$2.7 & 0.36$\pm$0.91 & 95.5$\pm$5.7 & 0.35$\pm$0.55 & {99.7} \\
\rowcolor[HTML]{F3F3F3} 
Bone\_CostalCartilages & 88.7$\pm$5.0 & 0.18$\pm$0.74 & 91.2$\pm$3.2 & 0.44$\pm$0.59 & 88.6$\pm$5.0 & 0.16$\pm$0.74 & {\cellcolor[HTML]{F3F3F3}99.5}
\\ \hline

Mean & \multicolumn{1}{c}{ 93.6 } & \multicolumn{1}{c|}{ 1.34 } & \multicolumn{1}{c}{ 91.9 } & \multicolumn{1}{c|}{ 0.77 } & \multicolumn{1}{c}{ 93.8 } & \multicolumn{1}{c|}{ 1.20 } & \multicolumn{1}{c|}{ 98.4 } \\ \hline
\end{tabular}
% }
\caption{\textbar~\textbf{Organ-wise performance comparison of \ac{nnu_e5}, MultiTalent, and \ac{cln_u5} on \ac{totalseg} `Rib' subgroup.} Organ-wise DSC (\%) and ASD (mm) of \ac{nnu_e5}, MultiTalent and \ac{cln_u5} are evaluated on \ac{totalseg} `Rib' subgroup. \ac{cln_u5} achieves the best mean DSC.}
\label{tab:pl_totalseg_ribs_ds5}
\end{table}
\clearpage

\begin{table}[htp]
\centering
\fontsize{10pt}{12pt}\selectfont
% \resizebox{\textwidth}{!}{
\begin{tabular}{|l|ll|ll|ll|r|}
\hline
 & \multicolumn{2}{c|}{\ac{nnu_e5}} & \multicolumn{2}{c|}{MultiTalent} & \multicolumn{3}{c|}{\ac{cln_u5}} \\ \cline{2-8} 
\multirow{-2}{*}{} & DSC$\uparrow$ & ASD$\downarrow$ & DSC$\uparrow$ & ASD$\downarrow$ & DSC$\uparrow$ & ASD$\downarrow$ & $\mathcal{T}\uparrow$ \\ \hline
\rowcolor[HTML]{F3F3F3} 
Bone\_Sacrum & 93.4$\pm$1.8 & 0.25$\pm$0.13 & 95.0$\pm$2.2 & 0.48$\pm$0.27 & 93.9$\pm$1.9 & 0.19$\pm$0.14 & {\cellcolor[HTML]{F3F3F3}99.0} \\
Bone\_Vert\_C1 & 94.5$\pm$20.3 & 0.25$\pm$0.53 & 94.3$\pm$4.2 & 0.42$\pm$0.53 & 95.3$\pm$10.2 & 0.23$\pm$0.50 & {96.0} \\
Bone\_Vert\_C2 & 98.0$\pm$20.7 & 0.31$\pm$3.04 & 93.2$\pm$20.1 & 0.09$\pm$0.08 & 97.9$\pm$10.5 & 0.14$\pm$0.47 &  \\
Bone\_Vert\_C3 & 98.1$\pm$23.4 & 0.16$\pm$3.22 & 94.8$\pm$10.3 & 0.28$\pm$0.64 & 98.0$\pm$15.5 & 0.16$\pm$3.18 &  \\
Bone\_Vert\_C4 & 97.0$\pm$23.4 & 0.34$\pm$3.60 & 92.1$\pm$20.0 & 0.60$\pm$1.99 & 96.6$\pm$19.6 & 0.30$\pm$3.54 &  \\
Bone\_Vert\_C5 & 96.8$\pm$17.5 & 0.22$\pm$1.62 & 88.7$\pm$22.3 & 0.26$\pm$0.62 & 97.8$\pm$17.4 & 0.21$\pm$1.63 &  \\
Bone\_Vert\_C6 & 95.0$\pm$10.6 & 0.60$\pm$0.93 & 96.9$\pm$4.6 & 0.16$\pm$0.33 & 96.0$\pm$10.5 & 0.45$\pm$0.93 &  \\
Bone\_Vert\_C7 & 95.2$\pm$17.5 & 0.41$\pm$0.86 & 96.5$\pm$5.8 & 0.19$\pm$0.45 & 95.3$\pm$13.1 & 0.18$\pm$1.00 &  \\
Bone\_Vert\_L1 & 95.7$\pm$12.1 & 1.27$\pm$0.98 & 97.1$\pm$3.1 & 0.14$\pm$0.30 & 94.9$\pm$12.1 & 1.19$\pm$0.98 &  \\
Bone\_Vert\_L2 & 94.8$\pm$18.2 & 0.29$\pm$1.09 & 97.4$\pm$1.7 & 0.11$\pm$0.21 & 93.7$\pm$12.3 & 0.18$\pm$0.95 &  \\
Bone\_Vert\_L3 & 96.0$\pm$16.1 & 0.24$\pm$1.09 & 97.0$\pm$1.8 & 0.13$\pm$0.16 & 96.9$\pm$16.2 & 0.18$\pm$1.06 &  \\
Bone\_Vert\_L4 & 94.1$\pm$14.3 & 0.43$\pm$0.41 & 95.7$\pm$6.4 & 0.22$\pm$0.39 & 94.2$\pm$14.2 & 0.23$\pm$0.41 &  \\
Bone\_Vert\_L5 & 92.9$\pm$17.3 & 0.21$\pm$1.49 & 94.6$\pm$9.5 & 0.31$\pm$0.70 & 93.9$\pm$17.4 & 0.19$\pm$1.43 &  \\
Bone\_Vert\_S1 & 96.3$\pm$23.8 & 0.20$\pm$2.10 & 92.0$\pm$14.2 & 0.42$\pm$0.96 & 97.3$\pm$18.9 & 0.19$\pm$1.22 &  \\
Bone\_Vert\_T1 & 96.3$\pm$23.5 & 0.15$\pm$1.28 & 92.5$\pm$13.0 & 0.41$\pm$0.94 & 95.9$\pm$13.5 & 0.15$\pm$1.37 &  \\
Bone\_Vert\_T2 & 98.0$\pm$5.2 & 0.48$\pm$0.74 & 88.6$\pm$22.0 & 0.59$\pm$1.38 & 97.7$\pm$5.4 & 0.32$\pm$0.76 &  \\
Bone\_Vert\_T3 & 97.0$\pm$2.7 & 0.48$\pm$0.20 & 90.4$\pm$17.1 & 0.64$\pm$1.58 & 97.9$\pm$2.6 & 0.46$\pm$0.21 &  \\
Bone\_Vert\_T4 & 89.6$\pm$17.0 & 1.12$\pm$0.21 & 91.2$\pm$16.9 & 0.51$\pm$1.50 & 90.5$\pm$8.9 & 1.05$\pm$0.20 &  \\
Bone\_Vert\_T5 & 90.7$\pm$22.9 & 0.92$\pm$0.38 & 93.6$\pm$15.8 & 0.42$\pm$1.73 & 90.4$\pm$23.1 & 0.88$\pm$0.39 &  \\
Bone\_Vert\_T6 & 86.3$\pm$28.3 & 1.15$\pm$0.07 & 92.8$\pm$15.8 & 0.48$\pm$2.13 & 85.4$\pm$28.1 & 1.04$\pm$0.06 &  \\
Bone\_Vert\_T7 & 96.8$\pm$1.0 & 1.58$\pm$0.04 & 89.9$\pm$17.5 & 0.37$\pm$1.00 & 97.1$\pm$1.0 & 1.48$\pm$0.06 &  \\
Bone\_Vert\_T8 & 91.5$\pm$13.0 & 1.08$\pm$0.40 & 84.9$\pm$24.3 & 0.18$\pm$0.22 & 91.6$\pm$13.0 & 0.93$\pm$0.16 &  \\
Bone\_Vert\_T9 & 94.6$\pm$44.2 & 0.53$\pm$0.23 & 94.4$\pm$4.6 & 0.08$\pm$0.03 & 94.3$\pm$14.4 & 0.42$\pm$0.24 &  \\
Bone\_Vert\_T10 & 97.6$\pm$29.9 & 0.36$\pm$2.18 & 95.8$\pm$0.9 & 0.10$\pm$0.07 & 97.5$\pm$13.0 & 0.38$\pm$1.32 &  \\
Bone\_Vert\_T11 & 95.1$\pm$19.4 & 0.25$\pm$1.64 & 94.7$\pm$5.5 & 0.22$\pm$0.46 & 95.1$\pm$12.4 & 0.22$\pm$1.67 &  \\
Bone\_Vert\_T12 & 97.2$\pm$16.3 & 0.32$\pm$1.30 & 84.3$\pm$27.3 & 0.24$\pm$0.34 & 97.5$\pm$16.0 & 0.26$\pm$1.35 &
\\ \hline

Mean & \multicolumn{1}{c}{ 94.9 } & \multicolumn{1}{c|}{ 0.52 } & \multicolumn{1}{c}{ 93.0 } & \multicolumn{1}{c|}{ 0.31 } & \multicolumn{1}{c}{ 95.1 } & \multicolumn{1}{c|}{ 0.45 } & \multicolumn{1}{c|}{ 97.5 } \\ \hline
\end{tabular}
% }
\caption{\textbar~\textbf{Organ-wise performance comparison of \ac{nnu_e5}, MultiTalent, and \ac{cln_u5} on \ac{totalseg} `Vertebrae' subgroup.} Organ-wise DSC (\%) and ASD (mm) of \ac{nnu_e5}, MultiTalent and \ac{cln_u5} are evaluated on \ac{totalseg} `Vertebrae' subgroup. \ac{cln_u5} achieves the best mean DSC.}
\label{tab:pl_totalseg_vert_ds5}
\end{table}
\clearpage

\begin{table}[htp]
\centering
\fontsize{10pt}{12pt}\selectfont
% \resizebox{\textwidth}{!}{
\begin{tabular}{|l|ll|ll|ll|r|}
\hline
 & \multicolumn{2}{c|}{\ac{nnu_e5}} & \multicolumn{2}{c|}{MultiTalent} & \multicolumn{3}{c|}{\ac{cln_u5}} \\ \cline{2-8} 
\multirow{-2}{*}{} & DSC$\uparrow$ & ASD$\downarrow$ & DSC$\uparrow$ & ASD$\downarrow$ & DSC$\uparrow$ & ASD$\downarrow$ & $\mathcal{T}\uparrow$ \\ \hline
\rowcolor[HTML]{F3F3F3} 
BrainStem & 93.7$\pm$1.5 & 0.32$\pm$0.10 & 87.2$\pm$2.5 & 0.84$\pm$0.26 & 92.9$\pm$1.6 & 0.23$\pm$0.11 & 99.5 \\
Eye\_L & 94.8$\pm$1.3 & 0.17$\pm$0.08 & 89.5$\pm$2.9 & 0.40$\pm$0.12 & 93.6$\pm$1.3 & 0.14$\pm$0.08 & 99.0 \\
Eye\_R & 93.2$\pm$4.2 & 0.24$\pm$0.28 & 90.0$\pm$2.5 & 1.43$\pm$2.58 & 95.3$\pm$4.3 & 0.28$\pm$0.29 & {} \\
\rowcolor[HTML]{F3F3F3} 
Lens\_L & 80.1$\pm$9.8 & 0.25$\pm$0.18 & 77.9$\pm$6.3 & 0.23$\pm$0.14 & 80.4$\pm$9.8 & 0.21$\pm$0.17 & 96.0 \\
\rowcolor[HTML]{F3F3F3} 
Lens\_R & 79.7$\pm$9.2 & 0.33$\pm$0.43 & 80.3$\pm$5.6 & 0.21$\pm$0.09 & 82.0$\pm$9.3 & 0.41$\pm$0.43 & {\cellcolor[HTML]{F3F3F3}} \\
OpticNerve\_L & 80.3$\pm$8.0 & 0.29$\pm$0.28 & 67.3$\pm$8.7 & 1.11$\pm$1.78 & 81.7$\pm$8.1 & 0.31$\pm$0.30 & 99.0 \\
OpticNerve\_R & 84.1$\pm$6.4 & 0.20$\pm$0.21 & 68.1$\pm$7.0 & 0.52$\pm$0.15 & 86.5$\pm$6.2 & 0.17$\pm$0.20 & {} \\
\rowcolor[HTML]{F3F3F3} 
Chiasm & 71.6$\pm$12.9 & 0.26$\pm$0.13 & 42.9$\pm$18.6 & 23.27$\pm$47.95 & 71.9$\pm$13.1 & 0.23$\pm$0.15 & 96.0 \\
TemporalLobe\_L & 94.7$\pm$1.2 & 0.24$\pm$0.08 & 85.6$\pm$7.0 & 1.93$\pm$0.94 & 95.3$\pm$1.3 & 0.22$\pm$0.09 & 99.5 \\
TemporalLobe\_R & 94.8$\pm$1.4 & 0.26$\pm$0.09 & 84.1$\pm$6.0 & 2.02$\pm$0.83 & 93.5$\pm$1.6 & 0.23$\pm$0.11 & {} \\
\rowcolor[HTML]{F3F3F3} 
Glnd\_Pituitary & 84.5$\pm$6.4 & 0.22$\pm$0.21 & 52.6$\pm$23.3 & 1.04$\pm$0.65 & 85.6$\pm$6.4 & 0.16$\pm$0.21 & 99.0 \\
Glnd\_Parotid\_L & 93.3$\pm$1.9 & 0.35$\pm$0.16 & 83.0$\pm$5.6 & 1.19$\pm$0.58 & 91.7$\pm$1.7 & 0.29$\pm$0.15 & 99.5 \\
Glnd\_Parotid\_R & 92.5$\pm$3.4 & 0.50$\pm$0.52 & 84.3$\pm$4.4 & 1.20$\pm$0.55 & 96.0$\pm$1.3 & 0.42$\pm$0.50 & {} \\
\rowcolor[HTML]{F3F3F3} 
Ear\_Inner\_L & 67.0$\pm$9.5 & 0.66$\pm$0.37 & 83.1$\pm$2.7 & 0.50$\pm$0.12 & 69.3$\pm$9.4 & 0.61$\pm$0.36 & 99.5 \\
\rowcolor[HTML]{F3F3F3} 
Ear\_Inner\_R & 64.3$\pm$9.3 & 0.66$\pm$0.30 & 86.8$\pm$2.7 & 0.33$\pm$0.11 & 63.8$\pm$9.4 & 0.75$\pm$0.31 & {\cellcolor[HTML]{F3F3F3}} \\
Ear\_Mid\_L & 90.4$\pm$3.7 & 0.07$\pm$0.04 & 75.2$\pm$15.7 & 1.30$\pm$1.14 & 90.6$\pm$3.6 & 0.07$\pm$0.03 & 99.5 \\
Ear\_Mid\_R & 90.2$\pm$3.3 & 0.08$\pm$0.06 & 81.2$\pm$7.0 & 1.69$\pm$2.60 & 92.1$\pm$3.4 & 0.07$\pm$0.06 & {} \\
\rowcolor[HTML]{F3F3F3} 
TMJ\_L & 84.4$\pm$9.5 & 0.31$\pm$0.23 & 78.1$\pm$6.8 & 0.79$\pm$0.34 & 82.5$\pm$9.3 & 0.24$\pm$0.22 & 98.0 \\
\rowcolor[HTML]{F3F3F3} 
TMJ\_R & 84.1$\pm$7.6 & 0.33$\pm$0.20 & 75.8$\pm$6.7 & 1.55$\pm$2.24 & 84.3$\pm$7.7 & 0.33$\pm$0.21 & {\cellcolor[HTML]{F3F3F3}} \\
SpinalCord & 91.5$\pm$2.9 & 0.36$\pm$0.19 & 81.4$\pm$3.2 & 0.61$\pm$0.14 & 90.8$\pm$3.1 & 0.33$\pm$0.21 & 99.0 \\
\rowcolor[HTML]{F3F3F3} 
Bone\_Mandible\_L & 95.7$\pm$2.1 & 0.17$\pm$0.08 & 91.4$\pm$1.9 & 0.37$\pm$0.13 & 97.5$\pm$2.0 & 0.15$\pm$0.07 & 99.7 \\
\rowcolor[HTML]{F3F3F3} 
Bone\_Mandible\_R & 95.7$\pm$2.5 & 0.17$\pm$0.08 & 91.3$\pm$1.1 & 0.36$\pm$0.10 & 95.6$\pm$2.6 & 0.13$\pm$0.09 & {\cellcolor[HTML]{F3F3F3}}

\\ \hline

Mean & \multicolumn{1}{c}{ 86.4 } & \multicolumn{1}{c|}{ 0.29 } & \multicolumn{1}{c}{ 79.0 } & \multicolumn{1}{c|}{ 1.95 } & \multicolumn{1}{c}{ 87.0 } & \multicolumn{1}{c|}{ 0.27 } & \multicolumn{1}{c|}{ 98.7 } \\ \hline
\end{tabular}
% }
\caption{\textbar~\textbf{Organ-wise performance comparison of \ac{nnu_e5}, MultiTalent, and \ac{cln_u5} on \ac{structseg}.} Organ-wise DSC (\%) and ASD (mm) of \ac{nnu_e5}, MultiTalent and \ac{cln_u5} are evaluated on \ac{structseg}. \ac{cln_u5} achieves the best mean DSC and ASD.}
\label{tab:pl_structseg_ds5}
\end{table}
\clearpage

\begin{table}[htp]
\centering
\fontsize{10pt}{12pt}\selectfont
% \resizebox{\textwidth}{!}{
\begin{tabular}{|l|ll|ll|ll|r|}
\hline
 & \multicolumn{2}{c|}{\ac{nnu_e5}} & \multicolumn{2}{c|}{MultiTalent} & \multicolumn{3}{c|}{\ac{cln_u5}} \\ \cline{2-8} 
\multirow{-2}{*}{} & DSC$\uparrow$ & ASD$\downarrow$ & DSC$\uparrow$ & ASD$\downarrow$ & DSC$\uparrow$ & ASD$\downarrow$ & $\mathcal{T}\uparrow$ \\ \hline
\rowcolor[HTML]{F3F3F3} 
Eso & 88.2$\pm$3.2 & 0.35$\pm$0.16 & 83.1$\pm$8.5 & 1.17$\pm$0.98 & 90.4$\pm$3.0 & 0.31$\pm$0.14 & 99.5 \\
Heart & 94.7$\pm$14.9 & 0.39$\pm$0.20 & 94.3$\pm$1.0 & 3.44$\pm$3.79 & 94.5$\pm$14.6 & 0.37$\pm$0.17 & 99.0 \\
\rowcolor[HTML]{F3F3F3} 
Trachea & 92.2$\pm$5.8 & 0.30$\pm$0.09 & 90.7$\pm$3.5 & 0.67$\pm$0.39 & 93.8$\pm$5.9 & 0.28$\pm$0.10 & 99.5 \\
A\_Aorta & 95.2$\pm$16.6 & 0.32$\pm$0.84 & 94.6$\pm$0.8 & 0.47$\pm$0.11 & 96.2$\pm$16.6 & 0.31$\pm$0.84 & 99.7
\\ \hline

Mean & \multicolumn{1}{c}{ 92.6 } & \multicolumn{1}{c|}{ 0.34 } & \multicolumn{1}{c}{ 90.7 } & \multicolumn{1}{c|}{ 1.44 } & \multicolumn{1}{c}{ 93.7 } & \multicolumn{1}{c|}{ 0.32 } & \multicolumn{1}{c|}{ 99.4 } \\ \hline
\end{tabular}
% }
\caption{\textbar~\textbf{Organ-wise performance comparison of \ac{nnu_e5}, MultiTalent, and \ac{cln_u5} on \ac{segthor}.} Organ-wise DSC (\%) and ASD (mm) of \ac{nnu_e5}, MultiTalent and \ac{cln_u5} are evaluated on \ac{segthor}. \ac{cln_u5} achieves the best mean DSC and ASD.}
\label{tab:pl_segthor_ds5}
\end{table}
\clearpage

\begin{table}[htp]
\centering
\fontsize{10pt}{12pt}\selectfont
% \resizebox{\textwidth}{!}{
\begin{tabular}{|l|ll|ll|ll|r|}
\hline
 & \multicolumn{2}{c|}{\ac{nnu_e5}} & \multicolumn{2}{c|}{MultiTalent} & \multicolumn{3}{c|}{\ac{cln_u5}} \\ \cline{2-8} 
\multirow{-2}{*}{} & DSC$\uparrow$ & ASD$\downarrow$ & DSC$\uparrow$ & ASD$\downarrow$ & DSC$\uparrow$ & ASD$\downarrow$ & $\mathcal{T}\uparrow$ \\ \hline
\rowcolor[HTML]{F3F3F3} 
Liver & 98.1$\pm$14.5 & 0.83$\pm$0.98 & 97.3$\pm$3.4 & 0.74$\pm$1.11 & 98.3$\pm$5.8 & 0.81$\pm$1.00 & 99.5 \\
Kidney\_R & 92.4$\pm$16.9 & 1.81$\pm$0.74 & 94.3$\pm$9.2 & 3.40$\pm$9.09 & 94.3$\pm$7.1 & 1.44$\pm$0.76 & 99.7 \\
Kidney\_L & 92.8$\pm$16.9 & 1.27$\pm$0.74 & 94.2$\pm$9.6 & 4.22$\pm$8.64 & 92.9$\pm$6.7 & 1.18$\pm$0.69 & {} \\
\rowcolor[HTML]{F3F3F3} 
Spleen & 97.9$\pm$15.6 & 0.30$\pm$0.37 & 97.1$\pm$3.4 & 0.45$\pm$0.89 & 98.0$\pm$6.6 & 0.19$\pm$0.37 & 99.5 \\
Pancreas & 91.6$\pm$17.2 & 0.80$\pm$0.69 & 89.0$\pm$4.8 & 0.98$\pm$0.68 & 92.4$\pm$7.0 & 0.72$\pm$0.38 & 99.5 \\
\rowcolor[HTML]{F3F3F3} 
A\_Aorta & 96.7$\pm$16.5 & 0.21$\pm$0.95 & 94.5$\pm$8.0 & 1.58$\pm$3.43 & 97.1$\pm$16.5 & 0.18$\pm$0.83 & 99.7 \\
V\_VenaCava\_I & 88.8$\pm$10.1 & 1.08$\pm$1.03 & 90.3$\pm$10.2 & 0.83$\pm$1.11 & 88.6$\pm$9.9 & 1.12$\pm$0.98 & 99.0 \\
\rowcolor[HTML]{F3F3F3} 
Glnd\_Adrenal\_R & 85.9$\pm$12.3 & 0.39$\pm$0.37 & 84.3$\pm$7.0 & 0.44$\pm$0.53 & 87.9$\pm$13.2 & 0.35$\pm$0.39 & 96.0 \\
\rowcolor[HTML]{F3F3F3} 
Glnd\_Adrenal\_L & 88.0$\pm$13.3 & 0.26$\pm$1.16 & 85.9$\pm$6.1 & 0.34$\pm$0.19 & 87.7$\pm$12.5 & 0.21$\pm$0.38 & {\cellcolor[HTML]{F3F3F3}} \\
GallBladder & 80.2$\pm$24.7 & 2.16$\pm$0.83 & 81.0$\pm$23.4 & 2.06$\pm$2.84 & 83.1$\pm$8.7 & 2.12$\pm$0.83 & 99.5 \\
\rowcolor[HTML]{F3F3F3} 
Eso & 84.4$\pm$3.1 & 1.70$\pm$0.15 & 84.1$\pm$15.4 & 1.93$\pm$3.34 & 86.0$\pm$3.0 & 1.61$\pm$0.14 & 99.7 \\
Stomach & 91.7$\pm$16.4 & 1.24$\pm$0.67 & 89.8$\pm$19.8 & 2.09$\pm$4.22 & 92.4$\pm$6.5 & 1.17$\pm$0.69 & 99.5 \\
\rowcolor[HTML]{F3F3F3} 
Duodenum & 79.8$\pm$25.9 & 2.71$\pm$1.29 & 79.0$\pm$16.9 & 4.77$\pm$8.97 & 81.0$\pm$16.1 & 2.39$\pm$1.18 & 99.5
\\ \hline

Mean & \multicolumn{1}{c}{ 89.9 } & \multicolumn{1}{c|}{ 1.14 } & \multicolumn{1}{c}{ 89.3 } & \multicolumn{1}{c|}{ 1.83 } & \multicolumn{1}{c}{ 90.7 } & \multicolumn{1}{c|}{ 1.04 } & \multicolumn{1}{c|}{ 99.2 } \\ \hline
\end{tabular}
% }
\caption{\textbar~\textbf{Organ-wise performance comparison of \ac{nnu_e5}, MultiTalent, and \ac{cln_u5} on \ac{flare}.} Organ-wise DSC (\%) and ASD (mm) of \ac{nnu_e5}, MultiTalent and \ac{cln_u5} are evaluated on \ac{flare}. \ac{cln_u5} achieves the best mean DSC and ASD.}
\label{tab:pl_flare_ds5}
\end{table}
\clearpage
\begin{table}[htp]
\centering
\fontsize{10pt}{12pt}\selectfont
% \resizebox{\textwidth}{!}{
\begin{tabular}{|l|ll|ll|ll|r|}
\hline
 & \multicolumn{2}{c|}{\ac{nnu_e5}} & \multicolumn{2}{c|}{MultiTalent} & \multicolumn{3}{c|}{\ac{cln_u5}} \\ \cline{2-8} 
\multirow{-2}{*}{} & DSC$\uparrow$ & ASD$\downarrow$ & DSC$\uparrow$ & ASD$\downarrow$ & DSC$\uparrow$ & ASD$\downarrow$ & $\mathcal{T}\uparrow$ \\ \hline

Kidney\_\Ac{gtv} & 86.7$\pm$9.0 & 1.18$\pm$1.06 & 79.6$\pm$20.8 & 9.94$\pm$17.66 & 87.2$\pm$9.0 & 1.03$\pm$0.98 & 96.0
\\ \hline
\end{tabular}
% }
\caption{\textbar~\textbf{Organ-wise performance comparison of \ac{nnu_e5}, MultiTalent, and \ac{cln_u5} on \ac{kits}.} Organ-wise DSC (\%) and ASD (mm) of \ac{nnu_e5}, MultiTalent and \ac{cln_u5} are evaluated on \ac{kits}. \ac{cln_u5} achieves the best mean DSC and ASD.}
\label{tab:pl_kits_ds5}
\end{table}
\clearpage

\begin{table}[htp]
\centering
\resizebox{\textwidth}{!}{%
\begin{tabular}{|l|lc|cc|ll|ll|ll|ll|}
\hline
 & \multicolumn{2}{c|}{SAT-pro} & \multicolumn{2}{c|}{Vista3D} & \multicolumn{2}{c|}{MultiTalent} & \multicolumn{2}{c|}{\ac{nnu_e36}} & \multicolumn{2}{c|}{\ac{cln_c36}} & \multicolumn{2}{c|}{\ac{cln_u36}} \\ \cline{2-13} 
\multirow{-2}{*}{} & \multicolumn{1}{c}{DSC$\uparrow$} & ASD$\downarrow$ & \multicolumn{1}{c}{DSC$\uparrow$} & ASD$\downarrow$ & \multicolumn{1}{c}{DSC$\uparrow$} & \multicolumn{1}{c|}{ASD$\downarrow$} & \multicolumn{1}{c}{DSC$\uparrow$} & \multicolumn{1}{c|}{ASD$\downarrow$} & \multicolumn{1}{c}{DSC$\uparrow$} & \multicolumn{1}{c|}{ASD$\downarrow$} & \multicolumn{1}{c}{DSC$\uparrow$} & \multicolumn{1}{c|}{ASD$\downarrow$} \\ \hline
\rowcolor[HTML]{F3F3F3} 
BrainStem & 62.1$\pm$20.8 & --- & --- & --- & 83.0$\pm$5.2 & 1.12$\pm$0.39 & 85.3$\pm$4.6 & 1.20$\pm$0.49 & 85.1$\pm$4.1 & 1.02$\pm$0.53 & 85.7$\pm$3.6 & 0.96$\pm$0.50 \\
Eye\_L & 76.5$\pm$13.5 & --- & --- & --- & 84.2$\pm$9.4 & 0.61$\pm$0.46 & 89.2$\pm$5.2 & 0.71$\pm$0.40 & 89.5$\pm$2.7 & 0.46$\pm$0.16 & 89.2$\pm$2.8 & 0.48$\pm$0.17 \\
Eye\_R & 76.7$\pm$9.9 & --- & --- & --- & 82.2$\pm$12.1 & 0.62$\pm$0.19 & 89.3$\pm$3.6 & 0.69$\pm$0.29 & 88.7$\pm$2.7 & 0.53$\pm$0.18 & 89.1$\pm$2.5 & 0.50$\pm$0.17 \\
\rowcolor[HTML]{F3F3F3} 
Lens\_L & 54.3$\pm$28.3 & --- & --- & --- & 71.3$\pm$14.5 & 0.63$\pm$0.31 & 74.5$\pm$7.9 & 0.48$\pm$0.24 & 75.1$\pm$10.1 & 0.57$\pm$0.39 & 75.8$\pm$8.6 & 0.53$\pm$0.32 \\
\rowcolor[HTML]{F3F3F3} 
Lens\_R & 60.3$\pm$16.7 & --- & --- & --- & 68.9$\pm$15.5 & 0.66$\pm$0.23 & 72.7$\pm$10.1 & 0.52$\pm$0.31 & 75.0$\pm$11.6 & 0.60$\pm$0.39 & 75.4$\pm$8.7 & 0.52$\pm$0.33 \\
Chiasm & 7.8$\pm$18.2 & --- & --- & --- & 39.8$\pm$18.5 & 2.15$\pm$0.95 & 50.5$\pm$13.1 & 1.42$\pm$0.66 & 54.6$\pm$12.3 & 0.82$\pm$0.59 & 54.6$\pm$12.2 & 0.82$\pm$0.63 \\
\rowcolor[HTML]{F3F3F3} 
OpticNerve\_L & 27.9$\pm$30.6 & --- & --- & --- & 65.6$\pm$13.9 & 0.74$\pm$0.28 & 68.0$\pm$9.3 & 0.79$\pm$0.38 & 68.2$\pm$10.4 & 0.56$\pm$0.34 & 69.7$\pm$9.7 & 0.58$\pm$0.35 \\
\rowcolor[HTML]{F3F3F3} 
OpticNerve\_R & 24.8$\pm$26.1 & --- & --- & --- & 67.6$\pm$13.5 & 0.63$\pm$0.26 & 67.4$\pm$11.1 & 0.84$\pm$0.48 & 69.4$\pm$8.7 & 0.59$\pm$0.41 & 69.2$\pm$7.9 & 0.61$\pm$0.47 \\
Glnd\_Parotid\_L & 62.3$\pm$21.6 & --- & --- & --- & 85.3$\pm$6.4 & 1.17$\pm$0.52 & 84.2$\pm$5.8 & 1.23$\pm$0.73 & 85.6$\pm$3.9 & 1.05$\pm$0.35 & 85.8$\pm$3.5 & 0.99$\pm$0.29 \\
Glnd\_Parotid\_R & 73.4$\pm$11.6 & --- & --- & --- & 86.4$\pm$6.5 & 1.24$\pm$0.42 & 83.9$\pm$5.6 & 1.27$\pm$0.53 & 85.7$\pm$3.5 & 1.09$\pm$0.39 & 85.8$\pm$3.5 & 1.05$\pm$0.39 \\
\rowcolor[HTML]{F3F3F3} 
TMJ\_L & 17.6$\pm$27.9 & --- & --- & --- & 71.9$\pm$16.2 & 0.91$\pm$0.76 & 69.8$\pm$11.5 & 1.07$\pm$0.53 & 72.2$\pm$12.0 & 0.89$\pm$0.57 & 72.4$\pm$9.3 & 0.63$\pm$0.48 \\
\rowcolor[HTML]{F3F3F3} 
TMJ\_R & 9.1$\pm$20.3 & --- & --- & --- & 71.1$\pm$7.5 & 0.99$\pm$0.26 & 69.6$\pm$12.3 & 1.10$\pm$0.58 & 71.5$\pm$12.6 & 0.94$\pm$0.58 & 71.9$\pm$10.7 & 0.67$\pm$0.52 \\ \hline
Mean & \multicolumn{1}{c}{46.1} & \multicolumn{1}{c|}{---} & --- & --- & \multicolumn{1}{c}{73.1} & \multicolumn{1}{c|}{0.96} & \multicolumn{1}{c}{75.4} & \multicolumn{1}{c|}{0.94} & \multicolumn{1}{c}{76.7} & \multicolumn{1}{c|}{0.76} & \multicolumn{1}{c}{77.1} & \multicolumn{1}{c|}{0.70} \\ \hline
\end{tabular}}
\caption{\textbar~\textbf{External testing: organ-wise segmentation performance comparison on 268 patients head \& neck dataset}. The organ-wise DSC (\%) and ASD (mm) are evaluated for comparison. Note that structures excluded from the officially released Vista3D model are indicated with a `-'. Furthermore, the SAT-pro inference framework removes image metadata headers during inference, making it impossible to calculate ASD. \ac{cln_u36} achieves the best mean DSC and ASD. }
\label{tab:external_hn}
\end{table}
\clearpage

\begin{table}[htp]
\centering
\resizebox{\textwidth}{!}{%
\begin{tabular}{|l|lc|ll|ll|ll|ll|ll|}
\hline
 & \multicolumn{2}{c|}{SAT-pro} & \multicolumn{2}{c|}{Vista3D} & \multicolumn{2}{c|}{MultiTalent} & \multicolumn{2}{c|}{\ac{nnu_e36}} & \multicolumn{2}{c|}{\ac{cln_c36}} & \multicolumn{2}{c|}{\ac{cln_u36}} \\ \cline{2-13} 
\multirow{-2}{*}{} & \multicolumn{1}{c}{DSC$\uparrow$} & ASD$\downarrow$ & \multicolumn{1}{c}{DSC$\uparrow$} & ASD$\downarrow$ & \multicolumn{1}{c}{DSC$\uparrow$} & \multicolumn{1}{c|}{ASD$\downarrow$} & \multicolumn{1}{c}{DSC$\uparrow$} & \multicolumn{1}{c|}{ASD$\downarrow$} & \multicolumn{1}{c}{DSC$\uparrow$} & \multicolumn{1}{c|}{ASD$\downarrow$} & \multicolumn{1}{c}{DSC$\uparrow$} & \multicolumn{1}{c|}{ASD$\downarrow$} \\ \hline
\rowcolor[HTML]{F3F3F3} 
BrachialPlex & \multicolumn{1}{c}{---} & --- & \multicolumn{1}{c}{---} & \multicolumn{1}{c|}{---} & \multicolumn{1}{c}{---} & \multicolumn{1}{c|}{---} & 40.5$\pm$11.8 & 3.35$\pm$3.69 & 49.0$\pm$8.8 & 2.52$\pm$2.88 & 49.9$\pm$8.8 & 2.54$\pm$2.09 \\
Eso & 80.5$\pm$7.5 & --- & 81.2$\pm$6.3 & 0.74$\pm$0.56 & 81.6$\pm$7.8 & 0.97$\pm$0.53 & 84.3$\pm$6.6 & 1.12$\pm$0.68 & 88.3$\pm$5.8 & 0.68$\pm$0.68 & 88.3$\pm$5.3 & 0.70$\pm$0.70 \\
\rowcolor[HTML]{F3F3F3} 
Lung\_L & 97.1$\pm$0.6 & --- & 97.2$\pm$0.6 & 0.60$\pm$0.22 & 98.0$\pm$0.0 & 0.64$\pm$0.05 & 98.0$\pm$0.4 & 0.85$\pm$1.07 & 98.1$\pm$0.4 & 0.84$\pm$0.99 & 98.2$\pm$0.6 & 0.68$\pm$1.35 \\
\rowcolor[HTML]{F3F3F3} 
Lung\_R & 94.7$\pm$0.4 & --- & 95.4$\pm$3.2 & 1.63$\pm$0.62 & 96.7$\pm$2.3 & 0.67$\pm$0.43 & 97.0$\pm$0.4 & 0.80$\pm$1.02 & 97.1$\pm$0.3 & 0.80$\pm$0.54 & 97.8$\pm$0.6 & 0.60$\pm$1.10 \\
Pericardium & 83.7$\pm$6.1 & --- & 73.6$\pm$4.1 & 3.00$\pm$1.36 & 87.6$\pm$2.4 & 39.75$\pm$6.56 & 92.2$\pm$1.8 & 1.62$\pm$3.84 & 92.2$\pm$1.4 & 1.62$\pm$3.69 & 92.4$\pm$1.8 & 1.54$\pm$4.45 \\
\rowcolor[HTML]{F3F3F3} 
SpinalCord & 77.9$\pm$5.5 & --- & 75.5$\pm$5.6 & 3.08$\pm$2.37 & 54.9$\pm$4.3 & 102.08$\pm$13.70 & 89.5$\pm$3.4 & 1.13$\pm$1.16 & 91.8$\pm$2.5 & 0.90$\pm$0.85 & 92.2$\pm$2.3 & 0.85$\pm$0.86 \\
ProximalBronchi & 13.6$\pm$3.3 & --- & 91.0$\pm$2.3 & 0.89$\pm$0.75 & 70.1$\pm$4.7 & 57.36$\pm$7.23 & 91.0$\pm$2.3 & 0.89$\pm$0.75 & 91.1$\pm$1.9 & 0.89$\pm$0.70 & 91.5$\pm$2.0 & 0.88$\pm$0.67 \\
\rowcolor[HTML]{F3F3F3} 
A\_Aorta & \multicolumn{1}{c}{---} & --- & 86.9$\pm$0.9 & 2.18$\pm$0.49 & 87.7$\pm$1.3 & 1.54$\pm$0.08 & 92.0$\pm$7.5 & 1.34$\pm$4.42 & 92.8$\pm$7.7 & 1.11$\pm$4.63 & 92.9$\pm$8.0 & 1.03$\pm$4.65 \\
A\_Pulmonary & 80.4$\pm$4.0 & --- & \multicolumn{1}{c}{---} & \multicolumn{1}{c|}{---} & \multicolumn{1}{c}{---} & \multicolumn{1}{c|}{---} & 87.8$\pm$4.0 & 1.12$\pm$1.47 & 87.8$\pm$4.0 & 1.10$\pm$1.09 & 89.0$\pm$3.8 & 0.88$\pm$1.44 \\
\rowcolor[HTML]{F3F3F3} 
V\_Pulmonary & 62.2$\pm$6.2 & --- & 66.0$\pm$6.4 & 1.29$\pm$0.62 & 67.4$\pm$5.9 & 54.81$\pm$9.14 & 69.7$\pm$9.5 & 2.21$\pm$1.24 & 71.2$\pm$8.0 & 1.94$\pm$0.99 & 73.3$\pm$8.2 & 1.80$\pm$1.43 \\
V\_Venacava\_I & \multicolumn{1}{c}{---} & --- & 41.6$\pm$9.7 & 8.16$\pm$1.07 & 42.5$\pm$10.0 & 41.78$\pm$3.19 & 88.0$\pm$7.6 & 1.22$\pm$1.10 & 86.3$\pm$7.5 & 1.16$\pm$1.10 & 88.6$\pm$8.1 & 0.95$\pm$1.15 \\
\rowcolor[HTML]{F3F3F3} 
V\_Venacava\_S & 81.8$\pm$4.1 & --- & 83.0$\pm$3.3 & 1.15$\pm$0.40 & 84.9$\pm$2.8 & 79.27$\pm$8.22 & 86.2$\pm$4.1 & 1.09$\pm$0.87 & 87.8$\pm$4.2 & 0.88$\pm$0.52 & 87.9$\pm$4.4 & 0.86$\pm$0.68 \\
ChestWall\_L & \multicolumn{1}{c}{---} & --- & \multicolumn{1}{c}{---} & \multicolumn{1}{c|}{---} & \multicolumn{1}{c}{---} & \multicolumn{1}{c|}{---} & 90.2$\pm$2.3 & 1.45$\pm$0.98 & 90.4$\pm$2.2 & 1.03$\pm$0.99 & 90.9$\pm$2.6 & 1.23$\pm$1.20 \\
ChestWall\_R & \multicolumn{1}{c}{---} & --- & \multicolumn{1}{c}{---} & \multicolumn{1}{c|}{---} & \multicolumn{1}{c}{---} & \multicolumn{1}{c|}{---} & 90.3$\pm$2.0 & 1.13$\pm$0.89 & 90.9$\pm$2.0 & 0.99$\pm$0.88 & 90.8$\pm$2.1 & 0.64$\pm$0.50 \\ \hline
Mean & \multicolumn{1}{c}{---} & --- & \multicolumn{1}{c}{---} & \multicolumn{1}{c|}{---} & \multicolumn{1}{c}{---} & \multicolumn{1}{c|}{---} & \multicolumn{1}{c}{85.5} & \multicolumn{1}{c|}{1.38} & \multicolumn{1}{c}{86.8} & \multicolumn{1}{c|}{1.18} & \multicolumn{1}{c}{87.4} & \multicolumn{1}{c|}{1.08} \\
 \hline
\end{tabular}}
\caption{\textbar~\textbf{External testing: organ-wise segmentation performance comparison on 60 patients chest dataset}. The organ-wise DSC (\%) and ASD (mm) are evaluated for comparison. Note that structures excluded from the officially released SAT-pro, Vista3D, and MultiTalent (trained on $\mathcal{C}_5$ datasets) models are indicated with a `-'. Furthermore, the SAT-pro inference framework removes image metadata headers during inference, making it impossible to calculate ASD. The \ac{cln_u36} achieves the best mean DSC and ASD. }
\label{tab:external_chest}
\end{table}
\clearpage

\begin{table}[htp]
\centering
\resizebox{\textwidth}{!}{%
\begin{tabular}{|l|cc|ll|ll|ll|ll|ll|}
\hline
 & \multicolumn{2}{c|}{SAT-pro} & \multicolumn{2}{c|}{Vista3D} & \multicolumn{2}{c|}{MultiTalent} & \multicolumn{2}{c|}{\ac{nnu_e36}} & \multicolumn{2}{c|}{\ac{cln_c36}} & \multicolumn{2}{c|}{\ac{cln_u36}} \\ \cline{2-13} 
\multirow{-2}{*}{} & \multicolumn{1}{c}{DSC$\uparrow$} & ASD$\downarrow$ & \multicolumn{1}{c}{DSC$\uparrow$} & ASD$\downarrow$ & \multicolumn{1}{c}{DSC$\uparrow$} & \multicolumn{1}{c|}{ASD$\downarrow$} & \multicolumn{1}{c}{DSC$\uparrow$} & \multicolumn{1}{c|}{ASD$\downarrow$} & \multicolumn{1}{c}{DSC$\uparrow$} & \multicolumn{1}{c|}{ASD$\downarrow$} & \multicolumn{1}{c}{DSC$\uparrow$} & \multicolumn{1}{c|}{ASD$\downarrow$} \\ \hline
\rowcolor[HTML]{F3F3F3} 
Spleen & --- & --- & 95.0$\pm$1.9 & 0.63$\pm$0.77 & 92.9$\pm$8.4 & 0.68$\pm$0.99 & 93.4$\pm$6.9 & 0.74$\pm$0.49 & 94.7$\pm$7.7 & 0.71$\pm$0.51 & 94.3$\pm$6.0 & 0.57$\pm$0.54 \\
Kidney\_R & --- & --- & \cellcolor[HTML]{F3F3F3}91.3$\pm$15.4 & \cellcolor[HTML]{F3F3F3}2.14$\pm$9.28 & 91.3$\pm$14.2 & 0.69$\pm$0.52 & 91.0$\pm$7.0 & 0.62$\pm$0.74 & 91.2$\pm$6.7 & 0.78$\pm$0.85 & 91.3$\pm$6.5 & 0.73$\pm$0.62 \\
Kidney\_L & --- & --- & \cellcolor[HTML]{F3F3F3}92.9$\pm$6.2 & \cellcolor[HTML]{F3F3F3}0.79$\pm$1.63 & 91.9$\pm$13.9 & 0.63$\pm$0.38 & 91.2$\pm$6.8 & 0.83$\pm$0.77 & 91.0$\pm$7.6 & 1.11$\pm$0.62 & 91.7$\pm$8.3 & 0.73$\pm$0.75 \\
\rowcolor[HTML]{F3F3F3} 
Gallbladder & --- & --- & 78.1$\pm$24.2 & 0.56$\pm$0.46 & 71.0$\pm$30.2 & 4.11$\pm$11.93 & 77.3$\pm$8.7 & 1.90$\pm$0.82 & 78.0$\pm$9.6 & 0.63$\pm$0.79 & 78.8$\pm$8.7 & 0.61$\pm$0.79 \\
Eso & --- & --- & \cellcolor[HTML]{F3F3F3}78.6$\pm$6.3 & \cellcolor[HTML]{F3F3F3}1.13$\pm$1.15 & 78.9$\pm$7.7 & 1.44$\pm$0.96 & 79.0$\pm$9.6 & 1.68$\pm$1.33 & 80.1$\pm$10.5 & 1.38$\pm$1.41 & 81.1$\pm$10.6 & 1.26$\pm$1.36 \\
\rowcolor[HTML]{F3F3F3} 
Liver & --- & --- & 96.7$\pm$0.8 & 0.46$\pm$0.17 & 96.8$\pm$0.7 & 0.85$\pm$0.35 & 97.0$\pm$0.5 & 0.79$\pm$1.03 & 96.4$\pm$0.4 & 0.41$\pm$0.10 & 97.3$\pm$0.7 & 0.44$\pm$0.09 \\
Stomach & --- & --- & \cellcolor[HTML]{F3F3F3}93.0$\pm$2.2 & \cellcolor[HTML]{F3F3F3}0.79$\pm$0.94 & 90.7$\pm$6.0 & 1.18$\pm$0.95 & 89.6$\pm$8.6 & 1.26$\pm$0.99 & 86.8$\pm$9.6 & 1.53$\pm$0.98 & 87.8$\pm$9.5 & 1.48$\pm$0.95 \\
\rowcolor[HTML]{F3F3F3} 
A\_Aorta & --- & --- & 91.7$\pm$2.2 & 0.81$\pm$0.94 & 92.5$\pm$2.3 & 0.76$\pm$0.81 & 91.1$\pm$3.6 & 1.51$\pm$0.36 & 92.2$\pm$4.2 & 0.61$\pm$0.34 & 92.5$\pm$4.2 & 0.61$\pm$0.28 \\
V\_VenaCava\_I & --- & --- & \cellcolor[HTML]{F3F3F3}86.9$\pm$5.2 & \cellcolor[HTML]{F3F3F3}1.38$\pm$1.97 & 87.6$\pm$5.8 & 1.22$\pm$0.50 & 88.0$\pm$7.0 & 2.03$\pm$0.49 & 86.8$\pm$6.4 & 1.22$\pm$0.56 & 86.3$\pm$6.4 & 1.21$\pm$0.53 \\
\rowcolor[HTML]{F3F3F3} 
V\_Portal\_Splenic & --- & --- & 74.8$\pm$7.3 & 0.57$\pm$0.48 & 76.5$\pm$11.3 & 2.19$\pm$2.29 & 77.0$\pm$19.0 & 1.48$\pm$0.79 & 75.9$\pm$19.9 & 0.71$\pm$0.73 & 75.8$\pm$19.2 & 0.74$\pm$0.74 \\
Pancreas & --- & --- & \cellcolor[HTML]{F3F3F3}83.9$\pm$4.6 & \cellcolor[HTML]{F3F3F3}0.77$\pm$0.99 & 81.4$\pm$7.2 & 1.61$\pm$1.71 & 83.4$\pm$9.2 & 1.53$\pm$0.37 & 82.2$\pm$10.1 & 0.56$\pm$0.33 & 81.9$\pm$7.3 & 0.48$\pm$0.38 \\
\rowcolor[HTML]{F3F3F3} 
Glnd\_Adrenal\_R & --- & --- & 72.0$\pm$5.7 & 0.60$\pm$0.30 & 73.0$\pm$5.2 & 0.88$\pm$0.31 & 72.1$\pm$14.2 & 1.30$\pm$0.43 & 75.1$\pm$14.9 & 0.69$\pm$0.46 & 76.0$\pm$16.0 & 0.66$\pm$0.50 \\
\rowcolor[HTML]{F3F3F3} 
Glnd\_Adrenal\_L & --- & --- & 72.3$\pm$12.7 & 0.48$\pm$0.14 & 69.9$\pm$18.0 & 1.80$\pm$3.13 & 72.0$\pm$16.4 & 1.67$\pm$0.45 & 75.0$\pm$17.3 & 0.69$\pm$0.46 & 75.4$\pm$17.1 & 0.65$\pm$0.51 \\ \hline
Mean & --- & --- & \multicolumn{1}{c}{85.2} & \multicolumn{1}{c|}{0.85} & \multicolumn{1}{c}{84.1} & \multicolumn{1}{c|}{1.39} & \multicolumn{1}{c}{84.8} & \multicolumn{1}{c|}{1.33} & \multicolumn{1}{c}{85.0} & \multicolumn{1}{c|}{0.85} & \multicolumn{1}{c}{85.4} & \multicolumn{1}{c|}{0.78}
 \\ \hline
\end{tabular}}
\caption{\textbar~\textbf{External testing: Organ-wise segmentation performance comparison on 30 patients BTCV~\cite{landman2015miccai} abdomen dataset}. The organ-wise DSC (\%) and ASD (mm) are evaluated for comparison. Note that structures not included in the officially released SAT-pro model and MultiTalent (trained on $\mathcal{C}_5$ datasets) are indicated with a `-'. \ac{cln_u36} achieves the best mean DSC and ASD. \\
*Note: BTCV is used for SAT-Pro training, therefore, it's not available for SAT-Pro external testing and no result is provided in the table. }
\label{tab:external_abdomen}
\end{table}
\clearpage

\begin{table}[htp]
\centering
\resizebox{\textwidth}{!}{%
\begin{tabular}{|l|cc|cc|cc|ll|ll|ll|}
\hline
 & \multicolumn{2}{c|}{SAT-pro} & \multicolumn{2}{c|}{Vista3D} & \multicolumn{2}{c|}{MultiTalent} & \multicolumn{2}{c|}{\ac{nnu_e36}} & \multicolumn{2}{c|}{\ac{cln_c36}} & \multicolumn{2}{c|}{\ac{cln_u36}} \\ \cline{2-13} 
\multirow{-2}{*}{} & \multicolumn{1}{c}{DSC$\uparrow$} & ASD$\downarrow$ & \multicolumn{1}{c}{DSC$\uparrow$} & ASD$\downarrow$ & \multicolumn{1}{c}{DSC$\uparrow$} & \multicolumn{1}{c|}{ASD$\downarrow$} & \multicolumn{1}{c}{DSC$\uparrow$} & \multicolumn{1}{c|}{ASD$\downarrow$} & \multicolumn{1}{c}{DSC$\uparrow$} & \multicolumn{1}{c|}{ASD$\downarrow$} & \multicolumn{1}{c}{DSC$\uparrow$} & \multicolumn{1}{c|}{ASD$\downarrow$} \\ \hline
\rowcolor[HTML]{F3F3F3} 
NPC\_\ac{gtv} & \multicolumn{1}{l}{\cellcolor[HTML]{F3F3F3}64.9$\pm$13.3} & --- & --- & --- & --- & --- & 70.3$\pm$13.1 & 2.02$\pm$1.46 & 70.5$\pm$14.2 & 2.01$\pm$1.59 & 70.3$\pm$9.9 & 1.98$\pm$1.58 \\
Eso\_\ac{gtv} & --- & --- & --- & --- & --- & --- & 75.0$\pm$18.4 & 8.85$\pm$14.23 & 74.9$\pm$16.7 & 8.87$\pm$12.24 & 75.1$\pm$18.8 & 8.67$\pm$14.50 \\
\rowcolor[HTML]{F3F3F3} 
Liver\_\ac{gtv} & \multicolumn{1}{l}{\cellcolor[HTML]{F3F3F3}33.7$\pm$33.3} & --- & --- & --- & --- & --- & 69.9$\pm$16.3 & 8.41$\pm$4.31 & 69.9$\pm$20.9 & 8.43$\pm$6.42 & 70.4$\pm$18.1 & 7.53$\pm$4.74 \\
Kidney\_\ac{gtv} & --- & --- & --- & --- & \multicolumn{1}{l}{81.3$\pm$11.4} & \multicolumn{1}{l|}{7.22$\pm$10.62} & 83.6$\pm$18.1 & 2.20$\pm$1.64 & 86.6$\pm$12.6 & 1.01$\pm$0.99 & 86.7$\pm$11.0 & 1.00$\pm$0.99 \\
 \hline
Mean & --- & --- & --- & --- & --- & --- & \multicolumn{1}{c}{74.7} & \multicolumn{1}{c|}{5.37} & \multicolumn{1}{c}{75.5} & \multicolumn{1}{c|}{5.08} & \multicolumn{1}{c}{75.6} & \multicolumn{1}{c|}{4.80}
 \\ \hline
\end{tabular}}
\caption{\textbar~\textbf{External testing: Organ-wise segmentation performance comparison on external 319 \ac{npc} \ac{gtv} patients, 148 Eso \ac{gtv} patients, 176 Liver \ac{gtv} patients, and 978 Kidney \ac{gtv} patients}. The organ-wise DSC (\%) and ASD (mm) are evaluated for comparison. Note that structures not included in the officially released SAT-pro model and MultiTalent (trained on $\mathcal{C}_5$ datasets) are indicated with a `-'. Furthermore, the SAT-pro inference framework removes image metadata headers during inference, making it impossible to calculate ASD. The \ac{cln_u36} achieves the best mean DSC and ASD. }
\label{tab:external_lesion}
\end{table}
\clearpage

\begin{table}[htp]
\centering
\fontsize{10pt}{12pt}\selectfont
% \resizebox{\textwidth}{!}{
\begin{tabular}{|l|ll|r|ll|r|}
\hline
\multicolumn{1}{|l|}{} &  \multicolumn{3}{c|}{\ac{cln_u36}} & \multicolumn{3}{c|}{\ac{cln_u36_ft}} \\ \cline{2-7} 
\multicolumn{1}{|l|}{\multirow{-2}{*}{}} &  DSC$\uparrow$ & ASD$\downarrow$ & \multicolumn{1}{r|}{$\mathcal{T}$$\uparrow$} & DSC$\uparrow$ & ASD$\downarrow$ & \multicolumn{1}{r|}{$\mathcal{T}$$\uparrow$} \\ \hline
\rowcolor[HTML]{F3F3F3} 
Spleen  & 97.8$\pm$6.9 & 0.29$\pm$0.41 & 99.5 & 97.2$\pm$6.8 & 1.67$\pm$0.40 & 99.5 \\
Kidney\_R  & 94.9$\pm$7.1 & 1.07$\pm$0.76 & 99.7 & 97.1$\pm$7.2 & 0.21$\pm$0.78 & 99.7 \\
Kidney\_L  & 94.2$\pm$6.9 & 1.46$\pm$0.72 & {} & 97.1$\pm$7.0 & 0.92$\pm$0.73 & {} \\
\rowcolor[HTML]{F3F3F3} 
GallBladder  & 86.6$\pm$8.8 & 1.96$\pm$0.84 & 99.5 & 93.9$\pm$8.9 & 0.91$\pm$0.85 & 99.0 \\
Liver  & 97.8$\pm$5.9 & 0.77$\pm$0.94 & 99.5 & 97.6$\pm$5.8 & 0.58$\pm$0.93 & 99.7 \\
\rowcolor[HTML]{F3F3F3} 
Stomach  & 92.4$\pm$6.2 & 1.07$\pm$0.65 & 99.5 & 96.1$\pm$6.1 & 1.08$\pm$0.64 & 99.5 \\
Pancreas   & 94.6$\pm$6.9 & 0.66$\pm$0.37 & 99.5 & 92.5$\pm$7.0 & 0.89$\pm$0.37 & 99.5 \\
\rowcolor[HTML]{F3F3F3} 
Glnd\_Adrenal\_R  & 90.6$\pm$13.1 & 0.22$\pm$0.38 & 98.0 & 85.2$\pm$13.5 & 0.43$\pm$0.41 & 98.0 \\
\rowcolor[HTML]{F3F3F3} 
Glnd\_Adrenal\_L  & 88.3$\pm$12.5 & 0.40$\pm$0.39 & {\cellcolor[HTML]{F3F3F3}} & 85.0$\pm$12.6 & 0.93$\pm$0.39 & {\cellcolor[HTML]{F3F3F3}} \\
Lung\_LUL  & 97.9$\pm$3.7 & 0.17$\pm$0.62 & 99.5 & 97.8$\pm$3.6 & 0.16$\pm$0.61 & 99.5 \\
Lung\_LLL  & 96.5$\pm$2.1 & 0.21$\pm$0.47 & {} & 97.4$\pm$2.0 & 0.19$\pm$0.46 & {} \\
\rowcolor[HTML]{F3F3F3} 
Lung\_RUL  & 97.7$\pm$2.5 & 0.20$\pm$0.98 & 99.5 & 97.8$\pm$2.6 & 0.21$\pm$1.00 & 99.5 \\
\rowcolor[HTML]{F3F3F3} 
Lung\_RML  & 92.3$\pm$3.3 & 0.22$\pm$1.50 & {\cellcolor[HTML]{F3F3F3}} & 93.1$\pm$3.3 & 0.23$\pm$1.37 & {\cellcolor[HTML]{F3F3F3}} \\
\rowcolor[HTML]{F3F3F3} 
Lung\_RLL  & 97.2$\pm$3.6 & 0.21$\pm$1.19 & {\cellcolor[HTML]{F3F3F3}} & 97.0$\pm$3.7 & 0.21$\pm$1.10 & {\cellcolor[HTML]{F3F3F3}} \\
Eso  & 85.4$\pm$3.0 & 0.64$\pm$0.15 & 98.0 & 94.5$\pm$3.1 & 0.46$\pm$0.16 & 99.0 \\
\rowcolor[HTML]{F3F3F3} 
Trachea  & 92.9$\pm$6.1 & 0.86$\pm$0.12 & 99.7 & 97.2$\pm$6.0 & 0.27$\pm$0.11 & 99.7 \\
Glnd\_Thyroid  & 82.7$\pm$14.5 & 0.95$\pm$0.96 & 94.0 & 93.3$\pm$14.6 & 0.26$\pm$0.97 & 99.0 \\
\rowcolor[HTML]{F3F3F3} 
SmallBowel  & 91.2$\pm$16.4 & 2.22$\pm$2.54 & 99.5 & 91.1$\pm$16.6 & 1.52$\pm$2.41 & 99.5 \\
Duodenum  & 82.9$\pm$16.0 & 2.29$\pm$1.18 & 99.5 & 89.4$\pm$15.9 & 1.74$\pm$1.19 & 99.5 \\
\rowcolor[HTML]{F3F3F3} 
Colon  & 92.9$\pm$15.9 & 2.72$\pm$1.13 & 99.5 & 90.7$\pm$15.7 & 2.49$\pm$1.13 & 99.7 \\
UrinaryBladder  & 93.6$\pm$9.7 & 1.82$\pm$0.94 & 98.0 & 89.7$\pm$9.5 & 2.99$\pm$0.96 & 99.5 \\
\rowcolor[HTML]{F3F3F3} 
Prostate  & 85.5$\pm$17.7 & 3.62$\pm$1.36 & 96.0 & 83.8$\pm$17.9 & 0.33$\pm$1.44 & 99.5 \\ \hline

Mean & \multicolumn{1}{c}{ 92.3 } & \multicolumn{1}{c|}{ 1.20 } & \multicolumn{1}{c|}{ 98.7 } & \multicolumn{1}{c}{ 93.4 } & \multicolumn{1}{c|}{ 0.85 } & \multicolumn{1}{c|}{ 99.4 }  \\ \hline

\end{tabular}
% }
\caption{\textbar~\textbf{Performance comparison of \ac{cln_u36} and \ac{cln_u36_ft} on \ac{totalseg} `Organ' subgroup.} Organ-wise DSC (\%), ASD (mm), and decoder-wise pruning rate $\mathcal{T}$ (\%) of \ac{cln_u36} and \ac{cln_u36_ft} are evaluated on \ac{totalseg} `Organ' subgroup. After fine-tuning, \ac{cln_u36_ft} achieves improved mean DSC and ASD, indicating that \ac{cln_u36} can further overfit its general label style learned from $\mathcal{U}_{36}$ datasets to the distinct label style of \ac{totalseg} `Organ' subgroup. }
\label{tab:pl_totalseg_organ_ds36}
\end{table}
\clearpage

\begin{table}[htp]
\centering
\fontsize{10pt}{12pt}\selectfont
% \resizebox{\textwidth}{!}{
\begin{tabular}{|l|ll|r|ll|r|}
\hline
\multicolumn{1}{|l|}{} &  \multicolumn{3}{c|}{\ac{cln_u36}} & \multicolumn{3}{c|}{\ac{cln_u36_ft}} \\ \cline{2-7} 
\multicolumn{1}{|l|}{\multirow{-2}{*}{}} &  DSC$\uparrow$ & ASD$\downarrow$ & \multicolumn{1}{r|}{$\mathcal{T}$$\uparrow$} & DSC$\uparrow$ & ASD$\downarrow$ & \multicolumn{1}{r|}{$\mathcal{T}$$\uparrow$} \\ \hline
\rowcolor[HTML]{F3F3F3} 
Heart  & 97.1$\pm$14.7 & 1.64$\pm$0.19 & 99.0 & 94.8$\pm$14.8 & 0.46$\pm$0.20 & 99.5 \\
A\_Aorta  & 95.0$\pm$16.9 & 0.75$\pm$0.86 & 99.0 & 96.0$\pm$16.8 & 0.42$\pm$0.86 & 99.7 \\
\rowcolor[HTML]{F3F3F3} 
V\_Pulmonary  & 75.5$\pm$15.3 & 1.37$\pm$0.14 & 99.0 & 90.7$\pm$15.4 & 0.30$\pm$0.14 & 99.0 \\
A\_BrachiocephalicTrunk  & 98.2$\pm$16.2 & 0.62$\pm$0.43 & 99.0 & 90.5$\pm$16.3 & 0.22$\pm$0.44 & 99.5 \\
\rowcolor[HTML]{F3F3F3} 
A\_Subclavian\_L  & 82.4$\pm$3.5 & 2.69$\pm$0.28 & 94.0 & 95.3$\pm$3.6 & 0.27$\pm$0.29 & 99.7 \\
\rowcolor[HTML]{F3F3F3} 
A\_Subclavian\_R  & 85.5$\pm$3.3 & 1.45$\pm$0.16 & {\cellcolor[HTML]{F3F3F3}} & 93.0$\pm$3.3 & 0.17$\pm$0.17 & {\cellcolor[HTML]{F3F3F3}} \\
A\_CommonCarotid\_L  & 88.3$\pm$7.6 & 0.89$\pm$3.83 & 99.7 & 89.6$\pm$5.5 & 0.23$\pm$0.29 & 99.7 \\
A\_CommonCarotid\_R  & 86.1$\pm$9.4 & 0.55$\pm$3.72 & {} & 87.2$\pm$3.5 & 0.26$\pm$1.01 & {} \\
\rowcolor[HTML]{F3F3F3} 
V\_Brachiocephalic\_L  & 92.5$\pm$5.2 & 0.36$\pm$0.14 & 99.5 & 92.2$\pm$6.2 & 0.33$\pm$1.91 & 99.5 \\
\rowcolor[HTML]{F3F3F3} 
V\_Brachiocephalic\_R  & 90.0$\pm$5.8 & 0.45$\pm$0.25 & {\cellcolor[HTML]{F3F3F3}} & 90.0$\pm$5.3 & 0.24$\pm$0.83 & {\cellcolor[HTML]{F3F3F3}} \\
LAA  & 91.7$\pm$3.3 & 0.16$\pm$0.15 & 99.7 & 92.5$\pm$3.4 & 0.18$\pm$0.16 & 99.7 \\
\rowcolor[HTML]{F3F3F3} 
V\_VenaCava\_S  & 92.0$\pm$21.4 & 0.83$\pm$2.23 & 99.0 & 93.7$\pm$21.5 & 0.17$\pm$2.23 & 99.5 \\
V\_VenaCava\_I  & 84.5$\pm$9.7 & 1.72$\pm$0.99 & 99.0 & 95.8$\pm$9.7 & 0.77$\pm$0.99 & 99.7 \\
\rowcolor[HTML]{F3F3F3} 
V\_Portal\_Splenic  & 86.7$\pm$29.1 & 1.12$\pm$0.63 & 98.0 & 89.0$\pm$27.1 & 0.18$\pm$0.63 & 98.0 \\
A\_Iliac\_L  & 95.1$\pm$3.3 & 0.50$\pm$0.20 & 99.0 & 96.0$\pm$3.3 & 0.50$\pm$0.20 & 99.7 \\
A\_Iliac\_R  & 91.9$\pm$19.6 & 0.39$\pm$0.18 & {} & 91.1$\pm$19.7 & 0.36$\pm$0.19 & {} \\
\rowcolor[HTML]{F3F3F3} 
V\_Iliac\_L  & 92.9$\pm$3.3 & 0.43$\pm$0.15 & 99.7 & 92.6$\pm$3.3 & 0.45$\pm$0.15 & 99.0 \\
\rowcolor[HTML]{F3F3F3} 
V\_Iliac\_R  & 93.6$\pm$4.8 & 0.45$\pm$0.27 & {\cellcolor[HTML]{F3F3F3}} & 94.4$\pm$4.9 & 0.47$\pm$0.28 & {\cellcolor[HTML]{F3F3F3}}
\\ \hline

Mean & \multicolumn{1}{c}{ 89.8 } & \multicolumn{1}{c|}{ 0.84 } & \multicolumn{1}{c|}{ 98.7 } & \multicolumn{1}{c}{ 92.5 } & \multicolumn{1}{c|}{ 0.33 } & \multicolumn{1}{c|}{ 99.4 }  \\ \hline
\end{tabular}
% }
\caption{\textbar~\textbf{Performance comparison of \ac{cln_u36} and \ac{cln_u36_ft} on \ac{totalseg} `Cardiac' subgroup.} Organ-wise DSC (\%), ASD (mm), and decoder-wise pruning rate $\mathcal{T}$ (\%) of \ac{cln_u36} and \ac{cln_u36_ft} are evaluated on \ac{totalseg} `Cardiac' subgroup. After fine-tuning, \ac{cln_u36_ft} achieves improved mean DSC and ASD, indicating that \ac{cln_u36} can further overfit its general label style learned from $\mathcal{U}_{36}$ datasets to the distinct label style of \ac{totalseg} `Cardiac' subgroup. }
\label{tab:pl_totalseg_cardi_ds36}
\end{table}
\clearpage

\begin{table}[htp]
\centering
\fontsize{10pt}{12pt}\selectfont
% \resizebox{\textwidth}{!}{
\begin{tabular}{|l|ll|r|ll|r|}
\hline
\multicolumn{1}{|l|}{} &  \multicolumn{3}{c|}{\ac{cln_u36}} & \multicolumn{3}{c|}{\ac{cln_u36_ft}} \\ \cline{2-7} 
\multicolumn{1}{|l|}{\multirow{-2}{*}{}} &  DSC$\uparrow$ & ASD$\downarrow$ & \multicolumn{1}{r|}{$\mathcal{T}$$\uparrow$} & DSC$\uparrow$ & ASD$\downarrow$ & \multicolumn{1}{r|}{$\mathcal{T}$$\uparrow$} \\ \hline
\rowcolor[HTML]{F3F3F3} 
Bone\_Humerus\_L  & 97.1$\pm$17.6 & 2.77$\pm$0.21 & 99.7 & 97.3$\pm$17.4 & 0.97$\pm$0.29 & 99.5 \\
\rowcolor[HTML]{F3F3F3} 
Bone\_Humerus\_R  & 97.1$\pm$17.0 & 3.88$\pm$0.17 & {\cellcolor[HTML]{F3F3F3}} & 97.2$\pm$17.0 & 0.19$\pm$0.17 & {\cellcolor[HTML]{F3F3F3}} \\
Bone\_Scapula\_L  & 93.8$\pm$21.4 & 0.97$\pm$0.17 & 99.7 & 94.0$\pm$21.5 & 0.98$\pm$0.19 & 99.5 \\
Bone\_Scapula\_R  & 94.4$\pm$1.9 & 1.63$\pm$0.08 & {} & 93.9$\pm$12.1 & 1.48$\pm$0.10 & {} \\
\rowcolor[HTML]{F3F3F3} 
Bone\_Clavicula\_L  & 95.2$\pm$0.9 & 0.45$\pm$0.11 & 99.7 & 96.0$\pm$1.0 & 0.36$\pm$0.13 & 99.5 \\
\rowcolor[HTML]{F3F3F3} 
Bone\_Clavicula\_R  & 98.3$\pm$16.0 & 1.37$\pm$0.20 & {\cellcolor[HTML]{F3F3F3}} & 98.4$\pm$15.8 & 0.29$\pm$0.18 & {\cellcolor[HTML]{F3F3F3}} \\
Bone\_Femur\_L  & 92.0$\pm$0.6 & 5.77$\pm$0.10 & 99.5 & 97.6$\pm$0.7 & 3.06$\pm$0.11 & 99.5 \\
Bone\_Femur\_R  & 94.6$\pm$0.2 & 5.95$\pm$0.06 & {} & 96.9$\pm$0.3 & 5.96$\pm$0.07 & {} \\
\rowcolor[HTML]{F3F3F3} 
Bone\_Hip\_L  & 97.0$\pm$1.9 & 0.29$\pm$0.14 & 99.5 & 97.1$\pm$2.0 & 0.31$\pm$0.14 & 99.7 \\
\rowcolor[HTML]{F3F3F3} 
Bone\_Hip\_R  & 97.0$\pm$31.9 & 0.27$\pm$0.11 & {\cellcolor[HTML]{F3F3F3}} & 96.5$\pm$31.8 & 0.30$\pm$0.10 & {\cellcolor[HTML]{F3F3F3}} \\
SpinalCord  & 94.1$\pm$16.7 & 0.84$\pm$0.26 & 99.7 & 93.1$\pm$16.4 & 0.27$\pm$0.23 & 99.5 \\
\rowcolor[HTML]{F3F3F3} 
Musc\_GluteusMax\_L  & 98.7$\pm$5.4 & 0.76$\pm$1.50 & 99.5 & 98.8$\pm$5.5 & 0.75$\pm$1.45 & 99.7 \\
\rowcolor[HTML]{F3F3F3} 
Musc\_GluteusMax\_R  & 97.9$\pm$2.2 & 0.71$\pm$0.34 & {\cellcolor[HTML]{F3F3F3}} & 97.7$\pm$2.3 & 0.66$\pm$0.35 & {\cellcolor[HTML]{F3F3F3}} \\
Musc\_GluteusMed\_L  & 98.4$\pm$1.5 & 0.45$\pm$0.23 & 99.5 & 98.3$\pm$1.5 & 0.47$\pm$0.23 & 99.5 \\
Musc\_GluteusMed\_R  & 97.3$\pm$7.5 & 0.47$\pm$0.21 & {} & 97.5$\pm$7.4 & 0.50$\pm$0.19 & {} \\
\rowcolor[HTML]{F3F3F3} 
Musc\_GluteusMin\_L  & 97.3$\pm$2.2 & 0.42$\pm$0.16 & 99.5 & 97.4$\pm$2.1 & 0.43$\pm$0.15 & 99.7 \\
\rowcolor[HTML]{F3F3F3} 
Musc\_GluteusMin\_R  & 98.2$\pm$1.1 & 0.36$\pm$0.10 & {\cellcolor[HTML]{F3F3F3}} & 98.3$\pm$1.0 & 0.34$\pm$0.09 & {\cellcolor[HTML]{F3F3F3}} \\
Musc\_Autochthon\_L  & 93.9$\pm$0.9 & 0.32$\pm$0.12 & 99.0 & 94.3$\pm$1.1 & 0.29$\pm$0.14 & 99.0 \\
Musc\_Autochthon\_R  & 94.8$\pm$0.8 & 0.41$\pm$0.11 & {} & 95.1$\pm$0.9 & 0.39$\pm$0.12 & {} \\
\rowcolor[HTML]{F3F3F3} 
Musc\_Iliopsoas\_L  & 93.5$\pm$19.6 & 0.36$\pm$0.19 & 99.0 & 92.5$\pm$19.8 & 0.33$\pm$0.20 & 99.0 \\
\rowcolor[HTML]{F3F3F3} 
Musc\_Iliopsoas\_R  & 94.9$\pm$9.1 & 0.32$\pm$0.25 & {\cellcolor[HTML]{F3F3F3}} & 95.6$\pm$9.2 & 0.30$\pm$0.26 & {\cellcolor[HTML]{F3F3F3}} \\
Brain  & 98.2$\pm$16.3 & 0.51$\pm$1.07 & 99.5 & 98.2$\pm$16.4 & 0.51$\pm$1.04 & 99.5 \\
\rowcolor[HTML]{F3F3F3} 
Skull  & 82.5$\pm$35.4 & 3.66$\pm$0.21 & 99.0 & 82.0$\pm$35.5 & 1.45$\pm$0.21 & 99.5
\\ \hline

Mean & \multicolumn{1}{c}{ 95.4 } & \multicolumn{1}{c|}{ 0.98 } & \multicolumn{1}{c|}{ 99.4 } & \multicolumn{1}{c}{ 95.8 } & \multicolumn{1}{c|}{ 0.90 } & \multicolumn{1}{c|}{ 99.5 }  \\ \hline
\end{tabular}
% }
\caption{\textbar~\textbf{Performance comparison of \ac{cln_u36} and \ac{cln_u36_ft} on \ac{totalseg} `Muscles' subgroup.} Organ-wise DSC (\%), ASD (mm), and decoder-wise pruning rate $\mathcal{T}$ (\%) of \ac{cln_u36} and \ac{cln_u36_ft} are evaluated on \ac{totalseg} `Muscles' subgroup. After fine-tuning, \ac{cln_u36_ft} achieves improved mean DSC, indicating that \ac{cln_u36} can further overfit its general label style learned from $\mathcal{U}_{36}$ datasets to the distinct label style of \ac{totalseg} `Muscles' subgroup. }
\label{tab:pl_totalseg_musc_ds36}
\end{table}
\clearpage

\begin{table}[htp]
\centering
\fontsize{10pt}{12pt}\selectfont
% \resizebox{\textwidth}{!}{
\begin{tabular}{|l|ll|r|ll|r|}
\hline
\multicolumn{1}{|l|}{} &  \multicolumn{3}{c|}{\ac{cln_u36}} & \multicolumn{3}{c|}{\ac{cln_u36_ft}} \\ \cline{2-7} 
\multicolumn{1}{|l|}{\multirow{-2}{*}{}} &  DSC$\uparrow$ & ASD$\downarrow$ & \multicolumn{1}{r|}{$\mathcal{T}$$\uparrow$} & DSC$\uparrow$ & ASD$\downarrow$ & \multicolumn{1}{r|}{$\mathcal{T}$$\uparrow$} \\ \hline
\rowcolor[HTML]{F3F3F3} 
Bone\_Rib1\_L  & 93.3$\pm$3.3 & 0.14$\pm$0.21 & {\cellcolor[HTML]{F3F3F3}96.0} & 92.4$\pm$3.1 & 0.14$\pm$0.19 & {\cellcolor[HTML]{F3F3F3}98.0} \\
\rowcolor[HTML]{F3F3F3} 
Bone\_Rib2\_L  & 90.2$\pm$1.5 & 0.74$\pm$1.05 &  & 91.5$\pm$1.6 & 0.59$\pm$0.92 &  \\
\rowcolor[HTML]{F3F3F3} 
Bone\_Rib3\_L  & 94.9$\pm$17.0 & 0.85$\pm$1.08 &  & 94.5$\pm$17.0 & 0.90$\pm$1.04 &  \\
\rowcolor[HTML]{F3F3F3} 
Bone\_Rib4\_L  & 93.9$\pm$18.7 & 0.63$\pm$1.77 &  & 94.1$\pm$18.8 & 0.61$\pm$1.67 &  \\
\rowcolor[HTML]{F3F3F3} 
Bone\_Rib5\_L  & 94.3$\pm$29.6 & 1.18$\pm$4.53 &  & 91.4$\pm$29.4 & 1.11$\pm$4.29 &  \\
\rowcolor[HTML]{F3F3F3} 
Bone\_Rib6\_L  & 92.8$\pm$24.1 & 1.51$\pm$3.35 &  & 92.8$\pm$24.3 & 1.55$\pm$3.15 &  \\
\rowcolor[HTML]{F3F3F3} 
Bone\_Rib7\_L  & 95.8$\pm$20.8 & 2.36$\pm$4.01 &  & 95.6$\pm$20.8 & 2.25$\pm$3.69 &  \\
\rowcolor[HTML]{F3F3F3} 
Bone\_Rib8\_L  & 95.2$\pm$21.1 & 4.15$\pm$4.04 &  & 95.5$\pm$20.9 & 3.82$\pm$3.80 &  \\
\rowcolor[HTML]{F3F3F3} 
Bone\_Rib9\_L  & 94.5$\pm$19.3 & 2.74$\pm$4.83 &  & 95.0$\pm$19.5 & 2.57$\pm$4.40 &  \\
\rowcolor[HTML]{F3F3F3} 
Bone\_Rib10\_L  & 91.9$\pm$13.2 & 1.48$\pm$6.45 &  & 91.5$\pm$13.1 & 1.46$\pm$6.42 &  \\
\rowcolor[HTML]{F3F3F3} 
Bone\_Rib11\_L  & 92.1$\pm$12.1 & 0.34$\pm$8.12 &  & 92.7$\pm$12.1 & 0.32$\pm$7.51 &  \\
\rowcolor[HTML]{F3F3F3} 
Bone\_Rib12\_L  & 94.7$\pm$16.3 & 0.33$\pm$0.41 &  & 94.5$\pm$16.3 & 0.31$\pm$0.40 &  \\
\rowcolor[HTML]{F3F3F3} 
Bone\_Rib1\_R  & 91.8$\pm$5.6 & 0.14$\pm$0.15 &  & 91.9$\pm$5.5 & 0.15$\pm$0.14 &  \\
\rowcolor[HTML]{F3F3F3} 
Bone\_Rib2\_R  & 94.6$\pm$3.4 & 0.09$\pm$0.91 &  & 94.7$\pm$3.2 & 0.09$\pm$0.89 &  \\
\rowcolor[HTML]{F3F3F3} 
Bone\_Rib3\_R  & 93.7$\pm$9.1 & 0.34$\pm$1.14 &  & 93.9$\pm$9.3 & 0.38$\pm$1.11 &  \\
\rowcolor[HTML]{F3F3F3} 
Bone\_Rib4\_R  & 90.4$\pm$18.6 & 0.73$\pm$1.92 &  & 91.1$\pm$18.4 & 0.73$\pm$1.85 &  \\
\rowcolor[HTML]{F3F3F3} 
Bone\_Rib5\_R  & 90.7$\pm$16.4 & 1.49$\pm$2.51 &  & 91.4$\pm$16.5 & 1.43$\pm$2.31 &  \\
\rowcolor[HTML]{F3F3F3} 
Bone\_Rib6\_R  & 93.4$\pm$13.0 & 1.38$\pm$2.54 &  & 93.0$\pm$13.1 & 1.46$\pm$2.45 &  \\
\rowcolor[HTML]{F3F3F3} 
Bone\_Rib7\_R  & 95.5$\pm$21.3 & 1.57$\pm$3.36 &  & 94.6$\pm$21.3 & 1.62$\pm$3.08 &  \\
\rowcolor[HTML]{F3F3F3} 
Bone\_Rib8\_R  & 95.0$\pm$18.0 & 1.78$\pm$4.22 &  & 95.5$\pm$17.9 & 1.67$\pm$4.06 &  \\
\rowcolor[HTML]{F3F3F3} 
Bone\_Rib9\_R  & 96.3$\pm$21.6 & 2.25$\pm$3.18 &  & 95.5$\pm$21.7 & 2.05$\pm$3.06 &  \\
\rowcolor[HTML]{F3F3F3} 
Bone\_Rib10\_R  & 95.9$\pm$23.3 & 2.30$\pm$7.41 &  & 95.9$\pm$23.4 & 2.27$\pm$7.24 &  \\
\rowcolor[HTML]{F3F3F3} 
Bone\_Rib11\_R  & 97.5$\pm$22.7 & 1.71$\pm$6.70 &  & 97.6$\pm$22.9 & 1.60$\pm$6.06 &  \\
\rowcolor[HTML]{F3F3F3} 
Bone\_Rib12\_R  & 97.4$\pm$27.5 & 0.46$\pm$0.10 &  & 97.2$\pm$27.6 & 0.41$\pm$0.12 &  \\
Bone\_Sternum  & 95.5$\pm$5.7 & 0.35$\pm$0.55 & {99.7} & 95.7$\pm$5.5 & 0.36$\pm$0.53 & {99.7} \\
\rowcolor[HTML]{F3F3F3} 
Bone\_CostalCartilages  & 88.6$\pm$5.0 & 0.16$\pm$0.74 & {\cellcolor[HTML]{F3F3F3}99.7} & 89.0$\pm$5.0 & 0.16$\pm$0.75 & {\cellcolor[HTML]{F3F3F3}99.5}
\\ \hline

Mean & \multicolumn{1}{c}{ 93.8 } & \multicolumn{1}{c|}{ 1.20 } & \multicolumn{1}{c|}{ 98.4 } & \multicolumn{1}{c}{ 93.8 } & \multicolumn{1}{c|}{ 1.16 } & \multicolumn{1}{c|}{ 99.1 }  \\ \hline
\end{tabular}
% }
\caption{\textbar~\textbf{Performance comparison of \ac{cln_u36} and \ac{cln_u36_ft} on \ac{totalseg} `Ribs' subgroup.} Organ-wise DSC (\%), ASD (mm), and decoder-wise pruning rate $\mathcal{T}$ (\%) of \ac{cln_u36} and \ac{cln_u36_ft} are evaluated on \ac{totalseg} `Ribs' subgroup. After fine-tuning, \ac{cln_u36_ft} achieves improved mean DSC, indicating that \ac{cln_u36} can further overfit its general label style learned from $\mathcal{U}_{36}$ datasets to the distinct label style of \ac{totalseg} `Ribs' subgroup. }
\label{tab:pl_totalseg_ribs_ds36}
\end{table}
\clearpage

\begin{table}[htp]
\centering
\fontsize{10pt}{12pt}\selectfont
% \resizebox{\textwidth}{!}{
\begin{tabular}{|l|ll|r|ll|r|}
\hline
\multicolumn{1}{|l|}{} &  \multicolumn{3}{c|}{\ac{cln_u36}} & \multicolumn{3}{c|}{\ac{cln_u36_ft}} \\ \cline{2-7} 
\multicolumn{1}{|l|}{\multirow{-2}{*}{}} &  DSC$\uparrow$ & ASD$\downarrow$ & \multicolumn{1}{r|}{$\mathcal{T}$$\uparrow$} & DSC$\uparrow$ & ASD$\downarrow$ & \multicolumn{1}{r|}{$\mathcal{T}$$\uparrow$} \\ \hline
\rowcolor[HTML]{F3F3F3} 
Bone\_Sacrum  & 93.9$\pm$1.9 & 0.19$\pm$0.14 & {\cellcolor[HTML]{F3F3F3}99.5} & 93.8$\pm$2.1 & 0.20$\pm$0.16 & {\cellcolor[HTML]{F3F3F3}99.5} \\
Bone\_Vert\_C1  & 95.3$\pm$10.2 & 0.23$\pm$0.50 & 96.0 & 94.6$\pm$10.0 & 0.24$\pm$0.49 & {96.0} \\
Bone\_Vert\_C2  & 97.9$\pm$10.5 & 0.14$\pm$0.47 &  & 98.0$\pm$10.7 & 0.15$\pm$0.49 &  \\
Bone\_Vert\_C3  & 98.0$\pm$15.5 & 0.16$\pm$3.18 &  & 98.2$\pm$15.4 & 0.15$\pm$2.86 &  \\
Bone\_Vert\_C4  & 96.6$\pm$19.6 & 0.30$\pm$3.54 &  & 95.9$\pm$19.5 & 0.27$\pm$3.36 &  \\
Bone\_Vert\_C5  & 97.8$\pm$17.4 & 0.21$\pm$1.63 &  & 97.8$\pm$17.3 & 0.19$\pm$1.49 &  \\
Bone\_Vert\_C6  & 96.0$\pm$10.5 & 0.45$\pm$0.93 &  & 95.1$\pm$10.5 & 0.44$\pm$0.93 &  \\
Bone\_Vert\_C7  & 95.3$\pm$13.1 & 0.18$\pm$1.00 &  & 96.1$\pm$13.0 & 0.19$\pm$0.98 &  \\
Bone\_Vert\_L1  & 94.9$\pm$12.1 & 1.19$\pm$0.98 &  & 94.4$\pm$12.0 & 1.13$\pm$0.97 &  \\
Bone\_Vert\_L2  & 93.7$\pm$12.3 & 0.18$\pm$0.95 &  & 92.8$\pm$12.1 & 0.18$\pm$0.93 &  \\
Bone\_Vert\_L3  & 96.9$\pm$16.2 & 0.18$\pm$1.06 &  & 97.3$\pm$16.3 & 0.20$\pm$0.96 &  \\
Bone\_Vert\_L4  & 94.2$\pm$14.2 & 0.23$\pm$0.41 &  & 94.6$\pm$14.0 & 0.24$\pm$0.39 &  \\
Bone\_Vert\_L5  & 93.9$\pm$17.4 & 0.19$\pm$1.43 &  & 93.9$\pm$17.2 & 0.18$\pm$1.40 &  \\
Bone\_Vert\_S1  & 97.3$\pm$18.9 & 0.19$\pm$1.22 &  & 97.3$\pm$19.0 & 0.18$\pm$1.12 &  \\
Bone\_Vert\_T1  & 95.9$\pm$13.5 & 0.15$\pm$1.37 &  & 96.8$\pm$13.6 & 0.14$\pm$1.25 &  \\
Bone\_Vert\_T2  & 97.7$\pm$5.4 & 0.32$\pm$0.76 &  & 97.5$\pm$5.6 & 0.32$\pm$0.77 &  \\
Bone\_Vert\_T3  & 97.9$\pm$2.6 & 0.46$\pm$0.21 &  & 97.9$\pm$2.5 & 0.44$\pm$0.20 &  \\
Bone\_Vert\_T4  & 90.5$\pm$8.9 & 1.05$\pm$0.20 &  & 91.4$\pm$8.9 & 0.95$\pm$0.20 &  \\
Bone\_Vert\_T5  & 90.4$\pm$23.1 & 0.88$\pm$0.39 &  & 91.0$\pm$22.9 & 0.97$\pm$0.37 &  \\
Bone\_Vert\_T6  & 85.4$\pm$28.1 & 1.04$\pm$0.06 &  & 85.4$\pm$28.2 & 0.94$\pm$0.06 &  \\
Bone\_Vert\_T7  & 97.1$\pm$1.0 & 1.48$\pm$0.06 &  & 97.0$\pm$1.2 & 1.62$\pm$0.08 &  \\
Bone\_Vert\_T8  & 91.6$\pm$13.0 & 0.93$\pm$0.16 &  & 91.9$\pm$13.0 & 1.00$\pm$0.17 &  \\
Bone\_Vert\_T9  & 94.3$\pm$14.4 & 0.42$\pm$0.24 &  & 93.3$\pm$14.4 & 0.46$\pm$0.25 &  \\
Bone\_Vert\_T10  & 97.5$\pm$13.0 & 0.38$\pm$1.32 &  & 97.4$\pm$13.0 & 0.37$\pm$1.21 &  \\
Bone\_Vert\_T11  & 95.1$\pm$12.4 & 0.22$\pm$1.67 &  & 95.6$\pm$12.4 & 0.23$\pm$1.64 &  \\
Bone\_Vert\_T12  & 97.5$\pm$16.0 & 0.26$\pm$1.35 &  & 97.7$\pm$16.1 & 0.24$\pm$1.26 & \\ \hline

Mean & \multicolumn{1}{c}{ 95.1 } & \multicolumn{1}{c|}{ 0.45 } & \multicolumn{1}{c|}{ 97.8 } & \multicolumn{1}{c}{ 95.1 } & \multicolumn{1}{c|}{ 0.45 } & \multicolumn{1}{c|}{ 97.8 }  \\ \hline
\end{tabular}
% }
\caption{\textbar~\textbf{Performance comparison of \ac{cln_u36} and \ac{cln_u36_ft} on \ac{totalseg} `Vertebrae' subgroup.} Organ-wise DSC (\%), ASD (mm), and decoder-wise pruning rate $\mathcal{T}$ (\%) of \ac{cln_u36} and \ac{cln_u36_ft} are evaluated on \ac{totalseg} `Vertebrae' subgroup. After fine-tuning, \ac{cln_u36_ft} achieves improved mean DSC and ASD, indicating that \ac{cln_u36} can further overfit its general label style learned from $\mathcal{U}_{36}$ datasets to the distinct label style of \ac{totalseg} `Vertebrae' subgroup. }
\label{tab:pl_totalseg_vert_ds36}
\end{table}
\clearpage

\begin{table}[htp]
\centering
\fontsize{10pt}{12pt}\selectfont
% \resizebox{\textwidth}{!}{
\begin{tabular}{|l|ll|r|ll|r|}
\hline
\multicolumn{1}{|l|}{} &  \multicolumn{3}{c|}{\ac{cln_u36}} & \multicolumn{3}{c|}{\ac{cln_u36_ft}} \\ \cline{2-7} 
\multicolumn{1}{|l|}{\multirow{-2}{*}{}} &  DSC$\uparrow$ & ASD$\downarrow$ & \multicolumn{1}{r|}{$\mathcal{T}$$\uparrow$} & DSC$\uparrow$ & ASD$\downarrow$ & \multicolumn{1}{r|}{$\mathcal{T}$$\uparrow$} \\ \hline
\rowcolor[HTML]{F3F3F3} 
BrainStem  & 93.6$\pm$1.7 & 0.76$\pm$0.11 & 99.5 & 93.1$\pm$1.7 & 0.24$\pm$0.12 & 99.0 \\
Eye\_L  & 86.6$\pm$1.4 & 0.69$\pm$0.09 & 99.0 & 93.7$\pm$1.3 & 0.13$\pm$0.08 & 99.0 \\
Eye\_R  & 86.7$\pm$4.3 & 0.57$\pm$0.29 & {} & 95.4$\pm$4.3 & 0.25$\pm$0.29 & {} \\
\rowcolor[HTML]{F3F3F3} 
Lens\_L  & 81.6$\pm$9.9 & 0.56$\pm$0.18 & 99.0 & 82.5$\pm$9.9 & 0.22$\pm$0.19 & 98.0 \\
\rowcolor[HTML]{F3F3F3} 
Lens\_R  & 84.3$\pm$9.4 & 0.37$\pm$0.44 & {\cellcolor[HTML]{F3F3F3}} & 82.1$\pm$9.3 & 0.43$\pm$0.43 & {\cellcolor[HTML]{F3F3F3}} \\
OpticNerve\_L  & 77.8$\pm$8.0 & 0.60$\pm$0.28 & 99.0 & 80.8$\pm$8.0 & 0.33$\pm$0.29 & 99.5 \\
OpticNerve\_R  & 79.3$\pm$6.1 & 0.59$\pm$0.19 & {} & 81.6$\pm$6.2 & 0.19$\pm$0.20 & {} \\
\rowcolor[HTML]{F3F3F3} 
Chiasm  & 74.5$\pm$13.0 & 0.62$\pm$0.14 & 96.0 & 69.1$\pm$13.3 & 0.29$\pm$0.16 & 99.5 \\
TemporalLobe\_L  & 88.2$\pm$1.3 & 1.41$\pm$0.09 & 99.5 & 95.5$\pm$1.1 & 0.20$\pm$0.07 & 99.7 \\
TemporalLobe\_R  & 87.6$\pm$1.7 & 1.73$\pm$0.13 & {} & 95.7$\pm$1.5 & 0.25$\pm$0.10 & {} \\
\rowcolor[HTML]{F3F3F3} 
Glnd\_Pituitary  & 85.2$\pm$6.5 & 0.49$\pm$0.21 & 98.0 & 88.7$\pm$6.3 & 0.15$\pm$0.20 & 99.5 \\
Glnd\_Parotid\_L  & 87.6$\pm$1.7 & 1.06$\pm$0.14 & 99.5 & 95.9$\pm$1.9 & 0.30$\pm$0.16 & 99.5 \\
Glnd\_Parotid\_R  & 86.6$\pm$1.4 & 1.09$\pm$0.51 & {} & 96.1$\pm$1.4 & 0.41$\pm$0.52 & {} \\
\rowcolor[HTML]{F3F3F3} 
Ear\_Inner\_L  & 58.3$\pm$9.5 & 0.60$\pm$0.37 & 99.5 & 65.5$\pm$9.3 & 0.61$\pm$0.35 & 98.0 \\
\rowcolor[HTML]{F3F3F3} 
Ear\_Inner\_R  & 57.1$\pm$9.4 & 0.57$\pm$0.32 & {\cellcolor[HTML]{F3F3F3}} & 65.0$\pm$9.4 & 0.79$\pm$0.32 & {\cellcolor[HTML]{F3F3F3}} \\
Ear\_Mid\_L  & 91.3$\pm$3.7 & 0.57$\pm$0.04 & 99.5 & 90.7$\pm$3.6 & 0.07$\pm$0.02 & 99.7 \\
Ear\_Mid\_R  & 90.4$\pm$3.4 & 0.64$\pm$0.06 & {} & 89.2$\pm$3.3 & 0.07$\pm$0.05 & {} \\
\rowcolor[HTML]{F3F3F3} 
TMJ\_L  & 78.6$\pm$9.4 & 0.65$\pm$0.23 & 94.0 & 78.7$\pm$9.5 & 0.36$\pm$0.23 & 96.0 \\
\rowcolor[HTML]{F3F3F3} 
TMJ\_R  & 76.8$\pm$7.9 & 0.71$\pm$0.23 & {\cellcolor[HTML]{F3F3F3}} & 80.5$\pm$7.5 & 0.37$\pm$0.20 & {\cellcolor[HTML]{F3F3F3}} \\
SpinalCord  & 94.4$\pm$3.0 & 0.67$\pm$0.20 & 99.7 & 94.9$\pm$3.2 & 0.35$\pm$0.22 & 99.0 \\
\rowcolor[HTML]{F3F3F3} 
Bone\_Mandible\_L  & 92.2$\pm$1.8 & 0.43$\pm$0.05 & 99.7 & 97.7$\pm$2.2 & 0.16$\pm$0.08 & 99.7 \\
\rowcolor[HTML]{F3F3F3} 
Bone\_Mandible\_R  & 91.5$\pm$2.4 & 0.47$\pm$0.07 & {\cellcolor[HTML]{F3F3F3}} & 97.7$\pm$2.6 & 0.14$\pm$0.09 & {\cellcolor[HTML]{F3F3F3}}
\\ \hline

Mean & \multicolumn{1}{c}{ 83.2 } & \multicolumn{1}{c|}{ 0.72 } & \multicolumn{1}{c|}{ 98.6 } & \multicolumn{1}{c}{ 86.8 } & \multicolumn{1}{c|}{ 0.29 } & \multicolumn{1}{c|}{ 98.9 }  \\ \hline

\end{tabular}
% }
\caption{\textbar~\textbf{Performance comparison of \ac{cln_u36} and \ac{cln_u36_ft} on \ac{structseg}.} Organ-wise DSC (\%), ASD (mm), and decoder-wise pruning rate $\mathcal{T}$ (\%) of \ac{cln_u36} and \ac{cln_u36_ft} are evaluated on \ac{structseg}. After fine-tuning, \ac{cln_u36_ft} achieves improved mean DSC, ASD and pruning rate, indicating that \ac{cln_u36} can further overfit its general label style learned from $\mathcal{U}_{36}$ datasets to the distinct label style of \ac{structseg}. }
\label{tab:pl_structseg_ds36}
\end{table}
\clearpage
% \begin{table}[]
% \centering
% \resizebox{\textwidth}{!}{
% \begin{tabular}{|l|ll|ll|r|ll|r|}
% \hline
% \multicolumn{1}{|l|}{} & \multicolumn{2}{c|}{nnUNet$^{E_{36}}$} & \multicolumn{3}{c|}{SUNSeg$^{\mathcal{U}_{36}}$} & \multicolumn{3}{c|}{SUNSeg$^{\mathcal{U}_{36}}_{\text{fine-tune}}$} \\ \cline{2-9} 
% \multicolumn{1}{|l|}{\multirow{-2}{*}{}} & DSC$\uparrow$ & \multicolumn{1}{l|}{ASD$\downarrow$} & DSC$\uparrow$ & ASD$\downarrow$ & \multicolumn{1}{r|}{$\mathcal{T}$$\uparrow$} & DSC$\uparrow$ & ASD$\downarrow$ & \multicolumn{1}{r|}{$\mathcal{T}$$\uparrow$} \\ \hline
% \rowcolor[HTML]{F3F3F3} 
% Eso & 86.0$\pm$3.2 & 0.66$\pm$0.16 & 89.9$\pm$2.8 & 0.53$\pm$0.12 & 98.0 & 90.5$\pm$2.9 & 0.32$\pm$0.13 & 99.5 \\
% Heart & 94.2$\pm$14.6 & 1.62$\pm$0.18 & 91.4$\pm$14.8 & 1.48$\pm$0.19 & 99.0 & 94.6$\pm$14.6 & 0.37$\pm$0.18 & 99.0 \\
% \rowcolor[HTML]{F3F3F3} 
% Trachea & 90.4$\pm$6.0 & 0.78$\pm$0.11 & 92.7$\pm$6.1 & 0.68$\pm$0.12 & 99.7 & 92.9$\pm$5.8 & 0.31$\pm$0.09 & 99.5 \\
% A\_Aorta & 94.3$\pm$16.5 & 0.90$\pm$0.82 & 95.3$\pm$16.8 & 0.75$\pm$0.86 & 99.0 & 95.9$\pm$16.7 & 0.34$\pm$0.85 & 99.5
% \\ \hline
% \end{tabular}
% }
% \caption{\ac{pl} performance details on $\mathcal{U}_{36}$ datasets: SegTHOR}
% \label{tab:pl_segthor_ds36}
% \end{table}

\begin{table}[htp]
\centering
\fontsize{10pt}{12pt}\selectfont
% \resizebox{\textwidth}{!}{
\begin{tabular}{|l|ll|r|ll|r|}
\hline
\multicolumn{1}{|l|}{} &  \multicolumn{3}{c|}{\ac{cln_u36}} & \multicolumn{3}{c|}{\ac{cln_u36_ft}} \\ \cline{2-7} 
\multicolumn{1}{|l|}{\multirow{-2}{*}{}} &  DSC$\uparrow$ & ASD$\downarrow$ & \multicolumn{1}{r|}{$\mathcal{T}$$\uparrow$} & DSC$\uparrow$ & ASD$\downarrow$ & \multicolumn{1}{r|}{$\mathcal{T}$$\uparrow$} \\ \hline
\rowcolor[HTML]{F3F3F3} 
Eso  & 89.9$\pm$2.8 & 0.53$\pm$0.12 & 98.0 & 90.5$\pm$2.9 & 0.32$\pm$0.13 & 99.5 \\
Heart  & 91.4$\pm$14.8 & 1.48$\pm$0.19 & 99.0 & 94.6$\pm$14.6 & 0.37$\pm$0.18 & 99.0 \\
\rowcolor[HTML]{F3F3F3} 
Trachea  & 92.7$\pm$6.1 & 0.68$\pm$0.12 & 99.7 & 92.9$\pm$5.8 & 0.31$\pm$0.09 & 99.5 \\
A\_Aorta  & 95.3$\pm$16.8 & 0.75$\pm$0.86 & 99.0 & 95.9$\pm$16.7 & 0.34$\pm$0.85 & 99.5
\\ \hline

Mean & \multicolumn{1}{c}{ 92.3 } & \multicolumn{1}{c|}{ 0.86 } & \multicolumn{1}{c|}{ 98.9 } & \multicolumn{1}{c}{ 93.5 } & \multicolumn{1}{c|}{ 0.34 } & \multicolumn{1}{c|}{ 99.4 }  \\ \hline
\end{tabular}
% }
\caption{\textbar~\textbf{Performance comparison of \ac{cln_u36} and \ac{cln_u36_ft} on \ac{segthor}.} Organ-wise DSC (\%), ASD (mm), and decoder-wise pruning rate $\mathcal{T}$ (\%) of \ac{cln_u36} and \ac{cln_u36_ft} are evaluated on \ac{segthor}. After fine-tuning, \ac{cln_u36_ft} achieves improved mean DSC, ASD and pruning rate, indicating that \ac{cln_u36} can further overfit its general label style learned from $\mathcal{U}_{36}$ datasets to the distinct label style of \ac{segthor}. }
\label{tab:pl_segthor_ds36}
\end{table}
\clearpage

\begin{table}[htp]
\centering
\fontsize{10pt}{12pt}\selectfont
% \resizebox{0.9\textwidth}{!}{
\begin{tabular}{|l|ll|r|ll|r|}
\hline
\multicolumn{1}{|l|}{} &  \multicolumn{3}{c|}{\ac{cln_u36}} & \multicolumn{3}{c|}{\ac{cln_u36_ft}} \\ \cline{2-7} 
\multicolumn{1}{|l|}{\multirow{-2}{*}{}} &  DSC$\uparrow$ & ASD$\downarrow$ & \multicolumn{1}{r|}{$\mathcal{T}$$\uparrow$} & DSC$\uparrow$ & ASD$\downarrow$ & \multicolumn{1}{r|}{$\mathcal{T}$$\uparrow$} \\ \hline
\rowcolor[HTML]{F3F3F3} 
Liver  & 93.7$\pm$5.9 & 0.81$\pm$0.95 & 99.5 & 97.9$\pm$5.6 & 0.79$\pm$0.99 & 99.5 \\
Kidney\_R  & 95.0$\pm$7.0 & 1.78$\pm$0.75 & 99.7 & 94.4$\pm$7.0 & 1.36$\pm$0.75 & 99.7 \\
Kidney\_L  & 91.1$\pm$6.7 & 1.29$\pm$0.70 & {} & 94.3$\pm$6.9 & 1.06$\pm$0.72 & {} \\
\rowcolor[HTML]{F3F3F3} 
Spleen  & 98.0$\pm$6.7 & 0.32$\pm$0.39 & 99.5 & 98.1$\pm$6.9 & 0.20$\pm$0.40 & 99.5 \\
Pancreas  & 94.9$\pm$7.0 & 0.86$\pm$0.38 & 99.5 & 92.6$\pm$7.0 & 0.74$\pm$0.38 & 99.5 \\
\rowcolor[HTML]{F3F3F3} 
A\_Aorta  & 92.2$\pm$16.7 & 0.23$\pm$0.85 & 99.0 & 97.1$\pm$16.6 & 0.18$\pm$0.84 & 99.5 \\
V\_VenaCava\_I  & 78.5$\pm$10.0 & 1.17$\pm$1.02 & 99.0 & 88.8$\pm$9.9 & 1.08$\pm$0.97 & 99.0 \\
\rowcolor[HTML]{F3F3F3} 
Glnd\_Adrenal\_R  & 90.3$\pm$13.4 & 0.43$\pm$0.41 & 98.0 & 89.0$\pm$13.2 & 0.33$\pm$0.39 & 99.0 \\
\rowcolor[HTML]{F3F3F3} 
Glnd\_Adrenal\_L  & 87.6$\pm$12.5 & 0.28$\pm$0.39 & {\cellcolor[HTML]{F3F3F3}} & 89.9$\pm$12.6 & 0.22$\pm$0.39 & {\cellcolor[HTML]{F3F3F3}} \\
GallBladder  & 85.5$\pm$8.4 & 2.15$\pm$0.80 & 99.5 & 83.2$\pm$8.5 & 2.27$\pm$0.82 & 99.5 \\
\rowcolor[HTML]{F3F3F3} 
Eso  & 81.8$\pm$3.2 & 1.66$\pm$0.16 & 98.0 & 86.1$\pm$2.8 & 1.50$\pm$0.13 & 99.5 \\
Stomach  & 92.3$\pm$6.5 & 1.17$\pm$0.68 & 99.5 & 92.5$\pm$6.2 & 1.05$\pm$0.65 & 99.5 \\
\rowcolor[HTML]{F3F3F3} 
Duodenum  & 82.6$\pm$15.9 & 2.60$\pm$1.26 & 99.5 & 83.2$\pm$15.8 & 2.36$\pm$1.19 & 99.5
\\ \hline

Mean & \multicolumn{1}{c}{ 89.5 } & \multicolumn{1}{c|}{ 1.13 } & \multicolumn{1}{c|}{ 99.2 } & \multicolumn{1}{c}{ 91.3 } & \multicolumn{1}{c|}{ 1.01 } & \multicolumn{1}{c|}{ 99.4 }  \\ \hline
\end{tabular}
% }
\caption{\textbar~\textbf{Performance comparison of \ac{cln_u36} and \ac{cln_u36_ft} on \ac{flare}.} Organ-wise DSC (\%), ASD (mm), and decoder-wise pruning rate $\mathcal{T}$ (\%) of \ac{cln_u36} and \ac{cln_u36_ft} are evaluated on \ac{flare}. After fine-tuning, \ac{cln_u36_ft} achieves improved mean DSC, ASD and pruning rate, indicating that \ac{cln_u36} can further overfit its general label style learned from $\mathcal{U}_{36}$ datasets to the distinct label style of \ac{flare}. }
\label{tab:pl_flare_ds36}
\end{table}
\clearpage
% \begin{table}[]
% \centering
% \resizebox{\textwidth}{!}{
% \begin{tabular}{|l|ll|ll|r|ll|r|}
% \hline
% \multicolumn{1}{|l|}{} & \multicolumn{2}{c|}{nnUNet$^{E_{36}}$} & \multicolumn{3}{c|}{SUNSeg$^{\mathcal{U}_{36}}$} & \multicolumn{3}{c|}{SUNSeg$^{\mathcal{U}_{36}}_{\text{fine-tune}}$} \\ \cline{2-9} 
% \multicolumn{1}{|l|}{\multirow{-2}{*}{}} & DSC$\uparrow$ & \multicolumn{1}{l|}{ASD$\downarrow$} & DSC$\uparrow$ & ASD$\downarrow$ & \multicolumn{1}{r|}{$\mathcal{T}$$\uparrow$} & DSC$\uparrow$ & ASD$\downarrow$ & \multicolumn{1}{r|}{$\mathcal{T}$$\uparrow$} \\ \hline

% Kidney GTV & 83.7$\pm$9.2 & 1.10$\pm$1.08 & 83.9$\pm$9.0 & 1.18$\pm$1.04 & 90.0 & 87.1$\pm$9.4 & 1.28$\pm$1.05 & 96.0
% \\ \hline
% \end{tabular}
% }
% \caption{\ac{pl} performance details on $\mathcal{U}_{36}$ datasets: KiTS19}
% \label{tab:pl_kits_ds36}
% \end{table}

\begin{table}[htp]
\centering
% \resizebox{0.9\textwidth}{!}{
\fontsize{10pt}{12pt}\selectfont
\begin{tabular}{|l|ll|r|ll|r|}
\hline
\multicolumn{1}{|l|}{} &  \multicolumn{3}{c|}{\ac{cln_u36}} & \multicolumn{3}{c|}{\ac{cln_u36_ft}} \\ \cline{2-7} 
\multicolumn{1}{|l|}{\multirow{-2}{*}{}} &  DSC$\uparrow$ & ASD$\downarrow$ & \multicolumn{1}{r|}{$\mathcal{T}$$\uparrow$} & DSC$\uparrow$ & ASD$\downarrow$ & \multicolumn{1}{r|}{$\mathcal{T}$$\uparrow$} \\ \hline

Kidney\_\Ac{gtv}  & 83.9$\pm$9.0 & 1.18$\pm$1.04 & 90.0 & 87.1$\pm$9.4 & 1.28$\pm$1.05 & 96.0
\\ \hline
\end{tabular}
% }
\caption{\textbar~\textbf{Performance comparison of \ac{cln_u36} and \ac{cln_u36_ft} on \ac{kits}.} Organ-wise DSC (\%), ASD (mm), and decoder-wise pruning rate $\mathcal{T}$ (\%) of \ac{cln_u36} and \ac{cln_u36_ft} are evaluated on \ac{kits}. After fine-tuning, \ac{cln_u36_ft} achieves improved DSC and pruning rate, indicating that \ac{cln_u36} can further overfit its general label style learned from $\mathcal{U}_{36}$ datasets to the distinct label style of \ac{kits}. }
\label{tab:pl_kits_ds36}
\end{table}
\clearpage

%% CSS order-wise results on 5 datasets
%% dataset-wise results on 5-dataset validation setting
\begin{table}[htp]
\centering
\resizebox{\textwidth}{!}{
\begin{tabular}{|l|ll|ll|c|ll|c|ll|c|ll|c|}
\hline
\multicolumn{1}{|l|}{} & \multicolumn{2}{c|}{nnUNet} & \multicolumn{3}{c|}{Order 1} & \multicolumn{3}{c|}{Order 2} & \multicolumn{3}{c|}{Order 3} & \multicolumn{3}{c|}{Order 4} \\ \cline{2-15} 
\multicolumn{1}{|l|}{\multirow{-2}{*}{}} & DSC$\uparrow$  & ASD$\downarrow$ & DSC$\uparrow$  & ASD$\downarrow$ & $\mathcal{T}$$\uparrow$  & DSC$\uparrow$  & ASD$\downarrow$ & $\mathcal{T}$$\uparrow$  & DSC$\uparrow$  & ASD$\downarrow$ & $\mathcal{T}$$\uparrow$  & DSC$\uparrow$  & ASD$\downarrow$ & $\mathcal{T}$$\uparrow$  \\ \hline
\rowcolor[HTML]{F3F3F3} 
Spleen & 97.6$\pm$15.6 & 1.95$\pm$0.40 & 97.6$\pm$6.8 & 1.87$\pm$0.40 & 99.5 & 98.4$\pm$6.9 & 1.81$\pm$0.41 & 99.7 & 97.1$\pm$7.0 & 1.87$\pm$0.41 & 99.0 & 97.6$\pm$6.8 & 1.87$\pm$0.40 & 99.5 \\
Kidney\_R & 95.9$\pm$16.9 & 0.19$\pm$0.73 & 96.8$\pm$7.0 & 0.25$\pm$0.75 & 99.7 & 95.4$\pm$6.9 & 0.17$\pm$0.74 & 99.7 & 96.0$\pm$7.1 & 0.22$\pm$0.76 & 99.0 & 96.8$\pm$7.0 & 0.25$\pm$0.75 & 99.7 \\
Kidney\_L & 96.1$\pm$26.7 & 1.15$\pm$6.45 & 95.9$\pm$6.8 & 1.07$\pm$0.71 & {} & 94.3$\pm$6.8 & 1.17$\pm$0.71 & {} & 97.2$\pm$6.6 & 1.10$\pm$0.69 & {} & 95.9$\pm$6.8 & 1.07$\pm$0.71 & {} \\
\rowcolor[HTML]{F3F3F3} 
GallBladder & 92.2$\pm$24.4 & 1.14$\pm$0.82 & 93.1$\pm$8.7 & 1.05$\pm$0.83 & 99.7 & 90.2$\pm$8.4 & 1.23$\pm$0.80 & 99.7 & 93.0$\pm$8.4 & 1.22$\pm$0.80 & 99.0 & 93.1$\pm$8.7 & 1.05$\pm$0.83 & 99.7 \\
Liver & 97.3$\pm$14.9 & 0.57$\pm$0.93 & 97.3$\pm$5.7 & 0.61$\pm$0.99 & 99.7 & 97.6$\pm$5.9 & 0.57$\pm$0.91 & 99.5 & 97.6$\pm$5.7 & 0.62$\pm$0.99 & 99.0 & 97.3$\pm$5.7 & 0.61$\pm$0.99 & 99.7 \\
\rowcolor[HTML]{F3F3F3} 
Stomach & 95.9$\pm$16.2 & 1.18$\pm$0.66 & 95.2$\pm$6.4 & 1.22$\pm$0.68 & 99.5 & 97.0$\pm$6.5 & 1.20$\pm$0.68 & 99.5 & 96.3$\pm$6.4 & 1.24$\pm$0.67 & 99.0 & 95.2$\pm$6.4 & 1.22$\pm$0.68 & 99.5 \\
Pancreas & 91.8$\pm$17.0 & 1.06$\pm$0.67 & 92.7$\pm$7.0 & 0.78$\pm$0.38 & 99.7 & 92.7$\pm$6.9 & 1.09$\pm$0.36 & 99.5 & 91.9$\pm$6.9 & 1.04$\pm$0.37 & 99.0 & 92.7$\pm$7.0 & 0.78$\pm$0.38 & 99.7 \\
\rowcolor[HTML]{F3F3F3} 
Glnd\_Adrenal\_R & 83.0$\pm$13.5 & 0.55$\pm$1.17 & 82.1$\pm$13.2 & 0.42$\pm$0.39 & 99.0 & 83.2$\pm$13.4 & 0.59$\pm$0.41 & 94.0 & 82.9$\pm$13.5 & 0.43$\pm$0.42 & 96.0 & 82.1$\pm$13.2 & 0.42$\pm$0.39 & 99.0 \\
\rowcolor[HTML]{F3F3F3} 
Glnd\_Adrenal\_L & 84.2$\pm$12.2 & 1.03$\pm$0.36 & 85.1$\pm$12.5 & 0.96$\pm$0.38 & {\cellcolor[HTML]{F3F3F3}} & 84.4$\pm$12.3 & 0.93$\pm$0.37 & {\cellcolor[HTML]{F3F3F3}} & 84.4$\pm$12.3 & 1.01$\pm$0.36 & {\cellcolor[HTML]{F3F3F3}} & 85.1$\pm$12.5 & 0.96$\pm$0.38 & {\cellcolor[HTML]{F3F3F3}} \\
Lung\_LUL & 97.8$\pm$3.6 & 0.31$\pm$0.93 & 97.7$\pm$4.0 & 0.19$\pm$0.64 & 99.5 & 99.3$\pm$3.8 & 0.34$\pm$0.63 & 99.5 & 98.0$\pm$3.7 & 0.27$\pm$0.62 & 99.0 & 97.7$\pm$4.0 & 0.19$\pm$0.64 & 99.5 \\
Lung\_LLL & 96.5$\pm$24.0 & 0.21$\pm$0.98 & 96.7$\pm$2.2 & 0.19$\pm$0.48 & {} & 97.1$\pm$2.3 & 0.22$\pm$0.50 & {} & 96.4$\pm$2.1 & 0.22$\pm$0.47 & {} & 96.7$\pm$2.2 & 0.19$\pm$0.48 & {} \\
\rowcolor[HTML]{F3F3F3} 
Lung\_RUL & 97.5$\pm$20.5 & 0.30$\pm$1.48 & 97.5$\pm$2.5 & 0.22$\pm$0.99 & 99.5 & 98.4$\pm$2.4 & 0.32$\pm$0.98 & 99.0 & 97.8$\pm$2.7 & 0.22$\pm$0.96 & 99.0 & 97.5$\pm$2.5 & 0.22$\pm$0.99 & 99.5 \\
\rowcolor[HTML]{F3F3F3} 
Lung\_RML & 92.0$\pm$6.2 & 0.47$\pm$1.41 & 92.0$\pm$3.3 & 0.21$\pm$1.45 & {\cellcolor[HTML]{F3F3F3}} & 92.0$\pm$3.0 & 0.48$\pm$1.47 & {\cellcolor[HTML]{F3F3F3}} & 92.1$\pm$3.2 & 0.21$\pm$1.40 & {\cellcolor[HTML]{F3F3F3}} & 92.0$\pm$3.3 & 0.21$\pm$1.45 & {\cellcolor[HTML]{F3F3F3}} \\
\rowcolor[HTML]{F3F3F3} 
Lung\_RLL & 97.4$\pm$26.8 & 0.16$\pm$2.45 & 97.5$\pm$3.6 & 0.22$\pm$1.21 & {\cellcolor[HTML]{F3F3F3}} & 96.9$\pm$3.7 & 0.15$\pm$1.18 & {\cellcolor[HTML]{F3F3F3}} & 97.2$\pm$3.8 & 0.26$\pm$1.22 & {\cellcolor[HTML]{F3F3F3}} & 97.5$\pm$3.6 & 0.22$\pm$1.21 & {\cellcolor[HTML]{F3F3F3}} \\
Eso & 94.8$\pm$2.9 & 0.61$\pm$0.15 & 94.4$\pm$2.9 & 0.50$\pm$0.13 & 99.7 & 95.4$\pm$2.9 & 0.64$\pm$0.13 & 99.5 & 94.2$\pm$3.0 & 0.66$\pm$0.14 & 98.0 & 94.4$\pm$2.9 & 0.50$\pm$0.13 & 99.7 \\
\rowcolor[HTML]{F3F3F3} 
Trachea & 96.7$\pm$5.8 & 0.36$\pm$0.12 & 96.9$\pm$6.1 & 0.30$\pm$0.12 & 99.5 & 95.9$\pm$5.9 & 0.33$\pm$0.10 & 99.0 & 97.2$\pm$6.0 & 0.30$\pm$0.11 & 99.0 & 96.9$\pm$6.1 & 0.30$\pm$0.12 & 99.5 \\
Glnd\_Thyroid & 91.7$\pm$14.7 & 0.41$\pm$0.96 & 92.4$\pm$14.8 & 0.32$\pm$0.99 & 99.0 & 90.6$\pm$14.6 & 0.40$\pm$0.97 & 99.0 & 92.5$\pm$14.8 & 0.42$\pm$0.99 & 99.7 & 92.4$\pm$14.8 & 0.32$\pm$0.99 & 99.0 \\
\rowcolor[HTML]{F3F3F3} 
SmallBowel & 91.2$\pm$26.4 & 1.63$\pm$2.53 & 90.7$\pm$16.5 & 1.55$\pm$2.77 & 99.7 & 93.0$\pm$16.2 & 1.66$\pm$2.61 & 99.5 & 91.1$\pm$16.3 & 1.82$\pm$2.71 & 98.0 & 90.7$\pm$16.5 & 1.55$\pm$2.77 & 99.7 \\
Duodenum & 87.3$\pm$26.0 & 1.87$\pm$1.26 & 87.0$\pm$16.2 & 1.94$\pm$1.20 & 99.0 & 87.0$\pm$16.0 & 2.04$\pm$1.26 & 98.0 & 87.4$\pm$15.9 & 2.29$\pm$1.20 & 98.0 & 87.0$\pm$16.2 & 1.94$\pm$1.20 & 99.0 \\
\rowcolor[HTML]{F3F3F3} 
Colon & 90.2$\pm$16.0 & 2.54$\pm$1.21 & 89.7$\pm$16.2 & 2.28$\pm$1.26 & 99.7 & 92.7$\pm$16.1 & 2.50$\pm$1.29 & 99.7 & 90.3$\pm$15.8 & 2.66$\pm$1.17 & 98.0 & 89.7$\pm$16.2 & 2.28$\pm$1.26 & 99.7 \\
UrinaryBladder & 89.1$\pm$9.5 & 2.98$\pm$1.08 & 89.2$\pm$9.4 & 2.91$\pm$1.04 & 99.0 & 87.9$\pm$9.8 & 2.99$\pm$1.05 & 99.5 & 89.2$\pm$9.6 & 2.84$\pm$1.00 & 98.0 & 89.2$\pm$9.4 & 2.91$\pm$1.04 & 99.0 \\
\rowcolor[HTML]{F3F3F3} 
Prostate & 81.7$\pm$38.0 & 0.52$\pm$2.06 & 81.7$\pm$17.8 & 0.36$\pm$1.42 & 94.0 & 83.4$\pm$17.8 & 0.54$\pm$1.43 & 94.0 & 81.5$\pm$18.0 & 0.36$\pm$1.53 & 96.0 & 81.7$\pm$17.8 & 0.36$\pm$1.42 & 94.0 \\ \hline

Mean & \multicolumn{1}{c}{92.6} & \multicolumn{1}{c|}{0.96} & \multicolumn{1}{c}{92.7} & \multicolumn{1}{c|}{0.88} & 99.1& \multicolumn{1}{c}{92.9} & \multicolumn{1}{c|}{0.97} & 98.7 & \multicolumn{1}{c}{92.8} & \multicolumn{1}{c|}{0.97} & 98.4 & \multicolumn{1}{c}{92.7} & \multicolumn{1}{c|}{0.88} & 99.1\\ \hline
\end{tabular}}
\caption{\textbar~\textbf{\Ac{css} order-wise performance details of \ac{cln_c5} on \ac{totalseg} `Organ' subgroup.} Organ-wise DSC (\%), ASD (mm), and decoder-wise pruning rate $\mathcal{T}$ (\%) of \ac{cln_c5} on \ac{totalseg} `Organ' subgroup are evaluated across all CSS orders. Notably, \ac{cln_c5} achieves mean DSC and ASD comparable to nnUNet upper bound and the average decoder pruning rates exceed 98\% across all orders, demonstrating high efficiency without compromising segmentation performance. }
\label{tab:css_totalseg_organ}
\end{table}
\clearpage

\begin{table}[htp]
\centering
\resizebox{\textwidth}{!}{
\begin{tabular}{|l|ll|ll|c|ll|c|ll|c|ll|c|}
\hline
\multicolumn{1}{|l|}{} & \multicolumn{2}{c|}{nnUNet} & \multicolumn{3}{c|}{Order 1} & \multicolumn{3}{c|}{Order 2} & \multicolumn{3}{c|}{Order 3} & \multicolumn{3}{c|}{Order 4} \\ \cline{2-15} 
\multicolumn{1}{|l|}{\multirow{-2}{*}{}} & DSC$\uparrow$  & ASD$\downarrow$ & DSC$\uparrow$  & ASD$\downarrow$ & $\mathcal{T}$$\uparrow$  & DSC$\uparrow$  & ASD$\downarrow$ & $\mathcal{T}$$\uparrow$  & DSC$\uparrow$  & ASD$\downarrow$ & $\mathcal{T}$$\uparrow$  & DSC$\uparrow$  & ASD$\downarrow$ & $\mathcal{T}$$\uparrow$  \\ \hline
\rowcolor[HTML]{F3F3F3} 
Heart & 95.2$\pm$14.7 & 0.42$\pm$0.20 & 95.1$\pm$14.6 & 0.44$\pm$0.17 & 99.5 & 96.0$\pm$14.8 & 0.40$\pm$0.19 & 99.5 & 94.8$\pm$14.6 & 0.54$\pm$0.18 & 99.0 & 95.1$\pm$14.6 & 0.44$\pm$0.17 & 99.5 \\
A\_Aorta & 96.0$\pm$16.7 & 0.44$\pm$0.95 & 96.6$\pm$16.5 & 0.43$\pm$0.82 & 99.7 & 93.7$\pm$16.7 & 0.45$\pm$0.85 & 99.0 & 96.4$\pm$16.7 & 0.46$\pm$0.85 & 99.0 & 96.6$\pm$16.5 & 0.43$\pm$0.82 & 99.7 \\
\rowcolor[HTML]{F3F3F3} 
V\_Pulmonary & 89.8$\pm$15.5 & 0.47$\pm$0.14 & 89.2$\pm$15.4 & 0.36$\pm$0.14 & 99.7 & 89.0$\pm$15.5 & 0.44$\pm$0.15 & 99.0 & 89.4$\pm$15.5 & 0.43$\pm$0.16 & 99.0 & 89.2$\pm$15.4 & 0.36$\pm$0.14 & 99.7 \\
A\_BrachiocephalicTrunk & 89.8$\pm$16.2 & 0.32$\pm$0.43 & 89.8$\pm$16.1 & 0.22$\pm$0.42 & 99.7 & 89.5$\pm$16.1 & 0.32$\pm$0.43 & 99.0 & 89.2$\pm$16.0 & 0.23$\pm$0.42 & 99.0 & 89.8$\pm$16.1 & 0.22$\pm$0.42 & 99.7 \\
\rowcolor[HTML]{F3F3F3} 
A\_Subclavian\_L & 94.6$\pm$3.5 & 0.44$\pm$0.28 & 94.9$\pm$3.4 & 0.29$\pm$0.27 & 99.7 & 94.6$\pm$3.3 & 0.47$\pm$0.26 & 98.0 & 94.8$\pm$3.5 & 0.59$\pm$0.28 & 98.0 & 94.9$\pm$3.4 & 0.29$\pm$0.27 & 99.7 \\
\rowcolor[HTML]{F3F3F3} 
A\_Subclavian\_R & 94.1$\pm$3.3 & 0.23$\pm$0.17 & 93.7$\pm$3.4 & 0.19$\pm$0.18 & {\cellcolor[HTML]{F3F3F3}} & 95.5$\pm$3.2 & 0.25$\pm$0.16 & {\cellcolor[HTML]{F3F3F3}} & 93.1$\pm$3.3 & 0.20$\pm$0.17 & {\cellcolor[HTML]{F3F3F3}} & 93.7$\pm$3.4 & 0.19$\pm$0.18 & {\cellcolor[HTML]{F3F3F3}} \\
A\_CommonCarotid\_L & 91.2$\pm$5.4 & 0.35$\pm$0.28 & 91.6$\pm$5.7 & 0.25$\pm$0.31 & 99.7 & 90.4$\pm$5.4 & 0.38$\pm$0.28 & 99.0 & 90.9$\pm$5.4 & 0.24$\pm$0.28 & 99.0 & 91.6$\pm$5.7 & 0.25$\pm$0.31 & 99.7 \\
A\_CommonCarotid\_R & 96.6$\pm$3.3 & 0.28$\pm$0.99 & 96.9$\pm$3.1 & 0.25$\pm$1.00 & {\cellcolor[HTML]{F3F3F3}} & 97.2$\pm$3.3 & 0.28$\pm$1.07 & {\cellcolor[HTML]{F3F3F3}} & 97.4$\pm$3.3 & 0.32$\pm$1.04 & {\cellcolor[HTML]{F3F3F3}} & 96.9$\pm$3.1 & 0.25$\pm$1.00 & {\cellcolor[HTML]{F3F3F3}} \\
\rowcolor[HTML]{F3F3F3} 
V\_Brachiocephalic\_L & 90.4$\pm$26.2 & 0.37$\pm$1.97 & 91.2$\pm$15.9 & 0.34$\pm$1.94 & 99.5 & 89.6$\pm$15.8 & 0.34$\pm$1.96 & 98.0 & 89.4$\pm$15.9 & 0.31$\pm$1.92 & 99.0 & 91.2$\pm$15.9 & 0.34$\pm$1.94 & 99.5 \\
\rowcolor[HTML]{F3F3F3} 
V\_Brachiocephalic\_R & 88.7$\pm$15.0 & 0.32$\pm$0.81 & 88.1$\pm$15.1 & 0.24$\pm$0.82 & {\cellcolor[HTML]{F3F3F3}} & 88.9$\pm$15.2 & 0.35$\pm$0.83 & {\cellcolor[HTML]{F3F3F3}} & 88.6$\pm$15.2 & 0.38$\pm$0.83 & {\cellcolor[HTML]{F3F3F3}} & 88.1$\pm$15.1 & 0.24$\pm$0.82 & {\cellcolor[HTML]{F3F3F3}} \\
LAA & 92.6$\pm$3.4 & 0.46$\pm$0.15 & 92.6$\pm$3.6 & 0.18$\pm$0.18 & 99.7 & 92.1$\pm$3.3 & 0.41$\pm$0.15 & 99.0 & 91.6$\pm$3.4 & 0.49$\pm$0.16 & 99.0 & 92.6$\pm$3.6 & 0.18$\pm$0.18 & 99.7 \\
\rowcolor[HTML]{F3F3F3} 
V\_VenaCava\_S & 93.8$\pm$21.4 & 0.23$\pm$2.35 & 93.8$\pm$21.4 & 0.19$\pm$2.31 & 99.7 & 95.6$\pm$21.4 & 0.23$\pm$2.47 & 99.0 & 93.6$\pm$21.5 & 0.16$\pm$2.49 & 98.0 & 93.8$\pm$21.4 & 0.19$\pm$2.31 & 99.7 \\
V\_VenaCava\_I & 93.9$\pm$9.8 & 1.03$\pm$1.05 & 94.5$\pm$9.9 & 0.87$\pm$0.96 & 99.5 & 95.5$\pm$10.1 & 1.13$\pm$1.04 & 99.0 & 95.1$\pm$10.0 & 0.88$\pm$0.95 & 99.0 & 94.5$\pm$9.9 & 0.87$\pm$0.96 & 99.5 \\
\rowcolor[HTML]{F3F3F3} 
V\_Portal\_Splenic & 85.5$\pm$29.3 & 1.29$\pm$0.62 & 86.1$\pm$29.3 & 1.28$\pm$0.65 & 99.0 & 86.8$\pm$29.0 & 1.27$\pm$0.62 & 98.0 & 85.7$\pm$29.3 & 1.28$\pm$0.64 & 98.0 & 86.1$\pm$29.3 & 1.28$\pm$0.65 & 99.0 \\
A\_Iliac\_L & 94.4$\pm$3.1 & 0.61$\pm$0.19 & 94.1$\pm$3.1 & 0.52$\pm$0.18 & 99.0 & 94.5$\pm$3.3 & 0.59$\pm$0.20 & 99.0 & 94.9$\pm$3.2 & 0.62$\pm$0.20 & 99.0 & 94.1$\pm$3.1 & 0.52$\pm$0.18 & 99.0 \\
A\_Iliac\_R & 91.9$\pm$19.5 & 0.45$\pm$0.15 & 91.0$\pm$19.5 & 0.40$\pm$0.18 & {\cellcolor[HTML]{F3F3F3}} & 93.2$\pm$19.4 & 0.45$\pm$0.16 & {\cellcolor[HTML]{F3F3F3}} & 91.7$\pm$19.3 & 0.45$\pm$0.16 & {\cellcolor[HTML]{F3F3F3}} & 91.0$\pm$19.5 & 0.40$\pm$0.18 & {\cellcolor[HTML]{F3F3F3}} \\
\rowcolor[HTML]{F3F3F3} 
V\_Iliac\_L & 92.2$\pm$3.0 & 0.48$\pm$0.13 & 91.7$\pm$3.4 & 0.47$\pm$0.16 & 99.7 & 91.7$\pm$3.1 & 0.47$\pm$0.14 & 99.7 & 92.8$\pm$3.3 & 0.59$\pm$0.15 & 99.0 & 91.7$\pm$3.4 & 0.47$\pm$0.16 & 99.7 \\
\rowcolor[HTML]{F3F3F3} 
V\_Iliac\_R & 94.2$\pm$4.9 & 0.44$\pm$0.31 & 93.4$\pm$5.1 & 0.46$\pm$0.30 & {\cellcolor[HTML]{F3F3F3}} & 95.3$\pm$4.8 & 0.43$\pm$0.27 & {\cellcolor[HTML]{F3F3F3}} & 93.5$\pm$4.9 & 0.69$\pm$0.28 & {\cellcolor[HTML]{F3F3F3}} & 93.4$\pm$5.1 & 0.46$\pm$0.30 & {\cellcolor[HTML]{F3F3F3}}\\ \hline

Mean & \multicolumn{1}{c}{92.5} & \multicolumn{1}{c|}{0.48} & \multicolumn{1}{c}{92.5} & \multicolumn{1}{c|}{0.41} & 99.6& \multicolumn{1}{c}{92.7} & \multicolumn{1}{c|}{0.48} & 98.9 & \multicolumn{1}{c}{92.4} & \multicolumn{1}{c|}{0.49} & 98.8 & \multicolumn{1}{c}{92.5} & \multicolumn{1}{c|}{0.41} & 99.6\\ \hline
\end{tabular}}
\caption{\textbar~\textbf{\Ac{css} order-wise performance details of \ac{cln_c5} on \ac{totalseg} `Cardiac' subgroup.} Organ-wise DSC (\%), ASD (mm), and decoder-wise pruning rate $\mathcal{T}$ (\%) of \ac{cln_c5} on \ac{totalseg} `Cardiac' subgroup are evaluated across all CSS orders. Notably, \ac{cln_c5} achieves mean DSC and ASD comparable to nnUNet upper bound and the average decoder pruning rates exceed 98\% across all orders, demonstrating high efficiency without compromising segmentation performance. }
\label{tab:css_totalseg_cardi}
\end{table}
\clearpage

\begin{table}[htp]
\centering
\resizebox{\textwidth}{!}{
\begin{tabular}{|l|ll|ll|c|ll|c|ll|c|ll|c|}
\hline
\multicolumn{1}{|l|}{} & \multicolumn{2}{c|}{nnUNet} & \multicolumn{3}{c|}{Order 1} & \multicolumn{3}{c|}{Order 2} & \multicolumn{3}{c|}{Order 3} & \multicolumn{3}{c|}{Order 4} \\ \cline{2-15} 
\multicolumn{1}{|l|}{\multirow{-2}{*}{}} & DSC$\uparrow$  & ASD$\downarrow$ & DSC$\uparrow$  & ASD$\downarrow$ & $\mathcal{T}$$\uparrow$  & DSC$\uparrow$  & ASD$\downarrow$ & $\mathcal{T}$$\uparrow$  & DSC$\uparrow$  & ASD$\downarrow$ & $\mathcal{T}$$\uparrow$  & DSC$\uparrow$  & ASD$\downarrow$ & $\mathcal{T}$$\uparrow$  \\ \hline
\rowcolor[HTML]{F3F3F3} 
Bone\_Humerus\_L & 96.9$\pm$17.5 & 2.84$\pm$0.20 & 97.0$\pm$17.7 & 2.80$\pm$0.21 & 99.5 & 96.8$\pm$17.9 & 3.00$\pm$0.23 & 99.5 & 97.1$\pm$17.8 & 2.85$\pm$0.23 & 99.0 & 97.0$\pm$17.7 & 2.80$\pm$0.21 & 99.5 \\
\rowcolor[HTML]{F3F3F3} 
Bone\_Humerus\_R & 96.6$\pm$17.3 & 4.31$\pm$0.17 & 97.2$\pm$17.2 & 4.23$\pm$0.19 & {} & 97.4$\pm$17.2 & 4.44$\pm$0.20 & {} & 97.2$\pm$17.2 & 4.57$\pm$0.19 & {} & 97.2$\pm$17.2 & 4.23$\pm$0.19 & {} \\

Bone\_Scapula\_L & 94.1$\pm$21.1 & 1.16$\pm$0.18 & 94.6$\pm$21.1 & 1.05$\pm$0.15 & 99.7 & 92.7$\pm$21.1 & 1.25$\pm$0.14 & 99.0 & 93.6$\pm$21.4 & 1.16$\pm$0.17 & 99.0 & 94.6$\pm$21.1 & 1.05$\pm$0.15 & 99.7 \\
 
Bone\_Scapula\_R & 95.1$\pm$1.7 & 1.95$\pm$0.07 & 95.2$\pm$1.9 & 1.67$\pm$0.08 & {\cellcolor[HTML]{F3F3F3}} & 93.5$\pm$1.6 & 1.78$\pm$0.05 & {\cellcolor[HTML]{F3F3F3}} & 94.2$\pm$1.6 & 2.12$\pm$0.05 & {\cellcolor[HTML]{F3F3F3}} & 95.2$\pm$1.9 & 1.67$\pm$0.08 & {\cellcolor[HTML]{F3F3F3}} \\
\rowcolor[HTML]{F3F3F3} 
Bone\_Clavicula\_L & 94.5$\pm$0.8 & 0.60$\pm$0.10 & 94.9$\pm$0.8 & 0.46$\pm$0.10 & 99.5 & 95.2$\pm$1.0 & 0.55$\pm$0.12 & 99.0 & 95.0$\pm$0.9 & 0.48$\pm$0.12 & 99.0 & 94.9$\pm$0.8 & 0.46$\pm$0.10 & 99.5 \\
\rowcolor[HTML]{F3F3F3} 
Bone\_Clavicula\_R & 98.4$\pm$16.1 & 1.52$\pm$0.19 & 98.6$\pm$16.1 & 1.43$\pm$0.21 & {\cellcolor[HTML]{F3F3F3}} & 97.5$\pm$16.1 & 1.50$\pm$0.21 & {\cellcolor[HTML]{F3F3F3}} & 98.3$\pm$15.8 & 1.60$\pm$0.18 & {\cellcolor[HTML]{F3F3F3}} & 98.6$\pm$16.1 & 1.43$\pm$0.21 & {\cellcolor[HTML]{F3F3F3}} \\

Bone\_Femur\_L & 97.4$\pm$15.4 & 3.51$\pm$0.29 & 97.4$\pm$0.5 & 3.39$\pm$0.09 & 99.5 & 98.7$\pm$0.6 & 3.82$\pm$0.10 & 99.0 & 97.5$\pm$0.4 & 3.17$\pm$0.08 & 99.0 & 97.4$\pm$0.5 & 3.39$\pm$0.09 & 99.5 \\

Bone\_Femur\_R & 95.8$\pm$0.5 & 6.85$\pm$0.08 & 96.3$\pm$0.3 & 6.22$\pm$0.06 & {\cellcolor[HTML]{F3F3F3}} & 96.0$\pm$0.5 & 6.53$\pm$0.09 & {\cellcolor[HTML]{F3F3F3}} & 96.1$\pm$0.3 & 7.59$\pm$0.06 & {\cellcolor[HTML]{F3F3F3}} & 96.3$\pm$0.3 & 6.22$\pm$0.06 & {\cellcolor[HTML]{F3F3F3}} \\
\rowcolor[HTML]{F3F3F3} 
Bone\_Hip\_L & 96.6$\pm$1.6 & 0.32$\pm$1.13 & 97.0$\pm$1.9 & 0.28$\pm$0.13 & 99.7 & 96.7$\pm$1.8 & 0.32$\pm$0.12 & 99.0 & 96.9$\pm$1.9 & 0.34$\pm$0.13 & 99.0 & 97.0$\pm$1.9 & 0.28$\pm$0.13 & 99.7 \\
\rowcolor[HTML]{F3F3F3} 
Bone\_Hip\_R & 95.9$\pm$31.6 & 0.34$\pm$0.10 & 95.6$\pm$31.8 & 0.29$\pm$0.10 & {\cellcolor[HTML]{F3F3F3}} & 97.3$\pm$31.6 & 0.32$\pm$0.08 & {\cellcolor[HTML]{F3F3F3}} & 96.8$\pm$31.9 & 0.44$\pm$0.11 & {\cellcolor[HTML]{F3F3F3}} & 95.6$\pm$31.8 & 0.29$\pm$0.10 & {\cellcolor[HTML]{F3F3F3}} \\

SpinalCord & 92.7$\pm$16.8 & 0.27$\pm$0.24 & 91.9$\pm$16.8 & 0.25$\pm$0.27 & 99.7 & 94.8$\pm$16.5 & 0.28$\pm$0.24 & 99.0 & 92.6$\pm$16.9 & 0.31$\pm$0.27 & 99.0 & 91.9$\pm$16.8 & 0.25$\pm$0.27 & 99.7 \\
\rowcolor[HTML]{F3F3F3} 
Musc\_GluteusMax\_L & 98.9$\pm$5.2 & 0.84$\pm$1.41 & 99.0$\pm$5.6 & 0.80$\pm$1.44 & 99.7 & 98.3$\pm$5.5 & 0.87$\pm$1.41 & 99.7 & 98.7$\pm$5.3 & 0.85$\pm$1.40 & 99.0 & 99.0$\pm$5.6 & 0.80$\pm$1.44 & 99.7 \\
\rowcolor[HTML]{F3F3F3} 
Musc\_GluteusMax\_R & 98.1$\pm$2.0 & 0.93$\pm$0.34 & 97.9$\pm$2.0 & 0.74$\pm$0.33 & {\cellcolor[HTML]{F3F3F3}} & 98.3$\pm$1.8 & 0.92$\pm$0.31 & {\cellcolor[HTML]{F3F3F3}} & 97.8$\pm$2.2 & 0.78$\pm$0.34 & {\cellcolor[HTML]{F3F3F3}} & 97.9$\pm$2.0 & 0.74$\pm$0.33 & {\cellcolor[HTML]{F3F3F3}} \\

Musc\_GluteusMed\_L & 98.2$\pm$1.5 & 0.43$\pm$0.22 & 98.2$\pm$1.2 & 0.45$\pm$0.20 & 99.7 & 98.6$\pm$1.6 & 0.38$\pm$0.23 & 99.7 & 98.4$\pm$1.3 & 0.48$\pm$0.21 & 99.0 & 98.2$\pm$1.2 & 0.45$\pm$0.20 & 99.7 \\
 
Musc\_GluteusMed\_R & 97.6$\pm$7.3 & 0.74$\pm$0.22 & 97.4$\pm$7.6 & 0.44$\pm$0.22 & {\cellcolor[HTML]{F3F3F3}} & 96.5$\pm$7.6 & 0.81$\pm$0.22 & {\cellcolor[HTML]{F3F3F3}} & 97.4$\pm$7.4 & 0.55$\pm$0.20 & {\cellcolor[HTML]{F3F3F3}} & 97.4$\pm$7.6 & 0.44$\pm$0.22 & {\cellcolor[HTML]{F3F3F3}} \\
\rowcolor[HTML]{F3F3F3} 
Musc\_GluteusMin\_L & 97.1$\pm$2.3 & 0.56$\pm$0.16 & 97.3$\pm$2.2 & 0.40$\pm$0.16 & 99.7 & 95.6$\pm$2.5 & 0.59$\pm$0.19 & 99.5 & 97.2$\pm$2.2 & 0.61$\pm$0.16 & 99.5 & 97.3$\pm$2.2 & 0.40$\pm$0.16 & 99.7 \\
\rowcolor[HTML]{F3F3F3} 
Musc\_GluteusMin\_R & 98.0$\pm$1.1 & 0.39$\pm$0.08 & 97.9$\pm$1.2 & 0.38$\pm$0.11 & {\cellcolor[HTML]{F3F3F3}} & 97.1$\pm$1.0 & 0.39$\pm$0.09 & {\cellcolor[HTML]{F3F3F3}} & 98.2$\pm$1.1 & 0.51$\pm$0.10 & {\cellcolor[HTML]{F3F3F3}} & 97.9$\pm$1.2 & 0.38$\pm$0.11 & {\cellcolor[HTML]{F3F3F3}} \\

Musc\_Autochthon\_L & 94.2$\pm$0.8 & 0.37$\pm$0.11 & 95.1$\pm$1.1 & 0.31$\pm$0.14 & 99.0 & 94.1$\pm$1.0 & 0.40$\pm$0.12 & 99.5 & 93.8$\pm$0.9 & 0.46$\pm$0.12 & 99.0 & 95.1$\pm$1.1 & 0.31$\pm$0.14 & 99.0 \\

Musc\_Autochthon\_R & 94.6$\pm$0.8 & 0.41$\pm$0.12 & 94.2$\pm$0.7 & 0.39$\pm$0.10 & {\cellcolor[HTML]{F3F3F3}} & 93.8$\pm$0.9 & 0.42$\pm$0.12 & {\cellcolor[HTML]{F3F3F3}} & 94.6$\pm$0.8 & 0.56$\pm$0.11 & {\cellcolor[HTML]{F3F3F3}} & 94.2$\pm$0.7 & 0.39$\pm$0.10 & {\cellcolor[HTML]{F3F3F3}} \\
\rowcolor[HTML]{F3F3F3} 
Musc\_Iliopsoas\_L & 94.3$\pm$19.7 & 0.37$\pm$0.18 & 94.9$\pm$19.7 & 0.35$\pm$0.20 & 99.7 & 93.4$\pm$19.7 & 0.36$\pm$0.20 & 98.0 & 93.3$\pm$19.6 & 0.67$\pm$0.19 & 99.0 & 94.9$\pm$19.7 & 0.35$\pm$0.20 & 99.7 \\
\rowcolor[HTML]{F3F3F3} 
Musc\_Iliopsoas\_R & 95.0$\pm$9.1 & 0.39$\pm$0.27 & 95.8$\pm$9.0 & 0.35$\pm$0.24 & {} & 93.4$\pm$9.1 & 0.40$\pm$0.25 & {} & 94.8$\pm$9.1 & 0.36$\pm$0.25 & {} & 95.8$\pm$9.0 & 0.35$\pm$0.24 & {} \\

Brain & 97.9$\pm$36.4 & 0.56$\pm$1.01 & 97.9$\pm$16.5 & 0.58$\pm$1.06 & 99.5 & 98.5$\pm$16.2 & 0.59$\pm$1.08 & 98.0 & 98.2$\pm$16.3 & 0.58$\pm$1.07 & 99.5 & 97.9$\pm$16.5 & 0.58$\pm$1.06 & 99.5 \\
\rowcolor[HTML]{F3F3F3} 
Skull & 93.1$\pm$35.4 & 3.91$\pm$0.20 & 92.3$\pm$35.5 & 3.59$\pm$0.21 & 99.7 & 93.0$\pm$35.5 & 3.86$\pm$0.21 & 99.0 & 92.4$\pm$35.3 & 3.96$\pm$0.20 & 99.0 & 92.3$\pm$35.5 & 3.59$\pm$0.21 & 99.7\\ \hline

Mean & \multicolumn{1}{c}{96.1} & \multicolumn{1}{c|}{1.46} & \multicolumn{1}{c}{96.2} & \multicolumn{1}{c|}{1.34} & 99.6& \multicolumn{1}{c}{96.0} & \multicolumn{1}{c|}{1.47} & 99.1 & \multicolumn{1}{c}{96.1} & \multicolumn{1}{c|}{1.52} & 99.1 & \multicolumn{1}{c}{96.2} & \multicolumn{1}{c|}{1.34} & 99.6\\ \hline

\end{tabular}}
\caption{\textbar~\textbf{\Ac{css} order-wise performance details of \ac{cln_c5} on \ac{totalseg} `Muscles' subgroup.} Organ-wise DSC (\%), ASD (mm), and decoder-wise pruning rate $\mathcal{T}$ (\%) of \ac{cln_c5} on \ac{totalseg} `Muscles' subgroup are evaluated across all CSS orders. Notably, \ac{cln_c5} achieves mean DSC and ASD comparable to nnUNet upper bound and the average decoder pruning rates exceed 99\% across all orders, demonstrating high efficiency without compromising segmentation performance. }
\label{tab:css_totalseg_musc}
\end{table}
\clearpage

\begin{table}[htp]
\centering
\resizebox{\textwidth}{!}{
\begin{tabular}{|l|ll|ll|c|ll|c|ll|c|ll|c|}
\hline
\multicolumn{1}{|l|}{} & \multicolumn{2}{c|}{nnUNet} & \multicolumn{3}{c|}{Order 1} & \multicolumn{3}{c|}{Order 2} & \multicolumn{3}{c|}{Order 3} & \multicolumn{3}{c|}{Order 4} \\ \cline{2-15} 
\multicolumn{1}{|l|}{\multirow{-2}{*}{}} & DSC$\uparrow$  & ASD$\downarrow$ & DSC$\uparrow$  & ASD$\downarrow$ & $\mathcal{T}$$\uparrow$  & DSC$\uparrow$  & ASD$\downarrow$ & $\mathcal{T}$$\uparrow$  & DSC$\uparrow$  & ASD$\downarrow$ & $\mathcal{T}$$\uparrow$  & DSC$\uparrow$  & ASD$\downarrow$ & $\mathcal{T}$$\uparrow$  \\ \hline
\rowcolor[HTML]{F3F3F3} 
Bone\_Rib1\_L & 93.6$\pm$3.3 & 0.16$\pm$0.23 & 93.8$\pm$3.4 & 0.15$\pm$0.22 & {\cellcolor[HTML]{F3F3F3}96.0} & 96.0$\pm$3.3 & 0.17$\pm$0.21 & {\cellcolor[HTML]{F3F3F3}94.0} & 93.1$\pm$3.4 & 0.27$\pm$0.22 & {\cellcolor[HTML]{F3F3F3}96.0} & 93.8$\pm$3.4 & 0.15$\pm$0.22 & {\cellcolor[HTML]{F3F3F3}96.0} \\
\rowcolor[HTML]{F3F3F3} 
Bone\_Rib2\_L & 89.9$\pm$1.7 & 0.95$\pm$0.95 & 89.5$\pm$1.7 & 0.88$\pm$0.96 &  & 89.7$\pm$1.4 & 0.85$\pm$0.99 &  & 89.6$\pm$1.5 & 0.84$\pm$0.97 &  & 89.5$\pm$1.7 & 0.88$\pm$0.96 &  \\
\rowcolor[HTML]{F3F3F3} 
Bone\_Rib3\_L & 94.7$\pm$17.0 & 0.91$\pm$1.06 & 95.0$\pm$17.2 & 0.90$\pm$1.06 &  & 96.3$\pm$17.0 & 0.87$\pm$1.04 &  & 94.8$\pm$16.9 & 0.81$\pm$1.04 &  & 95.0$\pm$17.2 & 0.90$\pm$1.06 &  \\
\rowcolor[HTML]{F3F3F3} 
Bone\_Rib4\_L & 94.5$\pm$18.9 & 0.76$\pm$1.82 & 94.8$\pm$19.0 & 0.67$\pm$1.80 &  & 94.6$\pm$18.9 & 0.69$\pm$1.89 &  & 93.8$\pm$18.9 & 0.66$\pm$1.89 &  & 94.8$\pm$19.0 & 0.67$\pm$1.80 &  \\
\rowcolor[HTML]{F3F3F3} 
Bone\_Rib5\_L & 94.2$\pm$29.7 & 1.52$\pm$4.61 & 94.8$\pm$29.3 & 1.18$\pm$4.70 &  & 94.3$\pm$29.3 & 1.49$\pm$4.41 &  & 94.2$\pm$29.4 & 1.08$\pm$4.49 &  & 94.8$\pm$29.3 & 1.18$\pm$4.70 &  \\
\rowcolor[HTML]{F3F3F3} 
Bone\_Rib6\_L & 93.2$\pm$23.9 & 1.64$\pm$3.32 & 92.7$\pm$23.9 & 1.42$\pm$3.24 &  & 93.8$\pm$23.8 & 1.59$\pm$3.35 &  & 92.6$\pm$24.2 & 1.75$\pm$3.23 &  & 92.7$\pm$23.9 & 1.42$\pm$3.24 &  \\
\rowcolor[HTML]{F3F3F3} 
Bone\_Rib7\_L & 94.5$\pm$20.7 & 2.71$\pm$4.08 & 93.6$\pm$20.6 & 2.68$\pm$3.98 &  & 96.6$\pm$20.7 & 2.47$\pm$4.10 &  & 95.6$\pm$20.8 & 2.72$\pm$4.11 &  & 93.6$\pm$20.6 & 2.68$\pm$3.98 &  \\
\rowcolor[HTML]{F3F3F3} 
Bone\_Rib8\_L & 94.6$\pm$21.1 & 4.29$\pm$4.00 & 95.3$\pm$20.9 & 4.59$\pm$3.81 &  & 94.0$\pm$20.8 & 4.08$\pm$3.82 &  & 95.1$\pm$21.0 & 4.12$\pm$4.10 &  & 95.3$\pm$20.9 & 4.59$\pm$3.81 &  \\
\rowcolor[HTML]{F3F3F3} 
Bone\_Rib9\_L & 94.0$\pm$19.4 & 2.83$\pm$5.21 & 93.5$\pm$19.4 & 2.93$\pm$5.24 &  & 95.4$\pm$19.2 & 2.59$\pm$5.01 &  & 94.3$\pm$19.4 & 2.73$\pm$4.94 &  & 93.5$\pm$19.4 & 2.93$\pm$5.24 &  \\
\rowcolor[HTML]{F3F3F3} 
Bone\_Rib10\_L & 93.0$\pm$22.9 & 1.51$\pm$6.63 & 92.8$\pm$12.9 & 1.38$\pm$6.61 &  & 92.4$\pm$13.2 & 1.61$\pm$6.77 &  & 91.8$\pm$13.0 & 1.41$\pm$7.00 &  & 92.8$\pm$12.9 & 1.38$\pm$6.61 &  \\
\rowcolor[HTML]{F3F3F3} 
Bone\_Rib11\_L & 93.0$\pm$22.4 & 0.36$\pm$8.01 & 92.8$\pm$12.5 & 0.31$\pm$8.12 &  & 95.2$\pm$12.4 & 0.34$\pm$7.89 &  & 92.0$\pm$12.4 & 0.39$\pm$8.20 &  & 92.8$\pm$12.5 & 0.31$\pm$8.12 &  \\
\rowcolor[HTML]{F3F3F3} 
Bone\_Rib12\_L & 94.2$\pm$26.2 & 0.54$\pm$0.41 & 93.5$\pm$16.3 & 0.34$\pm$0.41 &  & 93.3$\pm$16.1 & 0.54$\pm$0.39 &  & 94.5$\pm$16.2 & 0.34$\pm$0.39 &  & 93.5$\pm$16.3 & 0.34$\pm$0.41 &  \\
\rowcolor[HTML]{F3F3F3} 
Bone\_Rib1\_R & 91.2$\pm$5.6 & 0.11$\pm$0.17 & 91.0$\pm$5.8 & 0.13$\pm$0.17 &  & 92.7$\pm$5.6 & 0.11$\pm$0.15 &  & 91.7$\pm$5.7 & 0.21$\pm$0.16 &  & 91.0$\pm$5.8 & 0.13$\pm$0.17 &  \\
\rowcolor[HTML]{F3F3F3} 
Bone\_Rib2\_R & 94.5$\pm$3.5 & 0.23$\pm$0.94 & 93.6$\pm$3.4 & 0.09$\pm$0.91 &  & 94.1$\pm$3.6 & 0.22$\pm$0.93 &  & 94.4$\pm$3.6 & 0.15$\pm$0.93 &  & 93.6$\pm$3.4 & 0.09$\pm$0.91 &  \\
\rowcolor[HTML]{F3F3F3} 
Bone\_Rib3\_R & 92.8$\pm$9.1 & 0.48$\pm$0.75 & 92.9$\pm$9.0 & 0.34$\pm$1.05 &  & 92.5$\pm$9.1 & 0.47$\pm$1.05 &  & 93.6$\pm$9.1 & 0.38$\pm$1.14 &  & 92.9$\pm$9.0 & 0.34$\pm$1.05 &  \\
\rowcolor[HTML]{F3F3F3} 
Bone\_Rib4\_R & 90.5$\pm$18.8 & 0.83$\pm$2.02 & 91.3$\pm$18.8 & 0.69$\pm$1.88 &  & 88.3$\pm$18.6 & 0.83$\pm$2.02 &  & 90.3$\pm$18.8 & 0.77$\pm$1.82 &  & 91.3$\pm$18.8 & 0.69$\pm$1.88 &  \\
\rowcolor[HTML]{F3F3F3} 
Bone\_Rib5\_R & 90.3$\pm$26.5 & 1.70$\pm$2.42 & 90.5$\pm$16.5 & 1.64$\pm$2.53 &  & 89.5$\pm$16.4 & 1.84$\pm$2.61 &  & 90.6$\pm$16.6 & 1.98$\pm$2.52 &  & 90.5$\pm$16.5 & 1.64$\pm$2.53 &  \\
\rowcolor[HTML]{F3F3F3} 
Bone\_Rib6\_R & 93.4$\pm$18.9 & 1.52$\pm$2.45 & 93.4$\pm$13.0 & 1.65$\pm$2.32 &  & 93.0$\pm$13.0 & 1.48$\pm$2.54 &  & 93.3$\pm$13.0 & 1.74$\pm$2.42 &  & 93.4$\pm$13.0 & 1.65$\pm$2.32 &  \\
\rowcolor[HTML]{F3F3F3} 
Bone\_Rib7\_R & 94.2$\pm$21.0 & 1.56$\pm$3.51 & 93.8$\pm$21.2 & 1.61$\pm$3.66 &  & 92.5$\pm$21.1 & 1.56$\pm$3.36 &  & 95.4$\pm$21.2 & 1.70$\pm$3.40 &  & 93.8$\pm$21.2 & 1.61$\pm$3.66 &  \\
\rowcolor[HTML]{F3F3F3} 
Bone\_Rib8\_R & 95.5$\pm$18.3 & 1.88$\pm$4.08 & 95.4$\pm$18.1 & 1.81$\pm$4.16 &  & 95.4$\pm$18.2 & 1.79$\pm$3.92 &  & 94.8$\pm$18.2 & 1.72$\pm$4.08 &  & 95.4$\pm$18.1 & 1.81$\pm$4.16 &  \\
\rowcolor[HTML]{F3F3F3} 
Bone\_Rib9\_R & 95.9$\pm$21.4 & 2.48$\pm$5.16 & 95.1$\pm$21.6 & 2.31$\pm$3.20 &  & 96.8$\pm$21.4 & 2.58$\pm$3.35 &  & 96.2$\pm$21.3 & 2.93$\pm$3.41 &  & 95.1$\pm$21.6 & 2.31$\pm$3.20 &  \\
\rowcolor[HTML]{F3F3F3} 
Bone\_Rib10\_R & 96.3$\pm$23.3 & 2.50$\pm$7.23 & 97.1$\pm$23.1 & 2.56$\pm$7.10 &  & 97.0$\pm$23.3 & 2.60$\pm$7.30 &  & 95.8$\pm$23.3 & 2.67$\pm$7.41 &  & 97.1$\pm$23.1 & 2.56$\pm$7.10 &  \\
\rowcolor[HTML]{F3F3F3} 
Bone\_Rib11\_R & 96.4$\pm$22.8 & 1.85$\pm$6.95 & 96.1$\pm$22.7 & 1.88$\pm$7.02 &  & 97.0$\pm$23.0 & 1.90$\pm$7.27 &  & 97.5$\pm$23.0 & 1.82$\pm$6.86 &  & 96.1$\pm$22.7 & 1.88$\pm$7.02 &  \\
\rowcolor[HTML]{F3F3F3} 
Bone\_Rib12\_R & 97.1$\pm$27.5 & 0.63$\pm$0.07 & 97.2$\pm$27.4 & 0.45$\pm$0.10 &  & 97.6$\pm$27.2 & 0.57$\pm$0.08 &  & 97.3$\pm$27.2 & 0.51$\pm$0.08 &  & 97.2$\pm$27.4 & 0.45$\pm$0.10 &  \\
Bone\_Sternum & 94.1$\pm$5.8 & 0.63$\pm$0.55 & 94.8$\pm$5.9 & 0.38$\pm$0.57 & {99.0} & 95.6$\pm$5.7 & 0.64$\pm$0.55 & {99.0} & 95.3$\pm$5.9 & 0.39$\pm$0.57 & {99.5} & 94.8$\pm$5.9 & 0.38$\pm$0.57 & {99.0} \\
\rowcolor[HTML]{F3F3F3} 
Bone\_CostalCartilages & 88.7$\pm$5.0 & 0.18$\pm$0.74 & 89.6$\pm$5.1 & 0.18$\pm$0.76 & {\cellcolor[HTML]{F3F3F3}99.0} & 89.4$\pm$5.3 & 0.16$\pm$0.77 & {\cellcolor[HTML]{F3F3F3}98.0} & 88.3$\pm$5.1 & 0.17$\pm$0.75 & {\cellcolor[HTML]{F3F3F3}98.0} & 89.6$\pm$5.1 & 0.18$\pm$0.76 & {\cellcolor[HTML]{F3F3F3}99.0}\\ \hline

Mean & \multicolumn{1}{c}{93.6} & \multicolumn{1}{c|}{1.34} & \multicolumn{1}{c}{93.6} & \multicolumn{1}{c|}{1.28} & 98.0& \multicolumn{1}{c}{94.0} & \multicolumn{1}{c|}{1.31} & 97.0 & \multicolumn{1}{c}{93.7} & \multicolumn{1}{c|}{1.32} & 97.8 & \multicolumn{1}{c}{93.6} & \multicolumn{1}{c|}{1.28} & 98.0\\ \hline

\end{tabular}}
\caption{\textbar~\textbf{\Ac{css} order-wise performance details of \ac{cln_c5} on \ac{totalseg} `Ribs' subgroup.} Organ-wise DSC (\%), ASD (mm), and decoder-wise pruning rate $\mathcal{T}$ (\%) of \ac{cln_c5} on \ac{totalseg} `Ribs' subgroup are evaluated across all CSS orders. Notably, \ac{cln_c5} achieves higher mean DSC and ASD than nnUNet upper bound and the average decoder pruning rates exceed 97\% across all orders, demonstrating both high segmentation performance and efficiency. }
\label{tab:css_totalseg_ribs}
\end{table}
\clearpage

\begin{table}[htp]
\centering
\resizebox{\textwidth}{!}{
\begin{tabular}{|l|ll|ll|c|ll|c|ll|c|ll|c|}
\hline
\multicolumn{1}{|l|}{} & \multicolumn{2}{c|}{nnUNet} & \multicolumn{3}{c|}{Order 1} & \multicolumn{3}{c|}{Order 2} & \multicolumn{3}{c|}{Order 3} & \multicolumn{3}{c|}{Order 4} \\ \cline{2-15} 
\multicolumn{1}{|l|}{\multirow{-2}{*}{}} & DSC$\uparrow$  & ASD$\downarrow$ & DSC$\uparrow$  & ASD$\downarrow$ & $\mathcal{T}$$\uparrow$  & DSC$\uparrow$  & ASD$\downarrow$ & $\mathcal{T}$$\uparrow$  & DSC$\uparrow$  & ASD$\downarrow$ & $\mathcal{T}$$\uparrow$  & DSC$\uparrow$  & ASD$\downarrow$ & $\mathcal{T}$$\uparrow$  \\ \hline
\rowcolor[HTML]{F3F3F3} 
Bone\_Sacrum & 93.4$\pm$1.8 & 0.25$\pm$0.13 & 94.4$\pm$2.1 & 0.22$\pm$0.16 & {\cellcolor[HTML]{F3F3F3}99.7} & 93.4$\pm$1.9 & 0.26$\pm$0.14 & {\cellcolor[HTML]{F3F3F3}99.0} & 93.8$\pm$2.0 & 0.30$\pm$0.15 & {\cellcolor[HTML]{F3F3F3}99.0} & 94.4$\pm$2.1 & 0.22$\pm$0.16 & {\cellcolor[HTML]{F3F3F3}99.7} \\
Bone\_Vert\_C1 & 94.5$\pm$20.3 & 0.25$\pm$0.53 & 94.5$\pm$10.4 & 0.24$\pm$0.53 & {94.0} & 93.2$\pm$10.5 & 0.23$\pm$0.54 & {94.0} & 95.1$\pm$10.5 & 0.29$\pm$0.53 & {94.0} & 94.5$\pm$10.4 & 0.24$\pm$0.53 & {94.0} \\
Bone\_Vert\_C2 & 98.0$\pm$20.7 & 0.31$\pm$3.04 & 97.8$\pm$10.9 & 0.16$\pm$0.50 &  & 98.9$\pm$10.8 & 0.29$\pm$0.50 &  & 97.8$\pm$10.7 & 0.19$\pm$0.49 &  & 97.8$\pm$10.9 & 0.16$\pm$0.50 &  \\
Bone\_Vert\_C3 & 98.1$\pm$23.4 & 0.16$\pm$3.22 & 98.2$\pm$15.4 & 0.17$\pm$3.38 &  & 98.1$\pm$15.7 & 0.15$\pm$3.42 &  & 97.9$\pm$15.6 & 0.23$\pm$3.40 &  & 98.2$\pm$15.4 & 0.17$\pm$3.38 &  \\
Bone\_Vert\_C4 & 97.0$\pm$23.4 & 0.34$\pm$3.60 & 96.8$\pm$19.5 & 0.34$\pm$3.55 &  & 95.2$\pm$19.4 & 0.34$\pm$3.61 &  & 96.5$\pm$19.5 & 0.41$\pm$3.42 &  & 96.8$\pm$19.5 & 0.34$\pm$3.55 &  \\
Bone\_Vert\_C5 & 96.8$\pm$17.5 & 0.22$\pm$1.62 & 96.3$\pm$17.4 & 0.21$\pm$1.65 &  & 97.6$\pm$17.5 & 0.22$\pm$1.67 &  & 97.9$\pm$17.6 & 0.48$\pm$1.69 &  & 96.3$\pm$17.4 & 0.21$\pm$1.65 &  \\
Bone\_Vert\_C6 & 95.0$\pm$10.6 & 0.60$\pm$0.93 & 96.0$\pm$10.3 & 0.50$\pm$0.91 &  & 95.9$\pm$10.5 & 0.64$\pm$0.94 &  & 95.8$\pm$10.4 & 0.45$\pm$0.92 &  & 96.0$\pm$10.3 & 0.50$\pm$0.91 &  \\
Bone\_Vert\_C7 & 95.2$\pm$17.5 & 0.41$\pm$0.86 & 95.0$\pm$13.3 & 0.19$\pm$0.97 &  & 95.1$\pm$13.4 & 0.42$\pm$1.00 &  & 95.2$\pm$13.1 & 0.46$\pm$0.99 &  & 95.0$\pm$13.3 & 0.19$\pm$0.97 &  \\
Bone\_Vert\_L1 & 95.7$\pm$12.1 & 1.27$\pm$0.98 & 96.1$\pm$12.3 & 1.18$\pm$0.91 &  & 97.2$\pm$12.1 & 1.21$\pm$0.99 &  & 94.8$\pm$12.0 & 1.40$\pm$0.98 &  & 96.1$\pm$12.3 & 1.18$\pm$0.91 &  \\
Bone\_Vert\_L2 & 94.8$\pm$18.2 & 0.29$\pm$1.09 & 94.8$\pm$12.2 & 0.18$\pm$0.91 &  & 94.8$\pm$12.0 & 0.32$\pm$1.00 &  & 93.6$\pm$12.3 & 0.19$\pm$0.95 &  & 94.8$\pm$12.2 & 0.18$\pm$0.91 &  \\
Bone\_Vert\_L3 & 96.0$\pm$16.1 & 0.24$\pm$1.09 & 95.3$\pm$16.3 & 0.20$\pm$1.01 &  & 95.1$\pm$16.2 & 0.26$\pm$1.03 &  & 96.7$\pm$16.2 & 0.39$\pm$1.07 &  & 95.3$\pm$16.3 & 0.20$\pm$1.01 &  \\
Bone\_Vert\_L4 & 94.1$\pm$14.3 & 0.43$\pm$0.41 & 94.2$\pm$14.4 & 0.25$\pm$0.43 &  & 94.4$\pm$14.2 & 0.43$\pm$0.41 &  & 94.0$\pm$14.2 & 0.18$\pm$0.41 &  & 94.2$\pm$14.4 & 0.25$\pm$0.43 &  \\
Bone\_Vert\_L5 & 92.9$\pm$17.3 & 0.21$\pm$1.49 & 93.1$\pm$17.3 & 0.21$\pm$1.49 &  & 94.4$\pm$17.3 & 0.21$\pm$1.46 &  & 93.7$\pm$17.4 & 0.32$\pm$1.50 &  & 93.1$\pm$17.3 & 0.21$\pm$1.49 &  \\
Bone\_Vert\_S1 & 96.3$\pm$23.8 & 0.20$\pm$2.10 & 96.8$\pm$19.0 & 0.22$\pm$1.22 &  & 97.6$\pm$19.1 & 0.19$\pm$1.22 &  & 97.2$\pm$18.9 & 0.23$\pm$1.11 &  & 96.8$\pm$19.0 & 0.22$\pm$1.22 &  \\
Bone\_Vert\_T1 & 96.3$\pm$23.5 & 0.15$\pm$1.28 & 96.4$\pm$13.6 & 0.17$\pm$1.30 &  & 97.6$\pm$13.5 & 0.15$\pm$1.27 &  & 95.7$\pm$13.6 & 0.23$\pm$1.31 &  & 96.4$\pm$13.6 & 0.17$\pm$1.30 &  \\
Bone\_Vert\_T2 & 98.0$\pm$5.2 & 0.48$\pm$0.74 & 98.0$\pm$5.1 & 0.34$\pm$0.73 &  & 97.3$\pm$5.2 & 0.52$\pm$0.73 &  & 97.6$\pm$5.4 & 0.38$\pm$0.75 &  & 98.0$\pm$5.1 & 0.34$\pm$0.73 &  \\
Bone\_Vert\_T3 & 97.0$\pm$2.7 & 0.48$\pm$0.20 & 97.5$\pm$2.6 & 0.43$\pm$0.21 &  & 95.3$\pm$2.5 & 0.44$\pm$0.20 &  & 97.8$\pm$2.4 & 0.64$\pm$0.19 &  & 97.5$\pm$2.6 & 0.43$\pm$0.21 &  \\
Bone\_Vert\_T4 & 89.6$\pm$17.0 & 1.12$\pm$0.21 & 88.7$\pm$8.9 & 1.02$\pm$0.20 &  & 90.3$\pm$9.1 & 1.09$\pm$0.22 &  & 90.4$\pm$8.9 & 1.19$\pm$0.20 &  & 88.7$\pm$8.9 & 1.02$\pm$0.20 &  \\
Bone\_Vert\_T5 & 90.7$\pm$22.9 & 0.92$\pm$0.38 & 91.0$\pm$22.8 & 0.91$\pm$0.36 &  & 91.7$\pm$22.7 & 0.92$\pm$0.35 &  & 90.3$\pm$22.9 & 0.86$\pm$0.37 &  & 91.0$\pm$22.8 & 0.91$\pm$0.36 &  \\
Bone\_Vert\_T6 & 86.3$\pm$28.3 & 1.15$\pm$0.07 & 85.9$\pm$28.2 & 1.00$\pm$0.07 &  & 84.9$\pm$28.3 & 1.14$\pm$0.07 &  & 85.2$\pm$28.2 & 1.08$\pm$0.06 &  & 85.9$\pm$28.2 & 1.00$\pm$0.07 &  \\
Bone\_Vert\_T7 & 96.8$\pm$1.0 & 1.58$\pm$0.04 & 97.4$\pm$1.0 & 1.46$\pm$0.05 &  & 95.9$\pm$0.7 & 1.53$\pm$0.03 &  & 97.0$\pm$0.8 & 1.72$\pm$0.04 &  & 97.4$\pm$1.0 & 1.46$\pm$0.05 &  \\
Bone\_Vert\_T8 & 91.5$\pm$13.0 & 1.08$\pm$0.40 & 91.8$\pm$12.8 & 1.06$\pm$0.15 &  & 91.2$\pm$12.6 & 0.99$\pm$0.13 &  & 91.4$\pm$12.7 & 0.99$\pm$0.13 &  & 91.8$\pm$12.8 & 1.06$\pm$0.15 &  \\
Bone\_Vert\_T9 & 94.6$\pm$44.2 & 0.53$\pm$0.23 & 94.1$\pm$14.4 & 0.46$\pm$0.24 &  & 95.2$\pm$14.5 & 0.48$\pm$0.25 &  & 94.2$\pm$14.2 & 0.77$\pm$0.23 &  & 94.1$\pm$14.4 & 0.46$\pm$0.24 &  \\
Bone\_Vert\_T10 & 97.6$\pm$29.9 & 0.36$\pm$2.18 & 97.8$\pm$13.2 & 0.36$\pm$1.32 &  & 99.0$\pm$13.0 & 0.39$\pm$1.23 &  & 97.4$\pm$13.2 & 0.64$\pm$1.28 &  & 97.8$\pm$13.2 & 0.36$\pm$1.32 &  \\
Bone\_Vert\_T11 & 95.1$\pm$19.4 & 0.25$\pm$1.64 & 95.2$\pm$12.4 & 0.22$\pm$1.64 &  & 93.6$\pm$12.5 & 0.27$\pm$1.59 &  & 94.9$\pm$12.5 & 0.32$\pm$1.70 &  & 95.2$\pm$12.4 & 0.22$\pm$1.64 &  \\
Bone\_Vert\_T12 & 97.2$\pm$16.3 & 0.32$\pm$1.30 & 97.3$\pm$16.2 & 0.27$\pm$1.29 &  & 97.1$\pm$16.2 & 0.33$\pm$1.29 &  & 97.6$\pm$16.1 & 0.20$\pm$1.34 &  & 97.3$\pm$16.2 & 0.27$\pm$1.29 & 
\\ \hline

Mean & \multicolumn{1}{c}{94.9} & \multicolumn{1}{c|}{0.52} & \multicolumn{1}{c}{95.0} & \multicolumn{1}{c|}{0.46} & 96.9& \multicolumn{1}{c}{95.0} & \multicolumn{1}{c|}{0.52} & 96.5 & \multicolumn{1}{c}{95.0} & \multicolumn{1}{c|}{0.56} & 96.5 & \multicolumn{1}{c}{95.0} & \multicolumn{1}{c|}{0.46} & 96.9\\ \hline

\end{tabular}}
\caption{\textbar~\textbf{\Ac{css} order-wise performance details of \ac{cln_c5} on \ac{totalseg} `Vertebrae' subgroup.} Organ-wise DSC (\%), ASD (mm), and decoder-wise pruning rate $\mathcal{T}$ (\%) of \ac{cln_c5} on \ac{totalseg} `Vertebrae' subgroup are evaluated across all CSS orders. Notably, \ac{cln_c5} achieves mean DSC and ASD comparable to nnUNet upper bound and the average decoder pruning rates exceed 96.5\% across all orders, demonstrating high efficiency without compromising segmentation performance. }
\label{tab:css_totalseg_vert}
\end{table}
\clearpage
\begin{table}[htp]
\centering
\resizebox{\textwidth}{!}{
\begin{tabular}{|l|ll|ll|c|ll|c|ll|c|ll|c|}
\hline
\multicolumn{1}{|l|}{} & \multicolumn{2}{c|}{nnUNet} & \multicolumn{3}{c|}{Order 1} & \multicolumn{3}{c|}{Order 2} & \multicolumn{3}{c|}{Order 3} & \multicolumn{3}{c|}{Order 4} \\ \cline{2-15} 
\multicolumn{1}{|l|}{\multirow{-2}{*}{}} & DSC$\uparrow$ & ASD$\downarrow$ & DSC$\uparrow$ & ASD$\downarrow$ & $\mathcal{T}$$\uparrow$  & DSC$\uparrow$ & ASD$\downarrow$ & $\mathcal{T}$$\uparrow$  & DSC$\uparrow$ & ASD$\downarrow$ & $\mathcal{T}$$\uparrow$  & DSC$\uparrow$ & ASD$\downarrow$ & $\mathcal{T}$$\uparrow$  \\ \hline
\rowcolor[HTML]{F3F3F3} 
BrainStem & 93.7$\pm$1.5 & 0.32$\pm$0.10 & 92.9$\pm$1.4 & 0.28$\pm$0.09 & 99.0 & 92.2$\pm$1.5 & 0.25$\pm$0.10 & 98.0 & 92.8$\pm$1.4 & 0.25$\pm$0.09 & 99.0 & 92.9$\pm$1.4 & 0.28$\pm$0.09 & 99.0 \\
Eye\_L & 94.8$\pm$1.3 & 0.17$\pm$0.08 & 94.8$\pm$1.3 & 0.15$\pm$0.08 & 98.0 & 94.3$\pm$1.4 & 0.16$\pm$0.09 & 99.0 & 93.5$\pm$1.4 & 0.15$\pm$0.09 & 99.0 & 94.8$\pm$1.3 & 0.15$\pm$0.08 & 98.0 \\
Eye\_R & 93.2$\pm$4.2 & 0.24$\pm$0.28 & 94.1$\pm$4.1 & 0.28$\pm$0.28 & {} & 91.8$\pm$4.4 & 0.31$\pm$0.30 & {} & 95.2$\pm$4.3 & 0.30$\pm$0.29 & {} & 94.1$\pm$4.1 & 0.28$\pm$0.28 & {} \\
\rowcolor[HTML]{F3F3F3} 
Lens\_L & 80.1$\pm$9.8 & 0.25$\pm$0.18 & 79.6$\pm$10.0 & 0.23$\pm$0.20 & 98.0 & 78.7$\pm$9.9 & 0.24$\pm$0.18 & 99.0 & 80.2$\pm$9.6 & 0.21$\pm$0.16 & 98.0 & 79.6$\pm$10.0 & 0.23$\pm$0.20 & 98.0 \\
\rowcolor[HTML]{F3F3F3} 
Lens\_R & 79.7$\pm$9.2 & 0.33$\pm$0.43 & 78.0$\pm$9.1 & 0.35$\pm$0.41 & {\cellcolor[HTML]{F3F3F3}} & 78.6$\pm$9.0 & 0.40$\pm$0.41 & {\cellcolor[HTML]{F3F3F3}} & 80.4$\pm$9.3 & 0.42$\pm$0.43 & {\cellcolor[HTML]{F3F3F3}} & 78.0$\pm$9.1 & 0.35$\pm$0.41 & {\cellcolor[HTML]{F3F3F3}} \\
OpticNerve\_L & 80.3$\pm$8.0 & 0.29$\pm$0.28 & 80.9$\pm$8.0 & 0.29$\pm$0.28 & 94.0 & 81.1$\pm$8.0 & 0.31$\pm$0.28 & 98.0 & 81.0$\pm$7.9 & 0.33$\pm$0.27 & 96.0 & 80.9$\pm$8.0 & 0.29$\pm$0.28 & 94.0 \\
OpticNerve\_R & 84.1$\pm$6.4 & 0.20$\pm$0.21 & 85.0$\pm$6.5 & 0.17$\pm$0.22 & {} & 85.2$\pm$6.4 & 0.17$\pm$0.21 & {} & 85.6$\pm$6.3 & 0.18$\pm$0.20 & {} & 85.0$\pm$6.5 & 0.17$\pm$0.22 & {} \\
\rowcolor[HTML]{F3F3F3} 
Chiasm & 71.6$\pm$12.9 & 0.26$\pm$0.13 & 69.6$\pm$13.1 & 0.26$\pm$0.15 & 94.0 & 73.3$\pm$12.7 & 0.25$\pm$0.11 & 94.0 & 71.0$\pm$13.0 & 0.25$\pm$0.13 & 94.0 & 69.6$\pm$13.1 & 0.26$\pm$0.15 & 94.0 \\
TemporalLobe\_L & 94.7$\pm$1.2 & 0.24$\pm$0.08 & 93.7$\pm$1.2 & 0.21$\pm$0.08 & 99.7 & 93.2$\pm$1.3 & 0.22$\pm$0.08 & 98.0 & 95.1$\pm$1.2 & 0.23$\pm$0.08 & 99.0 & 93.7$\pm$1.2 & 0.21$\pm$0.08 & 99.7 \\
TemporalLobe\_R & 94.8$\pm$1.4 & 0.26$\pm$0.09 & 93.2$\pm$1.4 & 0.24$\pm$0.09 & {} & 94.4$\pm$1.6 & 0.24$\pm$0.11 & {} & 93.4$\pm$1.6 & 0.25$\pm$0.11 & {} & 93.2$\pm$1.4 & 0.24$\pm$0.09 & {} \\
\rowcolor[HTML]{F3F3F3} 
Glnd\_Pituitary & 84.5$\pm$6.4 & 0.22$\pm$0.21 & 84.5$\pm$6.3 & 0.21$\pm$0.19 & 99.5 & 83.5$\pm$6.5 & 0.19$\pm$0.22 & 98.0 & 83.7$\pm$6.5 & 0.18$\pm$0.22 & 99.5 & 84.5$\pm$6.3 & 0.21$\pm$0.19 & 99.5 \\
Glnd\_Parotid\_L & 93.3$\pm$1.9 & 0.35$\pm$0.16 & 94.9$\pm$1.9 & 0.32$\pm$0.16 & 96.0 & 95.3$\pm$2.0 & 0.34$\pm$0.17 & 96.0 & 91.5$\pm$2.0 & 0.32$\pm$0.17 & 94.0 & 94.9$\pm$1.9 & 0.32$\pm$0.16 & 96.0 \\
Glnd\_Parotid\_R & 92.5$\pm$3.4 & 0.50$\pm$0.52 & 95.3$\pm$1.6 & 0.40$\pm$0.54 & {} & 95.1$\pm$1.6 & 0.46$\pm$0.54 & {} & 95.9$\pm$1.4 & 0.44$\pm$0.52 & {} & 95.3$\pm$1.6 & 0.40$\pm$0.54 & {} \\
\rowcolor[HTML]{F3F3F3} 
Ear\_Inner\_L & 67.0$\pm$9.5 & 0.66$\pm$0.37 & 67.5$\pm$9.5 & 0.59$\pm$0.37 & 99.0 & 65.6$\pm$9.6 & 0.58$\pm$0.38 & 99.0 & 68.8$\pm$9.4 & 0.65$\pm$0.37 & 96.0 & 67.5$\pm$9.5 & 0.59$\pm$0.37 & 99.0 \\
\rowcolor[HTML]{F3F3F3} 
Ear\_Inner\_R & 64.3$\pm$9.3 & 0.66$\pm$0.30 & 65.6$\pm$9.3 & 0.67$\pm$0.31 & {\cellcolor[HTML]{F3F3F3}} & 62.9$\pm$9.4 & 0.81$\pm$0.31 & {\cellcolor[HTML]{F3F3F3}} & 63.7$\pm$9.4 & 0.83$\pm$0.31 & {\cellcolor[HTML]{F3F3F3}} & 65.6$\pm$9.3 & 0.67$\pm$0.31 & {\cellcolor[HTML]{F3F3F3}} \\
Ear\_Mid\_L & 90.4$\pm$3.7 & 0.07$\pm$0.04 & 90.0$\pm$3.9 & 0.07$\pm$0.05 & 99.5 & 91.4$\pm$3.7 & 0.07$\pm$0.04 & 99.0 & 88.6$\pm$3.9 & 0.07$\pm$0.06 & 98.0 & 90.0$\pm$3.9 & 0.07$\pm$0.05 & 99.5 \\
Ear\_Mid\_R & 90.2$\pm$3.3 & 0.08$\pm$0.06 & 90.2$\pm$3.2 & 0.08$\pm$0.04 & {} & 90.0$\pm$3.4 & 0.07$\pm$0.06 & {} & 91.9$\pm$3.4 & 0.07$\pm$0.06 & {} & 90.2$\pm$3.2 & 0.08$\pm$0.04 & {} \\
\rowcolor[HTML]{F3F3F3} 
TMJ\_L & 84.4$\pm$9.5 & 0.31$\pm$0.23 & 86.1$\pm$9.5 & 0.30$\pm$0.24 & 99.5 & 83.6$\pm$9.6 & 0.26$\pm$0.24 & 96.0 & 82.5$\pm$9.6 & 0.24$\pm$0.24 & 98.0 & 86.1$\pm$9.5 & 0.30$\pm$0.24 & 99.5 \\
\rowcolor[HTML]{F3F3F3} 
TMJ\_R & 84.1$\pm$7.6 & 0.33$\pm$0.20 & 83.4$\pm$7.4 & 0.33$\pm$0.18 & {\cellcolor[HTML]{F3F3F3}} & 82.5$\pm$7.4 & 0.33$\pm$0.19 & {\cellcolor[HTML]{F3F3F3}} & 82.9$\pm$7.6 & 0.34$\pm$0.20 & {\cellcolor[HTML]{F3F3F3}} & 83.4$\pm$7.4 & 0.33$\pm$0.18 & {\cellcolor[HTML]{F3F3F3}} \\
SpinalCord & 91.5$\pm$2.9 & 0.36$\pm$0.19 & 90.1$\pm$2.8 & 0.33$\pm$0.17 & 99.5 & 93.7$\pm$2.9 & 0.33$\pm$0.19 & 99.0 & 90.6$\pm$2.9 & 0.35$\pm$0.19 & 98.0 & 90.1$\pm$2.8 & 0.33$\pm$0.17 & 99.5 \\
\rowcolor[HTML]{F3F3F3} 
Bone\_Mandible\_L & 95.7$\pm$2.1 & 0.17$\pm$0.08 & 96.6$\pm$2.1 & 0.17$\pm$0.08 & 99.5 & 94.0$\pm$2.3 & 0.16$\pm$0.09 & 99.7 & 97.5$\pm$2.2 & 0.15$\pm$0.09 & 99.5 & 96.6$\pm$2.1 & 0.17$\pm$0.08 & 99.5 \\
\rowcolor[HTML]{F3F3F3} 
Bone\_Mandible\_R & 95.7$\pm$2.5 & 0.17$\pm$0.08 & 94.8$\pm$2.7 & 0.14$\pm$0.10 & {\cellcolor[HTML]{F3F3F3}} & 94.2$\pm$2.4 & 0.13$\pm$0.07 & {\cellcolor[HTML]{F3F3F3}} & 95.5$\pm$2.4 & 0.13$\pm$0.07 & {\cellcolor[HTML]{F3F3F3}} & 94.8$\pm$2.7 & 0.14$\pm$0.10 & {\cellcolor[HTML]{F3F3F3}}\\ \hline

Mean & \multicolumn{1}{c}{86.4} & \multicolumn{1}{c|}{0.29} & \multicolumn{1}{c}{86.4} & \multicolumn{1}{c|}{0.28} & 98.1& \multicolumn{1}{c}{86.1} & \multicolumn{1}{c|}{0.29} & 97.9 & \multicolumn{1}{c}{86.4} & \multicolumn{1}{c|}{0.29} & 97.5 & \multicolumn{1}{c}{86.4} & \multicolumn{1}{c|}{0.28} & 98.1\\ \hline

\end{tabular}}
\caption{\textbar~\textbf{\Ac{css} order-wise performance details of \ac{cln_c5} on \ac{structseg}.} Organ-wise DSC (\%), ASD (mm), and decoder-wise pruning rate $\mathcal{T}$ (\%) of \ac{cln_c5} on \ac{structseg} are evaluated across all CSS orders. Notably, \ac{cln_c5} achieves mean DSC and ASD comparable to nnUNet upper bound and the average decoder pruning rates exceed 97.5\% across all orders, demonstrating high efficiency without compromising segmentation performance. }
\label{tab:css_structseg}
\end{table}
\clearpage
\begin{table}[htp]
\centering
\resizebox{\textwidth}{!}{
\begin{tabular}{|l|ll|ll|c|ll|c|ll|c|ll|c|}
\hline
\multicolumn{1}{|l|}{} & \multicolumn{2}{c|}{nnUNet} & \multicolumn{3}{c|}{Order 1} & \multicolumn{3}{c|}{Order 2} & \multicolumn{3}{c|}{Order 3} & \multicolumn{3}{c|}{Order 4} \\ \cline{2-15} 
\multicolumn{1}{|l|}{\multirow{-2}{*}{}} & DSC$\uparrow$ & ASD$\downarrow$ & DSC$\uparrow$ & ASD$\downarrow$ & $\mathcal{T}$$\uparrow$  & DSC$\uparrow$ & ASD$\downarrow$ & $\mathcal{T}$$\uparrow$  & DSC$\uparrow$ & ASD$\downarrow$ & $\mathcal{T}$$\uparrow$  & DSC$\uparrow$ & ASD$\downarrow$ & $\mathcal{T}$$\uparrow$  \\ \hline
\rowcolor[HTML]{F3F3F3} 
Liver & 98.1$\pm$14.5 & 0.83$\pm$0.98 & 97.6$\pm$5.6 & 0.85$\pm$0.98 & 99.5 & 98.3$\pm$5.8 & 0.75$\pm$0.98 & 99.5 & 97.4$\pm$5.7 & 0.90$\pm$0.99 & 99.7 & 97.9$\pm$5.6 & 0.81$\pm$0.98 & 99.7 \\
Kidney\_R & 92.4$\pm$16.9 & 1.81$\pm$0.74 & 93.7$\pm$7.1 & 1.56$\pm$0.76 & 99.7 & 94.6$\pm$7.1 & 1.74$\pm$0.76 & 99.7 & 91.4$\pm$6.8 & 1.98$\pm$0.73 & 99.5 & 94.3$\pm$7.1 & 1.30$\pm$0.76 & 99.5 \\
Kidney\_L & 92.8$\pm$16.9 & 1.27$\pm$0.74 & 92.9$\pm$7.0 & 1.19$\pm$0.72 & {} & 94.0$\pm$6.8 & 1.27$\pm$0.70 & {} & 92.1$\pm$6.9 & 1.36$\pm$0.72 & {} & 92.2$\pm$6.7 & 1.26$\pm$0.69 & {} \\
\rowcolor[HTML]{F3F3F3} 
Spleen & 97.9$\pm$15.6 & 0.30$\pm$0.37 & 97.3$\pm$6.7 & 0.20$\pm$0.38 & 99.7 & 98.2$\pm$6.6 & 0.22$\pm$0.38 & 99.5 & 97.2$\pm$6.8 & 0.45$\pm$0.39 & 99.5 & 97.8$\pm$6.9 & 0.21$\pm$0.40 & 99.5 \\
Pancreas & 91.6$\pm$17.2 & 0.80$\pm$0.69 & 92.4$\pm$7.0 & 0.80$\pm$0.38 & 99.7 & 92.8$\pm$7.2 & 0.69$\pm$0.39 & 99.7 & 88.8$\pm$7.2 & 0.95$\pm$0.40 & 98.0 & 93.1$\pm$7.0 & 0.72$\pm$0.38 & 99.0 \\
\rowcolor[HTML]{F3F3F3} 
A\_Aorta & 96.7$\pm$16.5 & 0.21$\pm$0.95 & 97.9$\pm$16.7 & 0.19$\pm$0.85 & 99.7 & 97.0$\pm$16.6 & 0.18$\pm$0.84 & 99.5 & 96.4$\pm$16.5 & 0.37$\pm$0.82 & 99.7 & 97.2$\pm$16.8 & 0.22$\pm$0.86 & 99.7 \\
V\_VenaCava\_I & 89.8$\pm$10.1 & 1.08$\pm$1.03 & 88.4$\pm$10.0 & 1.03$\pm$1.03 & 99.0 & 87.9$\pm$9.8 & 1.18$\pm$1.04 & 98.0 & 90.5$\pm$10.0 & 1.25$\pm$0.99 & 99.0 & 88.1$\pm$9.9 & 1.14$\pm$0.96 & 99.5 \\
\rowcolor[HTML]{F3F3F3} 
Glnd\_Adrenal\_R & 85.9$\pm$12.3 & 0.39$\pm$0.37 & 86.4$\pm$13.5 & 0.39$\pm$0.42 & 99.0 & 87.0$\pm$13.4 & 0.35$\pm$0.41 & 98.0 & 85.2$\pm$13.2 & 0.62$\pm$0.39 & 96.0 & 86.8$\pm$13.4 & 0.42$\pm$0.41 & 98.0 \\
\rowcolor[HTML]{F3F3F3} 
Glnd\_Adrenal\_L & 88.0$\pm$13.3 & 0.26$\pm$1.16 & 89.9$\pm$12.5 & 0.21$\pm$0.39 & {\cellcolor[HTML]{F3F3F3}} & 93.2$\pm$12.4 & 0.30$\pm$0.38 & {\cellcolor[HTML]{F3F3F3}} & 87.8$\pm$12.4 & 0.37$\pm$0.37 & {\cellcolor[HTML]{F3F3F3}} & 90.0$\pm$12.5 & 0.18$\pm$0.39 & {\cellcolor[HTML]{F3F3F3}} \\
GallBladder & 80.2$\pm$24.7 & 2.16$\pm$0.83 & 81.8$\pm$8.4 & 2.07$\pm$0.80 & 94.0 & 82.3$\pm$8.5 & 2.03$\pm$0.81 & 98.0 & 81.2$\pm$8.5 & 2.33$\pm$0.81 & 99.0 & 82.1$\pm$8.4 & 2.29$\pm$0.80 & 94.0 \\
\rowcolor[HTML]{F3F3F3} 
Eso & 84.4$\pm$3.1 & 1.70$\pm$0.15 & 82.7$\pm$3.1 & 1.48$\pm$0.15 & 99.0 & 83.5$\pm$3.1 & 1.46$\pm$0.15 & 96.0 & 85.0$\pm$3.1 & 1.70$\pm$0.15 & 98.0 & 84.6$\pm$3.0 & 1.55$\pm$0.15 & 99.0 \\
Stomach & 91.7$\pm$16.4 & 1.24$\pm$0.67 & 93.0$\pm$6.5 & 1.14$\pm$0.69 & 99.5 & 94.9$\pm$6.4 & 1.29$\pm$0.67 & 99.0 & 91.0$\pm$6.2 & 1.47$\pm$0.66 & 99.7 & 92.9$\pm$6.3 & 1.29$\pm$0.66 & 99.5 \\
\rowcolor[HTML]{F3F3F3} 
Duodenum & 79.8$\pm$25.9 & 2.71$\pm$1.29 & 81.5$\pm$16.1 & 2.55$\pm$1.20 & 96.0 & 82.9$\pm$16.2 & 2.81$\pm$1.19 & 94.0 & 79.7$\pm$15.8 & 2.86$\pm$1.28 & 94.0 & 80.9$\pm$15.8 & 2.39$\pm$1.24 & 96.0 \\ \hline

Mean & \multicolumn{1}{c}{90.0} & \multicolumn{1}{c|}{1.14} & \multicolumn{1}{c}{90.4} & \multicolumn{1}{c|}{1.05} & 98.6& \multicolumn{1}{c}{91.3} & \multicolumn{1}{c|}{1.10} & 98.3 & \multicolumn{1}{c}{89.5} & \multicolumn{1}{c|}{1.28} & 98.4 & \multicolumn{1}{c}{90.6} & \multicolumn{1}{c|}{1.1} & 98.5\\ \hline

\end{tabular}}
\caption{\textbar~\textbf{\Ac{css} order-wise performance details of \ac{cln_c5} on \ac{flare}.} Organ-wise DSC (\%), ASD (mm), and decoder-wise pruning rate $\mathcal{T}$ (\%) of \ac{cln_c5} on \ac{flare} are evaluated across all CSS orders. Notably, \ac{cln_c5} achieves mean DSC and ASD comparable to nnUNet upper bound and the average decoder pruning rates exceed 98\% across all orders, demonstrating both high segmentation performance and efficiency. }
\label{tab:css_flare}
\end{table}
\clearpage
\begin{table}[htp]
\centering
\resizebox{\textwidth}{!}{
\begin{tabular}{|l|ll|ll|c|ll|c|ll|c|ll|c|}
\hline
\multicolumn{1}{|l|}{} & \multicolumn{2}{c|}{nnUNet} & \multicolumn{3}{c|}{Order 1} & \multicolumn{3}{c|}{Order 2} & \multicolumn{3}{c|}{Order 3} & \multicolumn{3}{c|}{Order 4} \\ \cline{2-15} 
\multicolumn{1}{|l|}{\multirow{-2}{*}{}} & DSC$\uparrow$ & ASD$\downarrow$ & DSC$\uparrow$ & ASD$\downarrow$ & $\mathcal{T}$$\uparrow$  & DSC$\uparrow$ & ASD$\downarrow$ & $\mathcal{T}$$\uparrow$  & DSC$\uparrow$ & ASD$\downarrow$ & $\mathcal{T}$$\uparrow$  & DSC$\uparrow$ & ASD$\downarrow$ & $\mathcal{T}$$\uparrow$  \\ \hline
\rowcolor[HTML]{F3F3F3} 
Eso & 88.2$\pm$3.2 & 0.35$\pm$0.16 & 89.2$\pm$3.1 & 0.34$\pm$0.16 & 99.7 & 89.1$\pm$2.8 & 0.48$\pm$0.12 & 99.5 & 90.0$\pm$3.0 & 0.33$\pm$0.15 & 99.7 & 88.6$\pm$2.9 & 0.33$\pm$0.14 & 98.0 \\
Heart & 94.7$\pm$14.9 & 0.39$\pm$0.20 & 95.9$\pm$14.6 & 0.40$\pm$0.17 & 99.5 & 94.6$\pm$14.5 & 0.53$\pm$0.16 & 99.7 & 95.5$\pm$14.8 & 0.36$\pm$0.19 & 99.0 & 94.4$\pm$14.5 & 0.37$\pm$0.17 & 99.5 \\
\rowcolor[HTML]{F3F3F3} 
Trachea & 92.2$\pm$5.8 & 0.30$\pm$0.09 & 92.6$\pm$6.1 & 0.28$\pm$0.12 & 99.5 & 92.8$\pm$5.7 & 0.48$\pm$0.08 & 99.7 & 93.2$\pm$5.9 & 0.30$\pm$0.10 & 99.7 & 93.7$\pm$5.9 & 0.30$\pm$0.10 & 99.7 \\
A\_Aorta & 95.2$\pm$16.6 & 0.32$\pm$0.84 & 95.2$\pm$16.4 & 0.30$\pm$0.82 & 99.7 & 95.7$\pm$16.6 & 0.43$\pm$0.83 & 99.7 & 95.1$\pm$16.7 & 0.30$\pm$0.85 & 99.7 & 96.0$\pm$16.5 & 0.32$\pm$0.83 & 99.7
\\ \hline

Mean & \multicolumn{1}{c}{92.6} & \multicolumn{1}{c|}{0.34} & \multicolumn{1}{c}{93.3} & \multicolumn{1}{c|}{0.33} & 99.6& \multicolumn{1}{c}{93.1} & \multicolumn{1}{c|}{0.48} & 99.7 & \multicolumn{1}{c}{93.5} & \multicolumn{1}{c|}{0.32} & 99.5 & \multicolumn{1}{c}{93.1} & \multicolumn{1}{c|}{0.33} & 99.2\\ \hline

\end{tabular}}
\caption{\textbar~\textbf{\Ac{css} order-wise performance details of \ac{cln_c5} on \ac{segthor}.} Organ-wise DSC (\%), ASD (mm), and decoder-wise pruning rate $\mathcal{T}$ (\%) of \ac{cln_c5} on \ac{segthor} are evaluated across all CSS orders. Notably, \ac{cln_c5} achieves higher mean DSC and ASD than nnUNet upper bound and the average decoder pruning rates exceed 98\% across all orders, demonstrating high efficiency without compromising segmentation performance. }
\label{tab:css_segthor}
\end{table}
\clearpage
\begin{table}[htp]
\centering
\resizebox{\textwidth}{!}{
\begin{tabular}{|l|ll|ll|c|ll|c|ll|c|ll|c|}
\hline
\multicolumn{1}{|l|}{} & \multicolumn{2}{c|}{nnUNet} & \multicolumn{3}{c|}{Order 1} & \multicolumn{3}{c|}{Order 2} & \multicolumn{3}{c|}{Order 3} & \multicolumn{3}{c|}{Order 4} \\ \cline{2-15} 
\multicolumn{1}{|l|}{\multirow{-2}{*}{}} & DSC$\uparrow$ & ASD$\downarrow$ & DSC$\uparrow$ & ASD$\downarrow$ & $\mathcal{T}$$\uparrow$  & DSC$\uparrow$ & ASD$\downarrow$ & $\mathcal{T}$$\uparrow$  & DSC$\uparrow$ & ASD$\downarrow$ & $\mathcal{T}$$\uparrow$  & DSC$\uparrow$ & ASD$\downarrow$ & $\mathcal{T}$$\uparrow$  \\ \hline
Kidney\_\Ac{gtv} & 86.7$\pm$9.0 & 1.18$\pm$1.06 & 86.9$\pm$9.4 & 1.32$\pm$1.06 & 98.0 & 87.1$\pm$9.3 & 0.92$\pm$1.01 & 94.0 & 86.6$\pm$9.1 & 1.13$\pm$1.07 & 94.0 & 86.9$\pm$9.4 & 1.32$\pm$1.06 & 98.0
\\ \hline
\end{tabular}}
\caption{\textbar~\textbf{\Ac{css} order-wise performance details of \ac{cln_c5} on \ac{kits}.} Organ-wise DSC (\%), ASD (mm), and decoder-wise pruning rate $\mathcal{T}$ (\%) of \ac{cln_c5} on \ac{kits} are evaluated across all CSS orders. Notably, \ac{cln_c5} achieves DSC and ASD comparable to nnUNet upper bound and the average decoder pruning rate exceeds 94\% across all orders, demonstrating high efficiency without compromising segmentation performance. }
\label{tab:css_kits}
\end{table}
\clearpage

\end{document}